\documentstyle[12pt]{article}

\newlength{\dinwidth}
\newlength{\dinmargin}
\def\lapproxeq{\lower .7ex\hbox{$\;\stackrel{\textstyle
<}{\sim}\;$}}
\def\gapproxeq{\lower .7ex\hbox{$\;\stackrel{\textstyle
>}{\sim}\;$}}

\def\nostrocostrutto#1\over#2{\mathrel{\mathop{\kern 0pt \rlap
  {\raise.2ex\hbox{$#1$}}}
  \lower.9ex\hbox{\kern-.190em $#2$}}}
\def\lsim{\nostrocostrutto < \over \sim}   
\def\gsim{\nostrocostrutto > \over \sim}   

\newcommand{\labl}[1]{\label{#1}}
\newcommand{\as}{\alpha_s}
\newcommand{\tij}{\vartheta_{ij}}
\newcommand{\vth}{\vartheta}

\def\t12{\vartheta_{12}}
\def\phi{\varphi}

\def\rtwo{r^{(2)}}

\def\Om{\Omega}

\def\vph{\varphi}

\catcode`@=11
\newcount\@tempcntc
\def\@citex[#1]#2{\if@filesw\immediate\write\@auxout{\string\citation{#2}}\fi
  \@tempcnta\z@\@tempcntb\m@ne\def\@citea{}\@cite{\@for\@citeb:=#2\do
    {\@ifundefined
       {b@\@citeb}{\@citeo\@tempcntb\m@ne\@citea\def\@citea{,}{\bf ?}\@warning
       {Citation `\@citeb' on page \thepage \space undefined}}%
    {\setbox\z@\hbox{\global\@tempcntc0\csname b@\@citeb\endcsname\relax}%
     \ifnum\@tempcntc=\z@ \@citeo\@tempcntb\m@ne
       \@citea\def\@citea{,}\hbox{\csname b@\@citeb\endcsname}%
     \else
      \advance\@tempcntb\@ne
      \ifnum\@tempcntb=\@tempcntc
      \else\advance\@tempcntb\m@ne\@citeo
      \@tempcnta\@tempcntc\@tempcntb\@tempcntc\fi\fi}}\@citeo}{#1}}
\def\@citeo{\ifnum\@tempcnta>\@tempcntb\else\@citea\def\@citea{,}%
  \ifnum\@tempcnta=\@tempcntb\the\@tempcnta\else
   {\advance\@tempcnta\@ne\ifnum\@tempcnta=\@tempcntb \else \def\@citea{--}\fi
    \advance\@tempcnta\m@ne\the\@tempcnta\@citea\the\@tempcntb}\fi\fi}
\catcode`@=12

\begin{document}
\noindent \hspace*{104mm} MPI-PhT/96-29\\
\hspace*{104mm} Durham DTP/96/36 \\
\hspace*{104mm} December 1996
\vfill
\begin{center}
{\Large\bf
 PERTURBATIVE QCD APPROACH TO MULTIPARTICLE PRODUCTION }\\
\mbox{ }\\
\vfill
{\bf Valery A. Khoze}\\
\mbox{ }\\
{\it Department of Physics\\
University of Durham, Durham DH1 3LE, UK}\\
and\\
{\it Institute for Nuclear Physics\\
St. Petersburg, Gatchina, 188350, Russia}\\
\mbox{ }\\
and\\
\mbox{ }\\
{\bf Wolfgang Ochs}\\
\mbox{ }\\
{\it Max Planck Institut f\"ur Physik \\
Werner Heisenberg Institut\\
F\"ohringer Ring 6, D-80805 Munich, Germany} 
\end{center}
\vfill

\begin{abstract}
In this review we discuss 
the analytical perturbative approach, based on perturbative
QCD and Local Parton Hadron
Duality (LPHD), and its application to 
multiparticle  production in jets in the semisoft region. 
Analytical formulae are
presented for various observables
within the accuracy of the Modified Leading Logarithmic Approximation
(MLLA), i.e. with terms of relative order $\sqrt{\alpha_s}$ 
taken into account systematically, 
and in some cases with even higher accuracy.
These predictions are confronted with existing  experimental
data. Many details of the perturbative approach to multiple hadroproduction
have been consolidated in recent years, and the overall picture is
remarkably healthy.
The prospects of future studies of the
semisoft jet physics are also discussed.
\end{abstract}

\vfill

\begin{center}
{\it To be published by International Journal of Modern Physics A}
\end{center}
\newpage
.
\newpage
\tableofcontents
\newpage

\section{Introduction}
\indent
The existence in hard processes of spectacular jets of hadrons
--- the footprints of QCD partons --- is one of the most
striking successes of QCD.

It was first observed by the MARK1 group at SPEAR in 1975
\cite{rfs} that the quarks, pair produced in $e^+ e^-$
collisions, began to appear as hadronic jets.  Five years
later the four PETRA experiments \cite{bbbb} proved the existence
of jets from gluon emission in quark pair production.  These
pioneering results opened up an important new area for
experimental tests of both perturbative and non-perturbative
QCD.  In the last decade hadron-jet physics has been intensively
studied in $e^+ e^-$, hadron-hadron and $ep$ scattering
processes (for reviews see e.g.\ 
\cite{h,bp,ms,brw1,al1,wjm1,my1,hm,wo,ts1}).
It will certainly
remain one of the main topics for investigation at the $e^+ e^-,
p\overline{p}$ and $ep$ colliders of the future.

$e^+ e^-$ annihilation into hadrons proves to be a
particularly wonderful laboratory for detailed experimental tests
of QCD.  A vast amount of data on multihadron production has been
accumulated in the experiments at PETRA, PEP and TRISTAN.  The
high statistics data collected from hadronic $Z^0$-decays
(more than 16 million events) have led to remarkable progress in
the understanding of QCD dynamics.  New QCD physics results
continue to pour out from the analysis of 
LEP-1 and SLD experiments \cite{ada1}. 
The record LEP-1 statistics 
allows one to perform detailed studies on 
subsamples of events of specific interest and to measure
relatively small effects of particular significance. 

As LEP-1 finishes its final analysis even more detailed results
will become available.  The full LEP-1 data will form the
priceless source for QCD studies.  It is an important task for
both LEP and QCD communities to fully exploit all record LEP-1
potentials stretching them to the limit.
From November 1995 LEP started operating at centre-of-mass energies 
above the $Z^0$-resonance.
The first experimental results at 130-140 GeV (LEP-1.5) 
and at higher energies agree well with the
perturbative predictions.
The experimental 
results from HERA have become available recently.
They allow, in
particular, to study the momentum transfer and 
energy dependence of various observables 
in the same apparatus. At the TEVATRON one 
is able to analyse jets of
 much higher energy than before. The prospective
laboratories for studying the various issues of QCD physics would
be provided by LEP-2 and future linear $e^+ e^-$ colliders (see
e.g.\ Refs.\ \cite{vak1,vak2}) where one enters the unexplored
energy regions. 

During the last few years the experiments have collected
exceedingly rich new information on QCD jets allowing one to
perform the important tests of the perturbative QCD and reduce
the domain of our ignorance on the physics of confinement. 
The data clearly show that the broad features of hadronic jet
systems (multiplicities and their correlations, angular
patterns of particle flows, inclusive spectra and correlation of
particles, etc.), calculated at the parton level agree
surprisingly well with the measured ones.  This convincingly
demonstrates the dominant role of the perturbative phase of jet
evolution and supports the hypothesis of Local Parton-Hadron
Duality (LPHD) \cite{dt,adkt1}.
The key assumption of this concept is that the
conversion of partons into hadrons occurs at a low virtuality
scale, of the order of hadronic masses,  
independent of the scale of the primary hard process, and
involves only low-momentum transfers.  
Such local duality is    
naturally connected with the \lq pre-confinement' properties
\cite{av} of the QCD cascades
which ensures that the colours are locally compensated. 
Only two essential parameters are involved in the
perturbative description: the effective QCD scale $\Lambda$ and a cut-off
parameter $Q_0$. 
The non-perturbative
effects are practically reduced to normalization constants
relating hadronic characteristics to partonic
ones\footnote{Certainly despite all its phenomenological
successes the LPHD is, by no means, a complete scheme but
rather the simplest model-independent approach.  However, there
exists a consistent field theoretical approach to the physics of
colour screening which seems to exhibit explicitly the LPHD
properties (Gribov's confinement scenario \cite{vng1}).}.

The LPHD approach tends to apply perturbative QCD as near as
possible
to the limits posed by confinement dynamics. On its
basis the detailed description of QCD jets in hard
collisions is possible.  Many important tests of the
underlying
QCD dynamics of multiple hadroproduction are allowed by the
studies of hadron distributions in the  semisoft
region
(small values of momentum fraction $x_p = p_h/E_{\rm jet}$).
In this region perturbative theory
can
be applied, provided that infrared enhanced terms are resummed
to
all orders (for reviews see 
\cite{bcm,dkt1,dkmt1,dkmt2}).
Formally $p_h \gg m_h$ in these calculations, but the results are smooth
down to small momenta $p_h$.
  As long as the process
of
colour blanching and non-perturbative hadronization of partons
is
local in the configuration space infrared
singularities can be factorized out of particle spectra, so
the
asymptotic shapes of these distributions are fully predicted
\cite{dkmt2}.  The sub-asymptotic corrections can be expressed
as
power series in $\sqrt{\alpha_s}$ (appearance of
$\sqrt{\alpha_s}$ terms is one of the surprising features of
the
semisoft physics).  All the terms of relative order
$\sqrt{\alpha_s}$ can be taken into account in a systematic
way. 
This is the so-called Modified Leading Logarithmic
Approximation
(MLLA) which allows one to make testable quantitative
predictions
with controllable accuracy \cite{dt,ahm1}.

The analytical perturbative approach (APA $\equiv$ PQCD +
LPHD)
to QCD jet physics attempts to describe the structure of the 
multi-hadronic final states with the minimal reference to the 
fragmentation dynamics. One of the goals of these studies 
is to find the areas of applicability and the limitations of this
perturbative approach. It is expected to work only for sufficiently inclusive
quantities.

Rediscovery of coherence in the context of QCD in the early
eighties \cite{ef,ahm2} has led to a dramatic revision of
theoretical expectations for semisoft particle distributions. 
Thus, the coherent effects in the {\bf intrajet} partonic
cascades, resulting on the average, in the angular ordering
($AO$) of sequential branching, gave rise to the {\it
hump-backed} shape of particle spectra --- one of the most
striking APA predictions \cite{adkt1,dfk1,bcmm}.  It is not the
softest particles, but those with the intermediate energies ($E_h
\sim E^{0.3-0.4}$) which multiply most effectively in QCD
cascades.  Due to the {\bf interjet} coherence which is
responsible for the string \cite{ags}/drag \cite{adkt2} effect
in the multijet hadronic events, a very important physical
phenomenon can be experimentally verified, namely, the fact that
it is the dynamics of the colour which governs the production of
hadrons in accordance with the QCD \lq\lq radiophysics" of
particle flows.

Without doubt, the hadronization effects could and should be
of importance in many cases.  After all, we observe jets of
hadrons in the detectors, not the quarks and gluons we are
dealing with in our perturbative calculations.  However, the
dynamics of hadronization is still not well understood from first
principles and one has to rely on the predictions of the
phenomenological models, for example, the LUND string model \cite{agis}
or  the cluster fragmentation model \cite{HERWIG}.
At the moment all the so-called WIG'ged $\equiv$ ({\bf W}ith {\bf
I}nterfering {\bf G}luons) Monte Carlo Models
(HERWIG \cite{HERWIG}, JETSET \cite{JETSET}, 
ARIADNE \cite{ARIADNE}) 
 are very successful in the representation of the existing
data and they are intensively used for the predictions of the
results of present and 
future measurements \cite{ts1,bba,ts4,ts2}. However, these 
models are far from perfect, e.g. \cite{brw2}.   
It is worthwhile to mention that
for many observables the LPHD concept (at least in its milder
formulation) is quantitatively realized within these algorithmic
schemes.

All existing phenomenological models are of a probabilistic and
iterative nature when the fragmentation process as a whole is
described in terms of simple underlying branchings.  Their
successfulness in representing the data to a certain extent
relies on the fact that in many cases it is possible to absorb
the quantum mechanical interference into the probabilistic
schemes.  However, the description of the evolution of a
multipartonic system in terms of classical Markov chains is of
limited value in principle.  For example, under special
conditions some subtle QCD phenomena, breaking the
probabilistic
picture, may even become dominant and observable, see Refs.\
\cite{dkmt2,dkt2} and section 6.

These days physics of QCD jets has entered a Renaissance age. 
So
it seems quite timely to realize where we are now and where we
are going.  In this review we shall focus on the properties of
QCD jets in the semisoft region.  The main goal is to survey
the
basic ideas and the recent impressive phenomenological
advances
of the APA 
\cite{bcm,dkt1,dkmt1,dkmt2}.
  The interest in the detailed
studies of QCD jets is twofold.  On the one hand, they are
important for testing both perturbative and nonperturbative
dynamics of multiple hadroproduction, for design of
experiments
and the analysis of the data.  On the other hand, the detailed
knowledge of the characteristic features of the multi-jet
states
could provide useful additional tools to study other physics. 
For instance, it could play a valuable role in digging out the
signals for new physics from the conventional QCD backgrounds.


\section{QCD Jets and the LPHD hypothesis}
\subsection{Development of Partonic Cascades}
\setcounter{equation}{0}
\indent

In the framework of APA  the
dominant source of multihadron production in hard processes is
gluon bremsstrahlung.  The produced hadrons bear information
about the underlying dynamics at small distances, so that the
distributions of the colour-singlet hadrons in the final state
are governed by the dynamics of the parton shower system, in
particular, by the flow of colour quantum numbers.

Since multiple gluon radiation plays the key role in
hadroproduction in QCD jets let us first recall the properties
of
the basic QCD process, namely the gluon bremsstrahlung off a
quark with momentum $p$.  The differential spectrum is given
by
the well-known formula
\begin{eqnarray}
d w^{q \rightarrow q + g} & = & \frac{\alpha_s
(k_\perp)}{4 \pi} \; 2C_F \; \left [ 1 + \left ( 1 -
\frac{k}{p} \right )^2 \right ] \; \frac{dk}{k} \:
\frac{dk_\perp^2}{k_\perp^2}, \nonumber\\
\alpha_s (k_\perp)& =&  2 \pi/b \ln (k_\perp/\Lambda),
\quad b = (11 N_C - 2n_f)/3
\labl{2.1}
\end{eqnarray}
where  $N_C=3$ is the number of colours,
$C_F  =  (N_C^2 \: - \: 1)/2 N_C \; = \; 4/3$
 and $k_{\mu}$ is the gluon 4-momentum.  
The effective coupling $\alpha_s$
 is given by its one-loop expression and 
runs with the gluon
transverse momentum $k_\perp$, which comes from higher-order
corrections to the Born probability \cite{gl,ddt2,abcmv};
$\Lambda$ is the QCD-scale and $n_f$ is the number of flavours.

Let us recall two important properties of the spectrum
(\ref{2.1}):  (i) the broad logarithmic distribution over
$k_\perp$ (typical for a theory with dimensionless coupling);
(ii) the broad logarithmic energy distribution (specific for
theories with massless vector particles).

At large emission angle and large energy $E$ one would get an
extra gluon jet with a small probability
\begin{equation}
{\rm Multijet \; topology:}  \;\; k_\perp \sim k \sim E
\rightarrow w \sim \frac{\alpha_s}{\pi} \ll 1.
\labl{2.2}
\end{equation}
\noindent At the same time, the bulk of radiation
(quasicollinear and/or soft partons) will not lead to the
additional visible jets but will instead populate the original
jet
with secondary partons influencing the particle multiplicity and
other jet characteristics.
\begin{equation}
{\rm Intrajet \; activity:} \;\; k_\perp \ll k \ll E
\rightarrow w \sim \alpha_s \: \log^2 \: E \sim 1.
\labl{2.3}
\end{equation}
\noindent The \lq\lq Double-Logarithmic" (DL) $q \rightarrow q
g$ process, together with two other basic parton splittings (DL
radiation $g \rightarrow g g$ and \lq\lq Single-Logarithmic"
(SL) decay $g \rightarrow q\overline{q}$), forms parton cascades.

APA aims to describe quantitatively the structure of
multipartonic systems produced by QCD branching and to gain
the
actual knowledge about physics of confinement by comparing the
calculable characteristics of quark-gluon ensembles with
measurable characteristics of the final multihadronic states.

To understand how the offspring partons govern the development
of
the hadronic yield, one has to find at first what is the
condition for a gluon to behave as an independent coloured
object.  Recall that the time interval needed for a gluon to
be
radiated (the so-called formation time) is
\begin{equation}
t_g^{\rm form} \; \sim \; \frac{1}{M_{\rm virt}} \;
\frac{E}{M_{\rm virt}} \; = \; \frac{E}{(p + k)^2} \; \simeq
\;
\frac{k}{k_\perp^2}.
\labl{2.4}
\end{equation}
Comparing Eq.\ (\ref{2.4}) with the hadronization time of a
gluon
\begin{equation}
t_g^{\rm had} \; \sim \; k \: R^2,
\labl{2.5}
\end{equation}
\noindent one concludes that it is the restriction
\begin{equation}
k_\perp > R^{- 1} \; = \; \mu \; = \; {\rm a \: few \; hundred
\;
MeV}
\labl{2.6}
\end{equation}
\noindent which guarantees the applicability of the
quark-gluon
language, i.e.\ of the perturbative QCD.  The partons with
$k_\perp \sim \mu$ can hardly be treated as gluons at all
since
they are forced to hadronize immediately after being formed. 
The
real strong interaction comes into play here $(\alpha_s \:
(k_\perp^2) \sim O (1))$ allowing a smooth transition to the 
nonperturbative
plateau of the old parton model.  It is worthwhile to
emphasize
that the very notion of $\alpha_s (Q^2)$ can hardly be
justified
at low $Q^2$ at all.  Nevertheless,
in some practical applications the procedure of treatment of the small $Q^2$
region does not influence the result, for example, for momentum spectra
\cite{adkt1}.
In other applications one is actually forced to put into the
theoretical
analysis, in one way or another, an infrared finite effective
coupling $\alpha_s^{\rm eff} (Q^2)$, see for details Ref.\
\cite{dkt3}.  The universal $\alpha_s^{\rm eff}$ becomes an
important ingredient of the perturbative LPHD scenario.  In
particular it allows one to keep under perturbative control
the
distributions of $c$- and $b$-flavoured hadrons, see section
9.  The hypothesis of the universal finite $\alpha_s^{\rm eff}$
can
also be applied to a good many interesting problems in the
light
quark sector \cite{dkt3,ms2}.  By this way one could obtain
valuable information on the integrated influence of the
large-distance phenomena upon the inclusive particle
distributions.  Thus, as shown in Ref.\ \cite{dkt3} (see also
section 9), the APA analysis of the heavy quark energy losses
provides one with the numerical value of the characteristic
integral
\begin{equation}
\int_0^{2 \: GeV} \: dk \: \frac{\alpha_s^{\rm eff} (k)}{\pi}
\;
\approx \; 0.36 \: {\rm GeV}.
\labl{2.7}
\end{equation}
\noindent Further phenomenological verifications of the fact
that
the effective QCD coupling remains numerically rather small
would
be of large practical importance.  For instance, Gribov's
scenario \cite{vng1} demonstrates how colour confinement can be
realized in a field theory of light fermions interacting with
comparatively small effective coupling --- a fact of
potentially
strong impact for enlarging the domain of applicability of the
APA analysis to the physics of hadrons and their interactions.

 The structure of the quantities considered below can be
expressed symbolically as
\begin{equation}
O(t) \; = \; C(\alpha_s (t)) \: * \: \exp \left \{ \int^t \:
\gamma
(\alpha_s (t^\prime)) \: dt^\prime \right \}.
\labl{2.8}
\end{equation}
\noindent This representation incorporates the fact that
independent sequential elementary processes exponentiate.  It
exhibits the property of locality inherent to the
probabilistic
shower picture.  Due to the $AO$ of sequential branching
the
evolution parameter (\lq time' $t$) is connected here with the
jet opening angle $dt = d \Theta/\Theta$.  The \lq time'
derivative of Eq.\ (\ref{2.8}) produces the factor $\gamma
(\alpha_s (t))$.  It is natural to call the symbolic
quantities
$\gamma$ and $C$, the {\it anomalous dimension} and the {\it
coefficient function}, respectively.

The exponent of the integrated anomalous dimension $\gamma$
incorporates the Markov chains of sequential angular ordered
parton decays.  The regular coefficient function $C$, being
free
of collinear (or mass) singularities, describes wide-angle
partonic configurations, i.e. {\it multijet contributions} to
the
evolution of the system.

Successive terms of symbolic series for $\gamma (\alpha_s)$,
\begin{equation}
\gamma \; = \; \sqrt{\alpha_s} \: + \: \alpha_s \: + \:
\alpha_s^{3/2} \: + \: \alpha_s^2 \: + \: \ldots ,
\labl{2.9}
\end{equation}
\noindent correspond to the increasing accuracy of the
description of elementary partonic decays at small angles
$\Theta_{i j} \ll 1$ and, thus, of the jet evolution.  Iterating
the coefficient function
\begin{equation}
C \; = \; 1 \: + \: \sqrt{\alpha_s} \: + \: \alpha_s \: + \:
\ldots ,
\labl{2.10}
\end{equation}
\noindent one accounts for the ensembles of increasing numbers
of
such jets with large relative angles $\Theta_{i j} \sim 1$.
An expansion in $\sqrt{\alpha_s}$ at first looks surprising. It arises in
an approximation with the
exponential accuracy to the exact result which is given as 
the perturbative expansion
in $\alpha_s$.\footnote{Consider, for example, the exact result for
multiplicity in double logarithmic approximation
$\overline{n}=\cosh(c\:\sqrt{\alpha_s}\:Y)$ with $c=\sqrt{2 N_C/\pi}$
for fixed $\alpha_s$
 which can be written as series in $\alpha_s$, in comparison
to the approximation 
 $\overline{n}=\exp(c\:\sqrt{\alpha_s}\:Y)/2$ for large $Y$. } 

Two alternative approaches were proposed to calculate SL
effects
in the development of QCD cascades:  the standard
renormalization
group technique \cite{ahm1} and the probabilistic shower
method
\cite{dt} based on the parton branching picture.  We shall
follow
below the latter approach which readily provides one with the
transparent physical interpretation.  The key idea is to
reorganize  the perturbative expansion in such a way that its
zero-order approximation includes all necessary contributions for an 
arbitrary
number of produced particles.  This zero-order approximation
can
be achieved through an iteration of basic $A \rightarrow B +
C$
parton branchings.  In principle, it should be possible to
include higher-order corrections to the basic branching along with
higher point branching vertices $A \rightarrow B + C + D
\ldots$
in order to improve the accuracy of a calculation.  It is
important to emphasize that the choice of an appropriate
evolution parameter ({\it i.e.\ the jet opening angle}) makes
it
possible to incorporate all substantial subleading SL terms
without such a complication.

Recall that the strong $AO$ \cite{ef,ahm2} provides one with the basis for
the
probabilistic interpretation of soft-gluon cascades in the double
logarithmic approximation (DLA):
\begin{equation}
k_s \; \ll \; k_f \; \ll \; k_g \: , \;\;\; \Theta_{s f} \;
\ll
\; \Theta_{f g},
\labl{2.11}
\end{equation}
\noindent where the subscripts denote the cascade genealogy
(first introduced by Fadin):  \lq grandpa', \lq father'
and
\lq son'.

The DLA is the simplest analytical scheme for the QCD parton cascade. It
takes into account the contributions from the infrared and collinear
singularities of gluon emission, Eq. (\ref{2.1}), but ignores the
energy-momentum balance (recoil effects). One can derive then the asymptotic
properties of the cascade at infinite energies. It also provides 
one with the
appropriate variables of a given problem and their scaling properties.
However, in many cases the DLA is too crude for making quantitative
predictions. In particular, 
it overestimates parton multiplicities in
cascading processes and the characteristic energy of the most
actively multiplying partons (the position of the maximum of
the
particle spectra). Quantitatively, in the DLA case one
keeps
track of the $\sim \sqrt{\alpha_s}$ term in the anomalous
dimension $\gamma$, disregarding $O (\alpha_s)$ contributions
which give rise to the essential pre-exponential
energy-dependent
factors.  For a more quantitative control over parton branching
processes one needs to take into full account SL
effects which are next-to-leading in the expansions (\ref{2.9}) and
(\ref{2.10}). This is achieved by the MLLA.

The MLLA can be viewed as
a straightforward generalization of the standard
leading logarithmic approximation
\cite{gl,ap,yld1}
 based on the
probabilistic scheme which takes into account both DL and
essential SL effects.  It is important to notice that beyond
the MLLA a probabilistic picture of the parton cascade evolution
does not exist (for instance, there are 
the so-called colour monsters: $\frac{1}{N_C^2}$
suppressed soft interference contributions, appearing only in
the 4th order of perturbation theory
\cite{dt,dkt4})\footnote{Another instructive example
is connected with the interjet collective phenomena which could
be reproduced in a probabilistic scenario only in the large $N_C$
limit \cite{adkt2,dkmt2,dkt4} (see also section 6).}.  Recall that 
the global features of the multiparticle final state, on one side, 
and the basic interaction cross section on the other are due to 
essentially different physical phenomena.  It is the accompanying 
soft gluon radiation that dominates the former and is completely 
unimportant for the latter.  As a result, the conventional evolution 
parton picture \cite{gl,ap,yld1}, by no means guarantees the very 
possibility of viewing the final state structure in a probabilistic 
way.  Each time one needs to reanalyse the structure of contributing 
Feynman amplitudes, accounting for the interference effects which 
are much more pronounced in the soft emission, to see whether it is 
still possible to present the final result in the probabilistic 
fashion.

The MLLA parton decay probabilities look as follows
\begin{eqnarray}
dw_A^{BC} & = & \frac{\alpha_s \: (k_\perp^2)}{2 \pi} \;
\Phi_A^{BC} (z) \: dz \: V(\vec{n}) \; \frac{d \Omega}{8 \pi},
\labl{2.12}  \\
& & \nonumber \\
V_{f (g)}^s (\vec{n}) & = & \frac{a_{sg} \: + \: a_{fg} \: -
\:
a_{sf}}{a_{sf} \: a_{sg}}, \nonumber
\end{eqnarray}
\noindent where subscripts refer to the \lq\lq soft gluon
family".  Hereafter we denote by $a_{ik}$ the angular factor
\begin{equation}
a_{ik} \;  = \; 1 \: - \: \vec{n}_i \: \vec{n}_k \; =
\; 1 \: - \: \cos \Theta_{ik}.
\labl{2.13}
\end{equation}
\noindent Eq.\ (\ref{2.12}) takes into full account the SL
effects.  Here the DGLAP ({\bf D}okshitzer-{\bf G}ribov-{\bf
L}ipatov-{\bf A}ltarelli-{\bf P}arisi
\cite{gl,ap,yld1}) splitting
functions $\Phi_A^{BC}$ include both soft-gluon emission and
terms corresponding to a loss of energy logs in $A \rightarrow
B + C$ decays.  The exact angular kernel $V (\vec{n})$,
depending on the directions of parton momenta of three sequential
generations, replaces the crude strong $AO$ (see Eq.\
(\ref{2.11})).  One can easily check that
\begin{equation}
\langle V_{f (g)}^s \: (\vec{n}) \rangle_{\rm azimuth \;
average}
\; = \; \int_0^{2 \pi} \; \frac{d \phi}{2 \pi} \; V_{f (g)}^s
\:
(\vec{n}) \; = \; \frac{2}{a_{sf}} \; \Theta (a_{fg} \: - \:
a_{sf})
\labl{2.14}
\end{equation}
\noindent with $\Theta$ the step function.

This means that the decay probability integrated over the
azimuth
of \lq son' around \lq father' results in the logarithmic
$\Theta$-distribution inside the parent cone $\Theta_{sf} \leq
\Theta_{fg}$ and vanishes outside.  It is important to
emphasize
that the \lq $V$-scheme', Eq.\ (\ref{2.12}), proves to
eliminate
not only $A \rightarrow A + g^\prime + g^{\prime\prime}$ but
also $A \rightarrow A + g^\prime + g^{\prime\prime} +
g^{\prime\prime\prime}$ elementary splitting processes,
factorizing them completely into the chains of two-parton
decays.
\subsection{Generating Functionals and their Evolution Equations.}
To describe the development of the QCD jets it was found very
convenient to apply the Generating Functional ({GF})
technique. 
The advantages of this method for the perturbative treatment
of the QCD partonic systems were first realized in Jet Calculus
of Ref.\ \cite{kuv}. After embodying the   QCD
coherence effects \cite{dfk1,dfk2,vsf}, this
technique can also be readily applied for studying soft partons.  
Since then GFs were used quite successfully allowing one to
fully account for the subleading corrections to the DL description
of the intrajet cascades in $e^+ e^-$ annihilation \cite{dt,bcm,dkt4}, 
see Ref.\ \cite{dkmt2}.  
The GF method is a very convenient tool for describing the structure of the 
final states in Deep Inelastic Scattering (DIS) as well.

\subsubsection{Inclusive Distributions and Multiplicity Moments}

GFs are the functionals over the space of the \lq\lq probing
functions" $u(k)$ and they accumulate the information about
the overall QCD cascade 
\begin{equation}
  Z(\{u\})=\sum_n \int d^3k_1 \ldots d^3k_n
  u(k_1)\ldots u(k_n) P_n(k_1,\ldots,k_n)
  \labl{gendef}
\end{equation}
where $P^{(n)}(k_1,\ldots, k_n)$ is the probability density for
exclusive production of particles with 3-momenta $k_1\ldots k_n$.

Taking the $n^{th}$ variational derivative of $Z$ over the
probing functions $u(k_i)$ at $u = 1$ one arrives at the
inclusive $n$-parton momentum distribution
\begin{equation}
 D^{(n)} (k_1,...,k_n)= \delta^n Z(\{u\})/\delta u(k_1)...
\delta u(k_n)\mid_{u=1} .     \labl{dndef}
\end{equation}
One also considers the connected (often called cumulant) correlation
function
\begin{equation}
\Gamma^{(n)} (k_1,\ldots,k_n)= \delta^n \ln Z(\{u\})/\delta u(k_1)\ldots
\delta u(k_n)\mid_{u=1}      \labl{gamdef}
\end{equation}
in which the reducible correlations are subtracted, e.g. for
$n=2$: $\Gamma^{(2)} (k_1,k_2)=D^{(2)}(k_1,k_2)-D^{(1)}(k_1)D^{(1)}(k_2)$.
The expansion of $Z$ at $u=0$ generates exclusive parton distributions and
correlations.

After integration over the momenta one obtains the generating
function for the multiplicities $Z(u)$, $u(k)\equiv u$
\begin{equation}
Z(u)=\sum^\infty_{n=0} P_n u^n
\labl{zmult}
\end{equation}
From the integral of the distributions $D^{(n)}$ and $\Gamma^{(n)}$
in (\ref{dndef}) and (\ref{gamdef}) over momenta one obtains the
factorial moments
\begin{eqnarray}
f^{(q)} &=& \sum_n n(n-1)\ldots (n-q+1)P_n\nonumber\\
        &=& \left(\frac{d}{du}\right)^q Z(u)\Big|_{u=1}   \labl{fmom}
\end{eqnarray}
and the cumulant moments
\begin{equation}
c^{(q)}=\left(\frac{d}{du}\right)^q \ln Z(u)\Big|_{u=1}.
\labl{cmom}
\end{equation}
The first moment is the average multiplicity $\bar n=f^{(1)}=c^{(1)}$.
In the applications the normalized moments
\begin{equation}
F^{(q)}=f^{(q)}/\bar n^q\/,\qquad C^{(q)}=c^{(q)}/\bar n^q\/.
\labl{momnor}
\end{equation}
are considered most frequently.
\subsubsection{Evolution Equation in MLLA}

When studying the global characteristics of parton systems,
such
as mean multiplicities and multiplicity fluctuations, energy
particle spectra and correlations, etc., one replaces the full
angular kernel $V (\vec{n})$ in Eq.\ (\ref{2.12}) by its {\it
azimuth averaged} analogue of Eq.\ (\ref{2.14}).  In this case
the exact $AO$ makes it possible to construct a simple
Evolution
Equation for jet GFs.

In MLLA the
system of two coupled equations for functionals $Z_q$ and
$Z_g$, which describe the parton content of quark and gluon jets with
initial momentum $p$ and upper limit  $\Theta$ on opening angles, reads
$(A,B,C = q,g)$: 
\begin{eqnarray}
Z_A (p, \Theta; \{u (k)\}) & = & e^{- w_A (p \Theta)} \: u_A (k =
p)
\; + \; \frac{1}{2!} \; \sum_{B,C} \; \int_0^{\Theta} \; \frac{d
\Theta^\prime}{\Theta^\prime} \; \int_0^1 \; dz \: e^{-w_A (p
\Theta) \: + \: w_A (p \Theta^\prime)} \nonumber \\
& & \labl{2.15}\\
& \times & \frac{\alpha_s (k_\perp^2)}{2 \pi} \; \Phi_A^{BC}
\:
(z) \: Z_B (zp, \Theta^\prime;\{ u\}) \: Z_C ((1 - z) p,
\Theta^\prime;\{ u\})  \Theta (k_\perp^2 - Q_0^2). \nonumber
\end{eqnarray}
where $\Phi_A^{BC}(z)$ are the splitting functions \cite{ap,yld1}
\begin{eqnarray}
\Phi_q^g(z)&=& \Phi_q^q(1-z)=2C_F\frac{1+(1-z)^2}{z}, \labl{ap1}\\
\Phi_g^q(z)&=& \Phi_g^q(1-z)=2T_R[z^2+(1-z)^2],       \labl{ap2}\\
\Phi_g^g(z)&=& \Phi_g^g(1-z)=4
N_C\left[z(1-z)+\frac{1-z}{z}+\frac{z}{1-z}\right],  \labl{ap3}
\end{eqnarray}
with $C_F$ as in (\ref{2.1}) and $T_R=1/2$.

These evolution equations (\ref{2.15})
correspond to the probabilistic picture of a
Markov chain of sequential parton decays.
The first term
in
the $r.h.s.$ corresponds to the case when the $A$-jet consists
of
the parent parton only.  The integral term describes the first
splitting $A \rightarrow B + C$ with angle $\Theta^\prime$
between the products.  The exponential factor guarantees this 
decay to be the first one:  it is the probability to emit
{\it nothing} in the angular interval between $\Theta^\prime$ and
$\Theta$.  The two last factors account for the further
evolution of the produced subjets $B$ and $C$ having smaller energies
and smaller $\Theta^\prime$ than the opening angle as required by angular
ordering.

The MLLA equation has first been written for small angles, then the jet
virtuality is given by the maximum transverse momentum 
$Q=p\Theta$ as in (\ref{2.15}). It may be desirable
to extend the evolution towards larger angles, in this case an appropriate
evolution variable is 
\begin{equation}
Q=2p\sin (\Theta/2) \labl{qscal}
\end{equation}
This scale is used, for example, 
in the topical application of the generating function
technique to the calculation of jet and subjet multiplicities 
using the popular
$k_\perp$/Durham clustering algorithm (see subsection 4.2).

Using the
normalization property of the GF
\begin{equation}
Z_A (p, \Theta; \{ u \})|_{u (k) \: \equiv \: 1} \; = \; 1
\labl{2.18}
\end{equation}
the MLLA formfactors can be found as 
\begin{eqnarray}
w_q & = & \int^\Theta \; \frac{d \Theta^\prime}{\Theta^\prime}
\;
\int_0^1 \; dz \; \frac{\alpha_s (k_\perp^2)}{2 \pi} \;
\Phi_q^q
(z), \;\;\;
\labl{2.19a}\\
& & \nonumber \\
w_g & = & \int^\Theta \; \frac{d \Theta^\prime}{\Theta^\prime}
\;
\int_0^1 \; dz \; \frac{\alpha_s (k_\perp^2)}{2 \pi} \; \left
[
\frac{1}{2} \; \Phi_g^g (z) \: + \: n_f \Phi_g^q (z) \right ].
\labl{2.19b}
\end{eqnarray}
\noindent Collinear and soft singularities in Eqs.\
(\ref{2.15}),%
(\ref{2.19a}) and (\ref{2.19b}) 
 are regularized (standard for APA) by the transverse
momentum restriction
\begin{equation}
k_\perp \; \approx \; z\:(1-z)\:p \: \Theta^\prime \; > \; Q_0.
\labl{2.20}
\end{equation}
\noindent where $Q_0$ is a cut-off parameter in the cascades.

Differentiating the product $Z_A \: \exp \: \left [w_A (p
\Theta)
\right ]$ with respect to $\Theta$ and using Eq.\ (\ref{2.15})
one arrives at
the
Master Equation \cite{dt,dkt4}\footnote{For earlier simplified versions of
this equation, see \cite{dfk1,bcm}} 
\begin{eqnarray}
\frac{d}{d \: \ln \: \Theta} \: Z_A (p, \Theta) & = &
\frac{1}{2}
\; \sum_{B,C} \; \int_0^1 \; dz \nonumber \\
& & \labl{2.23}\\
 \; &\times &\frac{\alpha_s (k_\perp^2)}{2 \pi} \: \Phi_A^{BC} (z)  
\left [Z_B (zp, \Theta) \: Z_C ((1 - z)p, \Theta) \: - \: Z_A
(p,\Theta) \right ]. \nonumber
\end{eqnarray}
The initial condition for solving this system of equations reads
\begin{equation}
Z_A (p, \Theta; \{ u \})|_{p \Theta \: = \: Q_0} \; = \; u_A
(k
\: = \: p).
\labl{2.16}
\end{equation}
\noindent The $A$-jet with the hardness parameter $p \Theta$
which is set to the boundary value $Q_0$ where the evolution
starts, only consists of the parent parton $A$.  This natural
condition is clearly seen from the integral Evolution Equation
(\ref{2.15}) where the second term disappears at $p \Theta =
Q_0$
together with the form factor in the Born term:
\begin{equation}
w_A (Q_0) \; = \; 0.
\labl{2.17}
\end{equation}

Results from these equations are complete within MLLA accuracy (i.e.
$\sqrt{\alpha_s}$ terms are taken systematically into account). In addition
they provide the solution including the boundary conditions.
Equation (\ref{2.23}) accumulates information about
{\it azimuth averaged} jet characteristics in the MLLA and
looks
similar to the DGLAP equations for single-inclusive parton
distributions 
\cite{gl,ap,yld1} but with a nonlinear kernel.

The global properties of a system of QCD jets can be described
by
applying variational derivatives to the proper product of GFs.
For instance, for $e^+ e^-$ annihilation into hadrons at the
total cms energy $W = 2E$ one obtains
\begin{equation}
Z_{e^+ e^-} \: (W; \{ u \}) \; = \; \left [ Z_q (E, \Theta =
\pi; \{ u \}) \right ]^2.
\labl{2.21}
\end{equation}
\noindent Correspondingly, the final particle distributions
generated by the two gluons in a colour singlet state at $W =
2E_g$ are described by
\begin{equation}
Z_{gg}^S \: (W; \{ u \}) \; = \; \left [ Z_g (E_g, \Theta =
\pi;
\{ u \}) \right ]^2
\labl{2.22}
\end{equation}
The equations (\ref{2.23}) are now actively exploited for studying
various properties of QCD jets, see e.g.\
\cite{brw1,dkmt2,cdotw,do,yld2,yld3,dkt5,lo,lo1}.

\subsubsection{Evolution Equation in DLA}
In the DLA one neglects the energy-momentum constraints and
approximates $1-z\approx 1$ in (\ref{2.23}). This yields a
simpler equation which can be integrated to give
(with $d^3k=d\omega d^2k_\perp$)
\begin{eqnarray}
  Z_A(p,\Theta;\{u\}) &=& u(p)\exp
     \biggl(\int\limits_{\Gamma(p,\Theta)}
     \frac{d\omega}{\omega}\frac{d^2k_\perp}{2\pi k^2_\perp} 
     \nonumber\\
&\times &  C_A \frac{2\alpha_s (k^2_\perp)}{\pi}\;
 [Z_g(k,\Theta_k;\{u\})-1]\biggr)  
\labl{dlaz}
\end{eqnarray}
The subscript $A$ refers to the $g$ and $q$ jet
GFs (A=g,q) and $C_A$ to the respective colour factors $N_C$ and $C_F$.
 The secondary gluon $g$ is emitted into the interval $\Gamma(p,\Theta)$:
$\omega<p^0=|\vec p|$ and $\Theta_k<\Theta$. Due to the angular ordering
constraint the emission of this gluon is  
limited by its angle
$\Theta_k$ to the primary parton $A$. 
This DLA Master Equation \cite{dfk1,dfk2,vsf} yields the correct results
in the high energy limit $p\to\infty$ and, therefore, allows one to study  the
asymptotic scaling properties of the observables under
consideration. 
At present energies the
results from the DLA are often unsatisfactory at a quantitative
level.
\subsection{The Concept of LPHD}
At present QCD has not yet evolved into a
theory which would describe directly the measurable observables 
of the hadronic final state. Therefore the application of PQCD
is not possible without additional assumptions about the
hadronization process  at large distances 
which is governed by the colour confinement forces.

The simplest approach is to neglect such
large distance  effects altogether and to compare directly
the perturbative predictions at the partonic level
 with the corresponding measurements at the hadronic level.
This idea can  be applied at first to the total cross sections, and then
to jet production for a given 
resolution whereby the partons are compared to hadronic jets at the same
angles and energies.
 This approach has led to spectacular successes
and has built up our present confidence 
in the correctness of QCD as the theory
of strong interactions, in particular, concerning the properties of
the basic vertices and the running coupling in terms of the QCD scale parameter 
$\Lambda$. In such applications the resolution or 
cut-off scale is normally a fixed fraction of the primary energy itself.

It is then natural to ask whether such 
a dual correspondence can be carried out further
to the level of partons and hadrons themselves.
To perform the necessary multiparton calculations one has to
introduce the additional cut-off parameter $Q_0$
as explained in the previous subsection. 

The answer is, in general, affirmative for ``infrared and collinear safe'' 
observables which do not change if a soft particle is added or one
particle splits into two collinear particles.
Such observables become insensitive to
the cut-off $Q_0$ for small $Q_0$ and have therefore a chance  to be
independent
of the final stage of the  jet evolution. Quantities of this type
are energy flows and correlations \cite{sw,bbel,ddt1} and
global event shapes like thrust, for a review, see e.g. \cite{lepv1}.

In the next step of comparison between
 partons and hadrons we consider observables  
which count individual particles, for example,
particle multiplicities, inclusive
spectra and multiparton correlations.
Such observables depend explicitly
on the cut-off $Q_0$ (the smaller the cut-off, the larger the particle
multiplicity). 
A possible correspondence between these infrared sensitive observables for
partons and hadrons has been considered in the DLA \cite{dfk1}
within the context of
``pre-confinement" \cite{av,bcm} or ``soft blanching" \cite{ddt2,yia2}.

According to the hypothesis of 
Local Parton Hadron Duality (LPHD) \cite{adkt1} the hadronic spectra
are proportional to the partonic ones if the cut-off $Q_0$ of the parton
cascade is decreased towards a small value of the order 
of the hadronic mass itself.
By making use of the increased accuracy 
of the MLLA, the LPHD hypothesis could
be subject to  meaningful tests \cite{adkt1}.

In this review we will concentrate on further tests of the
LPHD hypothesis concerning ``infrared sensitive quantities''.
To deal with the cut-off $Q_0$ one can proceed in different ways.
If the cut-off dependence factorizes (for example, multiplicity)
one can again get  ``infrared safe'' predictions after a proper normalization.
In other cases (for example, inclusive momentum spectrum) the observables become
insensitive to the cut-off at very high energies 
if  appropriately rescaled quantities are used.

Alternatively, one can try to relate the 
transverse momentum cut-off $Q_0$ to the hadronic mass. This is
suggested since this cut-off
 also sets a lower limit on the parton
energy $E_p \ge k_\perp \ge Q_0$, i.e. it acts like a parton mass.  
Then the general idea of LPHD is to compute hadronic observables 
from the partonic final state
\begin{equation}
O(x_1,x_2,...)|_{hadrons}\; \sim \; O(x_1,x_2,...;Q_0,\Lambda)|_{partons}
\labl{lphdeq}
\end{equation}
where $Q_0$ is an effective mass to be determined by experiment. For charged
particle hadronic final states $Q_0 \simeq \Lambda \simeq
250$ MeV is appropriate (see
section 3).

The hypothesis of LPHD  lies at the very heart of the APA 
but at the same time this key hypothesis could be considered as its
Achilles heel as it remains outside of what can be derived within the 
established framework of QCD today. On the other hand, LPHD fits naturally
into the space-time picture of the hadron formation in QCD jets 
\cite{yia2,dkt4,dkmt2}.
It is intriguing that the LPHD ideas look quite natural
within Gribov's confinement scenario \cite{vng1}.  In this
approach there is no need for high virtualities of gluons in
QCD
cascades. 
  This could provide  a clue for
the understanding of a conceptual problem of
 LPHD, namely the yield of
massive hadrons such as baryons in the QCD jets, see
\cite{cdgkr}.

The comparison of parton and hadron final states
can also be pursued by deriving parton level predictions from 
a parton Monte Carlo  calculation. This allows one to include more 
complex quantities in the discussion for which 
analytical calculations cannot be easily obtained with the necessary
precision.
 Some care is required,
however, in this case as typically
 the parameter $Q_0$ in the Monte Carlo
 cannot be made sufficiently small
to obtain a good fit to the data 
\footnote{The standard
JETSET \cite{JETSET} and HERWIG \cite{HERWIG} Monte Carlo's
are not appropriate in this connection. On the other hand, ARIADNE 
\cite{ARIADNE} allows for such parameters;
we thank G.~Gustafson for the discussion of this point.}.
Then only properly normalized
``infrared safe'' quantities should be compared.
 An early example of this type is the
 treatment of multiparton
correlations in rapidity windows \cite{gvh,ggsh}
which separates the ``safe'' normalized moments from the ``unsafe''
multiplicities (see subsection 4.1.3).

One may expect that LPHD works better with increasing energy 
as the sensitivity to the cut-off decreases. 
At realistic energies the LPHD is by no means an obvious
hypothesis.  For instance, the spectrum of each hadron depends on
the species of its \lq parent' particle.  The standard assumption
of the LPHD is that the convolution of these distributions is reproduced
by the spectrum as calculated for a parton shower with a properly
chosen cut-off.
Whether or not present energies are sufficiently large for the
perturbative approach to be applied is a question for experiment.


Nowadays, in the applications of the LPHD to the available data
there
are different interpretations by different authors.  For
instance, the \lq\lq hard-liners" follow literally the idea of
\lq\lq one parton --- one hadron" equivalence. 
We feel that a dual picture will only have a chance to work 
if some averages are taken and LPHD is not applied at an exclusive level.
A plausible
approach is based on the correspondence between the averaged local
phase-space densities of partons and hadrons \cite{dkmt2}.
Another possibility is to extend exclusive multi-jet analyses towards
smaller scales $\gapproxeq 1$ GeV such that an average over small mass
clusters is maintained \cite{cto}.
It is quite a delicate question of where to draw the line for
what precisely the LPHD is capable to predict at current energies
and what not\footnote{We are very grateful to T.\ Sj\"{o}strand
for the illuminating discussions of various aspects of the LPHD,
see also \cite{ts1}.}. To find out these lines will be a challenge to
experiment. 

Finally, let us recall that well-tuned Monte Carlo models are
vital for the practical purposes, for instance, for unfolding
parton distributions from hadron spectra.  
These models include exact energy-momentum conservation and contain further
physics assumptions concerning the hadronization stage and free parameters
that are not derivable within the purely perturbative framework.
In the WIG'ged Monte
Carlo the angular ordering is implemented through the so-called
coherent parton shower algorithms.  Contrary to these
phenomenological models, APA has no flexibility with the free
parameters and cannot challenge these phenomenological approaches
in the accuracy of the data description.  However, concerning
Monte Carlo algorithms there is always a \lq devil's advocate'
question of whether these models, with all their degrees of freedom,
lead to a better understanding of the hadronization dynamics, or
merely provide one with the better parametrization.

Concluding this subsection we would like to emphasize once
more that the LPHD is, by no means, a dogma and any of its
breakdown in the experiment may shed light on the details of
hadronization physics.

\section{One-Particle Distributions}

We first discuss the inclusive momentum spectra which 
were the starting point for the quantitative testing of the MLLA
predictions  (i.e., in  $\sqrt{\alpha_s}$ accuracy). 
Until today these calculations are the only ones which take  into full
account the constraints from the boundary conditions at the cut-off $Q_0$
and, therefore, play an especially important role in the 
conclusions concerning the
LPHD concept. Also we discuss here the differences between 
quark and gluon jets which requires Next-to-MLLA accuracy. 
The main emphasis in these studies is on the soft part of the spectrum
using the variable $\xi=\ln1/x_p$ with $x_p=p/E_{jet}$.
We come back to the quark gluon jet differences in section 4.2 on jet
multiplicities.

In a complementary approach
one considers the energy evolution of the
$x_p$ spectrum (``fragmentation function")
in two-loop accuracy for the running coupling (for a theoretical review
see \cite{nw} and for the recent analyses of $e^+e^-$ data, see Refs. 
\cite{bkk,alephx,delphix}). 
This is
expected to be more precise for large $x_p$ but does not take into 
full account the contributions important for the soft region.

Most experimental information on particle spectra has been collected in
$e^+e^-$ annihilation.
These data are discussed in this section. Recently first
results from HERA and TEVATRON have become available, see sections 6 and 8.

\subsection{Momentum Spectra and Multiplicity in MLLA}
\subsubsection{Evolution Equation and Analytical Prediction}
\setcounter{equation}{0}
\indent
The MLLA Evolution Equation for the inclusive distribution of partons $B$
\begin{equation}
x {D}_A^B (x)\;
=\; E_B \; \frac{\delta}{\delta
u(k_B)} \: Z_A\:(E_A,\: \Theta; \{u\})|_{u=1}
\labl{dbar}
\end{equation}
follows directly from Eq.\
(\ref{2.23}) and takes the form:
\begin{equation}
\frac{d}{d \: \ln \: \Theta} \: x {D}_A^B (x, \ln \:
E \Theta) \; = \; \sum_{C = q, \overline{q}, g} \; \int_0^1 \;
dz \; \frac{\alpha_s (k_\perp)}{2 \pi} \: \Phi_A^C (z) \: \left
[ \frac{x}{z} \: {D}_C^B \: \left ( \frac{x}{z}, \ln \:
z E \Theta \right ) \right ],
\labl{2.24}
\end{equation}
\noindent where $E, \Theta$ are respectively the energy and
opening polar angle of a
jet
$A$, and $\Phi_A^C$ stands for the regularized DGLAP kernels.  The
boundary condition for (\ref{2.24}) is
\begin{equation}
x {D}_A^B (x) |_{\ln \: E \Theta = \ln \: Q_0} \; =
\;
\delta (1 - x) \: \delta_A^B.
\labl{2.25}
\end{equation}
The QCD running coupling is given by its one-loop expression (\ref{2.1}).
  The scale of the coupling is given by the
transverse
momentum $k_\perp \simeq z (1 - z) E \Theta$.  The shower
evolution
is cut off by the parameter $Q_0$, such that $k_\perp \geq Q_0$.

The integral equation can be solved by Mellin transform:
\begin{equation}
D_\omega (Y, \lambda) \; = \; \int_0^1 \; \frac{dx}{x} \:
x^\omega \: \left [x {D} (x, Y) \right ] \; = \;
\int_0^Y \; d \xi e^{- \xi \omega} \: D (\xi, Y, \lambda)
\labl{2.26}
\end{equation}
\noindent with
\begin{equation}
Y \: = \: \ln \frac{E \Theta}{Q_0}, \;\;\; \lambda \: = \: \ln
\frac{Q_0}{\Lambda}, \;\;\; \xi \: = \: \ln \frac{E}{k}
\labl{2.27}
\end{equation}
\noindent and parton momentum $k$.

In flavour space the valence quark and ($\pm$) mixtures of sea
quarks and gluons evolve independently with different \lq\lq
eigenfrequencies".  At high energies and $x \ll 1$, the
dominant
contribution to the inclusive spectrum comes from the \lq\lq
plus"-term, which we denote by $D_\omega (Y, \lambda) \equiv
D_\omega^+ (Y, \lambda)$.  In an approximation where only the
leading singularity plus a constant term is kept, it satisfies
the following Evolution Equation:
\begin{eqnarray}
\left ( \omega + \frac{d}{d Y} \right ) \; \frac{d}{d Y} \:
D_\omega (Y, \lambda) & - & 4 N_C \: \frac{\alpha_s}{2 \pi} \:
D_\omega (Y, \lambda) \nonumber \\
& & \labl{2.28}\\
& = & - a \: \left ( \omega + \frac{d}{dY} \right ) \;
\frac{\alpha_s}{2 \pi} \: D_\omega (Y, \lambda) \nonumber
\nonumber
\end{eqnarray}
\noindent where
\begin{equation}
a =  \frac{11}{3}N_C + \frac{2n_f}{3 N_C^2}.
\labl{amldef}
\end{equation}
Taking $a=0$ and dropping the $r.h.s.$ would yield the DLA Evolution
Equation.

By defining now the anomalous dimension $\gamma_\omega$
according
to:
\begin{equation}
D_\omega (Y, \lambda) \; = \; D_\omega (Y_0, \lambda) \:
\exp \: \left ( \int_{Y_0}^Y \; dy \: \gamma_\omega \: \left [
\alpha_s (y) \right ] \right )
\labl{2.29}
\end{equation}
\noindent the Evolution Equation for the inclusive spectrum
can
be expressed in terms of a differential equation for the
anomalous dimension as follows:
\begin{equation}
(\omega + \gamma_\omega) \gamma_\omega \: - \: \frac{4 N_C
\alpha_s}{2 \pi} \; = \; - \beta (\alpha_s) \: \frac{d}{d
\alpha_s} \gamma_\omega \: - \: a (\omega + \gamma_\omega) \:
\frac{\alpha_s}{2 \pi}
\labl{2.30}
\end{equation}
\noindent where $\beta (\alpha_s) = \frac{d}{dY} \: \alpha_s
(Y)
\simeq - b \: \frac{\alpha_s^2}{2 \pi}$.  The first term on
the
right hand side of (\ref{2.30}) proportional to the $\beta$-
function keeps trace of the running coupling effects while the
second accounts for the \lq\lq hard SL correction" to the DL
soft
emission.  Both prove to be corrections (of relative order 
$\sqrt{\alpha_s}$) to the left hand side which is $O
(\alpha_s)$.
 Within the DL accuracy the $r.h.s.$ of (\ref{2.30}) is
negligible
and the well known DL anomalous dimension comes immediately:
\begin{equation}
(\omega + \gamma_\omega) \gamma_\omega \: - \: \gamma_0^2 \; =
\;
0, \;\;\; \gamma_\omega^{DL} (\alpha_s) \; = \; \frac{1}{2} \:
\left ( - \omega + \sqrt{\omega^2 + 4 \gamma_0^2} \right );
\labl{2.31}
\end{equation}
\noindent with
\begin{equation}
\gamma_0^2 \; = \; \gamma_0^2 (\alpha_s) \; \equiv \; 2 N_C \:
\frac{\alpha_s}{\pi}.
\labl{2.32}
\end{equation}
Choosing the (+) sign for the square root corresponds to the high energy
approximation.
 The next-to-leading order  result 
in MLLA follows from Eq.\ (\ref{2.30}):
\begin{equation}
\gamma_{\omega} \; = \; \gamma^{DL}_{\omega} \: + \: \frac{\alpha_s}{2 \pi} \:
\left [ - \frac{a}{2} \: \left ( 1 +
\frac{\omega}{\sqrt{\omega^2
+ 4 \gamma_0^2}} \right ) \: + \: b \frac{\gamma_0^2}{\omega^2
+
4 \gamma_0^2} \right ] \: + \: O (\alpha_s^{3/2}).
\labl{2.33}
\end{equation}
 
It is also possible, to find an analytical expression for the exact solution
 of the differential equation
(\ref{2.28}). It  can be expressed in terms of the confluent
hypergeometric functions \cite{ae}, see for details
\cite{dkmt2,dkt5},
\begin{eqnarray}
D_i^f (\omega, Y, \lambda) & = & \frac{\Gamma (A + 1)}{\Gamma (B +
2)} \: z_1 z_2^B \{ \Phi (- A + B + 1, B + 2, -z_1) \: \Psi
(A, B
+ 1, z_2) \nonumber \\
& & \labl{2.34}\\
& + & e^{z_2 - z_1} (B + 1) \: \Psi (A + 1, B + 2, z_1) \:
\Phi
(- A + B + 1, B + 1, - z_2) \} \:  C_i^f \nonumber
\end{eqnarray}
\noindent where we have used the notation
\begin{eqnarray}
A & = & \frac{4 N_C}{b \omega}, \;\;\;\;\;\;\;\;\;\; B \; = \;
\frac{a}{b},\nonumber \\
z_1 & = & \omega (Y + \lambda), \;\;\; z_2 \; = \; \omega
\lambda.
\labl{2.35}
\end{eqnarray}
\noindent Indices $i$ and $f$ in Eq.\ (\ref{2.34}) respectively 
stand for the initial parton generating a jet $(i = q, g)$ and for 
the final one whose spectrum is studied $(f = q, g)$.  In the leading
approximation the coefficient functions $C_i^g$ are simply
\begin{equation}
C_g^g \; = \; 1, \;\;\; C_q^g \; = \; \frac{C_F}{N_C} \; = \;
\frac{4}{9}.
\labl{2.36}
\end{equation}
\noindent To reconstruct the $x$-distribution one has to
perform
the inverse Mellin transformation
\begin{equation}
\left [ x {D}_i^f (x, Y) \right ] \; \equiv \;
{D}_i^f
(\xi, Y, \lambda) \; = \; \int_{\epsilon - i
\infty}^{\epsilon +
i \infty} \; \frac{d \omega}{2 \pi i} \: x^{- \omega} \: D_i^f
(\omega, Y, \lambda)
\labl{2.37}
\end{equation}
\noindent where the integral runs parallel to the imaginary
axis to the right of all singularities of the integrand in the
complex $\omega$-plane (if any).

After the chain of transformation the solution in the physical
region of particle energies $0 \leq \xi \leq Y$ can be written
as
\begin{equation}
x {D}_i^f (x, Y, \lambda) \; = \; \frac{4 N_C (Y +
\lambda)}{b B (B + 1)} \; \int_{\epsilon - i \infty}^{\epsilon
+
i \infty} \; \frac{d \omega}{2 \pi i} \: x^{- \omega} \Phi (-
A +
B + 1, B + 2, - \omega (Y + \lambda)) \: {\cal K}
\labl{2.38}
\end{equation}
\noindent where
$$
{\cal K} \; \equiv \; {\cal K} (\omega, \lambda) \; = \;
\frac{\Gamma (A)}{\Gamma (B)} \: (\omega \lambda)^B \: \Psi
(A, B
+ 1, \omega \lambda) \:  C_i^f.
$$
\noindent As we shall discuss below this restored x-spectrum
of
partons exhibits the celebrated hump-backed structure with a
maximum at particle energies approaching asymptotically
$\sqrt{E}$. Explixit analytical results can be obtained for the special case
$\lambda=0$ (``limiting spectrum") and for the moments of the distribution.
Alternatively one can evaluate (\ref{2.38}) numerically.

\subsubsection{Hadronization Effects and LPHD}
\indent
Let us recall the basic ideas of how to arrive at LPHD in the
analytical
predictions for the inclusive spectra of
hadrons, see Refs.\ \cite{adkt1,dkt6}.  In the leading twist
approximation this spectrum can be symbolically presented in the
general factorized form of Eq.\ (\ref{2.38}) as
\begin{equation}
x {D}^h (x, Y, \lambda) \; = \; \int \; \frac{d
\omega}{2 \pi i} \; x^{- \omega} \: \Phi (\omega, Y + \lambda)
\:
H (\omega, \lambda)
\labl{2.39}
\end{equation}
\noindent where 
$x \; = \; \frac{E_h}{E}$. 
Besides the jet energy $E$ and the QCD parameter
$\Lambda$ this expression also depends on the $k_\perp$ cut-off parameter
$Q_0$ which -- as already discussed -- also plays the role of
%
an \lq\lq effective mass" of a
parton.  The choice of the value of $Q_0$ sets a formal boundary
between
two stages of jet evolution:  the one of the parton branching
process, which is controlled by perturbative theory and then
the stage of non-perturbative transition into hadrons.  Thus, if there were
a theory of hadronization, the result would be
independent of the formal quantity $Q_0$ separating the two
stages.  

In Eq.\ (\ref{2.39}) only the $\Phi$ factor contains
the
$E$ dependence and is under perturbative control, while $H$
contains the contribution from hadronization. If this can be represented as
a convolution of the respective decay functions, we obtain
\begin{equation}
H (\omega, \lambda) \; = \; {\cal K} (\omega, \lambda) \: C^h
(\omega,
\lambda; m_h, J^{PC})
\labl{2.40}
\end{equation}
\noindent with $C^h$ the Mellin-transformed parton
hadronization function for the ``massive" parton.

We are interested in the kinematic region of relatively soft
particles $x \ll 1$.  In such a case the essential values of
$\omega$ under the integral in (\ref{2.39}) prove to be small,
$\omega \ll 1$ (near the maximum of the spectrum  $\omega \sim
1/\sqrt{Y}$).  To understand how hadronization affects the
spectrum shape one needs to know the behaviour of $C^h
(\omega)$
at $\omega \rightarrow 0$.

Let us compare two different possibilities \cite{dt,adkt1}:
\begin{itemize}
\item[(i)] The singular behaviour $C^h (\omega) \sim 1/\omega
+
\ldots$ would correspond to the picture in which each parton
produces hadrons with a plateau-like distribution.  In such a
case the depletion of particle yield at small $x$, which is a
characteristic manifestation of QCD coherence in the {\it
parton}
spectra, would never manifest itself in {\it hadron}
distributions.
\item[(ii)] The regular behaviour (namely, $C^h (\omega)
\rightarrow$ constant) corresponds to the {\it local} screening
of colour in phase-%
space and hadronization of partons.  In this
case
hadron and parton spectra prove to be similar and the LPHD is
realized.
\end{itemize}
Non-perturbative effects, for instance, resonance decays 
 could in principle smear the distributions over a
finite interval $\Delta \xi \sim 1$.  Therefore, as was
emphasized in Refs.\ \cite{dkt5,dkt6}, one should not
expect necessarily a one-to-one correspondence between partons and hadrons
at an exclusive level but
only in their average characteristics in the ensemble of
events
and in the fluctuations around the average.  
At high energies these smearing effects are small compared to the full
kinematic range $\Delta \xi \sim Y$ and can be neglected in the distribution
in the rescaled variable $\xi/Y$.
Then, the overall
normalization factor $C^h (0)$ remains the only arbitrary
parameter.  It may be fixed, for example, by fitting the
average
multiplicity while the effective value of $\Lambda$ could be
determined from the energy dependence of the spectrum.

Let us emphasize an important property of the spectrum, given
by
(\ref{2.39}).  Namely, at very high energies, when the typical
values of $\omega \lambda \sim \lambda/\sqrt{Y} \ll 1$, the
{\it
shape} of the $\xi$-spectrum appears to be insensitive to the precise value
of
$Q_0$
\begin{equation}
{\cal K} (\omega, \lambda)|_{\omega \lambda \ll 1} \; \approx \; {\cal K}
(\lambda) \; = \; {\rm const.}
\labl{2.41}
\end{equation}
\noindent Thus when the cascade is sufficiently developed, the
shape of the resulting distribution of particles becomes
insensitive to the last steps of evolution and approaches a limiting
behaviour.  
This observation
may
justify an attempt to provide the development of the cascade not with
an
increase of $E$ but with a decrease of $Q_0$, thus enhancing
the
responsibility of perturbative QCD at present energies.
The extreme case is the limit $\lambda \to 0$ (``limiting spectrum").

\subsubsection{Moments}
\indent
 It proves to be very
convenient (see e.g.\ \cite{fw,dkt5,lo}) to analyse inclusive
particle spectra in terms of the normalized moments
\begin{equation}
  \xi_q \equiv <\xi^q>
    = \frac{1}{N}\int d\xi \xi^q \overline D(\xi)
  \labl{ximomdf}
\end{equation}
 where $N$ is the mean
multiplicity in the jet, the integral of the spectrum. 
Also one defines the cumulant moments
$K_q (Y, \lambda)$ \cite{so}, or the reduced cumulants  $k_q \equiv
K_q/\sigma^q$,
which  are related by
\begin{eqnarray}
K_1 & \equiv & \overline{\xi} \; \equiv \; \xi_1 \nonumber\\
K_2 & \equiv & \sigma^2 \; = \; < (\xi - \overline{\xi})^2
>, \nonumber \\
K_3 & \equiv & s\sigma^3 \; \equiv \; 
< (\xi - \overline{\xi})^3 >, \nonumber \\
K_4 & \equiv & k \sigma^4 \; \equiv \; 
< (\xi - \overline{\xi})^4 > \: - \: 3 \sigma^4.
\labl{2.66}
\end{eqnarray}
where the third and forth reduced cumulant moments are
the skewness $s$ and the kurtosis $k$ of
the
distribution.

If the higher-order cumulants $(q > 2)$ are sufficiently
small, one can reconstruct the $\xi$-distribution from the
distorted Gaussian formula,  see Ref.\ \cite{fw}
\begin{equation}
{D}(\xi) \; \simeq \; \frac{N}{\sigma \:
\sqrt{2 \pi}} \; \exp \: \left [ \frac{1}{8} \: k \; - \;
\frac{1}{2} \: s \delta \; - \; \frac{1}{4} \: (2 + k)
\delta^2
\; + \; \frac{1}{6} \: s \delta^3 \; + \; \frac{1}{24} \: k
\delta^4 \right ]
\labl{2.54}
\end{equation}
where $\delta=(\xi-\bar\xi)/\sigma$.

A characteristic feature of the particle spectrum is
the
\lq\lq typical" value of the quantity $\xi$, which could be
represented by the various measures of the \lq\lq centre" of
the
distribution (see Ref.\ \cite{dkfb} for details). 
Statisticians
commonly employ three seemingly unrelated measures of the \lq
centre' of the probability distribution $P (\xi)$, namely
\begin{itemize}
\item[1.] The {\it mean} value, $\overline{\xi} = \int_0^\infty \:
\xi
P (\xi) d \xi$, already introduced in (\ref{2.66})
\item[2.] The {\it median} value $\xi_m$, which is such that
the
probabilities of lying above or below this value are equal,
i.e.
$\int_0^{\xi_m} \: P (\xi) \: d \xi = \frac{1}{2}$,
\item[3.] The {\it modal} or {\it peak} value $\xi^*$, at which $P
(\xi)$ reaches its maximum.
\end{itemize}

Higher-order cumulants can be found 
from the expansion of the Mellin-transformed spectrum
$D_\omega (Y, \lambda)$ in (\ref{2.26}):
\begin{equation}
\ln \: D_\omega (Y, \lambda) \; = \; \sum_{q = 0}^{\infty} \;
   K_q (Y, \lambda) \; \frac{(- \omega)^q}{q !},
\labl{2.68}
\end{equation}
\noindent or equivalently,
\begin{equation}
K_q (Y, \lambda) \; = \; \left . \left ( -
\frac{\partial}{\partial \omega} \right )^q \; \log \: D_\omega
(Y, \lambda) \right |_{\omega = 0}.
\labl{2.69}
\end{equation}
These moments can be calculated from the representation (\ref{2.38})
for $D_\omega$ which yields the expression \cite{dkt5}
\begin{equation}
< \xi^q > = \frac{1}{N} \sum_{k=0}^q \left({q\atop k}\right)
 (Y+\lambda)^{q-k}(-\lambda)^k ( N_1 L_k^{(q)}
+ N_2 R_k^{(q)} ) 
\labl{ximomk}
\end{equation}
where $N_1$, $N_2$, $L_k^{(q)}$ and $R_k^{(q)}$ are known functions of 
 $a$, $b$, $Y+\lambda$ and $\lambda$. So they depend on the two parameters
 $Q_0$ and $\Lambda$.

The high energy behaviour of moments in next-to-leading order can be
obtained from the appropriate expansion of (\ref{ximomk}) or 
directly from the high energy approximation of the anomalous dimension
$\gamma_\omega$ in (\ref{2.33}) using (\ref{2.29}). 
This yields for the cumulant moments 
\begin{equation}
K_q \; = 
 \; \int_0^{Y} \;
dy \: \left . \left ( - \frac{\partial}{\partial \omega} \right
)^q \; \gamma_\omega (\alpha_s (y)) \right |_{\omega = 0},
\labl{2.71}
\end{equation}
or $\bar \xi=K_1$, $\sigma^2=K_2$ and the reduced cumulants
$k_q=K_q/\sigma^q$. Eq. (\ref{2.71}) 
  shows the direct dependence of the
moments on $\alpha_s (Y)$.  For fixed $\alpha_s$, for example,
one obtains directly $K_q (Y) \: \propto \: Y$ for high
energies.

\subsubsection{Properties of the Limiting Spectrum}

\indent At $\lambda = 0$ the expression (\ref{2.38}) simplifies
drastically,
since in that case ${\cal K} (\omega, \lambda) \equiv 1$ at all
$\omega$.
Decreasing $Q_0$ means extending the responsibility of the
perturbative stage beyond its formal range of applicability.  
However, this limit is smooth: summing up the perturbative series yields a
finite result even if the coupling gets arbitrarily large.
At $\lambda = 0$ one obtains from 
(\ref{2.38})
the finite  expression for the limiting spectrum
\begin{equation}
{D}^{\lim} \: (\xi, \tau) \; \equiv \; \left \{ x \:
{D}_g^g \: (x, \tau) \right \}^{MLLA} \; = \; \frac{4
N_C \tau}{b B (B + 1)} \; \int_{\epsilon - i \infty}^{\epsilon
+
i \infty} \; \frac{d \omega}{2 \pi i} \: x^{- \omega} \: e^{-
\omega \tau} \: \Phi (A + 1, B + 2, \omega \tau)
\labl{2.43}
\end{equation}
 with $\tau = Y =  \ln \frac{E}{\Lambda}$.  Recall that in
the approximation (\ref{2.38})  the particle spectra in $q$
and
$g$-jets differ only by the factor $C_q^g = \frac{C_F}{N_C} =
\frac{4}{9}$.

Eq.\ (\ref{2.43}) is not so convenient for numerical
calculations.  Practically more useful is an expression first
derived in Ref.\ \cite{dt} using an integral representation
for
the confluent hypergeometric function $\Phi$,
\begin{eqnarray}
{D}^{\lim}(\xi,\tau) & = & \frac{4 N_C}{b} \: \Gamma (B) \;
\int_{-
\frac{\pi}{2}}^{\frac{\pi}{2}} \; \frac{d \ell}{\pi} \: e^{- B
\alpha} \; \left [ \frac{\cosh \: \alpha + (1 - 2 \zeta) \sin
h
\: \alpha}{\frac{4 N_C}{b} \; \tau \; \frac{\alpha}{\sinh \:
\alpha}}
\right ]^{B/2} \nonumber \\
& & \nonumber\\
& & \times \; I_B \; \left ( \sqrt{ \frac{16 N_C}{b} \; \tau \;
\frac{\alpha}{\sinh \: \alpha} \; [\cosh \: \alpha + (1 - 2
\zeta) \: \sinh \: \alpha ]} \right ),
\labl{2.44}
\end{eqnarray}
\noindent Here $\alpha = \alpha_0 + i \ell$ and $\alpha_0$ is
determined by $\tanh \: \alpha_0 = 2 \zeta - 1$ with $\zeta =
1
- \frac{\xi}{\tau}$.  $I_B$ is the modified Bessel function of
order $B$.

Characteristics of an individual quark jet are best studied
experimentally in the reaction $e^+ e^- \rightarrow$ hadrons.  An
inclusive hadron spectrum here is the sum of two $q$-jet
distributions, cf.\ Eq.\ (\ref{2.21}), and one obtains in the present
approximation 
\begin{equation}
\frac{1}{\sigma} \; \frac{d \sigma^h}{d \xi} \; 
    = \; 2K^h C_q^g {D}^{lim}(\xi,\tau)
\labl{2.42}
\end{equation}
 where $\xi = \ln \frac{1}{x}$ with $x =
\frac{E_h}{E}$ and $K^h$ is the hadronization constant.

Let us remind the reader that the basic equation (\ref{2.38})
has
been derived in a high energy approximation. 
However, even for
moderate energies it is expected to give reasonable
quantitative predictions because it represents the exact
solution
of the MLLA Evolution Equation (\ref{2.28}) 
which accounts for the main physical
ingredients of parton multiplication, namely, colour coherence
and the exact DGLAP kernels in 2-particle QCD branching, and it takes
 into account also the
boundary conditions for low virtuality $E\Theta$.

As a consequence of colour coherence
soft parton multiplication is suppressed and the $\xi$-distribution 
has the form
of
a hump-backed plateau \cite{dfk1,bcmm} which is asymptotically
Gaussian in the variable $\xi$ around the maximum ($\delta \lapproxeq 1$
in (\ref{2.54})).  
As was mentioned before, the
hump-backed plateau is among the fundamental predictions of
QCD. 
The asymptotic shape of the limiting distribution can be found
by
the saddle-point evaluation of the spectrum moment
representation
treating $\tau$ as a large parameter.  This Gaussian
distribution for a quark jet 
can be written 
\begin{equation}
{D}_q^{\lim} (\xi, \tau) \; \simeq \; \frac{N_q
(\tau)}{\sigma \: \sqrt{2 \pi}} \; \exp \: \left [ -
\frac{1}{2}
\: \delta^2 \right ].
\labl{2.45}
\end{equation}
 Asymptotically in the DLA
the multiplicity $N_q$, the average $\bar \xi$ and the 
dispersion $\sigma$ are given by
\begin{equation}
\hbox{DLA:}\qquad\qquad
N_q \; \sim \; \exp \sqrt{\frac{16N_C}{b}\tau},\qquad
\bar \xi \; = \; \frac{\tau}{2}, \qquad 
\sigma^2\; = \; \frac{1}{3}\sqrt{\frac{b\tau^3}{16N_C}}.
\labl{2.49}
\end{equation}


Next we turn to the moments in
next-to-leading order \cite{fw,dkt5}.
 In terms of the running coupling the asymptotic
behaviour of the limiting multiplicity can be written as, see
e.g.\
\cite{bw1,adkt1}
\begin{equation}
\ln \: N \; = \; \sqrt{ \frac{32 \pi \: N_C}{\alpha_s (\tau)}}
\;
\frac{1}{B} \: + \: \left ( \frac{B}{2} \: - \: \frac{1}{4}
\right ) \; \ln \: \alpha_s (\tau) \: + \: O (1).
\label{2.48}
\end{equation}
The reader is reminded that because of the
destructive
soft gluon interference, the first term in Eq.\ (\ref{2.48})
is
reduced by a factor $\frac{1}{\sqrt{2}}$ as compared to the
case
of incoherent cascades, see  \cite{dfk1}.
 
In the limiting case each of the quantities $\xi_i =
\overline{\xi}, \xi_m$ or $\xi^*$, introduced above, proves to have an energy
dependence of the form
\begin{equation}
\xi_i \; = \; \tau \: \left [ \frac{1}{2} \: + \:
\sqrt{\frac{C}{\tau}} \: + \: \frac{a_i}{\tau} \: + \: b_i
\tau^{- 3/2} \right ]
\labl{2.50}
\end{equation}
\noindent with
\begin{equation}
C \; = \; \frac{a^2}{16 \: N_C b} \; = \; 0.2915 \: (0.3513)
\;
{\rm for} \; n_f \: = \: 3 (5).
\end{equation}

\noindent The subleading coefficients $a_i$ have been computed
\cite{dkt5,dkfb,fw,dkt7} but the next term can be obtained
only
for certain combinations of the mean, median and peak.  For
peak
position $a_{\rm max} = - C$ and it appears that in the
available
energy range the expression
\begin{equation}
\xi^* \; = \; \tau \: \left [ \frac{1}{2} \: + \: \sqrt{
\frac{C}{\tau}} \: - \: \frac{C}{\tau} \right ]
\labl{2.51}
\end{equation}
\noindent leads to a nearly linear dependence of $\xi^*$ on
$\tau$. 
It is worthwhile to mention that in the large $N_C$ limit, when 
$11 N_C\gg 2n_f$ (cf. Eqs.(\ref{2.1}) and (\ref{amldef}))
the parameter $C$ becomes independent on both $n_f$ and $N_C$ and approaches
its asymptotical value of $C=\frac{11}{3}\frac{1}{2^4} \simeq 0.23$.
Therefore in this limit the effective gradient of the straight line is
determined by such a fundamental parameter of QCD as the celebrated 
$\frac{11}{3}$ factor (characterizing the gluon self interaction)
in the coefficient $b$. 

Note that the formula (\ref{2.51}) 
has been derived using the distorted Gaussian
approximation. The exact position of the maximum of the limiting spectrum
from the numerical evaluation of (\ref{2.44})
yields a slightly larger value.
In the energy range $\sqrt{s}=10$ GeV to 200 GeV the exact
position is well approximated by \cite{klo}
\begin{equation}
\xi^* \; = \; \tau \: \left [ \frac{1}{2} \: + \: \sqrt{
\frac{C}{\tau}} \: - \: \frac{C}{\tau} \right ]+0.10.
\labl{xistex}
\end{equation}

This formula describes suprisingly well the observed energy
evolution of the maximum of the spectrum, see section 3.3.
  The differences $\xi_m - \overline{\xi}$ and $\xi^* -
\overline{\xi}$,
revealing the net subleading MLLA effects, are asymptotically
constant
\begin{eqnarray}
\xi_m - \overline{\xi} & \simeq & \frac{a}{32 N_C},\labl{2.52a} \\
\xi^* - \overline{\xi} & = & \delta_{\rm max} \: \sigma \;
\simeq \; \frac{3a}{32 N_C}.
\labl{2.52b}
\end{eqnarray}
\noindent We cannot predict at the moment the first
non-leading
corrections to these results separately but the ratio
has been computed\footnote{The fact that the ratio
(mode-mean)/(median-mean) asymptotically equals to 3 for a
number
of \lq almost Gaussian' distributions was discovered
empirically
long ago by Pearson \cite{kp}.}:
\begin{equation}
\frac{\xi^* - \overline{\xi}}{\xi_m - \overline{\xi}} \; = \;
3
\: \left [ 1 \: + \: \frac{9}{10} \; \frac{1}{z} \right ], \qquad
 z \; \equiv \; \sqrt{
\frac{16 N_C}{b} \: \tau} \; = \; 2 \tau  \gamma_0
(\alpha_s).
\labl{2.53}
\end{equation}
Experimental studies of the quantities $\xi_i$ would
provide an interesting test of the APA picture.

At $N_C = 3$ the shape parameters $\sigma, k, s$ are given in 
next-to-leading order by 
\begin{eqnarray}
\sigma & = & \frac{\tau}{\sqrt{3 z}} \;  \left ( 1 \: - \:
\frac{3}{4 z} \right ) \; + \; O (\tau^{- 1/4})   \labl{2.56a} \\
s & = & - \: \frac{a}{16} \; \frac{1}{\sigma} \; + \; O (\tau^{-
3/4}) \labl{2.56b} \\
k & = & - \: \frac{27}{5 \tau} \; \left ( \frac{\tau}{\sqrt{z}}
\; - \; \frac{b}{24} \right ) \; + \; O (\tau^{- 3/2}) \labl{2.56c}
\end{eqnarray}
It is worthwhile to notice that the next-to-leading effects
are
very substantial at present energies.  In particular, the
spectrum significantly softens since the SL corrections take
into
full account energy conservation.  This affects the rate of
particle multiplication which is strongly overestimated by the DL
approximation\footnote{Based on the experience with fitting
the observed particle distributions in QCD jets into the MLLA
picture it looks surprising that the leading logarithmic
asymptotic predictions are phenomenologically successful
in the description of the data on DIS structure functions, see
e.g.\ \cite{xfe}.  The future data will show whether this
observation is only temporary or if it carries some important
message.}.

 Relative to the leading-order predictions, the MLLA
effects imply that the peak in the $\xi$-distribution is
shifted up (i.e.\ to lower $x$), narrowed, skewed towards higher
$x$, and flattened, with tails that fall off more rapidly than a
Gaussian, as shown in Fig.\ 3.1.

The higher order terms in the series expansion (\ref{2.56a}-\ref{2.56c}) left
out are still numerically sizable at LEP-1 energies
\cite{lo} ($\sim$ 10\% contribution from 
next-to-MLLA corrections to $\bar \xi$ and $\sigma^2$)
and increase towards lower energies. Therefore, it is appropriate in a
comparison with the data over a larger energy interval to use the exact result
(\ref{ximomk}) from the MLLA solution (\ref{2.34}) including the boundary
condition (\ref{2.25}) at low virtuality $Q_0=E\Theta$. In case of the
limiting spectrum the general result (\ref{ximomk}) simplifies 
and the moments can be expressed \cite{dkt5} in terms of the
parameter $B =  a/b$
and the variable $z$ as 
\begin{equation}
\frac{<\xi^q>}{\tau^q} = P_0^{(q)}(B+1,B+2,z) + \frac{2}{z}
\frac{I_{B+2}(z)}{I_{B+1}(z)} P_1^{(q)}(B+1,B+2,z)
\labl{momls}
\end{equation}
where $P_0^{(q)}$ and $P_1^{(q)}$ are polynomials
of order $2(q-1)$ in $z$. Explicit results for the full expressions  
 for $q < 3$ can be found in \cite{dkt5} and for $q$ =3,4 in
\cite{lo1}. For example, for the multiplicity and 
the first two moments one obtains 
\begin{equation}
N_q \; = \; C_q^g \: \Gamma (B) \: \left [ \frac{z}{2} \right
]^{- B + 1} \; I_{B + 1} (z), \;\;\; z \; \equiv \; \sqrt{
\frac{16 N_C}{b} \: \tau} \; = \; 2 \tau \: . \: \gamma_0
(\alpha_s),
\labl{2.47}
\end{equation}
\begin{equation}
\frac{\bar \xi}{\tau} = \frac{1}{2} + \frac{B}{z}
\frac{I_{B+2}(z)}{I_{B+1}(z)},
\labl{lavg:LS} 
\end{equation}
\begin{equation}
\frac{< \xi^2 >}{\tau^2} = \frac{1}{4} +   \frac{B(B+\frac{1}{3})}{z^2}
+ \frac{(B+ \frac{1}{3})}{z} \left( 1 - \frac{2 B (B+2)}{z^2} \right)
\frac{I_{B+2}(z)}{I_{B+1}(z)}.   
\labl{l2:LS}
\end{equation}
These moments fulfil the boundary conditions at threshold and approach
the high energy limits (\ref{2.50}) and (\ref{2.56a}).

Concerning the asymptotic behaviour $(n \geq 1)$
\begin{eqnarray}
\sigma^2 & \sim & Y^{3/2} \labl{2.72a}\\
k_{2 n + 2} & \sim & \left ( \sqrt{Y} \right )^{- n} \labl{2.72b}\\
k_{2 n + 1} & \sim & \left ( \sqrt{Y} \right )^{- n - 1/2}
\labl{2.72c}
\end{eqnarray}
\noindent one concludes that the higher cumulants $(n > 4)$
appear to be less significant for the shape of the spectrum in
the hump region $\delta \lapproxeq 1$.

It is interesting to investigate to what extent the predicted features of
the spectrum reflect directly  the QCD dynamics
beyond the more general constraints from the energy-momentum conservation. 
One such prediction is the suppression of soft particle production
due to colour coherence which leads to  the hump-backed structure already
discussed. Another feature is the running of the coupling
$\alpha_s(k_\perp^2)$ 
which yields an enhanced production of particles at low particle or low
cms energies as compared to the case with fixed coupling.

To this end the moments have been calculated in the same MLLA approximation
as above but with the coupling kept fixed \cite{lo}. In this case 
the evolution equation (\ref{2.28}) for the Mellin transform $D_\omega$ 
can be solved explicitely and one finds for the spectrum
\begin{equation}
D(\xi,Y,\lambda) = \gamma_0 \sqrt{\frac{Y-\xi}{\xi}}
I_1 \biggl(2 \gamma_0 \sqrt{\xi (Y-\xi)} \biggr)  e^{- 2 \eta (Y-\xi)}
\labl{fixed:solution}
\end{equation}
where $\eta=a\gamma_0^2/8N_C=a\alpha_s/4\pi$,
i.e. the MLLA correction modifies the known DLA expression \cite{dfk1} just
by an exponential factor \cite{dt,lo}.
In the same way the exact results
for the moments can be obtained \cite{lo,lo1}. 
As expected, the main difference between the fixed and running coupling
cases develops already at low cms energies (see next subsection). 
For high energies
one finds
\begin{equation}               
\bar N_{fix} = \frac{1}{2} \left( 1 + \frac{\eta}{\bar \gamma_0} \right)
 \exp \left( [\bar \gamma_0- \eta]  Y \right)
\labl{norm:fix}
\end{equation}               
\begin{equation}
\bar \xi_{fix}  =   \left[ 1 +
\frac{\eta }{\bar \gamma_0 }  \right] \frac{Y}{2}
\quad , \quad     
\sigma^2_{fix} = \frac{\gamma_0^2}{4 \bar \gamma_0^3} Y
\labl{momfa1}
\end{equation}
\begin{equation}
s_{fix} = - \frac{3 \eta}{\gamma_0} \frac{1}{\sqrt{\bar \gamma_0 Y}}
\quad , \quad     
k_{fix} = \frac{3 ( 4 \eta^2 - \gamma_0^2)}{\gamma_0^2 \bar \gamma_0}
\frac{1}{Y}  
\labl{momfa2}
\end{equation}
where $\bar \gamma_0 \equiv \sqrt{\gamma_0^2+\eta^2}$. 
This high energy behaviour has to be contrasted with Eqs. (\ref{2.49}),
(\ref{2.72b}) and (\ref{2.72c}) for the running $\alpha_s$ case. 

\subsubsection{Massive Particles from Truncated Cascades}

In the context of the LPHD logic the limiting formulae are
applied for dealing with the inclusive distributions of the
\lq\lq massless" hadrons ($\pi$'s) and, with some care, for
all
charged particle spectra.  To approximate the distributions of
\lq\lq massive" hadrons $(K, \rho, p \ldots)$ the partonic
formulae truncated at different cut-off values $Q_0 \: (Q_0
(m_h)
\: > \: \Lambda)$ could be used, see e.g.\
\cite{dkt5,dkt7,dkt9}.
 Within the framework of the LPHD-MLLA picture there is no
recipe
for relating $Q_0$ to the masses of the produced hadrons and
their quantum numbers.  Therefore, one could consider a
phenomenological study of the dependence of the hadron spectra
on
a parameter $Q_0$ as a specific simplified attempt to gain
information about confinement.

The expression (\ref{2.34}) for the truncated parton
distributions is not transparent for physical interpretation
and
is not easily suited to straightforward numerical
calculations\footnote{Only recently a code has been written by
the TOPAZ group \cite{my2} for the calculation of the contour
integral of the confluent hypergeometric functions in Eq.\
(\ref{2.34}) with $\Lambda$ and $Q_0$ both as free parameters.  
The coded results have been applied for comparison with the
experimental data \cite{topaz}}.  However, one can follow here
the route proposed in Ref.\ \cite{fw} for the case of the
limiting spectrum and use for $\delta \leq 1$ the distorted
Gaussian representation (\ref{2.54}).  The
MLLA
effects are encoded in terms of the analytically calculated
\cite{dkt5} (at $Q_0 \neq \Lambda$) shape parameters
$\overline{\xi}, \sigma, k, s$, see Eq.\ (\ref{2.66}).
Also one can compare directly these moments with the
experimental data.

Let us discuss some selected properties of truncated cascades.

The procedure for analytical calculations of the moments  $\xi_q (Y, \lambda)$
for arbitrary $Q_0$ and $\Lambda$ is described in \cite{dkt5}. The general
result can be expressed as in  (\ref{ximomk}).
The mean parton multiplicity can be written \cite{dt,adkt1} in
a
compact form in terms of modified Bessel (MacDonald) functions
$I_\nu (x)$ and $K_\nu (x)$, c.f.\ Eq.\ (\ref{2.47})
\begin{equation}
N_A (Y, \Lambda) \; = \; C_A^g \: x_1 \; \left (
\frac{z_2}{z_1}
\right )^B \; [I_{B + 1} (z_1) \: K_B (z_2) \: + \: K_{B + 1}
(z_1) \: I_B (z_2)],
\labl{2.64}
\end{equation}
\begin{equation}
z_1 \: = \: \sqrt{\frac{16 N_C}{b} \: (Y +
\lambda)}, \qquad \qquad z_2 \; = \; \sqrt{\frac{16 N_C}{b} \:
\lambda} \nonumber
\end{equation}
where A = q,g denotes the type of jet.
The first term in square brackets increases
exponentially with $\sqrt{Y}$ while the second term decreases.
Its role is to preserve the initial condition for the jet
evolution, namely, $N_g = 1$.  So, at $Y \gg \lambda$ it can, in
fact, be neglected.\footnote{Eq.(\ref{2.64}) does not fulfil the 
initial condition 
$dN_A/dY=0$. This can be achieved by adding further nonleading terms
\cite{lo1}.}

For the first moment, one obtains
\begin{equation}
\overline{\xi}=\frac{1}{2} Y + \frac{1}{2} \sqrt{\lambda(Y+\lambda)}
 \; \frac{ (I_{B+2}K_{B+1}-I_B K_{B-1}) + (K_{B+2}I_{B+1}-K_B I_{B-1})}
   {I_{B+1}K_B+K_{B+1}I_B}
\labl{xibarl}
\end{equation}
In this equation it is implied that in each of the products of the two
Bessel functions the argument of the first (second) one is $z_1$ ($z_2$).
The higher moments can be found in \cite{dkt5}. 

For a discussion of the main effect of truncating the cascade
at $Q_0>\Lambda$ we consider the high energy approximations which can also
be obtained readily from (\ref{2.71}):
\begin{equation}
\overline{\xi} \; = \; \frac{1}{2} Y\; + \; \frac{a}{4\sqrt{b N_C}} 
    (\sqrt{Y+\lambda}\; - \; \sqrt{\lambda}),
  \labl{xibnlo}
\end{equation}
\begin{equation}
\sigma^2 \; \approx \; \frac{1}{12} \; \sqrt{\frac{b}{N_C}}
\;
\left [ \left ( \sqrt{Y + \lambda} \right )^3 \: - \: \left (
\sqrt{\lambda} \right )^3 \right ] \; - \; \frac{b}{32 N_C} \;
\left [  (Y + \lambda) \: - \:
\lambda \right ],
\labl{2.73}
\end{equation}
(the limiting spectrum corresponds to $z_2 = 0$ $
 (\lambda = 0)$),
 c.f.\ Eqs.\ (\ref{2.50}) and (\ref{2.56a}).  Strictly
speaking this estimate is valid for large values of $\lambda$
only, since the asymptotic expansion (\ref{2.33}) relied upon
$\alpha_s (Q_0) \ll 1$.  However, the message we get
is clear: a stiffening of the spectrum is predicted for the truncated
cascade with $Q_0>\Lambda$, i.e. the average 
and the width of the spectrum become smaller with
$\lambda$ increasing. The same way one can find the higher order 
shape parameters $k_q$ explicitly and then examine the deviations from
the leading-order Gaussian spectrum which manifest the same
qualitative features as in the case of the limiting distribution,
see previous subsection and Fig.\ 3.1. 
 

Without going into detailed technical analysis of particle
spectra one can notice an interesting property which concerns
one of the most important characteristic --- the position of the
maximum.

\indent From the expression (\ref{2.54}) for the distorted Gaussian one
can see that the non-zero skewness leads to a calculable
splitting between $\xi^*$ and $\overline{\xi}$, namely
\begin{equation}
\delta_{\rm max} \; \approx \; - \: \frac{s}{2}.
\labl{2.74}
\end{equation}
\noindent As shown in Ref.\ \cite{dkt5} the difference $\xi^*
- \overline{\xi}$ is an energy independent constant.  Moreover,
this quantity proves to be $Q_0$ independent as well and, as a
result, formula (\ref{2.52b}) remains valid at $Q_0 \neq
\Lambda$.  Namely,
\begin{equation}
\xi^* - \overline{\xi} \; \approx \;  \frac{3 a}{32
N_C} \; = \; 0.351 \: (0.355) \;\;\; {\rm for} \: n_f = 3 (5).
\labl{2.75}
\end{equation}
\noindent Combining this result with Eq.\ (\ref{xibnlo}) we
conclude that MLLA predicts the energy-independent shift of
the peak position for truncated parton distributions compared to
the limiting spectrum.  This fact can be used to measure
effective $Q_0$ values for different hadron species by
comparative study of the energy evolution of the peak.
The observation of an increase 
 of $Q_0$ with $m_h$
(for hadrons with the same quark content and the same spin) would
evidence in favour of the APA expectation 
 that massive hadrons are produced effectively at
smaller space-time scales.  Fig.\ 3.2 illustrates the dependence
of $\xi^*$ on $Q_0$ at a number of energies 
 and on energy
for different $Q_0$ values as obtained 
from the numerical evaluation of Eq. (\ref{2.38}).
Also shown is the prediction of the asymptotical distorted Gaussian
result  (\ref{2.75}).

It is convenient for phenomenological applications (see e.g.\
\cite{dkt9,ncb}) to present $\xi^*$ in a form analogous to
Eqs.\ (\ref{2.51}),(\ref{2.52b}) for the limiting case
\begin{equation}
\xi^* \; \approx \; \tau \:  \left [ \frac{1}{2} \; + \;
\sqrt{\frac{C}{\tau}} \; - \; \frac{C}{\tau} \; + \; O
(\tau^{-3/2}) \right ] \; + \; F (\lambda)
\labl{2.76}
\end{equation}
\noindent with $F(0) = 0$.

Based on the results of Ref.\ \cite{dkt5} it was proposed in
\cite{ncb} to fit both function $F (\lambda)$ and its inverse
$\lambda (F)$ to polynomials as
\begin{equation}
F (\lambda) \; = \; - 1.46 \lambda \: + \: 0.207 \lambda^2 \:
\pm \: 0.06,
\labl{2.77}
\end{equation}
\vskip-.75cm
\begin{equation}
\lambda (F) \; = \; - 0.614 F \: + \: 0.153 F^2 \: \pm \: 0.06.
\labl{2.78}
\end{equation}
\noindent The given errors denote the maximum deviation from
the distribution plotted in Fig.\ 3.2 in the ranges $- 2 \: < \: F
\: < \: 0$ and $0 \: < \: \lambda \: < \: 2$.

To illustrate the effects of finite $Q_0$ on the parton spectrum
we present in Fig.\
3.3 the distribution ${D} (\xi, Y, \lambda)$ calculated
by numerical integration of $D (\omega)$ (see Eq.\ (\ref{2.37}))
in the complex $\omega$-plane and its energy evolution (from Ref.
 \cite{dkt5}.  
As example of a higher order cumulant we depict in Fig.\
3.4 the $Q_0$-dependence of kurtosis $k$ at the LEP-1 energy.
 Finally, let us note that
with increasing $W$ the shape characteristics of particle spectra
should become universal.

\subsection{Quark and Gluon Jet Differences}

Within the perturbative scenario for  the particle cascade the larger colour
charge of gluons ($C_A=N_C=3$) compared to quarks ($C_F=\frac{4}{3}$) 
 leads to various observable differences between
the two types of jets. In particular, the more intensive bremsstrahlung 
off a primary gluon yields asymptotically to the larger multiplicity in a
gluon jet by the factor $\frac{N_C}{C_F}=\frac{9}{4}$ \cite{bg}.
One expects also that the gluon jets are broader and contain
softer particles than quark jets of the same energy
\cite{ss,ew}.  
Such differences have been discussed already in the early eighties 
\cite{adk}. Meanwhile the calculations of increased accuracy became available.
Some predictions of this type will be discussed in the following.

It should be noted that these analytical results correspond to isolated
quarks and gluon jets emerging from pointlike colourless $q \bar q$ or $g g$
systems. Therefore, they describe the entirely inclusive quantities not
requiring the selection of jet topology or identification of jets within an
event. Some possible realizations of these requirements are discussed below
in subsection 3.3.5.

So far the bulk of
experimental information on the structure of a gluon jet has come
from the studies of 3-jet events in $e^+ e^-$ annihilation. 
Without special care, such an analysis is inherently ambiguous
and there is no direct correspondence between the data and the
properties of an individual gluon jet.  In particular, the
interjet coherence phenomena play an important role here
\cite{dkt1,dkmt2}.

The situation has been clarified a lot by the recent LEP-1 analyses which
complement the analytical predictions by the detailed Monte Carlo calculations at
the parton level to take into account the actual experimental situation 
 as will be discussed in  subsection 3.3.5.

\subsubsection{Multiplicity Ratio}

The ratio of multiplicities in quark and gluon jets has been calculated as
an expansion in $\sqrt{\alpha_s}$ or in
$\gamma_0=\sqrt{2N_C\alpha_s(Q)/\pi}$ at jet virtuality $Q$ \cite{mw,gm,dn1}
\begin{equation}
r\; \equiv \; \frac{N_g}{N_q}\; = \; 
    \frac{N_C}{C_F} \: (1-r_1\gamma_0-r_2\gamma_0^2) + O(\gamma_0^3)
    \labl{rqg}
\end{equation}
The leading term represents the asymptotic prediction which is given by the 
ratio of color charge factors \cite{bg}, the next terms describe the
approach to this limit. The result in order $\alpha_s$ has been reported as
\cite{dn1}
\begin{eqnarray}
r_1&=&2\left(h_1+\frac{n_f}{12N_C^3}\right)-\frac{3}{4},  \labl{r1qg}\\
r_2&=&\frac{r_1}{6}\left(\frac{25}{8}-\frac{3}{4}\frac{n_f}{N_C}-
     \frac{C_F}{N_C}\frac{n_f}{N_C}\right) +\frac{7}{8}-h_2
    -\frac{C_F}{N_C}h_3 + \frac{n_f}{12N_C}\frac{C_F}{N_C} h_4
   \labl{r2qg}
\end{eqnarray}
where $h_1=11/24$, $h_2=(67-6\pi^2)/36$, $h_3=(4\pi^2-15)/24$ and $h_4=13/3$.
Results for $r$ can be obtained in a diagrammatic approach \cite{gm} or by
the generating functional method \cite{mw,dh,dn1}. In the latter case one
starts from the coupled channel integro-differential equations (\ref{2.23})
for the generating function $Z_A(u)$  and obtains
the corresponding equations for the multiplicities $N_A=Z_A'(u)$.
The generating function which occurs at  displaced arguments
$Z(y+a)$ ($a=\ln x,\ln(1-x)$) is expanded into a Taylor series at large $y$
whereby terms of the second and higher derivatives are neglected \cite{imd,dn}
(see also section 4.1).
The $r_1$-term in (\ref{r1qg}) corresponds to the result of Ref.\ \cite{mw}. 
The first term
on the r.h.s of (\ref{r2qg}) has been obtained in \cite{gm},
whereas the other contributions according to  \cite{dn1}
come from the terms with  derivatives of
$Z(y)$.

Numerically one expects for 
the ratio $r$ in (\ref{rqg})  for $n_f=3$ and at the
$Z^0$ mass the value $r=2.25-0.197-0.214=1.84$ \cite{dn1}. Then,
 the asymptotic
prediction is reduced by about 20\%, but the expansion does not seem to be
rapidly convergent. A numerically similar result has been obtained for fixed
$\alpha_s$ \cite{dh}
in which case the evolution equation can be solved without Taylor
expansion of the generating function.

\subsubsection{Differential Distributions}
{\it $\xi$-distribution}\\
As was emphasized before, within the MLLA the momentum
fraction
distributions in quark and gluon jets differ only by the
normalization factor $\frac{C_F}{N_C} =
\frac{4}{9}$.
Taking account of the Next-to-MLLA effects,
namely
$\sim O (\sqrt{\alpha_s})$ terms in the \lq\lq
pre-exponential"
coefficient function factors $C_i^f$ in Eq.\ (\ref{2.38}),
leads  to some controllable change in the shape of
particle spectra in $q$ and $g$ jets \cite{fw,dkt7}.

This change can be described as a shift of the limiting
distribution (see Eqs.\ (\ref{2.43}),(\ref{2.44})) according to
$(A = q, g$ denotes the type of jet)
\begin{equation}
{D}_A (\xi, \tau) \; = \; \left [1 + \Delta_A \:
\left
(\frac{\partial}{\partial \xi} \: + \:
\frac{\partial}{\partial
\tau} \right ) \right ] \; {D}^{\lim} \: (\xi, \tau)
\;
\simeq \; {D}^{\lim} \: (\xi + \Delta_A, \tau +
\Delta_A),
\labl{2.59}
\end{equation}
\vskip-.75cm
\begin{equation}
\Delta_g \; = \;- \frac{n_f}{3} \:\: \frac{C_F}{N_C^2} \;
\left (
= \: - \frac{4}{27} \right );\qquad \Delta_q \; = \; \Delta_0 +
\Delta_g \; \left (= \: + \frac{1}{27} \right ),
\labl{2.60}
\end{equation}
\begin{equation}
\Delta_0\; \equiv \;r_1 \; = \; \frac{a-3N_C}{4N_C} \; \left(=\frac{5}{27}
\; {\rm for} \; n_f=3\right)
\labl{2.60a}                         
\end{equation}
This means that one obtains the particle distribution
for
a given jet energy $E$ with the $\sqrt{\alpha_s}$ corrections
taken into account by just evaluating the ${D}^{\lim}
\:
(\ln \: (E_{\lim}/E_h), \ln (E_{\lim}/\Lambda))$ spectrum
(\ref{2.44}) at
\begin{equation}
E_{\lim} \; = \; E \: \exp (\Delta_q) \; \simeq \; 1.04 \: E
\;\;\;\; {\rm for} \; q-{\rm jet}
\labl{2.61a}
\end{equation}
\noindent and 
\begin{equation}
E_{\lim} \; = \; E \: \exp (\Delta_g) \; \simeq \; 0.86 \: E
\;\;\;\; {\rm for} \; g-{\rm jet}
\labl{2.61b}
\end{equation}
\indent Thus, the distribution in a $q$-jet is fairly close to
the
limiting spectrum with a very small shift of the asymptotical
position of the maximum to higher $x$,
\begin{equation}
\delta \; \left ( \ln \: \frac{1}{x_{\rm max}} \right ) \; =
\; -
\: \left ( 1 \: - \: \frac{\ln \: 1/x_{\rm max}}{Y} \right )
\:
\Delta_q \; \simeq \; - 0.01.
\labl{2.62}
\end{equation}
The spectrum in a $g$-jet is shifted to lower $x$
compared with the limiting case
\begin{equation}
\delta \; \left ( \ln \: \frac{1}{x_{\rm max}} \right ) \; =
\; -
\: \left ( 1 \: - \: \frac{\ln \: 1/x_{\rm max}}{Y} \right )
\:
\Delta_g \; \simeq \; + 0.04.
\labl{2.63}
\end{equation}
Whereas the first moments for quark and gluon jets are different,
it is an immediate consequence of (\ref{2.59}) that the higher order 
 cumulant moments from
(\ref{2.66})
 are the same for both type of
jets in the given approximation.

The first moments of both $\xi$-distributions have also been computed in
Ref. \cite{fw}. The leading contribution in $\Delta_0$ is as in
(\ref{2.60}),
a contribution in higher order is presented as the relative shift in $\sigma$
and $k$.  

Strictly speaking, this consideration of
$\sqrt{\alpha_s}$ effects tells us about the relative features
of
$q$- and $g$-jets, i.e.\ about the correction to the
multiplicity
ratio and $\frac{5}{27}$ shift in $\xi$ between the spectra of
the two.  To derive absolute predictions with such an accuracy
one needs to go well beyond the MLLA scope to be able to
define
the energy scale precisely by fixing the normalization scheme.
Changing the scale of $\Lambda$ by a factor of $O (1)$ would
lead
to a correction of the relative order of $O \left (
\sqrt{\alpha_s} \right )$, which is subleading to the MLLA. 
The same remark holds true, e.g.\ for the next-to-leading
correction to the hump position.  Scaling the value of $\Lambda$
by a factor of $\rho$ would induce a shift $(\sim \ln \rho)$ in
the constant term in Eq.\ (\ref{2.51}).


\noindent{\it Particle and energy collimation}\\
The different angular spread of quark and gluon jets is illustrated
 in Fig.\ 3.5 which shows the
energy and multiplicity collimation in  isolated jets, i.e.\ when
the mutual influence of jets in their ensemble is neglected.
 As one clearly sees both the energy and
multiplicity collimation grows as jet energy increases, being
stronger for a quark jet than for a gluon jet.  With the jet
energy increasing the collimation of the multiplicity flows grows
more slowly than that of the energy flux, see for details
\cite{dkmt2,dkt10}.

\subsection{Experimental Data Confronted with Analytical Predictions}
\indent This subsection contains a selection of experimental tests of the
perturbative approach to particle distributions in QCD jets in the semisoft
region, based on MLLA calculations and LPHD.

There has been considerable progress in the experimental
studies of QCD jets due to several factors: 
the high statistics collected by the LEP-1 and TEVATRON
experiments; the improvements in detectors, especially in
particle identification and heavy quark tagging efficiency; new
tools for data analysis; and, of course, the advent of HERA,
which allows to perform the fundamental tests of the APA picture
in DIS 
\cite{al1} (see section 8).  A record precision in the studying of
identified final states has been reached at LEP-1 where about 20
particle species have been measured, see \cite{ada2,wjm2}.
The discussion of particle spectra presented in this subsection is based on
the experimental results reported in Refs. 
\cite{mark1,tasso1,tasso2,tpc,gdc,mark2,%
tassox,amyx,alephx2,delphix2,delphix3,delphix4,%
opal1,l31,l32,opal2,aleph1,delphi1,%
opal3,aleph2,delphi2,aleph3,opal5,%
delphi3,opal7,opal6,mark2n}. The recent LEP-1.5 data are presented 
in Refs. \cite{l3L15,delphiL15,alephL15,opalL15}; they are analysed in
Ref. \cite{klo}.  
For the pseudoscalar and vector meson nonets
and for the baryon octet and decuplet, at least one state per
isospin multiplet has been examined, plus the scalar $f_0$ (980)
and the tensors $f_2$ (1270) and $K_2^* (1430)^0$.

According to the LPHD approach \cite{adkt1}, 
the MLLA partonic predictions, 
when confronted with the experimental data,
are multiplied by the overall
normalization factors $K^h$
\begin{equation}
{D}_q^h \; = \; K^h \: {D}_q.
\labl{3.1}
\end{equation}
This prediction has been applied to the distribution of all charged
particles. The comparison of data with the limiting spectrum yields
parameters  $\Lambda_{\rm ch}
\simeq 250 - 270$ MeV \cite{lo,dkt7,dkt9,kdt}.

Alternatively, Eq.(\ref{3.1}) can be applied to various  
particle species separately, then the respective values of $K^h$ 
as well as  of the cut-off parameters $Q_0 \equiv Q_o
(m_h, J^{PC}, \ldots)$ should be determined phenomenologically
from comparison with the data.
The effective QCD
scale parameter $\Lambda_{\rm eff}$ could be found from the
fit to the $\pi$-spectra using the limiting distribution.  Once
determined, this quantity should remain fixed when fitting the
data for different particle species using the truncated parton
distributions.  Since we concentrate mainly on the semisoft
region the difference in the flavour composition of the
primary quarks is not so essential, but it could play a certain role in
fitting to data at current energies.
The values found for the QCD scale $\Lambda_{\rm eff} \gapproxeq$ 150 MeV
are a bit lower than those found for the charged particles.%
\footnote{In some MLLA-based fits
$\Lambda_{\rm eff}$ varies for different particle species, see
e.g.\ \cite{topaz}.}
 


The inclusive particle spectra have been studied vigorously in
the process $e^+ e^- \rightarrow$ hadrons since the early PETRA/PEP
days.  The first comparison \cite{adkt1} of the $\pi^\pm$ data
with the limiting spectrum showed quite satisfactory agreement. 
The new era of the detailed testing of the APA predictions has
been opened by the OPAL results on the charged particle momentum
fraction distribution \cite{opal1}.  The first measurement of
the identified particle ($\pi^0$) spectrum at LEP energy has been
performed by the L3 group \cite{l31}.  A wealth of new
experimental information on particle distributions in jets has
been collected at LEP, SLC and TRISTAN since then (see, e.g.,
reviews \cite{h,ms,ada1,ncb,wjm2}).

The comparison of charged particle distributions 
measured by the same experimental group 
at LEP-1, LEP-1.5 and then at LEP-2 will be very  useful 
for examining the relative normalization.

\subsubsection{Charged Particles and Limiting Spectrum}
{\it Energy evolution of $\xi$-spectrum}\\ 
To study the energy evolution, in Ref.\ \cite{opal1}
the $Z^0$ data were combined with the TASSO results \cite{tassox}
at lower
energies 9-44 GeV.  Fig.\ 3.6a shows the data on the $\xi_p = \ln
\left ( \frac{1}{x_p} \right )$ distribution together with the
predictions of the limiting spectrum (\ref{2.44}) and 
the distorted Gaussian (\ref{2.54}) with (\ref{2.56a}-\ref{2.56c}).
 Even now when we have the whole wealth of new data
on particle spectra at hand the agreement between the first
measurements \cite{opal1} and the MLLA-LPHD results (derived
five years prior to the data \cite{dt,adkt1}) looks very
impressive. In Fig. 3.6b we show the recent LEP-1.5 data which follow well the
predictions for the higher energy \cite{klo}.

All observed particle distributions,
for not too small momenta,  are in very good agreement
with the APA predictions, 
based on the QCD cascading picture of
multiple hadroproduction. The scale parameter used in the fit is
$\Lambda$ $\simeq$ 250 MeV. 
A closer inspection of the particle distributions shows a clear
deviation from the DLA predictions (\ref{2.45}),(\ref{2.49}):
the maximum of the $\xi$-distribution is shifted upwards from
$\tau/2\approx 2.6$, also the distribution is not symmetric around the
maximum.  

Let us make
a few comments concerning the charged particle distributions
exemplified in Fig.\ 3.6a, see Refs.\ \cite{dkt7,dkt9,kdt}.  It is
not surprising that the distorted Gaussian distribution
breaks down at large $x_p$.  More challenging is the agreement
of the data at $x_p \sim 1$ with Eq.\ (\ref{2.44}) representing
the true solution of the MLLA Evolution Equations.  Such a
coincidence stems from the fact that in terms of the Mellin-
transformed distribution (\ref{2.29}) the approximate expression
for anomalous dimension $\gamma_\omega (\alpha_s)$ at $\omega \ll
1$, that had been used to derive the semisoft spectrum
(\ref{2.44}), mimics reasonably well the behaviour of
$\gamma_\omega$ at $\omega \sim 1$ too, which is the region
responsible for $x_p \sim 1$.  At $\omega > 1$ it
becomes negative (as the true $\gamma$ does), thus imitating the
scaling violation at $x_p \rightarrow 1$.

The deviations from the limiting spectrum 
 at small $x_p$, seen in Fig.\ 3.6a, have a clear origin
and can be easily avoided.  The point is that in the MLLA one
makes no principal difference between the energy and momentum fractions
carried by a particle, i.e. $x_E$ and $x_p$.  Meanwhile, the MLLA
evolution equations, leading to (\ref{2.44}), in fact treat $x$ as 
$x_E=E_h/E$, so that parton distributions vanish below $x_E  =
Q_0/E$ $ (= \Lambda/E$ for the case of the limiting spectrum). 
If this kinematic effect is taken into account in a simple way
\cite{lo,klo} (see Eq.~(\ref{dual}) below), the distribution in $\xi_p$
acquires a tail for large $\xi_p$. This prediction
 is shown in Fig.~3.6b as the dashed
curve and fits the data well.

%

\noindent{\it Mean multiplicity}\\
Another piece of strong experimental evidence in favour of the
MLLA-LPHD cascading picture comes from measurements of
particle multiplicities.  The dynamics of the charged
multiplicity increase reflecting the size of the hump height
with the jet energy, is depicted in Fig.\ 3.7.  One finds
once more an excellent agreement between the data and the
analytical QCD expectation (\ref{2.48}). In particular, the recent LEP-1.5
data agree with the extrapolation of the fit to the lower energy data
\cite{ms}.
 
The experimental results are in 
distinct disagreement with the expectations following from the
incoherent QCD cascades.
In particular, neglecting the soft gluon interference would
increase the slope in Fig.\ 3.7 by $\sqrt{2}$ which is
inconsistent with the data for any reasonable $\Lambda_{\rm eff}$
parameter (see, for example, Ref. \cite{wo}).
A fit including lower energy data based on Eq. (\ref{2.47}) is shown in
Fig.\ 3.9, see discussion below.

\noindent{\it Peak position $\xi^*$}\\
Especially spectacular is the energy evolution of the peak
position $\xi^*$ 
 which
is independent of the normalization factors.  It is displayed in
Fig.\ 3.8 together with the exact prediction from the limiting spectrum
(\ref{xistex}).
These data distinctly demonstrate the essential role of the
next-to-leading corrections.  Thus the effective gradient $b_{\rm
eff}$ in the straight line fit
\begin{equation}
\xi^* \; = \; b_{\rm eff} \: \tau \: + \: {\rm const.}
\labl{3.3}
\end{equation}
\noindent appears to be about $\frac{5}{8}$ rather than
$\frac{1}{2}$ expected in the DL asymptotics (taking into account
the QCD coherence).
Moreover, the present accuracy in measurements of the energy
dependence of the hump position makes it possible to test
accurately expansion (\ref{xistex}) including the third,
next-to-MLLA, constant term. 

The existing data on $\xi^*$ are clearly incompatible with the
assumption of an incoherent branching picture or a simple
phase-space model $(\xi^* \simeq \tau +$ const.) \cite{ms}.  Notice that
the value of $\xi_p^*$ at the $Z^0$ corresponds to 
the rather low momenta $x_p
\approx 0.02$ or  $p \approx 1$ GeV. It is worthwhile to mention that the
energy dependence of the flavour composition of the primary quarks
could affect the energy evolution of $\xi^*$ in the wide energy range.

\noindent{\it Cumulant moments}\\
Instead of studying the energy evolution of the spectrum itself it proves
convenient to consider the cumulant moments of the distribution introduced
in (\ref{2.66}). Analytical predictions are available 
in the MLLA for the general case
of different $Q_0$ and $\Lambda$ parameters \cite{dkt5}. 
These moments are independent
of the normalization of the spectrum and their energy evolution 
is fully determined by the two parameters $Q_0$ and $\Lambda$,
see subsection 3.1.  

In the comparison of data with the theoretical predictions  one 
can try to avoid
the deviation seen for large $\xi$ in Fig. 3.6 by taking into
account the kinematical behaviour of parton and hadron spectra. 
Note that in the calculations the partons were taken as massless ($E_p=p_p$) and 
 $Q_0$ was the
cut-off in $k_\perp$, for hadrons $Q_0$ is taken as an effective mass and
$E_h=\sqrt{p_h^2+Q^2_0}$. A simple way to achieve a
consistent kinematical behaviour for
small energies $E_p$ and $E_h$ is to relate
parton and hadron spectra in the jet
through \cite{dfk1,lo} (for a primary parton $A$)
\begin{equation}
E_h \frac{dn_A(\xi_E)}{dp_h}=K^h E_p\frac{dn_A(\xi_E)}{dp_p}
\equiv K^h D_A^g(\xi_E,Y) \qquad
E_h = E_p \ge Q_0
\labl{dual}
\end{equation}
at the same energy $E$ or $\xi_E=\ln E_{jet}/E$
for partons and hadrons. 
This relation has the proper limit for $E\sim p \gg Q_0$ where parton and
hadron spectra become proportional, independent of $Q_0$\footnote{It
implies that parton and hadron momenta
are not the same for small momenta around $Q_0$.}.

With  relation (\ref{dual}) it is possible to have a finite limit of
the invariant distribution $E\frac{dn}{d^3p}$
 for $E\to Q_0$  in both cases \cite{lo}.
The experimental momentum spectra are then transformed into 
energy spectra for a given $Q_0$ and the moments are calculated. By
comparison with the MLLA predictions ($\bar \xi$ as in Eq. (\ref{xibarl})
and the higher moments as in (\ref{ximomk})\cite{dkt5}) the best agreement is
obtained for $Q_0 \simeq \Lambda \simeq 270 \ \hbox{MeV}$, i.e. for the
limiting spectrum \cite{lo}. 
The results for the moments at this $Q_0$ and the
predictions from the limiting spectrum (\ref{momls}) are shown in Fig.~3.9
as full lines. 
The multiplicity\footnote{The zero$^{\rm th}$ moment $N_E$ of the spectrum 
(\ref{dual}) approaches the particle multiplicity $N$ for high energies.} %
$N_E$ is fitted by Eq. (\ref{2.47}) with two further
parameters (a multiplicative and an additive one) to allow for a finite
value near threshold.
The number of flavours was taken  $n_f=3$ which is justified by the
low effective $k_\perp$ in the cascade.
Also shown are predictions with fixed coupling to which we
come back below. 

The MLLA predictions with running coupling are remarkably successful
considering the fact that there are only two parameters, actually
coinciding, for the four moments. A deviation is visible for $\bar \xi_E$ at
the lower energies which may indicate the influence of the leading primary
quark neglected in the calculation or other uncovered low energy effects. 
Otherwise, the predictions are confirmed
at an almost quantitative level down to the lowest cms energy of
3 GeV where charged particle spectra have been presented.
If a moment is given at a particular energy, then its further evolution 
is determined as in (\ref{2.71}). It turns out that the 
boundary condition (\ref{2.25}) for low virtuality $E\Theta=Q_0$
actually gives a satisfactory normalization
of the moments.   
 
\subsubsection{Identified Particles and Truncated Cascade}

Now let us come to the identified particle distributions.  All
existing data show that both the bell-shaped form of the spectra
and the energy evolution of the maximum $\xi^*$ 
are in a fairly good agreement with
the APA predictions.  For illustration we display in Fig.\ 3.10
the results of some recent measurements 
for different charged and neutral particles  
 at the $Z^0$   and from lower energies in Fig.\ 3.11.
Also shown is the comparison with the limiting spectrum for pions and
with the prediction (\ref{2.38})
from the truncated cascade with cut-off $Q_0$ for the
heavier particles.

To illustrate how the distorted Gaussian formalism of Ref.\
\cite{dkt5} (see subsection 3.1) works in the case of massive
particles we present in Fig.\ 3.10b the OPAL data of Ref.\
\cite{opal5} on the $K^0$ production together with the spectrum
of partons from the truncated MLLA cascades and its distorted
Gaussian approximation.  A fit to these data \cite{dkt9} gives
the values of the only free parameters, $Q_0 (m_K) = 300$ MeV,
$K^{K^0} = 0.38$.

Contrary to the case of the limiting distribution, one observes
here some deviation at the left wing of the spectrum.  This is
partly caused by a finite contribution of leading partons at $Q_0
> \Lambda$.  In the hard parton region $(x \sim 1)$ particle
yields in the quark and gluon jets become essentially different
and the soft approximation is no longer valid.  Note that this
region is also influenced by the heavy flavour decays.  The
flavour dependence does not only manifest itself at high momenta,
but it could also affect the low momentum range.  This could
cause shifts in $\xi^*$ which depend on the jet flavour.  As was
mentioned before, towards the right wing of the distribution one
should take into account the pure phase space effects.

The energy dependence of the peak position $\xi^*$ is
illustrated by Fig.\ 3.12 for some charged and neutral particles.
In Fig.\ 3.12b the data on $\pi^0$'s were combined with the
results for all charged particle production \cite{dkt9}.  The
latter were replotted on an effective energy scale
$(\sqrt{s})_{\rm ch} \; = \; \frac{\Lambda_{\rm
eff}}{\Lambda_{\rm ch}} \: W \; \approx \; 0.625 W$ (c.f.\
Eqs.\ (\ref{2.51}), (\ref{3.3})).  Comparison with the observed
energy dependence of $\xi^*$ allows further tests of the QCD
cascading picture.  In particular, Fig.\ 3.12b confirms the
perturbative universality of this dependence \cite{dkt5}, see
subsection 3.1.5.  The present data on the
identified particle spectra confirm the perturbative
expectation (\ref{2.76}) that for different particle species the energy
dependence of $\xi^*$ is universal. The data are completely incompatible
with the incoherent shower prescription (see Fig.\ 3.12a).

\subsubsection{Phenomenology of Mass Dependence}

The peak position of the $\xi$-spectra has been determined at the $Z^0$ for
a large variety of stable and unstable particles from Gaussian fits to the
central region.
For illustration we display in Fig.\ 3.13
the results of some recent ALEPH  \cite{aleph3} measurements of the
inclusive production of vector mesons.
One sees clearly the hump-backed shape of the particle
distributions, which is well represented by the Gaussian fits.


According to the LPHD concept it is natural to expect that two
particles with the same flavour content and the same spin and
nearly the same mass have practically the same
$\xi_p$-distributions.  This is confirmed by the
$\rho^0$ and $\omega$ mesons in  Fig.\ 3.13.  No evidence for
isospin dependence is seen in these data.  At the same time $K^*
(892)^o$ production is suppressed compared to the $\rho^0$ and
$\omega$ and its momentum distribution is harder.  This
qualitatively follows the MLLA-LPHD expectations.

Note also that, e.g.\ within Gribov's confinement picture
\cite{vng1} one would expect the double-strange $\Phi$ and $\Xi$
production to be strongly suppressed and their momentum spectra
to be stiffened as compared to $K^* (892)^o$ and $\Lambda$
respectively \cite{cdgkr}.  This seems to be in agreement with
the present data \cite{ada2,aleph3,opal6,opal7}.  In particular,
all experiments converge on a rather low $\Phi$ production rate.

The present LEP-1
results on $\xi^*$ are summarized for various particle species in Table 1.
\begin{table}[t]
\begin{center}
\begin{tabular}{|c|c|c|} 
\hline
Particle & $\xi^*$ & Exp \\ \hline
$\pi^+$  & 3.79 $\pm$ 0.02 & AOS \\ \hline
$\pi^0$  & 3.94 $\pm$ 0.13 & DL \\ \hline
$K^+$    & 2.64 $\pm$ 0.04 & ADO \\ \hline
$K^0$    & 2.68 $\pm$ 0.05 & ADLOS \\ \hline
$\eta$   & 2.52 $\pm$ 0.10 & L \\ \hline
$\eta^\prime (958)$ & 2.47 $\pm$ 0.49 & L \\ \hline
$\rho (770)^0$ & 2.80 $\pm$ 0.19 & A \\ \hline
$K^* (892)^0$ & 2.35 $\pm$ 0.07 & AO \\ \hline
$\phi (1020)$ & 2.21 $\pm$ 0.04 & ADO \\ \hline
$\omega (782)$ & 2.80 $\pm$ 0.36 & L \\ \hline
$p$ & 2.97 $\pm$ 0.09 & ADO \\ \hline
$\Lambda$ & 2.72 $\pm$ 0.04 & ADLOS \\ \hline
$\Xi^-$ & 2.58 $\pm$ 0.11 & DO \\ \hline
\end{tabular}
\caption{Maximum $\xi^*$ presented for various particles in review 
\protect\cite{ada2} based on data from ALEPH (A), DELPHI (D), L3 (L), 
OPAL (O) and SLD (S), see also discussion in text.}
\end{center}
\end{table}
Fig.\ 3.14 from Ref.\ \cite{delphi3} shows the measured
peak positions at $\sqrt{s} = M_z$ as a function of the particle
mass.  The lines are fits to the data using the functional form
$\xi_p^* = \ln \: \frac{m_0}{m_h}$ with changeable parameters
$m_0$ for mesons $(m_0 \simeq 6.6)$ and baryons $(m_0 \simeq
18)$.  In both cases the anticipated decrease of the maximum with
increasing particle mass is clearly seen.  The observation that
the simple logarithmic fits seem to describe the data reasonably
well looks quite intriguing.  It may well be that this analysis
has provided us with some (still) encoded messages concerning the
LPHD concept\footnote{A universal
mass dependence 
$\xi_{\rm prim}^* \; \propto \; \ln \: \frac{m_0}{m_h}$ 
which includes mesons and baryons has been found 
for the primary hadrons by taking into account
resonance decays \cite{ada2,delphi3,opal7} according to the 
hadronization scheme of the JETSET Monte Carlo \cite{JETSET}. 
Although interesting in its
own right, this result cannot be derived solely within the LPHD philosophy
 which
assumes resonance decays to be represented by the parton cascade in the
average.}.


\subsubsection{Sensitivity to Colour Coherence and Running Coupling}

It is sometimes argued, that the momentum spectrum is largely determined by
the global features of the final state. Here 
we want to discuss to which extent the momentum  spectra
reflect the  characteristic features of the QCD dynamics in the parton jet
evolution and what is the role of LPHD in comparison to hadronization models. 
To this end one may consider alternative models which miss an
important ingredient of the QCD description.  

\noindent{\it Hump-backed plateau}\\
Let us first consider the case of a cascade without the soft gluon coherence
taken into account.
This can be studied in Monte Carlo calculations where the angular ordering
constraint is ``switched off". Then the  $\xi$-distribution
does not show any more the Gaussian shape, the ``hump-backed plateau", 
but develops a shoulder towards
low momenta (large $\xi$), as has been shown in a numerical simulation
 \cite{marw1}. This is clearly seen also 
in Fig.\ 3.15 for the distribution of
gluons as obtained with the JETSET Monte Carlo \cite{JETSET}
at a cut-off $Q_0$=1 GeV
\cite{bcu}. Also shown in this figure are the spectra of charged particles
from both models after string hadronization of the partonic final state.
The parameters of the two Monte Carlo models
are tuned in order  to reproduce the main
global features of the hadronic final state. Both hadron distributions are then
found to coincide in Fig.\ 3.15
and also to agree with the measured spectrum in Fig.\ 3.6.
Therefore, it is not possible to claim 
evidence  in favour of the soft gluon interference within this hadronization
model \cite{bcu}:
the different treatment of the perturbative partonic interactions can be
compensated by a different treatment of the nonperturbative 
hadronization phase.
On the other hand, evolving the parton cascade further
according to 
perturbative QCD from $Q_0 \sim 1$ GeV to $Q_0 = m_h$, only the gluon
distribution for the coherent model would approach the data, as is asserted
by Fig.\ 3.15. This example nicely demonstrates the predictive power of the
LPHD hypothesis which predicts the similarity of parton and hadron spectra
at comparable scales without additional parameters. On the other hand,
hadronization models have considerable flexibility to bring
theoretical schemes, which are quite different 
at the parton level, into agreement with the data. In
particular, the parton shower with string hadronization corresponds the LPHD
predictions 
 only for a particular  set of model parameters.    

We should note another lesson from Fig. 3.15. The test of the LPHD
hypothesis through the comparison of data with Monte Carlo results at parton
level needs some care. The current Monte Carlo models use cut-off values
$Q_0$ greater than the hadronic scales 150-270 MeV obtained from the fits to
the analytical predictions whereas LPHD requires comparison at the same
scale. Therefore, the direct comparison of a parton level Monte Carlo with
the data makes only sense, if the observable considered is infrared safe, i.e.
does not depend essentially on the cut-off. This is not the case for the
$\xi$-spectrum which peaks in leading order at $\xi^*=Y/2\sim \ln(E/Q_0)/2$.

\noindent{\it Limiting behaviour for soft momenta}\\
Another consequence of soft gluon coherence is the behaviour of the spectrum
at small momenta. 
The coherence of the soft gluon emission from all harder partons
forbids the multiplication of the soft particles and one 
expects nearly an energy independent behaviour of the soft particle
rate \cite{adkt1}. Such a property has been pointed out to be present
indeed in the data
\cite{vakcar,lo}.

This problem has been studied recently in more detail
\cite{klo2}.
The analytical calculations both within the DLA and 
the MLLA have been shown to yield
the same limiting behaviour of the spectrum $D(\xi)$ 
for small particle energies, i.e. at  $\xi\approx Y$
($E\approx Q_0$),
independent of the $cms$ energy. 
In this limit the
energy conservation effects and large $z$ corrections from the splitting
functions which make up the differences between the DLA and MLLA
can be neglected. 
If the LPHD keeps some relevance towards
these low energies one expects the same to be true for the low momentum
hadrons. In the quantitative analysis it is convenient to work with the
invariant density $E \frac{dn}{d^3p}$ (= $E \frac{dn}{dy d^2p_T}$
with rapidity $y$ and transverse momentum $p_T$) and then one expects 
the energy independence of     
the invariant density $I_0$
of hadrons in the soft limit 
\begin{equation}
I_0 = \lim_{y \to 0, p_T \to 0} E \frac{dn}{d^3p} \quad = \quad
\frac{1}{2} \lim_{p \to 0} E \frac{dn}{d^3p}.
\labl{izero}
\end{equation}
The factor $\frac{1}{2}$ in this definition takes into account that both
hemispheres are included in the limit $p \to 0$. This energy independence
is a direct consequence of the coherence of the gluon
emission. Namely, the emission rate for the gluon of large wavelength
does not depend on the details of the jet evolution at smaller distances;
it is essentially determined by the colour charge of the hard initial
partons and is energy independent. The energy independent contribution comes
from the single gluon bremsstrahlung of order $\alpha_s$,
the higher order contributions 
generate the energy dependence but do not contribute in the
soft limit.

In Fig.~3.16 we show the experimental results on the invariant density
of charged particles
for $cms$ energies from 3 to 130 GeV in $e^+e^-$ annihilation.
An approximate energy independence of the soft limit (within about 20\%) is
indeed observed; the same is true for identified particles $\pi$,
$K$ and $p$ in the range from 1.6 to 91 GeV
\cite{klo2}.
It will be interesting to find out to which extent the residual energy
dependence between the lower energies and the LEP data is due to the known
incoherent sources, such as electromagnetic or weak decays (for example from
K$^0$'s or heavy quarks), or  instrumental systematic effects. In any case, 
the observed approximate 
energy independence of the soft particle production over two
orders of magnitude down to such a low energy is remarkable in its own right.
This result has to be compared with the overall rise of the rapidity plateau
$dn/dy$ at $y=0$ after $p_T$ integration.
Within a parton model description this
phenomenon can only be explained if the cut-off
scale is well below 1 GeV.   

The curves in Fig.~3.16 represent the MLLA results derived in a low energy
$E$ approximation. The different parton and hadron kinematics is taken into
account again as in (\ref{dual}) which implies
$E_h \frac{dn}{d^3p_h} = K^h D_A^g(\xi_E,Y)/4\pi (E_h^2-Q^2_0)$. 
This simple ansatz for
extrapolation of the predictions towards low particle energies is not unique
(see also \cite{klo3}), but the main prediction of the energy independence
of $I_0$ in (\ref{izero}) is not affected by any assumption of that type.  
The theoretical curves show the approach to the scaling limit and 
describe well the different slopes at larger particle energies. 
Except for the very small energies they are very close to the results from
the limiting spectrum \cite{klo2,sl}.

If the interpretation of the data in terms of the  underlying QCD process is
really correct, the soft particle production rate $I_0$ should depend on the 
colour charges $C_A$ of the primary hard partons.
For example, in  two
jet $gg$ events the low momentum particle rate should be larger 
than in $q\overline{q}$ events by the factor $9/4$. Realistic tests of this
type have been proposed in \cite{klo2} for several processes:
$e^+e^- \to$ 3 jets (study of the particle production perpendicular to the
event plane as a function of the relative jet angles), deep inelastic
scattering (study of $I_0(y)=\frac{dn}{d^2p_T}|_{p_T=0}$ as a function of
rapidity $y$ in $\gamma_V\gamma$ or $\gamma_Vp$ processes) and semihard
processes with gluon exchange. The rate $I_0$ 
in $e^+e^-$ annihilation can then be used as a standard
scale in comparison of various processes. In this way the soft particle
production could become a sensitive probe of the underlying partonic
processes and the contributions from the incoherent sources.                     

\noindent{\it Comparison of fixed and running $\alpha_s$ predictions}\\
Finally, let us  discuss  the sensitivity of the spectra to
the running of $\alpha_s$ \cite{lo}. The exact results for the 
 moments in MLLA at fixed $\alpha_s$
are compared in Fig.\ 3.9 with the data. Only the multiplicity
and the slope of the first moment could be fitted to the data choosing 
$\gamma_0=0.64$. The difference between both models
develops already at low energies where the
effect of the running coupling is largest. At high energies the two models
predict the different behaviour of Eqs. (\ref{norm:fix}-\ref{momfa1})
and (\ref{2.72a}-\ref{2.72c}). If the moments were not normalized at
threshold but choosen freely at some higher energy point, 
the fixed coupling model
would only work for some rather limited energy interval \cite{lo1}. 

To demonstrate
the difference between both models in the spectrum shape we show in
Fig.\ 3.17 the respective predictions for the Lorentz invariant
distributions
together with data at the low energy of 
$\sqrt{s}$ = 3 GeV, where the differences are expected to be largest. 
The fixed and running coupling predictions are obtained
using the exact result (\ref{fixed:solution}) and the distorted Gaussian
with the moments of Fig.\ 3.9 respectively. As for the decreasing particle
energy $E$ also the typical particle $k_\perp$ is necessarily decreasing,
the coupling $\alpha_s(k_\perp / \Lambda)$ is  increasing.
This yields the steeper slope in comparison to the fixed coupling model. 
Thus the running of the QCD coupling is visible in the particle spectra
even in the energy range of a
few hundred MeV  and this may be taken as further support
for LPHD. 

One may consider \cite{lo} some correspondence between the increasing
$\alpha_s$  towards small scales and resonance production which both yield
configurations with collimated particles in the final state. This view is
supported by the observation \cite{fb}, that a fixed coupling parton
evolution followed by string hadronization (based on the JETSET Monte Carlo
\cite{JETSET}) 
can fit the data throughout the PETRA/PEP energy range.
This indicates that hadron and resonance production in the final stage of
the evolution has a similar effect as the running of $\alpha_s$ at small
scales.

In conclusion, the agreement of the QCD-MLLA predictions based on coherent
shower evolution and running coupling with the momentum spectra is
nontrivial and constitutes in our view a strong support of the LPHD concept.

\subsubsection{Comparison of  Quark and Gluon Jets}

Many experimental studies of the gluon-quark differences have
been done in the period 1983-1995.  The measurements carried out in $e^+
e^-$ collisions below $Z^0$ (see e.g.\ \cite{jhmta} and
references therein) were not very encouraging.  Only qualitative
indications were reported for a slight difference between the
profiles of quark and gluon jets, and the measured value of
$\frac{N_g}{N_q}$ for charged particles was consistent with
unity.  However, in these studies the quark and gluon jets
to be compared often have different energies or belong to different
event types (2-jet, 3-jet), so their interpretation was quite
ambiguous, see for discussion Ref.\ \cite{jwg}.

Bearing in mind all the real-life circumstances ($q\overline{q}g$
environment, jet definition, heavy flavour effects, non-leading
and hadronization corrections, finite particle mass effects,
etc.) from the very beginning this result has not seemed a big
surprise for both the Monte Carlo community (e.g.\ \cite{ts4})
and the APA practitioners, e.g.\ \cite{dkt1}.  However, the QCD
theorists-purists (who have somehow become accustomed to
immediate successes of the asymptotic predictions in the \lq\lq
down-to-earth" studies) have taken this news as a quite unwelcome
fact.  Sometimes the previously observed small difference in
quark and gluon jet properties was even considered as the main
puzzle of perturbative QCD.  This has been a matter of hot debate 
for quite a long time.

\noindent{\it Methods of jet identification}\\
Luckily since then, with massive statistics, the
experimentalists have drastically improved both the quark tagging
methods and the procedures of the analysis and the recent
development looks quite optimistic, see e.g.\ reviews
\cite{ms,al1,my1}.

OPAL has firstly applied the heavy quark tagging method to study
the differences between quark and gluon jets
\cite{opal8,opal9,opal10}.  This is done in a model independent
way by comparing gluon jets in events with identified quarks to a
mixture of quark and gluon jets in a sample of 3-jet events
containing all flavours.  Detailed studies were performed also by
ALEPH \cite{aleph4,aleph3j,alephsj,aleph5} and DELPHI \cite{delphi4}.
Recently comparisons of separated
$b$ and light quark jets to gluon jets
became available as well \cite{opal11,aleph4,delphi4}.

Convincing experimental results concerning differences 
between quark and gluon jets have mainly come from the 
recent investigations at LEP-1. One method consists of studying  
the so-called $Y$-shaped symmetric 3-jet configuration where two 
lower energy jets 2,3 recoil against a high energy jet 1, $E_1 > E_2 \approx 
E_3$.  With an opening angle of $150^o$ between 
the high energy jet and the two low energy ones the high energy jet is a
quark jet in about $97\%$ of all cases.  Without further tagging
the sample of all low energy jets contains about equal amounts of
quark and gluon jets with comparable energy $E\approx 24$ GeV.  
Using lifetime
information or a lepton-tag to identify the quark jets allows one to
select an unbiased gluon-jet enriched sample without changing the
kinematical properties of the jets\footnote{As far as we are
aware, such a procedure was first discussed in Ref.\
\cite{adkt2}.}.  From those two samples which have different
quark-gluon compositions the properties of quarks and gluons can
be unfolded.  A similar method selects ``Mercedes Star" events 
\cite{delphi4} which yields gluon jets of energy $E\approx 30$ GeV.
In this case both quark jets have to be identified.

A new analysis using all 3-jet events at the $Z^0$ 
in comparison with the $q\bar q \gamma$ events has been
performed recently by DELPHI \cite{delphi4} which allows 
the comparison of gluon and quark jet properties over a range of energies.
More information on gluon jets is obtained
from the studies of subjet multiplicities (see subsection 4.2) and particle
flows in multi-jet events (see subsection 6.3.3).

\noindent{\it Multiplicity ratio}\\ 
First we discuss the ratio $r=N_g/N_q$ of gluon and quark jet 
multiplicities. This ratio has been determined by three groups 
from the symmetric, Y-shaped events. The results 
for charged particles are significantly different 
from unity, 
\begin{equation}
\left ( \frac{N_g}{N_q} \right )_{\langle E_j \rangle \: = \: 24
\: {\rm GeV}}^{ch} \; \simeq \; 1.2\; -\; 1.3,
\labl{3.8}
\end{equation}
see Table 2 and the open symbols in Fig. 3.18. The energy dependence of this
ratio has been obtained from the comparison of $q\bar q \gamma$ and $ q
\bar q g$ events. Note that these results refer to  different quark flavour 
compositions (``$\gamma$-flavour mix") from the above for symmetric events
which refer to the standard 
flavour composition at the $Z^0$ (``$Z^0$-flavour mix"), and
the difference is about 0.5 units (see also Fig. 3.18). One observes a clear
energy dependence of this ratio with positive slope \cite{delphi4}
\begin{equation}
\Delta r/ \Delta E = (86\pm 29\; (stat.) \pm 14\; (syst.)) \times 10^{-4}
\hbox{GeV}^{-1}.
\labl{drde}
\end{equation}

\begin{table}[ht]
\begin{center}
\begin{tabular}{|l|l|l|c|} \hline 
Experiment      & $r~\pm~ stat.~\pm~ syst.$ & MC-parton & MC-hadron \\ \hline
ALEPH \cite{aleph3j}  & 1.19 $\pm$ 0.04 $\pm$ 0.002   & 1.29 (J) & 1.28 \\
       &                         & 1.24 (H) &  \\ \hline
DELPHI \cite{delphi4} & 1.279 $\pm$ 0.021 $\pm$ 0.020 & 1.4 (J) & 1.3 \\ \hline
OPAL \cite{opal9,opal10}
   & 1.25 $\pm$ 0.02 $\pm$ 0.03 & 1.29 (J)$^{(*)}$ & 1.26\\
       &                           & 1.19 (H)$^{(*)}$ & 1.21 \\ \hline
\end{tabular}
\caption{Experimental results on the charged particle 
multiplicity ratio $r=N_g/N_q$   
obtained from symmetric, Y-shaped 
 3-jet events in comparison with Monte Carlo (MC)
predictions from JETSET (J) \protect\cite{JETSET} and HERWIG (H)
\protect\cite{HERWIG} performed
at parton and hadron level (data$^{(*)}$ refer to
ratio
$r'=(N_g -1)/(N_q-1) > r)$.}
\end{center}
\end{table}

As already emphasized these results depend on the quark flavours. 
The charged particle multiplicity ratio when selecting 
 $b$-quark jets is close to unity
\cite{opal11,aleph3j,delphi4} whereas the same ratio  
using the light (uds) quark jet sample  is larger
than that observed between the gluon jet and a normal mixture of quark jets 
\cite{opal11,delphi4}
\begin{equation}
\left ( \frac{N_g}{N_{\rm uds \: quark}} \right )_{\langle E_j
\rangle \: = \: 24 \: {\rm GeV}}^{ch} \; \simeq \; 1.3\; - \; 1.4.
\labl{3.9}
\end{equation}

The reported results for the multiplicity
ratios depend on the jet finding algorithms, see e.g.\ \cite{jwg,opal10}.
 Thus, the values (\ref{3.8}), (\ref{3.9}) correspond to the
analysis based on the $k_\perp$ jet finder \cite{rep,cdotw,do}. 
The ratio $\left (\frac{N_g}{N_q} \right )^{ch}$ is found to be essentially
independent on the $y_{\rm cut}$ value.  The other results on
quark-gluon differences are also only modestly affected by the $y_{\rm
cut}$ choice.  Employing a cone definition \cite{jeh} for the
jets with a cone size of $R = 30^o$ leads to a lower value 
$\left ( \frac{N_g}{N_q} \right )^{ch}=1.10$ which, however, becomes
larger as $R$ increases.  This behaviour is expected as the multiplicities
depend on the virtuality $Q\approx ER$ 
rather than on the energy alone;
it is also in  agreement with the general
perturbative expectation \cite{dkt10,dkt11,dkmt2} that the memory
of the colour charge of the parent parton weakens as the
aperture of the registered cone decreases.

Let us now discuss these results in comparison with the theoretical
expectations.
Asymptotically the multiplicity of an isolated gluon initiated
jet should be  larger
than the multiplicity of a quark jet by a factor 
$r= \frac{N_C}{C_F} = 2.25$,
see subsection 3.2.1.  For
the energy range of interest here one expects $r\approx 2$ in MLLA  and 
$r\approx 1.8$ in next-to-MLLA  (see Eq. (\ref{rqg})). 
 The reader is reminded that this analytical result
corresponds to the ratio of the mean number of particles radiated
from a point-like colourless $gg$ system to that produced from a
colour-singlet $q\overline{q}$ point source under the same
circumstances.  Therefore it is entirely inclusive, not
requiring the selection of jet topology or the identification of
jets within an event.

The recent experimental findings clearly confirm the anticipated
differences in multiplicity between quark and gluon jets. 
The observed values in Fig. 3.18 are, however, considerably smaller than
these analytical results.
Let us recall that in 3-jet events soft hadrons cannot 
in principle be assigned
to a given jet and the experimental
definitions of jets normally do not exactly match those of
theory, see e.g.\ \cite{dkt1,dkt4,dkt10,jwg}.  
Perturbative calculations which match directly the experimental conditions 
have been performed using the JETSET \cite{JETSET} or HERWIG \cite{HERWIG}
Monte Carlo at the parton level.
These results are also presented in Table 2. As can be seen, these
perturbative predictions are considerably smaller than the fully inclusive
next-to-MLLA results, and are close to both the actual experimental findings
and to the results of the hadronization models quoted in the 
table.\footnote{A larger hadronization correction has been reported 
\cite{jwg} for the final $gg$ state. Part of this  effect may come from
using the larger ratio $(N_g-1)/(N_q-1)$ on the parton level.}
We, therefore, conclude that the multiplicity difference of both types of jets
can be largely explained by the perturbative approach. Also the flavour
dependence of the multiplicity ratio follows the old expectations
(see e.g. \cite{dkt1}).

In order to bring the analytical predictions into agreement with experiment,
it seems one has to include higher order terms in the $\sqrt{\alpha_s}$
expansion (\ref{rqg}), as is done, for example, 
for the inclusive particle spectrum, and/or to take into account the
influence of the nearby jets in the event using methods discussed in section
6. Alternatively, one may try to approach in the experiment the conditions
which are assumed in the theoretical analysis (see below in this subsection).

\noindent{\it Particle spectra}\\
Besides the particle multiplicities the experimental groups have 
also presented
detailed results on the inclusive distributions of fractional momenta and
angles. No comparison with perturbative predictions has been carried out 
unlike the
 case of particle multiplicity. However, one can say that
the differences between quark and gluon jets
 prove to be in a qualitative agreement with the perturbative
expectations (see subsection 3.2.2)
and it is well described quantitatively by the
WIG'ged Monte Carlo models.  In particular, one observes the following
\cite{opal10,aleph3j,delphi4}:
\begin{itemize}
\item[(i)] Being consistent with the higher multiplicity, the
fragmentation function of gluon jets is found to be markedly
softer than for quark jets; for high momenta the multiplicity in a gluon jet
is about one order of magnitude smaller than in the quark jet;
\item[(ii)] Studying the energy and multiplicity fractions
contained in fixed angular cones around the jet axis one finds
that gluon jets are indeed substantially less collimated than
quark jets of equivalent energy;
\item[(iii)] Corresponding differences have been observed in the rapidity
distributions and in some topological observables like ``jet broadness".
\end{itemize}

\noindent{\it Future measurements of gluon jet properties}\\
To bring the experimental conditions for the studies of gluon jets
closer to the theoretical ones the following possibilities have been
proposed.

One could sharpen the
manifestation of an individual gluon jet in the 3-jet ensemble 
\cite{dkt11,dkt4} (see also
\cite{vak1,jwg}) by studying  the $q\overline{q}g$ events in which
the hard gluon jet is observed in the hemisphere recoiling
against (preferably quasi-collinear) $q$ and $\overline{q}$-jets.
 The colour structure of such events resembles that of the $gg$
events produced from a point-like colour singlet source, see also
section 6.  Since the $q\overline{q}g$ events with the identified
$q$ and $\overline{q}$ jets in the same hemisphere are quite
rare, this proposal requires massive statistics such as those
available at the $Z^0$ peak.  The detailed examination 
in Ref.\ \cite{jwg} has demonstrated the promising prospects of
such a technique for testing the differences between quark and
gluon jets in the analysis of LEP-1 data.  This would allow us to
perform inclusively the experimental selection of the gluon jet
and the value of the observed ratio $\left ( \frac{N_g}{N_q}
\right )_{\langle E_j \rangle \: = \: 44 \: {\rm GeV}}$ obtained
in this manner is expected to be on the level around 1.4.

As was pointed out in Ref.\ \cite{vak1}, a prospective
way of studying experimentally the gluon jet properties
could arise from the process $\gamma\gamma \rightarrow gg$. 
This
reaction provides a unique environment in which two isolated
high
energy gluon jets are produced in a colour singlet state (the
only counterpart to the celebrated process $e^+ e^-
\rightarrow
q\overline{q}$)\footnote{One can often meet in the literature
a statement that the properties of a gluon jet could be well
studied at hadronic colliders by separating the contribution
of the $gg \rightarrow gg$ subprocess.  This argument may be quite
misleading since at all realistic energies the overall
structure of the final state is strongly affected by the whole
complex of interjet collective effects, see e.g.\
\cite{dkmt2,dkt8,emw}.  In particular, the colour structure of
the $gg$ events does not correspond to a colour singlet.  For 
experimental studies of gluon jets in $p\overline{p}$ collisions
see e.g.\ Ref.\ \cite{ua2}.}.  At large angles its cross section
is lower, approximately by an order of magnitude, than the cross
section of background process $\gamma\gamma \rightarrow
q\overline{q}$, see e.g.\ Refs.\ \cite{bkf,bfkk}.  This
background could be substantially reduced by using the polarized
$\gamma\gamma$ facility (which is \lq\lq for free" at future
linear $e^+ e^-$ colliders) to ensure that the colliding photons
have the same helicities.

\noindent{\it Hadronization phenomena}\\
It is clear that there are hadronization phenomena not 
quantitatively accessible within the
perturbative approach to particle production that we discuss in this review.
Nevertheless, we mention here some
further aspects of quark and gluon jets which could be important for our
overall understanding of the phenomena.     

In the fragmentation region $(x_p \sim 1)$ some qualitative
differences between quark and gluon jets are anticipated,
see e.g.\ \cite{adk,pw,im}.  In particular, one might expect
glueball production or at least a surplus of isosinglet hadrons
$(\eta, \eta^\prime \ldots)$ in the gluon fragmentation.

Some important news has been reported by the L3 Collaboration
\cite{l33} which has observed enhanced $\eta$ production at
large $x_p$ in gluon jets compared with quark jets.  This
enhancement increases with $x_p$ and it is not reproduced
within the existing Monte Carlo models.  At the same time no
anomaly is observed in the $\pi^0$ distribution.  
It could well be that the long awaited
\cite{adk,pw,im} specific manifestation of the gluon jet
fragmentation has been finally found.  If so, even more
spectacular discrepancies could be expected for the $\eta^\prime$.

There also appears to be some problems with strangeness 
production (mainly $K$ and $\Lambda$), see \cite{ts1} for further 
details.  A deficit of strange particles seems to be seen in the 
extreme two-jet events of $e^+ e^-$ annihilation.  There are some 
other indications for the possible breakdown of jet universality 
from HERA data.  One may start worrying whether these indications 
are related to the specific properties of the hadronization of the 
gluon jet. 
No doubt, this area should be watched 
closely in the future.

As indicated in Ref.\ \cite{cdgkr} the $f_0 (975)$ and $a_0
(980)$ mesons could play a special role in the dynamics of quark
confinement, and the studying of their production in quark and gluon
jets may allow some important phenomenological tests of Gribov's
theory \cite{vng1}.  One possibility is to measure the yield of
$f_0$ production at large $x_p$ in quark
jets \cite{cdgkr}.


\section{Particle and Jet Multiplicities}
\setcounter{equation}{0}
\subsection{Distribution of Particle Multiplicities}

We turn now to the more complex characteristics of the multiparticle
final state. Let us consider first the multiplicity distribution $P_n$
or the moments of this distribution which are the integrals of the
respective multiparticle correlation functions (see section~2).~

\subsubsection{KNO-Scaling and its violation}

Analytical results can be obtained from the evolution equations in MLLA
(\ref{2.23}) or DLA (\ref{dlaz}) of the multiplicity generating
function $Z(\tau,u)$ in (\ref{zmult}). This yields a coupled system of
equations for $Z_g$ and $Z_q$. Important features of the solutions
can already be obtained in ``gluodynamics'', i.e.\ when quarks are
neglected. Then the MLLA evolution equation in $\tau =\ln (Q/\Lambda)$
reads
\begin{equation}
Z'_g(\tau,u)=\int\limits^1_0 dx \gamma^2_0 (k_\perp)P(x)[Z_g(\tau+\ln x,u)Z_g
(\tau +\ln (1-x),u)-Z_g(\tau,u)]\Theta (k_\perp^2-Q_0^2).  \labl{zgev}
\end{equation}
Here $k_\perp\approx x(1-x)Q$ and $P(x)$ represents the gluon splitting
function $\Phi^g_g(x)$. Due to the symmetry of
(\ref{zgev}) one can also
write
\begin{equation}
P(x)=(1-x)\Phi^g_g (x)=x^{-1}-(1-x)[2-x(1-x)]\/.\labl{pxgg}
\end{equation}

The dominant contribution comes from $x<< 1$ and
one obtains the DLA equation \cite{dfk2}
\begin{equation}
(\ln Z_A(\tau))'=\frac{C_A}{N_C}\int\limits^\tau d\tau'
\gamma_0^2(\tau')[Z_g(\tau')-1] \labl{zerdla}\/.
\end{equation}
In this limit one can derive \cite{bcm1} a scaling property for
the multiplicity moments or the probabilities $P_n$ in the
rescaled multiplicity $n/\bar n$, the famous ``KNO-scaling''
\cite{kno,amp}
\begin{equation}
\bar n(\tau)P_n(\tau) =f(n/\bar n (\tau))   \labl{knosc}
\end{equation}
independent of energy variable $\tau$. The normalized factorial moments
(\ref{momnor}) for the gluon jet fulfil the recursion
relation
\begin{equation}
F_g^{(q)}=\frac{q^2}{2(q^2-1)}\sum^{q-1}_{m=1}\left({q\atop m}\right)
\frac{F^{(m)}_g F_g^{(q-m)}}{m(q-m)},\qquad F_g^{(1)}=1\/, \labl{frecur}
\end{equation}
e.g.\ $F_g^{(2)}=\frac{4}{3}$. For a general superposition of $N_q$
and $N_g$ quark and gluon jets respectively the generating function
is rescaled according to $Z_g\to Z^\rho_g$ with $\rho=\frac{C_q}{N_C}
N_q+N_g$
or the cumulant moments (\ref{cmom}) $c_g\to \rho c_g$, e.g.
$F^{(2)}_{q\bar q}=\frac{11}{8}$. Remarkably, these normalized
moments are pure numbers, i.e.\ they are infrared safe and even
independent of the coupling. For large $q$ one finds \cite{dfk2}
$F^{(q)}_g\approx 2qq!/C^q$ where $C\approx 2.5527$. This
corresponds with a good accuracy
to the asymptotic KNO distribution \cite{yld}
\begin{eqnarray}
f(x) &=&\frac{2}{x}\; \frac{(Cx)^2}{1+(Cx)^{-1}}\; \exp(-Cx)\;{\cal L}(x)\/,
     \nonumber\\
     &&{\cal L}(x)=\exp\:(-\frac{1}{2}\ln^2[1+(Bx)^{-1}])\labl{knoas}
\end{eqnarray}
with $B\approx 2.04$. 

The approach to this asymptotic limit is very slow, as
the recoil effects play an important role at present energies.
This has been studied analytically and numerically
\cite{yld,ct,og}. As an example we show in Fig.~4.1 a comparison
at LEP-1 energies of the DLA
limit (\ref{knoas}) with an analytical formula \cite{yld} 
obtained from the
equation (\ref{zgev}) in gluodynamics, keeping track of the
specific next-to-MLLA effects responsible for energy-momentum
conservation, with some further
simplifications $(\alpha_s$ fixed, $P(x)\approx x^{-1})$. 
As can be seen, the
distribution at finite energies is much narrower. Whereas the high
multiplicity tail in DLA behaves like $f(x)\sim \exp (-Cx)$, see
Eq.~(\ref{knoas}), one obtains at finite energies
the much faster decrease \cite{yld}
\begin{equation}
f(x)\sim exp (-[Dx]^\mu),\quad \mu=(1-\gamma)^{-1}>1,\quad
D\approx C(2\gamma/\pi)^\gamma\/,\labl{knofe}
\end{equation}
with the anomalous dimension $\gamma \sim\sqrt{\alpha_s}$
($\gamma\approx 0.40$ at LEP-1).
The experimental data confirm the approximate
scaling behaviour in a large energy region (see, e.g.
\cite{delphi6,aleph8}). The broadening of the KNO distribution
expected from the above considerations would become clearly visible
when increasing the $e^+e^-$ energy from $\sqrt{s}=90$~GeV to
$\sqrt{s}=2000$~GeV as suggested in
a Monte Carlo (JETSET) \cite{JETSET} study \cite{ugl}.

\subsubsection{Multiplicity Moments in Next-to-Leading Order}

A convenient way to analyse multiplicity distributions is in terms of
their moments (see section~2).
Analytical results have been derived for the realistic cases of gluon and
quark jets (and the application to
$e^+e^-$ collisions) by solving the coupled equations for the
generating functions $Z_g,Z_q$ in next-to-leading order \cite{dt,mw}.
For the factorial moments one finds
\begin{equation}
F^{(q)}_A(Q^2)\sim a^{(q)}_A[1+(b^{(q)}_A+c^{(q)}_A n_f)
\sqrt{\alpha_s (Q^2)}]\labl{fqmw}
\end{equation}
where the constants $a^{(q)}_A,b^{(q)}_A$  and $c^{(q)}_A$
have been calculated for various processes. For example, one obtains
\begin{equation} 
F^{(2)}_{e^+e^-}=\frac{11}{8} (1\; - \;
\frac{4255}{1782\sqrt{6\pi}} \; \sqrt{\alpha_s})
\labl{f2eplm}
\end{equation}
 for $n_f=5$.
This prediction is compared to experimental data in Fig.~4.2. It is
found to be 
smaller than the asymptotic result 
$F^{(2)}_{e^+e^-}=\frac{11}{8}$ by $\sim$30\%
but still differs from the data by $\sim$10\%. 
The scale parameter $\Lambda$
in the coupling $\alpha_s(Q^2)$ in (\ref{fqmw}) 
has been adjusted to fit the rise of the 
multiplicity $\bar n$ in the same approximation 
and energy region and cannot be found to fit
both $F^{(2)}$ and $\bar n$ at the same time \cite{aleph8}.

A good agreement with the data is finally achieved 
\cite{wo} by calculating
this moment with the Monte Carlo method (using HERWIG \cite{HERWIG}
at the parton
level\footnote{For the perturbative cascade without the nonperturbative
$g\to q\bar q$ splitting at the end and with
parameters $\Lambda =0.15 GeV$,
$m_q=m_g=0.32$~GeV.}) which fully takes into account the energy-momentum
conservation constraints. At this level LPHD is verified for the
normalized moments.

These moments have another interesting property (at least up to order
$q=4$): they obey approximately the recursion relation
\begin{equation}
F^{(q+1)}=F^{(q)}(1+\frac{q}{k})\/,\qquad F^{(1)}=1\/, \labl{nbrel}
\end{equation}
characteristic for the phenomenologically successful negative
binomial distribution with parameter $k$.

This distribution -- originally derived for a gluon jet as a coherent
superposition of $q$ Bose-Einstein sources
\cite{ag1} -- fits reasonably well the multiplicity
distribution in full phase space and also in restricted rapidity
intervals of various processes. Recently the more precise measurements
at LEP-1 
and SLD \cite{delphim,opalm,alephm,sld} have shown a small shoulder in the
multiplicity distribution in deviation from the simple negative
binomial distribution which will be discussed further below.

\subsubsection{Multiplicity Moments in Rapidity Windows}
The comparison of multiplicity distributions of partons and hadrons
for different central rapidity intervals from the JETSET Monte Carlo
\cite{JETSET}
has revealed an interesting regularity \cite{gvh,ggsh}. Good fits have
been obtained with the negative binomial distributions 
for the multiplicities of both partons and hadrons.
The
parameters $\bar n$ and $k$ obtained at parton and hadron level are connected
by the formulae:
\begin{equation}
k_h\approx k_p\/,\quad \bar n_h=\rho n_p,\quad \rho \approx 2 \labl{glphd}
\end{equation}
The first relation also implies $(F^{(q)})_p\approx (F^{(q)})_h$
because of (\ref{nbrel}), consistent with the conclusions from
Fig. 4.2, but now it is
also valid in restricted rapidity intervals $\Delta y$.
The results (\ref{glphd}) can be obtained if one assumes in rapidity space
$D^{(q)}_h(y_1,\ldots, y_n)=\rho^q D^{(q)}_p (y_1,\ldots,y_n)$
(``generalized local parton hadron duality'' \cite{gvh}).
We note that $k$ or $F^{(q)}$ are infrared safe quantities but
$\bar n$ in (\ref{glphd}) is not and the factor $\rho$ depends on the
cut-off $Q_0$.

\subsubsection{Oscillating Cumulant Moments }
New insights into the correlations of higher order have emerged from
the recent studies of the generating function evolution equations
in which the contributions beyond MLLA 
are taken into
account  \cite{yld,imd}. The
importance of going beyond this approximation of \cite{mw} originates from
the observation that the relevant expansion
parameter is not the anomalous dimension $\gamma_0\sim\sqrt{\alpha_s}$
itself but the combination $q\gamma_0$. Therefore, at finite
energies higher order terms $(q\gamma_0)^n$ are not necessarily small
and their role has to be investigated properly.

The main consequences for the multiplicity moments is
that these higher order terms generate
negative cumulant moments $C^{(q)}$ with a minimum in the
ratio \cite{imd}
\begin{equation}
H^{(q)} =\frac{C^{(q)}}{F^{(q)}}\/,\labl{hqdef}
\end{equation}
followed by oscillations \cite{dn,dh},
whereas the factorial moments $F^{(q)} $ are rising rather smoothly with
$q$,
see also the review \cite{imd1}. Such
oscillations are actually found in the data \cite{dab,sld} although
with smaller amplitude than predicted by the QCD calculations.

The main qualitative features of these results can be
understood in gluodynamics starting from the evolution equation
(\ref{zgev}) \cite{imd}. One replaces in the nonsingular parts of the
integrand in this equation
$Z(\tau +a)$ by $Z(\tau)+a Z'(\tau)$ and neglects terms
$\sim (\ln Z)' Z'$ of yet higher order.\\
Then one arrives at the equation
\begin{equation}
(\ln Z)'' = \gamma_0^2(Z-1-2 h_1Z'+h_2Z'')  \labl{lnzeq}
\end{equation}
with the derivatives with respect to $\tau$ and 
$h_1=11/24 = b/8N_C$, $ h_2=(67 - 6\pi^2)/36$ $\approx$ 0.216.
The $q$-fold differentiation with respect to
the second variable $u$ (c.f.\ Eqs.~~(\ref{fmom}), (\ref{cmom}))
yields an equation between a cumulant moment on the $l.h.s.$ 
and factorial moments on the $r.h.s.$ of (\ref{lnzeq}).
With the KNO ansatz $f^{(q)}=F^{(q)}\bar n^q$ and $c^{(q)}=C^{(q)}
\bar n^q$ and introducing the anomalous dimension
$\gamma$ according to
\begin{equation}
\bar n(\tau)=\exp\int\nolimits^\tau d\xi\gamma (\xi)  \labl{anom}
\end{equation}
one obtains for $H^{(q)}$ in (\ref{hqdef})
\begin{equation}
H^{(q)}=\frac{\gamma_0^2 [1-2h_1q\gamma +h_2(q^2\gamma^2+q\gamma')]}
{q^2\gamma^2 +q\gamma'}. \labl{hq1}
\end{equation}

From this equation for $q=1$ an expansion of $\gamma$
in powers of $\gamma_0$ can be found. The result (\ref{hq1})
clearly shows the leading dependence of the higher moments on
$q\gamma\approx q\gamma_0$ 
(whereas $q\gamma'\simeq qh_1\gamma_0^3\ll q\gamma_0$).
One obtains again the DLA result $H^{(q)}\simeq q^{-2}$
-- equivalent to relation (\ref{frecur}) -- for
$\gamma^2\approx\gamma_0^2\;\to\;0$. The MLLA results \cite{mw} are found
for $h_2=0$, if only  terms of relative order $\gamma_0$ are kept; this
yields $H^{(q)}\lapproxeq 0$ for larger $q$. With all terms included
one finds for
large~$q$
\begin{equation}
H^{(q)}=h_2\gamma_0^2 +{1\over q^2} 
[1-(2q-1)h_1\gamma_0-q(h_1\gamma_0)^2]\/,
\labl{hqlq}
\end{equation}
which develops a minimum with
\begin{eqnarray}
q_{min} &=&\frac{1}{h_1\gamma_0}+\frac{1}{2}\approx 5\nonumber\\
H^{(q)}_{min} &=&(h_2-h^2_1)\gamma^2_0+O(\gamma_0^4)\;\approx\; 1.4\times
10^{-3}\/.\labl{qmin}
\end{eqnarray}
and reaches $H^{(q)}=h_2\gamma^2_0\approx 0.05$ for large $q$
(using $\gamma_0\approx 0.48$). The ratio $H^{(q)}$ is of the order
$10^{-2}$ - $10^{-3}$ for $q$ changing from 3 to 10.
A calculation which takes into account the difference 
between  quark and gluon
jets has been performed
as well \cite{dln} in the same linear approximation for
$Z(\tau+a)$. One finds results qualitatively similar to
(\ref{hqlq}) but with $q_{min}$ shifted to a larger value $q_{min}
\approx 7-8$.

If $Z(\tau +a)$ is expanded to higher orders one obtains a sequence
of oscillations of $H^{(q)}$ \cite{dn} in the order $q$.
The coupled equation for $Z_g$ and $Z_q$ can actually be solved
exactly in the high energy KNO approximation if $\alpha_s$ is kept
constant \cite{dh}. Then one finds again an oscillatory structure.

\subsubsection{Experimental Evidence for Oscillations}
An analysis of existing experimental multiplicity
distributions \cite{dab} shows the predicted trend of the data.
In a dedicated study of moments up to $q\approx 16$ 
the SLD collaboration \cite{sld} concludes
that the steep decrease of $H^{(q)}$ for 
$q\lapproxeq 5$, the negative minimum at
$q\approx 5$ and the quasi-oscillation around
zero for $q\gapproxeq 8$ are
well established features of the data, see Fig.~4.3.
Whereas these qualitative features follow the QCD predictions, the
actual size of the $H^{(q)}$ moments as computed in \cite{dln} 
for the $e^+e^-$ reaction is more than an order of magnitude
 larger than that measured in \cite{sld}.
This discrepancy could be due either to the limited
accuracy of the theoretical calculation or to the influence of
hadronization effects.

Also the effective cut-off of high multiplicities may induce an
oscillatory behaviour \cite{ugl1}, although
with smaller amplitude and not just as observed
\cite{sld}. 

In another approach \cite{glu2} the oscillations of the
cumulant moments are related to the shoulder observed in an
intermediate multiplicity range \cite{delphim,opalm,alephm}.
This shoulder has been explained by the DELPHI Collaboration
\cite{delphimj} as the result of superposition of multiplicity
distributions coming from events with 2, 3 and 4 jets, corresponding
to a certain jet resolution criterion $y_{\rm cut}$.  
The multiplicity
distributions for each class of events are well reproduced by  negative
binomial distributions. A superposition of two such distributions
with the known fractions of 2 and $\geq$ 3 jet events can
reproduce \cite{glu2} both
the full phase space multiplicity distribution and the
$H^{(q)}$ moments. This would imply that the oscillations in the  cumulant
moments come from large angle hard gluon emissions. A correct description
of such processes also requires contributions of NLO terms and,
therefore,both effects, the shoulder and the
oscillating moments could have the same QCD origin.

In any case, it is interesting that the simple consequence of
such a peculiar higher order QCD effect has actually been observed.
It remains to be seen whether more complete analytical calculations can
provide also a quantitative description of the observed  multiparticle
phenomena.

\subsection{Jet Multiplicities}

By studying the multiparticle final state with 
variable resolution one can
learn about the earlier stages of the jet evolution. Looking at an $e^+e^-$
event with  
low resolution one observes only a few objects (jets). The rate
and the distributions of these jets can be calculated from the first terms 
of the 
perturbative expansion in absolute normalization. 
The resolution parameter(s) can be introduced
in such a way that  
the perturbative results are collinear and infrared safe. These results are
then directly compared with the experimental data at the same resolution. 
In this application it is implicitly assumed that the hadronization process
does not change the perturbative result. 

If the resolution is increased, more and more jets are resolved, until finally
all individual hadrons are distinguished. In this case it is not sufficient to
consider the first terms of the perturbative expansion but a resummation of the
full series is required to obtain reliable results. Several resolution criteria
and jet finding algorithms have been applied but the perturbative resummation
has not been achieved in all cases. If the  transverse
 momentum separation 
between the two jets is taken as the resolution measure
($k_T$/Durham algorithm, see below), then there is a
matching with the parton cascade results  discussed in the previous section
in the limit of small transverse momentum separation $Q_0\sim 250$ MeV.
It is an interesting question, of down to which
resolution scale the perturbative predictions agree with 
the experimental data at the 
inclusive or exclusive level. As discussed in section 2.3, in the limit when all
hadrons are resolved, the parton hadron duality 
can only be expected for the sufficiently
inclusive quantities. Also in this limit the overall 
normalization of jet (hadron)
rates is treated as an adjustable parameter. 
Therefore, the studies of jet rates
at variable resolution may provide us with
further  information about  the transition from
the perturbative to  the non-perturbative region. 
Of special interest here are the differences between
quark and gluon jets 
and the role of the soft gluon coherence in the
final state.

\subsubsection{Resolution criteria and jet algorithms}
Finite results for the exclusive final states are obtained in 
perturbation
theory by introducing an appropriate resolution criterion. Then the
cross section for all final states which are indistinguishable
according to this criterion are finite. This result was first derived in
QED and is expressed by the  Kinoshita-Lee-Nauenberg
theorem \cite{kln} but it applies also for QCD \cite{sw}
(for a review, see \cite{guk}).
Several resolution criteria have been considered:\\
(a) energy-angle $(\varepsilon,\delta)$ cutoff \cite{sw}\\
Two partons (jets) with fractional energies $x_i$ and relative angles
$\Theta_{kl}$ are counted as distinguishable objects if the following
conditions are fulfilled
\begin{equation}
x_k>\varepsilon,\quad x_l>\varepsilon \qquad {\rm and} \qquad 
\Theta_{kl}>\delta
\label{epsdel}
\end{equation}
for given resolution parameters $\varepsilon$ and $\delta$.
In case of $e^+e^-\to q\bar qg$ the configurations 
satisfying (\ref{epsdel})
correspond to 3-jet events; since the soft and collinear singularities
are excluded the cross section is finite. In the opposite  case
the singularities in the $q\bar qg$ final states cancel against the
singularities from the virtual corrections to the $q\bar q$ final
state and the rate for these indistinguishable configurations is finite
as well.

\noindent(b)
invariant mass cut-off (``JADE-algorithm'' \cite{JADE}).\\
In this case both collinear and soft divergencies are avoided by a
single cut-off 
\begin{equation}
y_{kl}=M^2_{kl}/s > y_c \label{jadecond}
\end{equation} 
where $M^2_{kl}$ and $s$ are the
invariant mass squared of two partons and of the full event respectively.
This resolution criterion has been applied  to the detailed computation
of jet rates in $e^+e^-$ annihilation in ${\cal O}(\alpha_s^2)$
\cite{kl}.

\noindent(c)
relative transverse momentum $k_T$-cutoff 
(``Durham-algorithm''\footnote{This 
algorithm was discussed within the framework of analytical calculations 
at the  Durham Workshop \cite{rep}.
The very idea of $k_T$ clustering has been
already applied since the early eighties, see e.g. 
\cite{sjoclu}.} \cite{bs,cdotw,bkss}).\\
In this case two particles (jets) of energies $E_k, E_l$ and relative angle
$\Theta_{kl}$ are distinguished if
\begin{equation}
y_{kl}=2\; (1-\cos \Theta_{kl})\;  {\rm min} (E_k^2,E_l^2)/s\; > \;y_c.
\label{jade}
\end{equation}

For each resolution criterion a jet finding algorithm can be defined by
an iterative procedure.
For every pair of particles one computes the corresponding distance
$y_{kl}$ defined by the resolution criterion
 (restricting here to the case of a single resolution parameter). 
If the smallest distance
obtained in this way is smaller than 
the resolution parameter $y_c$ the two particles are combined into
a single jet according to a recombination scheme. 
In the simplest prescription (E-scheme) the 4-momenta of
jets are added; alternatively, one may require that either the momenta
or the energies are rescaled in such a way that massless jets are obtained.
This procedure is repeated
until all pairs satisfy $y_{kl} > y_c$. The remaining objects are the jets
at resolution $y_c$.

In this way one can study the configurations of variable resolution between
the two extreme limits, the jet limit for  $y_c \to 1$ 
where the multiplicity of resolved objects 
approaches its minimal value ($N=2$ in $e^+e^-$ annihilation) and 
the particle limit $y_c\to 0$
where the jet final states coincide with the particle final states.

Whereas all cut-off procedures yield finite results for the multi-jet rates
there are some advantages of using the $k_T$ algorithm. In this case a summation
of the leading $\alpha_s^n\ln^{2n} y_c$ and next-to-leading 
$\alpha_s^n\ln^{2n-1} y_c$ terms of the perturbative expansion into an
exponential form is possible which is important for small 
values of the parameters $y_c$
\cite{bs,cdotw,bkss,cdfw,cwdf,do}.
Another interesting feature of the $k_T$ algorithm is the small size of
``hadronization corrections" which are 
obtained from the most popular parton shower
Monte Carlo's \cite{bkss}. In the following we discuss some results using
this algorithm with special attention to the high resolution limit.

\subsubsection{Average Jet Multiplicity}

For low resolution (large parameter $y_c\sim {\cal O}(1)$) one can derive the jet
rates $R_k=\sigma_{k-jets}/\sigma_{tot}$ and the average multiplicity 
$N=\Sigma kR_k$ from the exact matrix element calculations \cite{ert}.
 In $e^+e^-$ annihilation the jet rates have been
calculated for the $k_T$-algorithm \cite{bkss} in two loop order and the 
result has the general form
\begin{eqnarray}
N(y_c,Q)\;=\; 2 &+& \left(\frac{\alpha_s(\mu)}{2\pi}\right) A(y_c) \;+ \;
 \left( \frac{\alpha_s(\mu)}{2\pi} \right)^2 [B(y_c)+2C(y_c)\nonumber\\
  &\;&\qquad -\; (2+b\ln(Q/\mu))A(y_c)] \ + \ {\cal O}(\alpha_s^3)
   \label{jetmul}
\end{eqnarray}
where $\mu$ is the renormalization scale. In this approximation for large
$y_c$ there is no strong dependence on $\mu$ and one sets conveniently
$\mu=Q=\sqrt{s}$. As can be seen from (\ref{jetmul}) the $Q$-dependence enters only 
through the running coupling constant and for fixed coupling the jet
multiplicity would depend only on $y_c$. The coefficient functions include
the expected logarithmic terms with powers of $L=\ln(1/y_c$) which dominate
for small $y_c$; one finds for the leading terms up to ${\cal O}(\alpha_s^3)$
\cite{cdotw,cdfw}
\begin{eqnarray}
N\; = \; 2 & + & C_F \frac{\alpha_s}{2\pi} \;[L^2-3L+{\cal O}(1)]\nonumber\\
  & + & C_F \left(\frac{\alpha_s}{2\pi}\right)^2 \;
   \left[\frac{1}{12} N_CL^4 + \frac{1}{9} 
          N_C-n_f)L^3 +{\cal O}(L^2)\right]\label{npert}\\
 & + & C_F \left(\frac{\alpha_s}{2\pi}\right)^3 \;
     \left[\frac{1}{360} N_C^2L^6 +
          \left(\frac{7}{72}N_C^2 - \frac{1}{30}N_C n_f + \frac{1}{90} C_F
          n_f\right)L^5 +{\cal O}(L^4)\right]\nonumber
\end{eqnarray}
At small $y_c$ the double log terms $\alpha_s^nL^{2n}$ dominate and with
decreasing $y_c$ the full perturbative series has to be summed. The double
log terms in (\ref{npert}) just correspond to the expansion of
\begin{equation} 
N\; = \; 2 \; + \; 2 \;\frac{C_F}{N_C}\; 
     [\cosh \sqrt{\alpha_s N_CL^2 /2\pi} \;-\;
1];
\label{nfal}
\end{equation}
this corresponds to the DLA result for fixed $\alpha_s$ \cite{dfk1}\footnote{
The coupling is kept fixed at a given $Q$ in this calculation if the
resolution is varied.}, which follows from the 
master equation (\ref{dlaz}) 
for
the multiplicity generating functions $Z_A(p,\Theta,u_a)$ with constant
$u_A$,
see also section 3.1.

The result in the next-to-leading accuracy (summing also the terms $\alpha_s^n
L^{2n-1}$) can be obtained by solving 
the corresponding MLLA master equation (\ref{2.23}).
To this end one simplifies the $r.h.s.$ of (\ref{2.23}) and keeps only
${\cal O}(\sqrt{\alpha_s})$ and    ${\cal O}(\alpha_s)$ contributions
\cite{mw,cdotw,cdfw,do}. Note that within the required accuracy one can replace 
$Z_A((1-z)p,\Theta)$ by $Z_A(p,\Theta)$; furthermore, one separates the
singular $1/z$ parts in the splitting functions from the regular parts
and then neglects the $z$-dependence in the arguments of $Z_A$ and
$\alpha_s$ of the regular $z$-integrals. This yields the 
modified master equation for
$Z_A$ within the MLLA accuracy. 
The equations for jet multiplicities in single 
quark and gluon jets at resolution
$Q_0$ and jet virtuality $Q$ are obtained by differentiating 
$Z_A$ over $u_A$ 
\begin{eqnarray}
N_q(Q,Q_0) &=& 1\; + \; \int_{Q_0}^Q \; dq\ \Gamma_q(Q,q)\ N_g(q,Q_0)
              \nonumber\\
N_g(Q,Q_0) &=& 1\; + \; \int_{Q_0}^Q \; dq\ [\Gamma_g(Q,q)\ N_g(q,Q_0)
              \nonumber\\
   & \; & \qquad \qquad 
    +\;\Gamma_f(Q,q)\ (2N_q(q,Q_0)-N_g(q,Q_0))] \label{jetmeq}
\end{eqnarray}
where the initial condition (\ref{2.25}) has been used (at threshold, 
i.e. at $Q=Q_0$ or $y_c=1$, there
is only one particle in the jet). For the single jet the virtuality of the
jet is taken as $Q\approx p\Theta$ or as in (\ref{qscal}) (see, for example,
\cite{do}).
The emission probabilities $\Gamma_i$ in (\ref{jetmeq}) 
are given within MLLA accuracy by
\begin{eqnarray}
\Gamma_q(Q,q)= \frac{C_F}{N_C} \frac{\gamma_0^2(q)}{q} 
\left(\ln\frac{Q}{q} -
       \frac{3}{4}\right),& &\qquad 
\Gamma_g(Q,q)=  \frac{\gamma_0^2(q)}{q} \left(\ln\frac{Q}{q} -
       \frac{11}{12}\right) \nonumber \\
\Gamma_f(q) & = & \frac{n_f T_R \gamma_0^2(q)}{3N_Cq}. \label{gammaq}    
\end{eqnarray} 
The equations (\ref{jetmeq}) can be solved in terms of Bessel functions
\cite{cdfw} and the leading high energy contribution comes from a term as
given in (\ref{2.64}). 

In this approximation the proper limits of the $z$-integration have been
replaced by (0,1) which is consistent with the MLLA accuracy at small $y_c$. 
At $y_c\sim 1$, on the other hand, this approximation leads to unphysical
results with $N<1$. This deficiency can be avoided using the improved master
equation \cite{do} in which the proper integration limits are restored.
Another type of improvement of the perturbative expansion can be achieved
by matching the result with
the matrix element calculations (\ref{jetmul}) including the
nonleading terms up to ${\cal O}(\alpha_s^2)$.

Experimental results on jet multiplicities $N$ in $e^+e^-$ annihilation
have been presented by L3 \cite{L3jmul} and OPAL \cite{opaljmul}.
A comparison of the L3 data on $N-2$ 
with the next-to-leading order QCD prediction, improved by the matching with
the exact matrix element results,
 is shown in Fig.~4.4.
The data on this quantity had been corrected for ``hadronization effects" as
suggested by hadronization models by an amount 
found to be less than 5\%. A very good
agreement of the perturbative prediction 
with the experimental data can be seen both in shape and normalization
for $y_c\gsim 10^{-3}$ (corresponding to $Q_0\gsim 2.8$ GeV) after adjusting
the only parameter $\alpha_s(M_Z)$. 
For yet smaller $y_c$ the data would
rise above the analytical predictions. 
The DLA result (\ref{nfal}) on $N-2$ is still larger by a factor of 2 at
$y_c\sim 0.1$ in comparison to the prediction in Fig.~4.4 but 
both theoretical results approach each other for larger multiplicities at
smaller  $y_c$.

\subsubsection{Exclusive jet cross sections}

The same MLLA master equation for the generating functional
can be used to derive exclusive
jet cross sections
by differentiation over the probing function $u$ at $u = 0$. 
In the simplest case of the
exclusive channel ($N = 1$) one obtains the  Sudakov
form factor for quark and gluon jets
\begin{eqnarray}
F_q&=& \exp\left(-\int_{Q_0}^Q dq\; \Gamma_q(Q,q)\right)  \nonumber\\
F_g&=& \exp\left(-\int_{Q_0}^Q dq\; 
[\Gamma_g(Q,q)\;+\; \Gamma_f(q)]\right) 
\label{sudakov}
\end{eqnarray}
with the parton emission densities given by (\ref{gammaq}).
These form factors describe the probability for no
emission between the scales $Q$ and $Q_0$, so at resolution
$Q_0$ there is no other parton resolved than the original one.
Explicit expressions in improved MLLA are presented in
\cite{do}.

The jet rates of higher order 
can then be calculated iteratively and are given as nested
integrals. The exclusive jet rates of the lowest order in $e^+e^-$ annihilation
are given within the MLLA by \cite{cdotw} (for improved MLLA, see \cite{do})
\begin{eqnarray}
R_2&=& F_q(Q)^2 \label{r2jet}\\
R_3&=& 2 F_q(Q)^2 \int_{Q_0}^Q dq \Gamma_q(Q,q) F_g(q)  \label{r3jet}\\
R_4&=& 2 F_q(Q)^2
[\left(\int_{Q_0}^Q dq \Gamma_q(Q,q) F_g(q) \right)^2 \nonumber\\
  &\;& \qquad  + \;\int_{Q_0}^Q dq \Gamma_q(Q,q) F_g(q) 
      \int_{Q_0}^q dq' \{\Gamma_g(q,q') F_g(q')+\Gamma_f(q') F_f(q')\}
   ]
 \label{r4jet}
\end{eqnarray}
with $F_f=F_q^2/F_g$.
Whereas the form factors or jet rates $R_2$ are falling off exponentially
with increasing resolution $Q$ the jet rates of higher order
$R_n$ first rise from zero at $Q=Q_0$  
and after reaching a maximum fall off like products of
the Sudakov form factors
$R_n\sim F_g^{n_g} F_q^{n_q+n_{\bar q}}$
for a final state with $n_g$ gluons and $n_q + n_{\bar q}$
quarks and antiquarks \cite{do}.

It would be interesting to study experimentally these jet rates
down to small $y_c$.
In the transition region to the final hadronic states (small $y_c$)
one can investigate  the possible deviation from the perturbative
predictions due to hadronization effects.
An instructive example has been presented 
\cite{webaach}
for the 2-jet  and 3-jet rates in $e^+e^-$ annihilation where large
differences appear between parton and hadron level results as
obtained from the HERWIG Monte Carlo
\cite{HERWIG}.

\subsubsection{Subjet Multiplicities}

The study of jet rates can be carried further by considering
the subjet structure of multijet events or of individual resolved jets
\cite{cwdf}. This allows one  to study the
differences of quark and gluon jets at variable resolution. To this end, in
a process with the primary hardness $Q$, one first combines the hadrons
to clusters corresponding to the resolution
$y_0=Q_0^2/Q^2$, also called subjets, and then combines these clusters to
jets at the resolution scale $y_1=Q_1^2/Q^2$ with $y_1>y_0$.  
In this way one can define for $e^+e^-$ annihilation the average
number of subjets $M_2 (y_0$, $y_1)$ and $M_3 (y_0$, $y_1)$
in 2- and 3-jet events
respectively.

The subjet rates can be calculated perturbatively for
$Q_0\gg \Lambda$
in absolute normalization, and in this way a new information on the
difference between quark and gluon jets can be obtained.
In the limit $Q_0 \sim m_{\pi}$ the subjets correspond to the
final state hadrons and we may perform comparison to the results in section 3.

Naively one would expect that the ratio of subjet multiplicities 
in 3-jet 
and 2-jet events approach
\begin{equation}  
\frac{M_3}{M_2} = \frac{2C_F+N_C}{2C_F}=\frac{17}{8}.
\label{rnaiv}
\end{equation}
according to the ratio $C_F/N_C$ of subjet multiplicities in quark and gluon
jets.
The actual calculation, however, in the region of
parameters relevant today yields much  smaller values of less than 1.5 as a
consequence of the destructive soft gluon interferences.

The theoretical analysis of $e^+e^-$ annihilation
\cite{cwdf} 
proceeds in analogy to the previous case 
of jet multiplicities.
One can derive from the exact matrix
element calculations
\cite{ert}
the perturbative results of order $\alpha_s$ for the subjet rates in
terms of the logarithms
$L_0=\ln(1/y_0)$
and
$L_1=\ln(1/y_1)$ in the next-to-leading order \cite{cwdf},
 i.e. including double
and single logarithmic terms in each order in $\alpha_s$, 
\begin{eqnarray}
M_2&=& 2\; + \; \left(\frac{\alpha_s}{2\pi}\right)
    C_F(L_0-L_1)(L_0+L_1-3) \label{M2}\\
M_3&=& 3\; + \; \frac{1}{2} \left(\frac{\alpha_s}{2\pi}\right)
    (L_0-L_1) [ 2C_F (L_0+L_1-3) \nonumber \\
   &\;& \qquad\qquad + \; N_C \left(L_0-\frac{1}{3}L_1-\frac{14}{3}\right) 
   \; +\; \frac{2}{3}N_f].
    \label{M3}
\end{eqnarray}
The ratio $M_3/M_2$ with increasing subjet resolution $L_0$ first decreases
from the limiting value $3/2$ at $L_0=L_1$ because of the additional
contributions in $M_3$ from the gluon jet, before it rises towards the
asymptotic limit (\ref{rnaiv}) for $L_0 \gg L_1$.  

These calculations can be generalized to all orders in $\alpha_s$ using the
generating functional technique for two scales $Q_0$ and $Q_1$. 
 The results from the
next-to-leading calculation can be written \cite{cwdf}
\begin{eqnarray}
M_2(Q,Q_1,Q_0)&=&2N_q(Q,Q_0)\; +\; 2C_F\ln(Q/Q_1)H(Q_1,Q_0)
\label{sjm2}\\
M_3(Q,Q_1,Q_0)&=& M_2(Q,Q_1,Q_0)\; + \; N_g(Q_1,Q_0)\; + \; 
    N_C<\eta>_g H(Q_1,Q_0).
\label{sjm3}
\end{eqnarray}
The subjet multiplicities $M_i$ obtain the contributions from 
the intrinsic subjet multiplicities $2N_q$ and $N_g$ 
at scale $Q_1$  as expected; in addition, there are the contributions from the
interjet production of gluons at angles larger than the typical
opening  angle of the jet 
of virtuality 
$Q_1$ but with energy small enough not to be resolved as a new
jet. 
The additional term in (\ref{sjm2}) for $M_2$ can be interpreted as the 
multiplicity of such interjet gluons within the rapidity range $2\ln
(Q/Q_1)$ with density $C_F H(Q_1,Q_0)$ where
\begin{equation}
H(Q_1,Q_0)\;= \; \frac{2}{\pi} \int_{Q_0}^{Q_1} \frac{dq}{q} \alpha_s(q) 
     N_g(q,Q_0).   \label{Hqq}
\end{equation}
The last term in (\ref{sjm3}) for $M_3$ corresponds to the emission
of gluons with density $N_C H(Q_1,Q_0)$ into the interjet rapidity interval
of the gluon jet of length
\begin{equation}
<\eta>_g\;=\; \frac{
   \int_{Q_1}^Q dq \Gamma_q(Q,q) F_g(Q_1,q) \ln(q/Q_1)}
   { \int_{Q_1}^Q dq \Gamma_q(Q,q) F_g(Q_1,q)}. 
   \label{etag}
\end{equation}
If the subjet scale $y_0$ is decreased from threshold $y_0=y_1 \ll 1$
more subjets are resolved in the quark jets where the interjet rapidity
range is rising with total energy $Q$ like $\ln (Q/Q_1)$ in comparison to the
gluon jet whose rapidity interval for fixed $Q_1$ does not increase with
$Q$.  This reduced gluon production from the gluon jet is a consequence of the
angular ordering which forbids emissions at angles larger than the
production angle of the gluon jet itself. This effect of soft gluon
coherence leads to the suppression of $M_3/M_2 < 1.5$ for scales 
$ y_0\lsim y_1$.   

These analytical results have been compared with experimental data
from  L3
\cite{L3sj}, OPAL \cite{opalsj} and AMY \cite{amysj}
collaborations.
In Fig.~4.5 we show the average subjet multiplicities
$M_2-2$ and $M_3-3$ as measured by OPAL in comparison
to the theoretical predictions. The calculation of order
$\alpha_s$ as in (\ref{M2}) and (\ref{M3}) 
is close to the data only near threshold for
$2 \times 10^{-3} < y_0 < y_1= 10^{-2}$. The all order $\alpha_s$, 
next-to-leading order ($\ln y_{0,1}$) results 
(\ref{sjm2}), (\ref{sjm3}) follow the data well 
in the  region considered with $y_0>10^{-4}$ within 10-20\%. 
Again the theoretical predictions are given
in absolute normalization and depend only on the scale $\Lambda $,
chosen here to be $\Lambda$ = 0.35 GeV.
The ratio $M_3/M_2$ follows the prediction to decrease initially for
decreasing $y_0$ as expected from the colour coherence and then
to increase towards the asymptotic value of 17/8 for $y_0 \ll y_1$.

The measured ratio ($M_3-3)/(M_2-2$) is found to 
follow the predictions rather well down to
$y_0 \sim 4 \times 10^{-4}$.
At this value of $y_0$ the data indicate a sudden change
of slope whereas the theoretical prediction behaves
smoothly. This discrepancy around the observed kink 
in the data has been interpreted as a
possible sign of the onset of hadronization
corresponding to the scale $Q_0 \sim $1.6~GeV \cite{opalsj,amysj}.
One should observe, however, that the discrepancy
between theory and data in the unnormalized
quantities starts already before,
around $y_0 \sim 2\times 10^{-3}$ 
or $Q_0 \sim $4~GeV (see Fig. 4.5) which is a bit large for a
hadronization effect and this moderate discrepancy
could be within the accuracy of the calculation.
In any case, it would be interesting to investigate in more detail the
origin of the kink in the data.

Another interesting result with 3-jet events has been
presented by the ALEPH collaboration
\cite{alephsj}. 
They identified the gluon jet by
tagging the heavy quarks in two other jets. Studying 
the ratio 
$r' \equiv <N_g-1>/<N_q-1>$
of subjet multiplicities minus one in quark and gluon
jets (see Fig. 4.6) they find that with decreasing $y_0$ 
this ratio first rises towards a
maximum of 
\begin{equation}
r'_{max} 
     =  1.96 \pm 0.15\qquad 
   ({\rm ALEPH}) \label{rmalep}
\end{equation} 
at $y_0 \sim 2\times 10^{-3}$
and then approaches the lower value $r' = 1.29 \pm 0.03$
for
 $y_0 \sim 1.6\times 10^{-5}$.

This result shows that there is indeed the large difference
between quark and gluon jet multiplicities close to the
expected asymptotic ratio of 9/4 at intermediate or low resolution
$y_0 \lsim y_1$\footnote{
The drop for $y_0\to y_1$ has been related \cite{alephsj} to the different
energies of quark and gluon jets in the 3-jet sample considered.}.
For small $y_0 \sim 10^{-5}$ on the other hand 
the experimental ratio $r'$ approaches the
value obtained for hadrons as expected. Such small value of $r'$  
was found to be in a
rough agreement with the Monte Carlo predictions at the
parton level (see table 2 in section 3)\footnote{ 
In \cite{alephsj} the result of
a parton level calculation is quoted with
$r'\sim 1.6$ 
rather than the measured $r' =  1.29$. 
Here the discrepancy is also clearly visible in the
unnormalized multiplicities $N_{q,g}-1$: the multiplicities of the partons
tend to  saturate below a resolution of
$y_0\sim 10^{-4}$,
earlier than the multiplicities of the hadrons.
This is expected because of the larger parton mass around 1 GeV used in the
Monte Carlo.}.

In the analytical description of these measurements
\cite{sey} 
one has to take proper care of the definition
of quark and gluon jets in the experimental analysis.
The result of a next-to-leading calculation describes
well the region around the maximum of $r'$ 
but does not reproduce the smaller values of $r'$ at
small scales $y_0$. It would be clearly desirable to arrive at a common
understanding of the analytical and Monte Carlo results in the small $y_0$
region.

We conclude that 
the results on jets and subjets are generally in a very
good agreement with perturbative predictions for sufficiently
large scales $Q_0$ both for the shape and for the 
absolute normalization by adjusting
only one free parameter, namely the QCD scale $\Lambda$.
In particular, the large differences between quark and gluon jets and the
influence of colour coherence are demonstrated. 
So far no detailed comparison of the
predictions on exclusive jet rates have become available.
There is a tendency of the data to disagree with perturbative
predictions at scales $Q_0\lsim 2$ GeV.
This is an interesting problem which deserves further study.
In the moment it is not clear whether such discrepancies
are the first signs of non-perturbative hadronization effects
or merely a consequence of the limitation in the accuracy
of the theoretical calculations.


%
\section{Particle Correlations inside Jets}
\setcounter{equation}{0}
\subsection{Momentum Correlations}

\indent
The normalized correlation function is defined by
\begin{equation}
R_2(\xi_1, \xi_2) =
   \frac{d^2N/d\xi_1 d\xi_2}
 {(dN/d\xi_1)(dN/d\xi_2)}
\labl
{r2def}
\end{equation}
where again the variables $\xi_i=\ln(E/\omega_i)$ are used with $E$ and
$\omega_i$ denoting the jet and parton energies (momenta) respectively.
The QCD prediction for the leading contribution at high energies to this
correlation has 
been derived in the DLA as
 \cite{dfk2}
\begin{equation}
          R_2(\xi_1, \xi_2) =
           1 + \frac {1}{6r} \left[1+{4 \over 3} \sinh^2
\left(\frac {\mu(\zeta_1,\eta)-\mu(\zeta_2,\eta)}{2}\right)\right]^{-1}
\labl{r2dla}
\end{equation}
which depends only on the scaling variables $\zeta_i = y_i/Y $ and
$\eta=\lambda/Y$
with $y_i = \ln(\omega_i/Q_0)\; = \;Y -\xi_i$ and also on 
the constant $r$ which for the process  $e^+e^- \to q \bar q$
is given by $r=C_F/N_C=4/9$.
The function $\mu(\zeta,\eta)$ is determined
implicitly through the equations
\begin{eqnarray}
       2 \zeta -1&=&\frac{(\sinh(2\mu)-2\mu)\;-\;(\sinh(2\nu)-2\nu)}
       {2(\sinh^2\mu-\sinh^2\nu)}
\nonumber\\
        \sinh\nu&=&\frac{\sinh\mu}{\sqrt{1+1/\eta}}.
\labl{mudef}
\end{eqnarray}
For high energies $Y\gg\lambda$ 
the correlation function (\ref{r2dla}) depends 
only on the scaling variables
$\zeta_i$ and 
peaks for equal momenta $(\zeta_1=\zeta_2)$.
At the maximum it coincides with the second multiplicity moment
$F^{(2)} = 1 + 1/6 r$ from (\ref{frecur}).

Next-to-leading order corrections of relative order
$\sqrt{\alpha_s}$ have been obtained
for the kinematic region around the maximum correlation
keeping only linear and quadratic terms in the variables
$\xi_i$ \cite{fw}
\begin{equation}
R_2(\xi_1,\xi_2,\tau)=1+\frac{1}{6r}-\frac{1}{2r}
\left(\frac{\xi_1-\xi_2}{\tau}\right)^2 - \frac{1}{216 r}\left[A-B 
\left(\frac{\xi_1+\xi_2}{\tau}\right)\right]\sqrt{\frac{48}{b\tau}}
\labl{r2fw}
\end{equation}
\begin{equation}
A=44+9t+5rt-30r^2t,\qquad B=3 (11+t+rt-6r^2t).
\labl{r2fwc}
\end{equation}
Here the expansion variable is chosen as $\tau=\ln(E/\Lambda)$, 
with $\tau\approx Y$ for high energies, also
$t = n_f/2N_C$ and $b$ from (\ref{2.1}).
The dependence on momenta again comes only through the
rescaled variables $\xi_i/\tau \approx 1-\zeta_i$.
The leading terms at high energies depend only on the difference 
$(\zeta_1-\zeta_2)^2$ and  
coincide at this order with
Eq. (\ref{r2dla}) of the DLA (for $\eta\ll 1$, $\nu \ll \mu$). 
The second, next-to-leading order, term depends not only
on the difference but also on the absolute size of the
$\xi_i$'s. The range of these partonic correlations is large,
with $\tau$ of order  $ \ln(Q/\Lambda)$,
whereas the hadronic correlations, say from resonance decays,
are expected to be much smaller, of
order $\ln(M_h/\Lambda)\ll \tau$.

Experimental data on the correlation $R_2$ have been presented
by the OPAL Collaboration
\cite{opalmc}
for different bands in the $(\xi_1,\xi_2)$ space. In Fig.~5.1
we show two examples exhibiting the dependence of the correlation
$R_2$ on $\xi_1-\xi_2$ and $\xi_1+\xi_2$. Also shown are 
the predictions from Eq. (\ref{r2fw})
for different values of the scale parameter $\Lambda$
and the leading order DLA contribution.
The approximately linear and quadratic dependences 
are well reproduced for 
the parameter $\Lambda \sim 250$ MeV which fits best the single
particle spectra.
The normalization is off by an amount of the order of the neglected
next-to-next-to-leading terms \cite{webdal}, as one might expect.

In order to see whether this discrepancy is really due to the
neglect of the higher order terms one may compare it to a Monte Carlo
calculation at the parton level which includes the most important 
nonleading corrections.
In Fig.~5.2 we show $R_2(\xi_1, \xi_2)$ for
$\xi_1=\xi_2$ as a function of the rescaled variable
$\xi/\xi^*$ (where $\xi \equiv \xi_1$ and $\xi^*$ is the
$\xi$ value at the peak, so $\xi/\xi^*
\approx 2\zeta$) at the parton and hadron level
as obtained from the HERWIG Monte Carlo \cite{HERWIG} at two different
primary quark energies \cite{wo}.

Contrary to the case of the global multiplicity moment $F^{(2)}$
the discrepancy between the parton model prediction and the
experimental data (which in turn agree with the hadron level Monte Carlo)
remains. Apparently LPHD is not realized here quantitatively,
however, with
increasing energy both distributions approach each other.
This may be related to the fact that the momentum spectra and correlations
are in general dependent on the cut-off
$Q_0$ as can be seen from Eqs. (\ref{r2dla}),(\ref{mudef}) involving
$Y \approx \ln(E/Q_0$) and $y = \ln(\omega/Q_0$),
and this is different from the global moments $F^{(n)}$,
which are infrared safe.

So far only the central region around $\xi$ = $\xi^*$
has been studied experimentally. For small $\xi$ (large $x$) the Monte Carlo
calculation predicts
an anticorrelation with $R_2 < 1$ (see Fig.~5.2). This nonleading effect 
is a consequence of energy-momentum 
conservation, as two particles cannot have large $x$
at the same time.
Asymptotically, one expects $R_2 \geq 1$ from Eq. (\ref{r2dla})
and for $\xi_1 = \xi_2$
the approach to the constant
$R_2=1+1/6r$, i.e. $R_2=11/8\approx 1.375$ for 
 $e^+e^- \to q \bar q$.
Indeed, the Monte Carlo data on $R_2$ in Fig.~5.2 show a rise at small
$\xi$ for increasing energy $\sqrt{s}$, but the approach to the
asymptotic limit is very slow. On the other hand, for the very soft
particles with large $\xi/\xi^*$ the correlation $R_2$ comes much closer to
the asymptotic value 11/8. This agrees with the expectation that very soft
particles are not much influenced by the energy conservation constraints. 
It will be interesting to test this prediction experimentally.

We conclude that
the dominant nonleading effects of                    
momentum correlations are generated by energy
conservation constraints.
The role of colour coherence
in these correlations has also been investigated \cite{opalmc}.
From the comparison of QCD based hadronization models
with and without colour coherence included no important 
effect could be established.

\subsection{Azimuthal Angle Correlations}

Angular correlations promise a higher sensitivity to colour
coherence and to
the predicted angular ordering of subsequent partons in the
cascade. 
We consider first the azimuthal correlation, especially those
between two gluons emitted from a $q \bar q$ pair in $e^+e^-$ annihilation
\cite{dkmw1,dmo}.
This problem is closely related to the classic string/drag effect to be
discussed in section~6.2 which concerns the soft gluon emission
from a hard $q \bar q g$ system.
The production of the soft gluon opposite to the hard gluon is
suppressed because of a negative interference effect. In the limit in
which the hard gluon becomes soft as well, we expect the same kind of
suppression for two gluons in a hard $q \bar q$ event 
emitted in opposite directions, i.e.
with the azimuthal difference 
$\Delta \phi = \pi$\cite{dkmw1}.

\subsubsection{The Energy-Multiplicity-Multiplicity Correlation in Leading
Order}
The distribution of two gluons 
in the process $e^+e^-\to q(1) \: \bar q(2) \: g(3) \: g(4)$
in directions $\vec n_3,\vec n_4$ is obtained by adding another gluon
to the 1-gluon emission process   \cite{dkmw1}, cf. Eq.  (\ref{4.1}).
\begin{equation}
\frac{d^2N^{q\bar q}}{d\Omega_{\vec n_3} d\Omega_{\vec n_4}} \; \propto
    \; \left(\frac{a_{12}}{a_{13}a_{34}a_{42}}+
         \frac{a_{12}}{a_{23}a_{34}a_{41}}
         -\frac{1}{N_C^2} \frac{a_{12}^2}{a_{13}a_{32}a_{14}a_{42}}\right)
   \labl{qqgg}
\end{equation}
with $a_{ij}=1-\vec n_i \vec n_j \equiv 1-\cos\vartheta_{ij}$.
We are interested here again in the normalized correlation
\begin{equation}
R^{q\bar q}(\vec n_3,\vec n_4) =
     \frac{d^2N^{q\bar q}/d\Omega_{\vec n_3} d\Omega_{\vec n_4}}
         {(dN/d\Omega_{\vec n_3})\;(dN/d\Omega_{\vec n_4})} 
\labl{cazid}
\end{equation}
which is obtained using (\ref{qqgg}) and (\ref{4.1}) as
\begin{equation}
   R^{q\bar q}(\vec n_3,\vec n_4) = 1+ \frac{N_C}{2C_F}
    \left(\frac{a_{13}a_{24}+a_{14}a_{23}}{a_{12}a_{34}}-
       1\right).
\labl{cazid1}
\end{equation}
In the soft gluon limit the quark and antiquark will be anti-collinear
and we may introduce the gluon pseudorapidities
$\eta_{3},\eta_{4}$ and azimuthal angles $\phi_{3},\phi_{4}$ with respect
to the quark direction $\vec n_1\simeq -\vec n_2$ and one obtains for 
Eq. (\ref{cazid1}) simply
\begin{equation}
C^{q\bar q}(\eta_{34},\phi_{34}) = 1+\frac{N_C}{2C_F}
   \frac{\cos \phi_{34}}{\cosh \eta_{34}-\cos \phi_{34}}.
\labl{caziang}
\end{equation}
This quantity depends only on the rapidity and angle differences
$\eta_{34}$ = $\eta_3-\eta_4$ and
$\phi_{34}$ = $\phi_3-\phi_4$.
Note that this result is infrared stable, i.e. it is independent of the cut-off
$Q_0$, 
 and also independent of the primary energy $E$. For
 back-to-back gluons with $\phi_{34}$ = $\pi$ there is a
destructive interference with
\begin{equation}
C^{q\bar q}(0,\pi) = \frac{N_C^2-2}{2(N_C^2-1)}=\frac{7}{16} \approx 0.44.
\labl{cazi}
\end{equation} 
This value of the correlation function -- depending only on the number of
colours -- has been found also 
in the analysis of the string
effect, see below Eq.~(\ref{4.6}).
On the other hand for orthogonal gluon directions one finds
$C^{q \bar q}(0,\pi/2)\: =\: 1$, i.e. there is no correlation.

A convenient way to measure this correlation is through the
Energy-Multiplicity-Multiplicity Correlation (EM$^2$C)
\cite{dkmw1,dkmt2}. This avoids the selection of 2-jet
events and the definition of a jet axis. Each particle $i$ of the
event defines in turn an axis with respect to which $\eta_j$
and $\phi_k$ for any pair of particles are defined. Each
correlation as in 
(\ref{cazid}) is then weighted by the energy $E_i$ of the selected particle.

The azimuthal correlation is then defined as
\begin{equation}
C(\phi) = \frac{C_{EMM}(\eta_{min},\eta_{max},\phi)\cdot  C_E}
{[C_{EM}(\eta_{min},\eta_{max})]^2}
\labl{cphi}
\end{equation}
where
\begin{eqnarray}
C_E &=& \frac{1}{\sigma}\int dE_iE_i\;\frac{d\sigma}{dE_i}
\labl{ce}\\
C_{EM}(\eta_{min},\eta_{max}) &=& \frac{1}{\sigma}\int dE_iE_i\:dE_j
                                 \int_{\eta_{min}}^{\eta_{max}} d\eta_j
   \; \int_0^{2\pi} d\phi_j \frac{d\sigma}
             {dE_idE_jd\eta_jd\phi_j}
              \labl{cem}\\
C_{EMM}(\eta_{min},\eta_{max},\phi)&=&\frac{1}{\sigma}\int dE_i E_i dE_j dE_k
                        \nonumber \\
           \times   \int_{\eta_{min}}^{\eta_{max}}
    d\eta_j  d\eta_k & &
   \int_0^{2\pi} d\phi_j d\phi_k \delta (\phi-\phi_j+\phi_k) \frac{d\sigma}
             {dE_idE_j dE_k d\eta_j d\eta_k d\phi_j d\phi_k}
\labl{cemm}
\end{eqnarray}
In order to obtain
the distribution of soft gluons from the $q \bar q gg$-antenna
one has to replace the gluons 3 and 4 by their associated parton cascades.
This can be achieved by multiplying the unnormalized correlations $C_{EM}$
and  $C_{EMM}$ by a cascading factor $N_g'(\ln(E))$ for each gluon where
$N_g(\ln(E))$ is the mean multiplicity in the gluon jet, see 
subsection 5.2.2. 
For normalized correlations these
factors cancel and at leading order (DLA) the
EM$^2$C is given by the corresponding primary parton correlation
(\ref{caziang}). Therefore for a sufficiently narrow rapidity interval
$[\eta_{min},\eta_{max}]$
\begin{equation}
C(\phi) \approx C^{q \bar q} (0,\phi).
\labl{cfin}
\end{equation}
This is an important result \cite{dkmw1} which relates the 
energy  correlation to
the underlying antenna pattern. It supports
the idea that the particle
whose energy is used as the
weight effectively determines the jet direction.

\subsubsection{Predictions in Next-to-Leading Order and Monte Carlo Results}
The NLO corrections of O($\sqrt{\alpha_s})$ \cite{dmo} to these results
take into account not only
the overestimation of parton cascading in the
DLA known already from the inclusive energy spectra. It  also accounts
for the influence of the hard matrix elements at large angles
and the screening of the singularity of small relative angles $\vartheta_{34}$
from intermediate partonic processes.
In what follows we will describe the main consequences of these
corrections.

{\it i) Corrections to the multiplicity flows}\\
As already mentioned above
the perturbative corrections to the
hard gluon emission are provided by the cascading factors.
They take into account that the triggered gluon is a part
of a jet. For the  
1-particle flow from a hard parton one obtains 
\begin{equation}
\frac{dN}{d\Omega_{\vec n_3}} = \frac{W_{12}^3}{2\pi} \int_0^1
    \frac{dx}{x}\frac{\alpha_s}{2\pi} N_g(xQ_3)
    \quad \hbox{with} \quad W^3_{12} = 2 C_F \frac{a_{12}}{a_{13}a_{23}}.
 \labl{mfl}   
\end{equation}
The contribution of lowest order would be like Eq.~(\ref{mfl})
but without the $N_g$ factor in the integral.
The multiplicity factor $N_g$ accounts for the intermediate parton
cascades up to the scale $xQ_3 = xE \sin\vartheta_{13}$,
given by the angular ordering constraint.

In order to obtain the next-to-leading corrections
one has to take into
account the evolution for multiplicity
to order $\sqrt{\alpha_s}$ which reads
\begin{eqnarray}
N'_g(Q) &=& \int_0^1 \frac{dx}{x}\left(2N_C \frac{\alpha_s(xQ)}{\pi}\right)
            N_g(xQ)-\bar a\left(2N_C \frac{\alpha_s(Q)}{\pi}\right)N_g(Q)
         \nonumber\\
        &=& \gamma(\alpha_s(Q))N_g(Q)
\labl{nglu}
\end{eqnarray}
with
\begin{equation}
\gamma \simeq \gamma_0-\frac{1}{2} \bar a
\gamma_0^2(1-\frac{1}{2B}) +\cdots, \qquad
\gamma_0=(2N_C\alpha_s/\pi)^{1/2}, \qquad 
\bar a=\frac{11}{12}+\frac{n_f}{6N_C^3}=\frac{a}{4N_c}\/
\labl{gam0a}
\end{equation}
(and $B$ from (\ref{2.35}). Then,
one obtains for the 1-particle flow from a hard parton at NLO
\begin{equation}
\frac{dN}{d\Omega_{\vec n_3}}
        =\frac{W_{12}^3}{8\pi N_C} N'_g(Q_3)(1+\bar a\gamma)
\labl{mflow}
\end{equation}

For 2-gluon emission one has to consider in addition 
the mutual influence
of both flows. Neglecting running $\alpha_s$ effects at small
angles $\vartheta_{34}$ one can exponentiate 
such contributions and obtain
the improved formula for the full two particle correlation
\begin{equation}
R^{q\bar q}(\vec n_3,\vec n_4) \; = \;
   1+\frac{N_C}{2C_F}
  \left[
   \left(\frac{a_{13}a_{24}}{a_{12}a_{34}}\right)^{\rho}   
+\left(\frac{a_{14}a_{23}}{a_{12}a_{34}}\right)^{\rho}
   -1\right]
\labl{rrho}
\end{equation}
with the exponent $\rho~ = ~1-\frac{3}{4}\gamma$,
or, in terms of the relative angles $\eta_{34},~\phi_{34}$
\begin{equation}
R^{q\bar q}(\vec n_3,\vec n_4) \; = \;
   1+\frac{N_C}{2C_F}\left[\frac{2\cosh (\rho\eta_{34})}{(2
(\cosh\eta_{34}-\cos\phi_{34}))^{\rho}}-1\right].
\labl{rrhoa}
\end{equation}
For $\rho=1$ ($\gamma=0$ at infinite energies) one would  
restore the leading results in Eqs.
(\ref{cazid1}), (\ref{caziang}) but since $\rho < 1$ the peak near small
angles $\vartheta_{34}$ or $\phi_{34}$ will be reduced
because of the intermediate parton emissions.

{\it ii) Hard matrix elements} \\
Subleading corrections also occur if
both gluons are not soft and strongly ordered in energy.
These corrections are particularly important for larger relative angles
and have to be extracted from the matrix
elements. The corresponding diagrams considered 
in the calculation of $C_{EM}$ and $C_{EMM}$
are shown in Fig.~5.3.

{ \it iii) Small relative angles $\vartheta_{34}$}\\
 The various contributions to the correlation functions
$C_{EM}$ and $C_{EMM}$ are obtained as product of a matrix
element contribution with cascading factors $N_{g}'(Q)$.
The singularity $\vartheta^{-2}_{34}$ is screened by summing the
$\alpha_s \ln^2 \vartheta_{34}$ terms which leads to the replacement
\begin{equation}
R^{q\bar q}(\vartheta_{34})
      \; \sim \;\frac{1}{\vartheta^{2}_{34}}
    \; \to \; \frac{1}{\vartheta^{2}_{34}} \;
     \left[\frac{N_g(E\vartheta_{34})}{N_g(E\Theta)}\right]^{\nu_{\rm{eff}}}
\labl{screen}
\end{equation}
where the anomalous dimension $\gamma_0$ is absorbed into the
multiplicity $N_g$. This agrees with Eq.~(\ref{rrho}) for
$\nu_{\rm eff} = \frac{3}{2}$, the value for fixed coupling.
For running coupling
$\nu_{\rm eff}$ is a smooth function which increases from
$\frac {3}{2}$ for $\vartheta_{34} \lapproxeq \Theta$ to the value
$\nu_{\rm eff}\: =\: 2$ for $\vartheta_{34} \to \Lambda/E$.
Approximate formulae for $\nu_{\rm eff}$ will be given below, 
see Eqs. (\ref{nueff}), (\ref{omln}).

These analytical results for the general 2-particle angular
correlation $R(\vec{n_3},\: \vec{n_4}$)
are explicitly evaluated for the azimuthal correlations.
The LO and NLO results for $C(\phi$) are shown in Fig.~5.4
together with the predictions from the HERWIG Monte Carlo \cite{HERWIG}
at parton and
hadron level. The leading contribution to $C_{EM}$ --
the normalization of $C(\phi)$ -- comes from diagram (a) in Fig.~5.3
($q$ triggered, $g$ registered) and the NLO corrections of order
$\gamma_0 C_F/N_C$ amount to only 10-20\%.
On the other hand, $C_{EMM}$ gets very large corrections, in
particular, at small angles $\phi$ from the intermediate
partons which screen the $1/\vartheta^2_{34}$ singularity.

At large angles $\phi\approx\pi$ 
the negative soft gluon 
interference, which at LO leads to $C(\pi$) = 7/16
$\approx$ 0.44, as
in the string effect, is largely
removed and one finds $C(\pi) \:\approx\: 0.93$ at NLO.
This comes from the fact that the non-leading contribution (C) in
Fig.~5.3 with a gluon trigger becomes comparable for
$\phi \approx \pi$ to the leading contribution (A) with a quark
trigger. The soft gluon coherence effect is therefore better
seen in the ``classic" string effect in 3 jet-events where the
identity of the jets is better known.

The large contribution from the gluon jet could be avoided if a heavy quark
trigger is applied. Then one can
relate the particle whose energy (E) is measured to the heavy
quark particle and the particles whose multiplicities are counted (MM)
to the particles in a rapidity interval in the same hemisphere.
Alternatively, one could think of making a cut for the fast 
heavy quark to diminish the gluon bremsstrahlung contribution.
Then the negative interference effect should become more pronounced
and approach the magnitude expected in the string effect.

For not too small angles $\phi$
the analytical results agree within $\sim $10\% 
with the results from the parton Monte Carlo (Fig. 5.4) which
takes into account the most important subleading effects.
The remaining discrepency can be attributed to neglected terms of even
higher order.
Furthermore, by comparing Monte 
Carlo results at parton and hadron level (see
Fig. 5.4) one observes that
hadronization effects are negligible in the region
$\phi \gapproxeq \pi/4$ at LEP-1 energies.

\subsubsection{Comparison with Experimental Data}
The azimuthal correlation  $C(\phi)$ has been measured by OPAL \cite{opalph}
and ALEPH \cite{aleph7} at LEP-1 and by AMY at TRISTAN \cite{amy}.
The LEP-1 results are found to be in good agreement (within a few percent)
with the JETSET and HERWIG Monte Carlo's which take the colour
coherence in parton branching into account (see OPAL data in Fig.~5.5,
for example), 
in particular
\begin{eqnarray}
C(\pi)\;&=&\; 0.787 \pm 0.002 \pm 0.004 \quad \hbox{(OPAL)}\nonumber\\
C(\pi)\;&=&\; 0.783 \pm 0.001 \pm 0.016 \quad \hbox{(ALEPH)} 
\labl{copal}
\end{eqnarray}
where the first error is statistical and the second is systematic.
The results are found consistent 
also with the Monte Carlo predictions at parton level \cite{opalph} and
the data points are slightly
below the analytical NLO prediction of $C(\pi)\approx 0.93$ (see Fig. 5.4). 

The AMY data at the lower CM energy approach about the same value
for large angles $C(\phi\approx \pi)\approx 0.78$
but the correlation is rising more slowly towards small angles
with  $C(\phi\approx 0)\approx 2.75$ as compared to
$C(\phi\approx 0)\approx 3.25$ at LEP-1. This behaviour for small angles
violates the prediction of scaling with energy
at leading order, Eq. (\ref{caziang}), but
is expected from the screening effect, Eq. (\ref{rrho}), 
(increasing $\rho$-parameter for increasing energy).
Another effect is the influence of the cut-off $Q_0$ at small angles.
This can be seen from the comparison \cite{opalph}
of parton level Monte Carlo's
with different cut-off: the larger cut-off in JETSET corresponds to 
a stronger suppression of the correlation at small angles in comparison to
the HERWIG Monte Carlo  without affecting the large angle correlation.

The prediction from a model without coherence
(``JETSET incoherent (J3)", see Fig.~5.5) 
deviates from the data significantly.
However, in absolute terms the difference from $\phi \sim \pi$
is rather small, only about 5\%, i.e. of the same order as the
difference from the analytical NLO calculation. Therefore,
one has to conclude that this
observable is not particularly sensitive to the soft gluon
coherence effect. This is expected to be changed if the energy trigger is
related to the heavy quark as discussed above.

The results, however, represent another remarkable success of the LPHD concept
in that a hadronic particle correlation can be derived from a parton
level calculation analytically within $\sim$ 10\% for
$\phi \gapproxeq \pi$/4 (from the parton Monte Carlo
 even better). The discrepancy
at small angles is apparently due to cut-off effects 
which can be concluded from the cut-off dependence of the
 Monte Carlo's at the parton
level.

\subsection{Solid Angle Correlations:
     Observables and Generic Results}

A consequence of soft gluon coherence is the ``angular ordering"
of subsequent gluon emission
in the cascade evolution, see section 2.
These constraints are expected to affect most directly 
the correlations of the solid  angles in a jet.
Such correlations will be discussed in the remaining part of
this section. 

\subsubsection{Observables in Analytical Calculations}
The following observables have been considered in analytical calculations
\cite{ow1,ow2,dd,bmp}:

{\it i) Relative solid angle}\\
The simplest observable of this type, partially integrated, 
 is the distribution
of the relative angle
$\vartheta_{12}$ between two particles with solid angles
$\Omega_1$ and $\Omega_2$ in a forward cone of half opening angle
$\Theta$ around the primary parton of energy E
\begin{equation}
\frac{dN^{(2)}(E, \Theta)}{d \vartheta_{12}}
      = \int_{\Theta} \frac{dN^{(2)}(E, \Theta)}{d\Omega_1 d\Omega_2}\:
      \delta(\vartheta_{12}-\vartheta_{\Omega_1,\Omega_2}) \:d\Omega_1
      d\Omega_2\/.
 \labl{d2def} 
\end{equation}
A related quantity is the ``Particle-Particle Correlation" (PPC) and its
asymmetry which has been studied experimentally 
\cite{aleph7,ppc1,ppc2,ppc3}, but is not
analytically computed.

{\it ii) Fully inclusive $n$-particle distributions}\\
 The particle densities
$D^{(n)}(\Omega_1\ldots\Omega_n,E,\Theta) 
=dN^{(n)}/d\Omega_1,\ldots,d\Omega_n$
 of general order $n$
or the connected (cumulant) correlation functions 
$\Gamma^{(n)}(\Omega_1,\ldots,\Omega_n,E,\Theta)$, introduced in Sect.
2.2.1, for particles emitted from a primary parton of energy $E$ within
a forward cone of half-opening angle $\Theta$.
 

{\it iii) Differential moments (``star integrals")}\\
 By partial integrations of fully differential distributions one
obtains the differential
moments, which occur naturally in the leading pole approximation \cite{ow2}.
Properties of such moments, also called ``star
integrals" 
have been discussed in detail in 
\cite{star}. One defines, e.g. for cumulant moments of rank $n$,
\begin{equation}
h^{(n)}(\Omega,\delta) = {1\over n} \sum_{i=1}^n \int_{\gamma(\Omega,\delta)}
d\Omega_1\dots d\Omega_n \Gamma^{(n)} (\Omega_1, \dots \Omega_n)
 \delta (\Omega-\Omega_i), 
\labl{hdef}
\end{equation}
i.e. one particle is kept fixed at solid angle $\Omega$
 and all others
are counted in the cone $\delta$ around it. In practical applications one 
has to integrate 
$h^{(n)}(\Omega,\delta) $ over an area in $\Om$.

{\it iv) Multiplicity moments in angular intervals}\\
 The integral of the correlation functions
$D^{(n)}$ or $\Gamma^{(n)}$ 
over selected angular intervals $\gamma(\vartheta,\delta)$
 yields the respective unnormalized factorial and cumulant 
multiplicity moments
\begin{eqnarray}
f^{(n)}(\vartheta,\delta)\;&=&\; \int_{\gamma(\vartheta,\delta)}
D^{(n)}(d\Omega_1\ldots d\Omega_n)
d\Omega_1\ldots d\Omega_n
\labl{fmomg}\\
c^{(n)}(\vartheta,\delta)\; &=&\; \int_{\gamma(\vartheta,\delta)}
\Gamma^{(n)}(d\Omega_1\ldots d\Omega_n)       
d\Omega_1\ldots d\Omega_n. 
\labl{cmomg}
\end{eqnarray}
Usually one considers
moments $F^{(n)}$ and $C^{(n)}$ normalized by the total multipliciy 
$N\equiv\bar n$
as in (\ref{momnor}).
Calculations have been performed for two types of 
 angular intervals $\gamma(\vartheta,\delta)$: (a)
the symmetric ring where the
polar angles $\vartheta_i$ with respect to the initial parton 
(jet axis) are
within the range $\vartheta -\delta\leq\vartheta_i\leq\vartheta +\delta$,
and (b) the sidewise cone, centered at a polar angle $\vartheta$ and
half opening $\delta$. We refer to these configurations according to the
dimensions $D=1$ and $D=2$  of the respective phase space  
volume $\delta^D$.

A study of multiplicity fluctuations at two resolution levels has been
suggested using the ``Double Trace Moments" \cite{ratti} as an extension of
the above one level moment analysis. Predictions have been obtained in the
DLA and can be found in \cite{dd}.

\subsubsection{Leading Order Results and Critical Angle}
Complete analytical results for these observables
have been derived in the DLA to
exponential accuracy which
provides the asymptotic limit at high energies, in particular the
scaling properties in the proper variables
of the various observables and
relations between them. Certain subleading corrections have been computed as well.

Two different methods have been applied to obtain results on
the solid angle correlations:

{\it i) Partial integration of $n$-particle distribution} \cite{ow1,ow2}\\  
Starting from the generating
functional in DLA, Eq.~(\ref{dlaz}) one
obtains the integral equation for the general
n-th order correlation function
or the above partially integrated correlations.
In the leading pole approximation 
they can all be built from functions
$h^{(n)}(\delta, E, \Theta)    $\footnote{For the observable listed
in {\sl (i)} above $\delta\equiv
\vartheta_{12}$, in {\sl (ii)} there is a sum 
over $n$ such terms involving
the relative angles $\vartheta_{ij}$ instead of the single angle $\delta$.}
which fulfill the integral equation
\begin{equation}
h^{(n)}(\delta, E, \Theta)=d^{(n)}(\delta, E)
      +\int_{Q_0/\delta}^E\frac{dK}{K}\;\int_{\delta}^{\Theta}
 \frac{d\Psi}{\Psi}\gamma_0^2(K\Psi) h^{(n)}(\delta, K, \Psi).
\labl{hnint}
\end{equation}
Here the inhomogeneous term 
 $d^{(n)}$ corresponds to the
contribution from the direct emission of the
considered particles off the primary parton, the integral corresponds
to the
emission from an intermediate gluon of momentum $K$ and angle
$\Psi$ to the primary parton. The direct term
is typically given by the $n^{th}$
power of the angular distribution $D^{(1)}(\delta,E)$
or by the event multiplicity $N(E,\delta)$
in the forward cone $\delta$ 
\begin{equation}
d^{(n)}(\delta, E)\; \sim \; [ N(E,\delta)]^n \; \sim \;
\exp({2n\beta\sqrt{\ln(E\delta/\Lambda)}})
\labl{dinh}
\end{equation}
with $\beta^2=12/b$, i.e. $\beta^2=\frac{4}{3}$ for $n_f=3$.
Eq. (\ref{hnint}) can be solved by the resolvent (Greens function)
method or by the WKB approximation to the corresponding 
partial differential equation.  
For $n$=2 the consistency between the fully
differential distribution $\Gamma^{(2)}(\Omega_1,\Omega_2)$
and the global moment $F_2(E,\Theta)$ is
demonstrated by subsequent integrations,
for fixed $\alpha_s$ also for the exact solution 
at finite energies in the DLA.

{\it ii) Jet calculus diagrams}\\
Considering the appropriate jet calculus diagrams \cite{kuv} 
the results for the multiplicity moments of arbitrary order are built
from the relevant 1-particle distributions and cascading factors
\cite{dd} following the procedure applied for 2-particle
azimuthal correlations \cite{dmo} as explained in the previous
subsection. Certain MLLA corrections have been evaluated as well in this
case.
This procedure is followed also in Ref.~\cite{bmp} where, in addition,
a description in terms of a 
random cascading process has been presented and further phenomenological
applications have been discussed.

Available results from both methods are found to agree 
in the high energy limit.
They are derived for a restricted region of phase space
away from kinematical boundaries
where recoil effects can be neglected, 
formally for $\frac{\Lambda}{E}\ll \delta \ll\Theta$.

The results can be obtained from an integral representation \cite{dmo,ow1,ow2}
\begin{equation}
h^{(n)}(\delta, E, \Theta)\; \sim \;
     \int^1_{Q_0/q} \frac{dx}{x} D^{\delta}(x,Q,q) [ N(xq)]^n
     \labl{hresol}
\end{equation}
with the two momentum scales $Q=E\Theta$ and $q=E\delta$.
The resolvent $D^{\delta}(x,Q,q)$ is the inclusive momentum
distribution of the intermediate partons
as discussed in Sect.~3, but these partons are required
with angular cut-off $\delta$ from below and not only with the usual
transverse momentum cut-off $k_\perp=xE\delta>Q_0$. 
This more
severe restriction is a further consequence of angular ordering:
final partons of angular separation $\delta$ can only fragment
from an intermediate parton which is produced with minimum angle $\delta$
itself, i.e. with virtuality $xE\delta$. This distribution is given by 
the inverse Laplace transform
\begin{equation}
D^{\delta}(x,Q,q)=\int \frac{d\omega}{2\pi i}
      \exp\left(\omega \ln\frac{1}{x}+\int_{xq}^Q
      \frac{dK_\perp}{K_\perp}\gamma_{\omega}
      (\alpha_s(K_\perp))\right).
          \labl{delgen}
          \end{equation}
where the anomalous dimension $\gamma_{\omega}$ is given in Sect.~3
for different approximation schemes.
In the DLA for fixed $\alpha_s$, for example,
one finds an explicit result which
can be easily discussed:
\begin{eqnarray}
D^{\delta}(x,Q,q)&=&\delta(x-1)+\gamma_0\sqrt{\frac{\ln Q/q}{\ln 1/x}}
   I_1\left(2\gamma_0\sqrt{\ln\frac{1}{x}\ln\frac{Q}{q}}\right)
   \labl{dtheta}\\
          &\sim&
    \exp\left(2\gamma_0\sqrt{\ln\frac{1}{x}\ln\frac{Q}{q}}\right).
 \labl{dtheta1}
\end{eqnarray}
This distribution drops for fast partons with $x\approx 1$ but
peaks for soft partons with small $x$ and is
different from the usual momentum
spectrum obtained for 
cut-off $K_\perp>Q_0$ in Eq.~(\ref{delgen}), the ``hump-backed plateau"
discussed in Sect.~3, which drops towards small $x$ as well. The appearence of
the distribution $D^{\delta}$ in Eq.~(\ref{hresol})
makes the above observables
$h^{(n)}$ sensitive
to the angular ordering constraint with increasing
$\delta$.

The integrand in Eq.~(\ref{hresol}) is sharply peaked 
at high energies ($N$ vanishes for
small $x$, $D^{\delta}$ for large $x$). If the lower limit
of the integral $Q_0/q$
is below the maximum ($\delta$ sufficiently large) one
obtains the integral (\ref{hresol}) by a saddle point method.
If the lower limit goes beyond the $x$-value at the maximum -- corresponding
to the ``critical angle" $\delta=\vartheta_c$ -- this method cannot be
applied any longer but the integral is well
approximated by the integrand, i.e. the 1-particle spectrum, at the lower
limit. For fixed $\alpha_s$ and $n=2$ this critical angle is
at \cite{dmo,ow1,ow3}
\begin{equation}
\vartheta_c =\frac{Q_0}{E} \left(\frac{E\Theta}{Q_0}\right)^{1/5},
\labl{thcrit}
\end{equation}
for running $\alpha_s$ one finds in a high energy approximation \cite{ow3}
\begin{equation}
    \vartheta_c=\frac{\Lambda}{E}
    \left(\frac{\ln (E\Theta/\Lambda)}{\lambda}\right)^{\frac{4}{9}\lambda}.
\labl{thcritra}
\end{equation}
The critical angle limits the regime of scaling from below. For smaller
angles the correlations become sensitive to the cut-off $Q_0$. It is
interesting that the perturbative QCD alone reveals the existence of these 
two regimes with quite different physical characteristics 
 (see below). For the critical
angle $\vartheta_c$ the correlation function in the high energy limit
develops a nonanalytic behaviour (discontinuous second derivative). This
behaviour is quite unique for angular correlations and does not occur, for
example, for momentum spectra.

We consider first the important case of larger angles $\delta>\vartheta_c$
where the integral in Eq. (\ref{hresol}) can be obtained from the
saddle point method.
The high energy limit in DLA for general order $n$ is obtained from 
(\ref{hresol}) and (\ref{delgen}) by
applying a saddle point approximation and
can be written as 
\begin{equation}
h^{(n)}(\delta, E, \Theta)\; \sim \;
        \exp \left( 2\beta 
\sqrt{\ln (E\Theta/\Lambda)}\omega(\epsilon,n)\right )
\labl{hsol}
\end{equation}
or, after normalization by the multiplicity $ N(E,\Theta)$  
with $H^{(n)} \; \equiv \; h^{(n)}/N^n$
\begin{equation}
H^{(n)}(\delta, E, \Theta) \; \sim \;
        \exp \left( 2\beta  
\sqrt{\ln (E\Theta/\Lambda)}(\omega(\epsilon,n)-n)\right)\/.
\labl{Hsol}  
\end{equation}
The dependence of the exponent on the scale $E\Theta$ comes from the 
anomalous dimension $\gamma_0$ of the multiplicity evolution,
see Eq.~(\ref{dinh}), therefore, the normalized correlation is naturally
expressed by
\begin{eqnarray}
H^{(n)}(\delta, E, \Theta) \;& \sim & \;
     \left[\frac{N(E\delta)}{N(E\Theta)}\right]^{\nu_{eff}}
     \labl{Heff}\\  
 \nu_{\rm eff}\;&=&\;
\frac{n-\omega(\epsilon,n)}{1-\sqrt{1-\epsilon}}.
\labl{nueff}
\end{eqnarray}
Remarkably, the dependence on the small angle $\delta$ occurs
only through the 
scaling variable
\begin{equation}
\epsilon={\ln{(\Theta/\delta)}\over \ln{(E\Theta/\Lambda)}
}.
\labl{epsdef}
\end{equation}
As the angle $\delta$ varies  typically in the range 
$\frac{\Lambda}{E} \lapproxeq \frac{Q_0}{E} 
\leq \delta \leq \Theta $,
the corresponding bound for $\epsilon$ is then 
$0\leq \epsilon \leq 1$. The region $\Theta\leq \delta \leq 2\Theta$ is
kinematically allowed but is not reliably calculated until now because
of important recoil effects in this domain.

The function $\omega(\epsilon, n)$ is given by\footnote{
This function has been introduced in Ref.~\cite{ow2} where more details are
given; the equivalent function $\nu_{eff}$, see Eq.~(\ref{nueff}), is
defined in Ref.~\cite{dmo} where it appears in the 
analysis of the azimuthal angle correlations, see  Eq.~(\ref{screen}) above
for $R^{q\bar q}$. The results on $\omega(\epsilon)$ which follow
 are found to agree numerically with
$\nu_{eff}(\tau),\; \tau=1-\epsilon$ in Fig.~3 of Ref.~\cite{dmo}, 
see also Ref. \cite{ow2}.}

\begin{eqnarray}
\omega(\epsilon,n)&=&\gamma(z(\epsilon,n))+\epsilon
z(\epsilon,n),\labl{omege}\\
 \gamma(z)&=&{1\over 2} (\sqrt{z^2+4}-z),\labl{gamz}
\end{eqnarray}
where $z(\epsilon, n)$ is the solution of the algebraic
equation
\begin{equation}
\gamma^2(z)-
2\ln{\gamma(z)} - \epsilon z^2 = n^2 - 2\ln{n}. \labl{alg}
\end{equation}
Equations (\ref{omege})-(\ref{alg}) determine completely the scaling
function $\omega(\epsilon,n)$. 
For small $\epsilon $ this function has a linear behaviour
\begin{equation}
\omega(\epsilon,n)\simeq  n-{n^2-1\over 2 n} \epsilon,\quad
\nu_{eff}\simeq\frac{n^2-1}{n}
\labl{omlin}
\end{equation}
whereas  $\omega\to 0$ or $\nu_{eff}\to n$ for
$\epsilon\to1$.
In the limit $\epsilon $ fixed, $n \to \infty$ one finds the
approximation
\begin{equation}
\omega(\epsilon, n)=n\sqrt{1-\epsilon}\;
(1-\frac{1}{2n^2}\ln (1-\epsilon)+\cdots).
\labl{omln}
\end{equation}
Remarkably, this approximation has an accuracy of better than 1\%
for $\epsilon < 0.95$ already for $n = 2$ and is, therefore, well
suited for numerical analysis.

\subsubsection{Characteristic predictions for High Energies}
There are several important aspects of these asymptotic results of the
DLA:

{\it i) Angular scaling}\\
The dependence on the scaling variable $\epsilon$ implies
a finite high energy limit of the quantity 
\begin{equation}
\frac{\ln h^{(n)}(\delta, E, \Theta)}{ 2n\beta\sqrt{\ln
(E\Theta/\Lambda)}}
\to \omega(\epsilon,n)\/.
\labl{epscal}
\end{equation}
This asymptotic scaling 
property has been proposed for various observables  and tested
with Monte Carlo models whereby it has been already confirmed in some
cases already for energies $E \gapproxeq 10$ GeV, and not too small angles
$\delta$ \cite{ow2,wo1}.
The relevance of asymptotic DLA
scaling predictions at present
energies has been already found
in case of multiplicity scaling (KNO). Namely, at present energies the
KNO scaling is observed, but the scaling function is still far away
from the asymptotic expectation.
The first experimental results support the scaling in the jet opening
angle $\Theta$ \cite{mb}. An interesting feature 
implied by Eq.~(\ref{epscal}) is the scaling in two
external variables $E$ and
$\Theta$. This prediction has not yet been explored experimentally and 
can be tested most conveniently
with jets of variable energy (e.g. at HERA and TEVATRON).

The normalizing factor in Eq.~(\ref{epscal}) can also be expressed
in terms of  $\ln N$ in leading exponential accuracy, see Eq.~(\ref{dinh}), 
similarly the variable
$\epsilon$ can be given
in terms of $\ln N(E\delta)/\ln N(E\Theta)$ \cite{ow4}. 
If written in
this way, Eq.~(\ref{epscal}) is valid also for fixed $\alpha_s$.
In this formulation the parameter $\Lambda$ in the definitions of
observables is avoided but it may be difficult to determine $N(E,\delta)$.
A similar scaling law  has been noted
already \cite{dfk2}
for rescaled logarithms of momenta, in the variable 
$\zeta=\ln (E/\omega)/\ln (E\Theta/Q_0)$. 

Distributions in these 
rescaled logarithmic variables approach finite asymptotic limits in QCD.
Note that this is not true for rescaled quantities like $x=\omega/E$:
Bjorken- or Feynman-scaling is violated and there is only the trivial
asymptotic limit $D(x)\to \delta(x)$ for a review, see Ref. \cite{wozak}.

{\it ii) Universality}\\
As the various observables $h^{(n)}$ introduced above
approach the same function 
$\omega(\epsilon, n)$ in Eq.~(\ref{epscal}), nontrivial relations appear
between seemingly quite different observables.

{\it iii) Power behaviour for large angles}\\
For sufficiently large angles $\delta$ (small $\epsilon$) the linear
approximation Eq.~(\ref{omlin}) applies and one obtains the power behaviour
\begin{equation}
H^{(n)}(\delta, E, \Theta) \sim
\left(\frac{\delta}{\Theta}\right)^{\frac{n^2-1}{n}\gamma_0(E\Theta)}.
\labl{Hpow}
\end{equation}
Power laws of this type are characteristic
for  selfsimilar structures or ``fractals"
\cite{mbrot}. The fractal behaviour of jets has 
 been noted already in the 
beginning of QCD jet studies
\cite{kuv,gv}.
There has been a renewed interest in the role of power laws and fractal
behaviour with
intensive investigations in the field of multiparticle production,
following the discussion of ``intermittency" \cite{bip}.
The evolution of a QCD 
jet in a fractal phase space
has been discussed within the LUND dipole cascade model
\cite{lufractal}.  
For a review of various experimental and theoretical results
see, e.g., Ref.~\cite{wdk}.

The selfsimilarity of the QCD cascade
is mildly broken by the logarithmic scale
dependence of $\gamma_0(E\Theta)$ from the running $\alpha_s$. The same
power laws (\ref{Hpow}) with constant exponents and therefore with global
selfsimilarity are obtained in case of
fixed $\alpha_s$.
These results, valid for 
sufficiently large angles $\delta$, are 
infrared safe as they do not depend on the cut-off $Q_0$. This 
occurs because the
integrand in Eq.~(\ref{hresol}) is independent of $Q_0$ except through the 
prefactor of $N$ which cancels after normalization; the integration limit
disappears in the saddle point approximation for high energies.

{\it iv) Correlations below the critical angle}\\
The perturbative analysis \cite{ow3} reveals two different kinematic regimes 
which are separated by
the critical angle $\vartheta_c$ or $\epsilon_c$
\begin{eqnarray}
\epsilon_c(\rho)&=&\left(\gamma^2(z)-n^2-\ln(\gamma^2(z)/n^2)\right)/z^2
\labl{epscex}\\
   \rho&=&\sqrt{\frac{\ln(Q_0/\Lambda)}{\ln(E\Theta/\Lambda)}} \qquad
    z\: =\: -\frac{n^2-1}{n\rho}
\labl{rhoz}
\end{eqnarray}
with $\gamma$ given by (\ref{gamz}). It is found that
$\epsilon_c\to 1$ for $E\to \infty$.
The
correlation functions vanish for $\delta \to \delta_{min}=Q_0/E$
or for $\epsilon\to \epsilon_{max}=1-\rho^2$, with
$\epsilon_{max}<1$ for $Q_0/\Lambda>1$.
At LEP-1 energies this  angle corresponds to $\epsilon_c\gapproxeq 0.7$
or $\delta_c \lapproxeq 1^o$. 

There are two characteristic differences in the behaviour of correlations
in the two regimes:

\noindent {\it a) dependence on the cut-off $Q_0$.} 
In the region $\epsilon<\epsilon_c$ (large angles $\delta$)
the correlation function is $Q_0$-independent and develops 
the angular scaling behaviour
 with the only scale available being $\Lambda$.
In the complementary region
($\epsilon>\epsilon_c$) this scaling is broken 
and the correlation function
drops to zero at $\epsilon_{max}$ which depends on a second scale $Q_0$.
If the cut-off $Q_0$ is related to the particle mass according to  LPHD,
one expects the following pattern for correlations between particles of
different species (say $\pi\pi$, $KK$, $pp$): for small $\epsilon$ the
correlations should be the same, and
for large $\epsilon$ the heavier particles
drop faster according to their smaller $\epsilon_{max}$.

\noindent {\it b) dependence on the order $n$ of the correlation.} 
The rescaled quantities in Eq.~(\ref{epscal}) depend on $n$ 
in the small $\epsilon$ region; beyond $\epsilon_c$ they
depend only on the
1-particle spectrum and are independent of $n$.

It should be noted that the separation of these regimes is incomplete 
at finite energies and the characteristic features are realized away from
the critical angle. 
In the next subsections more details on specific observables are reported.

\subsection{Distribution in  the Relative Solid Angle $\vartheta_{12}$}

This is the simplest correlation variable.
There are two observables for which results have been obtained.

\subsubsection{Normalized Correlation Function and Test of Angular Scaling}
Besides the unnormalized correlation defined in Eq.~(\ref{d2def})
one also considers the normalized correlations
\begin{equation}
r(\vartheta_{12})=\frac{dN^{(2)}/d \vartheta_{12}} 
                  {dN_{\rm prod}^{(2)}/d\vartheta_{12}}, \;\;\qquad 
\hat r(\vartheta_{12})=\frac{dN^{(2)}/d \vartheta_{12}}
                       {N^2}
\labl{r2}
\end{equation}
In the first observable $r$ (``correlation integral"
\cite{lceb}) the normalizing quantity is defined as in
(\ref{d2def}) but with $dN^{(2)}/d\Omega_1 d\Omega_2$
replaced by $(dN^{(1)}/d\Omega_1) \; (dN^{(1)}/d\Omega_2)$\footnote{
In the experiment one takes the angle $\vartheta_{12}$
between particles of different events (``event mixing").};
in the second observable $\hat r$ the square 
of the particle multiplicity $N^2(E,\Theta)$ in the
forward cone defines the normalizing factor. The first quantity
is more sensitive to the non-trivial correlations as it
measures the deviations from the distribution of uncorrelated
pairs, but it depends more critically on the choice of the jet
axis in the definition of the angles $\Omega_i$; the
second quantity depends only weakly on the jet axis through the opening
angle $\Theta$.

The direct term in Eq.~(\ref{hnint}) is
given by $d^{(2)}(\vartheta_{12},E)=dN^{(2)}_{\rm prod} /d\vartheta_{12}$
 as in Eq. (\ref{r2}).
The leading contribution in pole approximation is given by
Eq. (\ref{dinh}) for $n = 2$, $\delta = \vartheta_{12}$.
The correlation functions are obtained from Eq.~(\ref{hsol})
\begin{eqnarray}
 r(\vartheta_{12})&=&\exp \left(2\beta
 \sqrt{\ln(E\Theta/\Lambda)}
\left(\omega(\epsilon,2)-2\sqrt{1-\epsilon}\right)
    \right)
     \labl{rres},\\
 \hat r(\epsilon)&=&\bar b \exp \left(\bar b
     \left(\omega(\epsilon,2)-2\right)
    \right), \quad \bar b=2\beta\sqrt{\ln(E\Theta/\Lambda)}.
     \labl{rhres}
\end{eqnarray}
Note that in the correlation $r(\vartheta_{12})$
the leading term in the exponent 
from the uncorrelated pairs is canceled, see Eq.~(\ref{omln}),
whereas it contributes to the correlation $\hat{r}(\epsilon)$.
The correlation $\hat{r}(\epsilon)$ in
(\ref{rhres}) is defined as in Eq.~(\ref{r2}) but in terms of the
variable $\epsilon$ instead of $\vartheta_{12}$.


In the kinematic region of large $\vartheta_{12}$ or small $\epsilon$,
 when the linear
approximation Eq.~(\ref{omln}) for $\omega (\epsilon)$  applies,
 the correlation functions show power behaviour
\begin{equation}
 r(\vartheta_{12}) \simeq 
 \left({\Theta\over\vartheta_{12}}\right)^{{\gamma_0\over 2}},\qquad
\hat r (\vartheta_{12}) \simeq  {2\gamma_0 \over \vartheta_{12}}
\left({\Theta\over\vartheta_{12}}\right)^{-{3\gamma_0\over 2}},
  \labl{rpow}
\end{equation}
as expected from the selfsimilar structure of the fully developed cascade.

The first experimental
results on the $\vartheta_{12}$ correlation became available from
the DELPHI Collaboration
\cite{mb}.
Fig.~5.6 shows the quantity $-\ln( \hat r(\epsilon)/\bar b)$ from
Eq.~(\ref{rhres}) after rescaling by $2\sqrt{\ln E\Theta/\Lambda}$.
The data are in reasonable agreement with the
asymptotic prediction $\sim$ 2$\beta(1-\omega(\epsilon$, 2)/2).
Better agreement is obtained with the more precise Monte Carlo calculation
at partonic level (HERWIG \cite{HERWIG})
 in the full angular region supporting
once again the LPHD hypothesis. The same 
parton level Monte Carlo also reveals nicely
the angular scaling 
 property for not too small angles
($\epsilon \lapproxeq 0.5$) if one compares the $\sqrt s$ = 90 
and $\sqrt s$ =
1800 GeV results. The scaling region is below $\epsilon_c<0.7$, in
agreement with the general discussion above.
Very good agreement with 
the experimental data has been obtained by the
JETSET Monte Carlo \cite{JETSET} which includes hadronization.

An experimental  test of angular scaling 
 by varying the 
jet opening angle $\Theta $ is presented in Fig. 5.7 for the 
correlation $ r(\vartheta_{12})$. 
The rescaling of the horizontal
and vertical axis yields the approximate scaling for angles 
$\Theta \gapproxeq 30^0$.
The deviation at smaller angles most likely reflects the sensitivity 
to the jet axis (sphericity)
determination. This sensitivity is weak for the 
correlation $\hat r(\epsilon)$ through the cone opening angle $\Theta$
but proves to be more severe for the correlation
$r(\vartheta_{12})$ through the 1-particle angular distribution which enters
the normalizing factor. 
Whereas the prediction of scaling is sufficiently well satisfied the
prediction on the asymptotic shape from (\ref{rres}) reproduces only the
qualitative trend of the data. A computation which evaluates the integral
equation (\ref{hnint}) numerically at the LEP-1 energy and thereby takes
into account the cut-off $Q_0$ properly gives an improved agreement in the
shape.  It still differs though in the absolute normalization which enters
in NLO in the expansion of  $\sqrt{\ln E\Theta /\Lambda}$.

Results with less sensitivity to the jet axis could be obtained by 
using 
the corresponding $\rm EMM$-Correlations between 3
particles 
instead of the correlations $r$ and $\hat r$ 
-- as discussed for the azimuthal angle correlation
in Sect.~5.2 -- whereby 
two particles (1,2) define the correlations in $\vartheta_{12}$
as above and the third particle in turn with energy weight represents the jet   
axis.

\subsubsection{PPCA and Test of Color Coherence}

An interesting aspect of the relative 
spherical angle correlation is its sensitivity to the
soft gluon interference. This has been investigated experimentally
with the  observable  ``Particle-Particle Correlation''
(PPC) and its asymmetry (PPCA) in the relative angle $\chi \equiv \vartheta_{12}$
\cite{aleph7,ppc1,ppc2,ppc3}
which is closely related but not identical
to the above correlation $\hat r(\vartheta_{12})$:
\begin{eqnarray}
PPC(\chi)&=& \frac{1}{N_{ev}}\frac{1}{\Delta\chi}
\sum_1^{N_{ev}}\sum_{i=1}^{N_{ch}} \sum_{j=1,j\neq i}^{N_{ch}}
    \frac{1}{N_{ch}^2} \delta_{bin}(\chi-\chi_{ij})
    \labl{PPC}\\
PPCA(\chi)&=&PPC(180^0-\chi)-PPC(\chi)
\labl{PPCA}
\end{eqnarray}
with $N_{ch}$ and $N_{ev}$ denoting the number of 
charged tracks in an event
and the number of events respectively,
$\Delta \chi$  is the bin width. If the multiplicity $N_{ch}$ per event in
the definition (\ref{PPC}) were replaced by the average 
$\bar N_{ch}$ we had just obtained $\hat r (\vartheta_{12})$ from
the last subsection; also, one has the approximate relation 
$PPCA(\vartheta_{12}) 
\approx-(dN^{(2)}/d\vartheta_{12}-dN^{(2)}_{prod}/d\vartheta_{12})/\bar N$
if the forward-backward correlations of two jets across the hemispheres
are neglected.
The quantities
(\ref{PPC}),(\ref{PPCA}) are built in close analogy to the
well known Energy-Energy Correlations and their asymmetry
\cite{bbel,ddt1}
but without the energy weight factors in the respective sums.

These new quantities
have not been calculated analytically yet,
so only comparison with Monte Carlo
calculations are available. As an example, we show in
Fig.~5.8 the experimental results obtained by the L3 collaboration
\cite{ppc3} in comparison with 
the JETSET model \cite{JETSET} predictions with and without
soft gluon coherence included in the jet algorithm.
The models which include the coherence are 
generally in agreement with the experimental data,
except for very small angles $\chi \lapproxeq 6^0$ where the
Bose-Einstein effects are important as well.
The sharp dip visible for $\chi < 20^0$ can be  related to the peak in
$dN^{(2)}/d\vartheta_{12} \sim 1/\vartheta_{12}$ in the DLA
(note that the PPCA and $dN^{(2)}/d\vartheta_{12}$ have opposite sign).

The angular ordering condition implies that a soft gluon 
is emitted from parton 1 only
within an angular cone limited by the next colour connected parton
 ($\vartheta_{12}\leq\vartheta_{{\rm 1~next}}$).
One may estimate roughly \cite{wo}
$\vartheta_{1~next}\sim\bar\vartheta_{12}\sim\Theta/
\sqrt{\bar N}$ which corresponds to about
$1/\sqrt{15}{~\rm rad}\sim 15^0$.
One expects the emission within this angular region to be enhanced,        
and outside to be suppressed in comparison with the non-ordered case.
Interestingly, such an effect is indeed seen 
in Fig.~5.8 a,b, if ``JETSET coherent'' and ``JETSET incoherent'' are compared.
In the coherent case the correlation disappears almost completely 
for angles around 25$^0$. Remarkably, the interference effect expected for
gluons is actually visible at the  hadronic level.

These results suggest that the effect of angular ordering is most
clearly seen in the correlations in polar angles and less pronounced
in correlations in azimuthal angles or momenta.

\subsection{Fully Differential Multiparton Correlations}

The general $n$-particle cumulant correlation function 
$\Gamma^{(n)}(\Omega_1,\ldots,\Omega_n)$
has also
been studied in the DLA \cite{ow2}.
Consider first the simplest case $n=2$ for fixed $\as$, for which exact
results have been obtained. Starting again from Eq.~(\ref{hnint}) with
inhomogenous term $D^{(1)}(\Omega_1)D^{(1)}(\Omega_2)$ and angular
distribution 
$D^{(1)}(\Omega)=\gamma_0/2\pi
\vartheta^2\ \sinh (\gamma_0 \ln(E\vartheta/Q_0))$
one finds using Eqs.~(\ref{hresol}),(\ref{dtheta}) the exact result
\cite{ow2,jw}
\begin{eqnarray}
\lefteqn{\Gamma^{(2)}(\Omega_1,\Omega_2) =}
\nonumber \\& &
  = \frac{\gamma_0^2}{(2\pi)^2 \vartheta_1^2 \vartheta_{12}^2 }
\sum_{m=0}^\infty (2^{2m}-1) \left( \frac{l}{L_1} \right)^{m+1}
I_{2m+2}(2\gamma_0\sqrt{l L_1}) +(1\leftrightarrow 2).\labl{cora2}
\end{eqnarray}
Here $l=\ln{(E\vartheta_{12}/Q_0)},
L_1=\ln{(\vartheta_1/\vartheta_{12})}$, furthermore there is the constraint
 $\vartheta_1 (\vartheta_2) > \vartheta_{12}$ required by angular ordering.
At high energies one finds again the characteristic power behaviour
\begin{equation}
\Gamma^{(2)}(\Omega_1,\Omega_2)
\simeq \frac{\gamma_0^2}{2(4\pi)^2} \frac{1}{\vartheta_{12}^2}
\left( \frac{E\vartheta_{12}}{Q_0} \right)^{2\gamma_0}
\left ( \frac{1}{\vartheta_{1}^2}
 \left (\frac{\vartheta_1}{\vartheta_{12}}\right )^{\gamma_0/2}
+ \frac{1}{\vartheta_{2}^2}
\left (\frac{\vartheta_2}{\vartheta_{12}}\right )^{\gamma_0/2} \right).
\labl{Gas}
\end{equation}
By appropriate integrations one can obtain the corresponding results for
$dN^{(2)}/d\vartheta_{12},\; h^{(2)}(\Omega,\delta)$ and, finally, the
well known result for the global moment $F_2=4/3$.

For general $n$ and running $\as$ an expression has been derived
for the typical configurations where the relative angles  $\vartheta_{ij}$
between partons are of comparable order and much larger
than the asymptotically small angle $\Lambda/E$, so that
\begin{equation}  
\delta_{lm}^{ij}= \ln(\vartheta_{ij}/\vartheta_{lm})/ 
\ln(E\vartheta_{lm}/\Lambda)\gg 1\/.
\labl{delij}
\end{equation}
On the other hand, the quantity $\epsilon^i_{lm}=
\ln(\vartheta_i/\vartheta_{lm})/\ln
(E\vartheta_i/\Lambda)$ involving the polar angle $\vartheta_i$ is allowed to
vary within the full range 
$0\leq\epsilon^i_{lm}<1$. The correlation function is then given
as a sum of terms singular in the production angles $\vth_i$
\begin{equation}
\Gamma^{(n)}(\{\Omega\})\simeq
\left({f\over 4\pi}\right)^n {1\over n}
\sum_{i=1}^n {\gamma_0^{n\over 2}(E\vartheta_i)\over \vartheta_i^2}
\exp{\biggl(2\beta \omega(\epsilon_i,n)\sqrt{\ln E\vth_i/\Lambda}\biggr)}
   F_i^n (\{\chi\})\/.
\labl{gamom}
\end{equation}
Here $\epsilon_i=\ln(\vth_i/(\tij)_{{\rm min}})/\ln (E\vth_i/\Lambda)$
and $F_i^n(\{\chi\})$, the $i^{th}$ term for the $i^{th}$ particle,
is homogeneous of degree $p=-2(n-1)$ in
the relative angles built from factors $1/\tij^2$. In term $i$
the particle $i$ is connected with the initial parton direction
(factor $\vth_i^{-2})$ and all other partons $j,l$ are either
connected to $i$ (factor $\tij^{-2}$) or among themselves
(factor $\vth_{jl}^{-2}$). Again a power behaviour
results for sufficiently small $\epsilon_i$.

\subsection{Multiplicity Moments in Angular Bins}

In order to study genuine multiparticle angular correlations it is
convenient to integrate the correlation function over the
angular intervals as
in Eqs.~(\ref{fmomg}), (\ref{cmomg}) which yields the respective multiplicity
moments.

\subsubsection{Scaling Predictions and Universality}
Due to the singularity structure of the connected
correlation function just mentioned,
it is most convenient to consider the differential
 moments defined in Eq.~(\ref{hdef}),
where one particle is kept fixed at angle $\Om$ and the others
are counted in the cone $\delta$ around it.
The functions $\bar h^{(n)}(\Om,\delta)=\vartheta^2 h^{(n)}(\Om,\delta)$ 
fulfill the generic
integral equation (\ref{hnint}) with (\ref{dinh}) in the high energy limit
and the asymptotic solution is again given by Eq.(\ref{hsol}). 
In taking the direct term (\ref{dinh}) one applies KNO scaling which relates
the $n^{th}$ global moment to the $n^{th}$ power of multiplicity.

The multiplicity moments in the 
sidewise angular bins as defined in Eqs. (\ref{fmomg}),(\ref{cmomg})
are then obtained approximately as an integral over $\Om$ in
the appropriate interval $\gamma$:
\begin{equation}
c^{(n)} (\vth,\delta)=\int_{\gamma(\vth,\delta)} h^{(n)}
(\Omega,\delta) d \Omega \approx c_D\:
h^{(n)}(\Omega,\delta)\/,
\labl{c2app}
\end{equation}
where $c_1=2\pi \delta\vartheta$ for the ring (D=1)
and $c_2= \pi\delta^2$ for the cone (D=2).
Although in the 1D case all partons are counted in the ring,
in the calculation at leading order the full area is considered only
for the first parton, whereas the relevant phase space for the other partons
in the leading pole approximation
is the bin width $\delta$ and not the ring (see, e.g. \cite{dd}).
The latter approximation requires formally
$\Lambda/E \ll\delta\ll \vartheta$ and deteriorates for
$\delta\to \vartheta$ where the bremsstrahlung singularities
$1/\vartheta_i$ present for all partons become important.
For the normalized moments one obtains
after division by $N(\vth,\delta)^n$ with
multiplicity $N(\vth,\delta)\sim (\delta/\vth)^D 
\exp(2\beta\sqrt{\ln E\vth/\Lambda})$ \cite{ow2,dd,bmp}
\begin{equation}
M^{(n)}(\vth,\delta) \sim\left({\vth\over\delta}\right)^{\phi_n}, \qquad
\phi_n=D(n-1)-2\gamma_0(E\vth)(n-\omega(\epsilon,n))/\epsilon
\labl{cmomno}
\end{equation}
with the usual $\gamma_0=\beta/\sqrt{\ln E\vth/\Lambda}$.
This result for normalized moments has been given for the cumulant moments
\cite{ow2} or for the factorial moments \cite{dd,bmp}. The difference
comes from a different treatment of nonleading effects in the DLA and
vanishes in the high energy limit
(for $\epsilon$ fixed $M^{(n)}$ grows
$\sim \exp(c(n-1)\sqrt{E})$ whereas $F^{(n)}=C^{(n)}+O(C^{(n-1)})$). 
The more precise parton Monte Carlo results indeed show
the convergence of
both moments at higher energies \cite{ow2}. Furthermore,
these Monte Carlo results as
well as the first experimental data \cite{mb} give better agreement 
with the theoretical prediction  (\ref{cmomno}) 
at present energies if $M^{(n)}$ is taken to be the
factorial moment rather than the cumulant moment. 
%

The dependence of the moments in the DLA on the various
parameters can best be seen from the moments $\tilde F^{(n)}$,
 rescaled as in
(\ref{epscal}), which project out the $\epsilon$-dependent part of the
exponent of $F^{(n)}$.
They approach for $\epsilon$ fixed, $E\to\infty$ a scaling limit  
\begin{eqnarray}
\tilde F^{(n)} =-{\ln \lbrack(\delta/\vartheta)^{D(n-1)}
F^{(n)}\rbrack \over n\sqrt{\ln {E\vartheta\over \Lambda}}}
&\to&  2\beta (1-{\omega(\epsilon ,n)\over n})
\labl{rnall}\\
&\approx& 2 \beta(1-\sqrt{1-\epsilon})
\labl{rnallf}
\end{eqnarray}
independent of dimension $D$, and within the large $n$
approximation of Eq.~(\ref{omln}) 
also independent of the order
$n$ in Eq.~(\ref{rnallf}). 

In Fig.~5.9 (a) experimental data from DELPHI Collaboration 
\cite{mb} on the
second and third factorial moments are compared with the predictions 
(\ref{cmomno}) and in (b) on the rescaled moments  compared with
(\ref{rnall}). The qualitative trend of the data is well reproduced, and
the
agreement could be further improved by adjustment of the normalization which
is nonleading in DLA. 

The moments have also been computed with the parton Monte Carlo. These
results are in close agreement with the data of Fig.~5.6 and also with the
results from the hadron level Monte Carlo for not too small angles
$\epsilon\lapproxeq 0.5$. Also in this region 
the approximate angular scaling 
of $\tilde F^{(n)}$
in (\ref{rnall}) has been verified by the Monte Carlo for
different primary energies \cite{ow2}. In this region 
($\epsilon < \epsilon_c$) 
the theoretical predictions are independent of the cut-off $Q_0$ and it is in
this kinematic regime that the
hadronization effects are negligible and LPHD
works. As $\epsilon_c\to 1$ with increasing energies (for running $\as$)
the range of validity of LPHD is expected to increase as well in this limit.

The comparison of Figs. 5.6 and 5.9b supports the universality of angular
correlations according to Eq.~(\ref{epscal}) which relates quite different
observables: it is to be noted
that in the measurement of $\hat r$ and $F^{(2)}$ particle pairs  
from different regions of phase space are gathered, 
with the first quantity being
differential, and the second cumulant in the angles.

\subsubsection{Intermittency and Fractal Structure}
The behaviour of moments is qualitatively similar to the one
of $\rtwo(\t12)$ discussed above.
For small $\epsilon$, i.e.\ sufficiently large opening angle
$\delta$ (or for constant $\alpha_s$) we can apply the linear
approximation (\ref{omlin})
to $\omega (\epsilon,n)$ and find the power law
for moments
\begin{equation}
M^{(n)}(\vartheta,\delta)
\sim \left({\vth\over\delta}\right)^{\phi_n},\qquad
\phi_n=D(n-1)-(n-{1\over n})\gamma_0(E\vth)\/.
\labl{cmomn}
\end{equation}
For larger $\epsilon$ (smaller angles $\delta$) the full result
(\ref{cmomno}) implies
a deviation from the power law, larger for $D=1$ than
for $D=2$, which eventually leads to a saturation and downward
bending of the moments. The above formulae cease to be valid beyond $\epsilon
\approx \epsilon_c$.
The dimension D enters the result only through 
the kinematic factor $(\vth/\delta)^{D(n-1)}$.

The power behaviour  of moments 
is also refered to as ``intermittency" \cite{bip}.
There have been detailed studies of the intermittency property of
factorial moments both experimentally and theoretically
in the different types of collision processes, motivated
originally by analogies to selfsimilar structures in dynamical statistical
systems, for reviews see, e.g., \cite{wdk,intref}.
Various qualitative trends of the data, e.g. dependence on the resolution scale
and on the dimension (D=1,2,3)
had been predicted from simple models of branching
processes \cite{bip,intmod}.
However, the above QCD results (for hard processes) are more
specific: they select the angles ($\vartheta,\varphi$) as the relevant
variables (rather than rapidity and $\varphi$) as a consequence of angular
ordering (see e.g. \cite{bmp}). Also the power law occurs only for the fully
developed cascade at sufficiently large angles ($\epsilon$ small) whereas
most phenomenological studies discuss the power law at small angles in
the limit $\delta\to 0$. Finally, the intermittency exponents $\varphi_n$
became quantitatively calculable in QCD in terms of the multiplicity anomalous
dimension $\gamma_0\sim \sqrt{\alpha_s}$, see Eq.~(\ref{cmomn}); this
implies a logarithmic variation of the intermittency exponent 
with the jet virtuality $E\Theta$ and, therefore,
a weak breaking of selfsimilarity.

The power law for moments can also be interpreted in terms
of the fractal structure of the jets. The fractal dimension
$D_F$ is derived from the result (\ref{cmomn}) for  $\vph_n$,
 see e.g. \cite{wdk,bl,cello}, as
\begin{equation}
\vph_n=(n-1)(D-D_F^n).
\labl{dfdef}
\end{equation}
 The dimension $D_F^n$
has a simple geometrical interpretation \cite{cello}.
An equipartition of points in $D$ dimensions yields a Poisson
distribution for a subinterval of any size. Therefore one obtains for
the
factorial moments $F^{(n)}=1$ and $\vph_n=0$, i.e.\
the fractal dimension equals the topological dimension $D^n_F=D$.
On the other hand, if in a subdivision into smaller intervals only
one interval is occupied and the others are empty then
$\vph_n=(n-1)D$ and $D_F^n=0$ as intuitively required.
>From (\ref{cmomn}) for $\vph_n$ one finds for the
fractal dimension independent of dimension $D$ the simple
result
\begin{equation}
D^n_F={n+1\over n} \gamma_0\/.
\labl{dfrac}
\end{equation}
For large $n$ the fractal dimension approaches the anomalous
dimension $D^n_F\approx \gamma_0$, and so there is a geometrical
interpretation of the QCD quantity $\gamma_0$. 

More analytical results on the fractal structure of QCD jets concerning
the multifractal approach and Levy indices have been presented in \cite{dd}.
The fractal properties of multiplicity fluctuations have been addressed
also in \cite{gn} in a study of
related observables (``x-curve").
The results for unnormalized moments are found as in
(\ref{dfrac}) in the large $n$ approximation;
the numerical results on normalized quantities
are consistent with (\ref{dfrac}).
 
\subsubsection{MLLA Prediction on Fractal Dimension}
Corrections to the leading DLA results on moments
have been computed in the 
MLLA \cite{dd}. One important effect is the softening of the particle
spectrum because of energy conservation. This can be taken into account by
calculating the distribution $D^{\delta}$ and multiplicity $N$ in
Eqs.~(\ref{hresol}) and (\ref{delgen}) in MLLA accuracy with the help of the
anomalous dimension $\gamma_{\omega}$ from Eq.~(\ref{2.33}).

Explicitly the fractal dimension in (\ref{dfdef}) has been derived.
The result is as in  
(\ref{dfrac}) but with $\gamma_0$ replaced by
\begin{equation}
\gamma=\gamma_0+\gamma_0^2\frac{b}{4 N_C}
  \left[-B\frac{n-1}{2(n+1)}+\frac{1}{4}+
   \frac{n-1}{2(n+1)(n^2+1)}\right]+O(\gamma_0^3).
\labl{gamma}
\end{equation}
The $n_f$-dependent constants $B$ and $b$ 
are as in (\ref{2.35}). For large ranks
$n\gg1$ one obtains a negative shift
\begin{equation}
\gamma=\gamma_0+\gamma_0^2\frac{b}{4 N_C}
    \left[-\frac{B}{2}+\frac{1}{4}\right].
\labl{gamoa}
\end{equation}
This corresponds to the MLLA correction to the multiplicity 
growth Eq.~(\ref{gamoa}) which
dominates at large $n$ in the convolution (\ref{hresol}) over the energy
loss correction in $D^{\delta}$. The numerical coefficients of $\gamma_0^2$
in (\ref{gamma}) for $n_f=3$ light quark flavours 
have been computed as  (+0.057, -0.028,
-0.107, -0.153) for $n$=2...5 and -0.280 for $n=\infty$. The corresponding
corrections for the slopes $\varphi_n$ are of the order of 10\%.

Finally, a word of caution on the accuracy of the
 analytical results for angular correlations reported in this section.
There are some effects which are not yet included
in the present treatment of moments. There is the neglect of differences
between the factorial moments and ``normal" moments $<N^n>$ 
and the KNO scaling assumption which enter Eq.~(\ref{hresol}). This
formally requires the multiplicity factors $N(E\delta)$ to be large.
Another uncertainty comes from the definition of the jet axis which is
particularly important for the correlation $r(\vartheta_{12})$ in Sect.~5.3
and can, in principle, be avoided by considering the corresponding EMM
Correlations \cite{dkmw1,dmo}. 
On the other hand, if the above quantities are calculated with
a parton level Monte Carlo good agreement is found with hadronic data for
not too small relative angles where $Q_0$ cut-off effects are absent (for
$\epsilon<\epsilon_c$). Therefore,
angular correlations for sufficiently large
relative angles provide evidence in favour of LPHD, despite the remaining
uncertainties in the analytical results.

\section{QCD Collective Effects in Multi-jet Events}
\setcounter{equation}{0}

\subsection{Radiophysics of Particle Flows}

The colour structure of QCD is now well tested by the LEP-1 and
Tevatron experiments, see e.g.\ Refs.\ \cite{bp,ms,brw1,hm}.  As we
have already discussed, in the framework of APA an intimate
connection is expected between the multi-jet event structure and
the underlying colour dynamics at small distances.  The detailed
features of the parton-shower system, such as the flow of colour
quantum numbers, influence significantly the distribution of
colour-singlet hadrons in the final state (see e.g.\
\cite{dkmt2}).  Such a phenomenon was first observed about 15
years ago by the JADE Collaboration \cite{jade} studying the
angular flows of hadrons in three-jet $(q\overline{q}g)$ events
from $e^+ e^-$ annihilation, the string \cite{ags} or drag
\cite{adkt2} effect.  The data clearly show in favour of QCD
coherence.  In particular, less particles were found to be
produced in between the $q$ and $\overline{q}$ jets as compared
to the other two inter-jet regions.  This observation was later
confirmed by other $e^+ e^-$ experiments \cite{tpcc}.  The
PETRA/PEP data have convincingly demonstrated that the wide-angle
particles do not belong to any particular jet but have emission
properties dependent on the overall jet ensemble.

The interjet coherence phenomena were then successfully studied
at LEP-1 (e.g.\ \cite{h,ms,al1,wjm1,ts1,wjm2}), TRISTAN (e.g.\
\cite{my1,amy,amysj}) and TEVATRON (e.g.\ \cite{pb1,cdf1,d0}).  These
experimental findings have beautifully demonstrated the connection
between colour and hadronic flows. Surely, it is entirely
unremarkable that the quantum mechanical interference effects are
observed in QCD.  Of real importance is that the experiment
proves that these effects survive the hadronization phase.  The
current experimental situation is briefly discussed in the next
section.

The interjet coherence deals with the angular structure of
particle flow when three or more hard partons are involved.  The
hadron distribution proves to depend upon the geometry and colour
topology of the hard parton skeleton.  The clear observation of
interjet interference effects gives another strong evidence in
favour of the LPHD concept.  The collective nature of
multiparticle production reveals itself here via the QCD wave
properties of the multiplicity flows.

The detailed experimental studies of the colour-related effects
are of particular interest for better understanding of the
dynamics of hadroproduction in the multi-jet events.  For
instance, under special conditions some subtle interjet
interference effects, breaking the probabilistic picture, may
even become dominant, see Refs.\ \cite{adkt2,dkt2}.  We remind
the reader that QCD radiophysics predicts both attractive and
repulsive forces between the active partons in the event
\cite{dkmt1,dkmt2}.  Normally the repulsion effects are small,
but in the case of colour-suppressed $O \left (\frac{1}{N_C^2}
\right )$ phenomena they may play a leading role.  It should be
noted that within the APA picture the interjet collective
effects are viewed only on a completely inclusive basis, when
all the colour emitters are simultaneously active in particle
production.  There is an important difference between the
perturbative radiophysics scenario and the parton-shower Monte
Carlo models.  The latter not only allow but even require a
completely exclusive probabilistic description (e.g.\ the
negative-sign interference effects are, as usual, 
absent)\footnote{New ideas of how to incorporate the colour-suppressed 
terms into the Lund dipole cascade picture \cite{gg1} were advocated 
in Ref.\ \cite{gg2}.  This opens a possibility \cite{gg3} to include 
the so-called \lq\lq negative dipoles" into the ARIADNE Monte Carlo 
model \cite{ARIADNE}.}.  Normally (such as in the case of $e^+ e^- 
\rightarrow q\overline{q}g$, see next subsection) the two pictures 
work in a quite peaceful coexistence; the difference only becomes drastic 
when one deals with the small colour-suppressed effects.

Let us emphasise that the relative smallness of the non-classical
effects by no means diminishes their importance.  This
consequence of QCD radiophysics is a serious warning against the
traditional ideas of independently evolving partonic subsystems. 
So far (despite the persistent pressure from the APA side) no
clear evidence has been found experimentally in favour of the
non-classical colour-suppressed effects in jets, and the peaceful
coexistence between the perturbative interjet coherence and
colour-topology-based fragmentation models remains unbroken. 
However, these days the colour suppressed interference effects
attract increased attention.  This is partly boosted by the
findings (e.g.\ \cite{sk1,vak3,guha} that the QCD interference
(interconnection) between the $W^+$ and $W^-$ hadronic decays
could affect the $W$ mass reconstruction at LEP-2.  Clear
understanding of the dynamics of the colour-suppressed phenomena
is of importance also for elucidation of the role of perturbative
mechanism of the so-called rapidity gap
events in $e^+ e^-$ annihilation (e.g.\ \cite{bbl,er}),
 see for more details Ref.\
\cite{kms}.  These new developments make the systematic detailed
analysis of the colour-related effects based on the massive LEP-1
statistics even more desirable.

Finally, let us recall that the colour-related collective
phenomena could become a phenomenon of large potential value as a
new tool helping to distinguish the new physics signals from the
conventional QCD backgrounds (e.g.\ \cite{dkt1,dkmt2}).  The reader 
is reminded that when the energy (hardness) of a process is increased, 
the amount of cascading is sharply rising.  This leads to high particle 
multiplicities, and thus larger experimental challenges and complexities.  
For instance, for the LHC the expected mean charged multiplicity per event 
is above 100.  This means that at nominal luminosity, any physics will 
have to be dug out among 4000 charged or neutral particles, e.g. \cite{ts1}.  
Therefore, the new ideas providing one with the analysis strategy that cuts 
away these backgrounds are very desirable.

\subsection{String/Drag Effect in $q\overline{q} g$ Events}

The first (and still best) example of the interjet colour-related
phenomena is the string/drag effect in $e^+ e^- \rightarrow
q\overline{q}g$.  Since the new results based on the refined 
analysis of $q\overline{q}g$ events continue to pour out from the
LEP-1 collaborations \cite{wjm2,ts1,al1} it seems useful to
recall the main ideas \cite{adkt2} of the perturbative
QCD explanation of this
bright coherence phenomenon.

\subsubsection{Interjet Particle Flows}

We consider first the angular distribution of particle flows at
large angles to the jets in $e^+ e^- \rightarrow q\overline{q}g$.
The more general case including the intrajet particle flow will be discussed
below in subsection 6.3.3.

Let all the angles
$\Theta_{ij}$ between jets and the jet energies $E_i$ be large
($i=\{+-1\}\equiv \{q \overline q g\}$)\ : 
$\Theta_{+ -} \sim \Theta_{+1} \sim \Theta_{-1} \sim 1, E_1 \sim
E_+ \sim E_- \sim E \sim W/3$.  As was discussed above, within
the perturbative picture the angular distribution of soft interjet
hadrons carries information about the coherent gluon radiation
off the colour antenna formed by three emitters $(q, \overline{q}
\: {\rm and} \: g)$.  The wide-angle distribution of a secondary
soft gluon $g_2$ displayed in Fig.\ 6.1 can be written as
\begin{equation}
\frac{8 \pi d N_{q\overline{q}g}}{d \Omega_{\vec{n}_2}} \; = \;
\frac{1}{N_C} \; W_{\pm 1} \: (\vec{n}_2) \: N_g^\prime (Y_m) \;
= \; \left [ ( \widehat{1 +}) + ( \widehat{1 -}) \: - \:
\frac{1}{N_C^2} \; ( \widehat{+ -}) \right ] \; N_g^\prime (Y_m).
\label{4.1}
\end{equation}
\indent Here the \lq\lq antenna" $(\widehat{i j})$ is represented as
\begin{equation}
(\widehat{i j}) \; = \; \frac{a_{ij}}{a_i a_j}, \;\;\;\; a_{ij}
\; = \; (1 \: - \: \vec{n}_i \vec{n}_j), \;\;\;\; a_i \; = \; (1
\: - \: \vec{n}_2  \vec{n}_i)
\label{4.2}
\end{equation}
and  $N_g^\prime (Y_m)$ is the so-called cascading factor
taking into account that a final soft particle is a part of
cascade (see \cite{adkt2,dkmt2} and subsection 5.2); furthermore, $Y_m = \ln E
\Theta_m/\Lambda$, where one defines the angle
$\Theta_m = {\rm min} \{\Theta_+, \Theta_-
, \Theta_1 \}$ with $\cos \Theta_i = \vec{n}_2 \vec{n}_i$ for $i =  
\{+,-,1\}$.

The radiation pattern corresponding to the case when a photon
$\gamma$ is emitted instead of a gluon reads (cf. eq.~\ref{mfl})
\begin{equation}
\frac{8 \pi d N_{q\overline{q}\gamma}}{d \Omega_{\vec{n_2}}} \; =
\;
\frac{1}{N_C} \; W_{+ -} (\vec{n}_2) \: N_g^\prime (Y_m) \; = \; 
\frac{2C_F}{N_C} \; (\widehat{+ -}) \:  N_g^\prime (Y_m).
\label{4.3}
\end{equation}
\noindent The dashed line in Fig.\ 6.2 displays the corresponding
\lq\lq directivity diagram", which represents the particle density
(\ref{4.3})
projected onto the
$q\overline{q}\gamma$
plane:
\begin{eqnarray}
W_{+ -} (\phi_2) & = & 2C_F \; \int \; \frac{d \cos \Theta_2}{2}
\; 
(\widehat{+ -}) \; = \; 2C_F \: a_{+ -} \: V (\alpha, \beta),
\label{4.4a} \\ 
V (\alpha, \beta) & = & \frac{2}{\cos \alpha - \cos \beta} \:
\left ( 
\frac{\pi - \alpha}{\sin \alpha} \: - \: \frac{\pi - \beta}
{\sin \beta} \right); \: \alpha \: = \: \phi_2,\;\; \beta \: = \:
\Theta_{+ -} - \: \phi_2.
\nonumber
\end{eqnarray}
\noindent Expression $W_{+ -} (\vec{n}_2)$ is simply related to
the particle
distribution in two-jet events $e^+ e^- \rightarrow q(p_+) +
\overline{q}
(p_-)$, Lorentz boosted from the quark {\it cms} to the lab system.

Replacing $\gamma$ by $g_1$ changes the directivity diagram
essentially because
the antenna element $g_1$ now participates in the emission as
well.  However,
this change leads not only to an appearance of an additional
particle flow in 
the $g_1$ direction.  Integrating Eq.\ (\ref{4.1}) over
$\Theta_2$ one obtains
$(\gamma = \Theta_{1+} + \phi_2)$:
\begin{equation}
W_{\pm 1} (\phi_2) = N_C \: \left [ a_{+1} V (\alpha, \gamma) +
a_{-1} V (\beta,
\gamma) - \frac{1}{N_C^2} \: a_{+ -} V (\alpha, \beta) \right ].
\label{4.5}
\end{equation}
\noindent Fig.\ 6.2 illustrates that the particle flow in the
direction opposite 
to $\vec{n}_1$ appears to be {\it considerably lower} than in the
photon case.  
So, the destructive interference diminishes radiation in the
region between the 
quark jets giving a surplus of radiation in the $q-g$ and
$\overline{q} - g$ 
valleys.  One easily sees that the colour-coherence phenomena
strongly affect the
total three-dimensional shape of particle flows in three-jet
events, practically 
excluding the very possibility of representing it as a sum of
three parton 
contributions.

Let us recall here that, owing to coherence, the
radiation of a soft 
gluon $g_2$ at angles larger than the characteristic angular size
of each parton 
jet proves to be insensitive to the jet internal structure: 
$g_2$ is emitted 
by a colour current which is conserved when the jet splits.  This
is the reason 
why one may replace each jet by its parent parton with $p_i^2
\approx 0$.

For a clean experimental verification of the drag phenomenon it
was found 
\cite{adkt2} to be convenient to examine the fully symmetric
$q\overline{q}g$ 
events (Mercedes-type topology), when $\vec{n}_+ \vec{n}_- =
\vec{n}_+ \vec{n}_1 
= \vec{n}_- \vec{n}_1$.  Contrary to the asymmetric
configurations, the Mercedes 
events are practically unbiased due to kinematic selections. 
For illustration 
let us consider the fully symmetric topology and take $\vec{n}_2$
pointing in the  
direction opposite to $\vec{n}_1$, that is, midway between
quarks.  Then, 
neglecting the weak dependence $N_g^\prime$ on $\Theta$, one
arrives at the 
ratio (numbers for $N_C=3$)
\begin{equation}
\frac{d N_{q\overline{q}g}/d \vec{n}_2}{d
N_{q\overline{q}\gamma}/
d \vec{n}_2} \; = \; \frac{N_C^2 - 2}{2 (N_C^2 - 1)} \; \approx
\; 0.44.
\label{4.6}
\end{equation}
The corresponding ratio for the projected particle flows  obtained from 
(\ref{4.4a}) and (\ref{4.5}) reads
\begin{equation}
\frac{d N_{q\overline{q}g}/d \varphi_2}{d
N_{q\overline{q}\gamma}/
d \varphi_2} \; \approx \; \frac{9 N_C^2 - 14}{14 (N_C^2 - 1)} \; \approx
\; 0.60.
\label{rphigg}
\end{equation}
 We emphasize that Eq.\ (\ref{4.1}) provides not only the
planar
picture, but the global three-dimensional wide-angle pattern of
particle flows.
  
 For $q\overline{q}g$ events, owing to the
constructive 
interference, the radiation is strongly enhanced in the $q - g$ and
$\overline{q} - g$ 
valleys:  for the Mercedes-like configuration midway between the
jet directions 
one arrives at the asymptotic ratio
\begin{equation}
\frac{d N_{\langle qg \rangle}/d\vec{n}_2}{d N_{\langle q\overline{q}
\rangle}/d\vec{n}_2} \; = \; 
\frac{5 N_C^2 - 1}{2 N_C^2 - 4} \; \approx \; 3.14.
\label{4.9}
\end{equation}
The projected particle flow ratio for the same configuration is given
approximately by
\begin{equation}
\frac{d N_{\langle qg \rangle}/d\varphi_2}{d N_{\langle q\overline{q}
\rangle}/d\varphi_2} \; = \; 
\frac{15( 5N_C^2 - 1)}{4(9 N_C^2 - 14)} \; \approx \; 2.46.
\label{rphiqg}
\end{equation}
As can be seen from (\ref{rphigg}) and (\ref{rphiqg}) the effect in the
projected flows is smaller than in the differential flows of
(\ref{4.6}) and (\ref{4.9}).

It is worth noting that the destructive interference proves to be
strong enough 
to dump the particle flow in the direction opposite to the gluon
to even smaller 
values than that in the most kinematically \lq\lq unfavourable"
direction, which 
is transverse to the event plane.  The asymptotic ratio of these
flows in the 
case of Mercedes-like $q\overline{q}g$ events is
\begin{equation}
\left ( \frac{d N_{\perp}}{d N_{\langle
q\overline{q} \rangle}}
\right )^g \; = \; \frac{N_C + 2C_F}{2 (4C_F - N_C)} \; \approx
\; 1.2.
\label{4.7}
\end{equation}
\noindent In the case of $q\overline{q} \gamma$ events this ratio
is
\begin{equation}
\left ( \frac{d N_{\perp}}{d N_{\langle
q\overline{q} \rangle}}
\right )^\gamma \; = \; \frac{1}{4}.
\label{4.8}
\end{equation}
\noindent It is of interest to test the ratios (\ref{4.7}),
(\ref{4.8}) 
experimentally.  

 For hadron states associated with the $\overline{q}qg$
and 
$\overline{q}q\gamma$ events the ratios of particle flows given
by Eqs.\ (\ref{4.6}-\ref{4.8})
 should remain asymptotically correct, since
non-perturbative 
hadronization effects are expected to cancel at very
high energies.  
It is an interesting question, to what extent the  non-perturbative effects
are important at
 present energies
 for a  quantitative description of the
data.

\subsubsection{Perturbative and Nonperturbative Effects and Colour
topology} 

Thus, the analysis of the bremsstrahlung pattern clearly
demonstrates particle 
\lq\lq drag" by the gluon jet $g_1$.  If one drops the colour
suppressed 
contribution, the two remaining terms in Eq.\ (\ref{4.1}) may be
interpreted 
as the sum of two independent $(\widehat{1+})$ and
$(\widehat{1-})$ antenna 
patterns, boosted from their respective rest frames into the
overall 
$q\overline{q}g$ {\it cms}.  The point is, that by neglecting the
$1/N_C^2$ terms, 
the hard gluon can be treated as a quark-antiquark pair.  In this
approximation 
each external quark line is uniquely connected to an external
antiquark line of 
the same colour, forming colourless $q\overline{q}$ antennae.  In
the general 
case, when calculating the resultant soft radiation pattern, one
can only deal 
with a set of such colour-connected $q\overline{q}$ pairs because
the 
interference between gluons emitted from {\it
non-colour-connected} lines proves 
to be suppressed by powers of $1/N_C^2$ \cite{dkmt2}.

Along this line of reasoning the depletion of the
$q-\overline{q}$ valley is a 
direct consequence of Lorentz boosts.  Such a scenario reproduces
literally the 
explanation of drag/string effect given by the standard Lund
string model 
\cite{ags,agis} where particles are created from the breakup of a
colour 
flux-tube stretching from the quark to the gluon and from the
gluon to the 
antiquark, see Fig.\ 6.3. Notice that within the latter model
there is no string 
piece spanned directly between the quark and antiquark, no
particles are produced 
in between them, except by some minor \lq\lq leakage" from the
other two regions.

So, one sees that for $q\overline{q}g$ physics, as
well as in many 
other cases, the Lund string scenario provides an excellent
picture for reproducing in the large $N_C$ limit the
QCD phenomena that reflects radiophysics of particle flows.  Let
us emphasise 
that the interjet coherence \cite{adkt2} is a purely perturbative phenomenon
and the string 
effect \cite{ags} is a non-perturbative one, but both have one
important 
phenomenon in common, namely a respect for the colour flow
topology of events.

However, at some point significant differences should
certainly manifest themselves 
(e.g.\ \cite{dkmt2}).  First of all, one might expect a serious
discrepancy 
between QCD and the string model when studying some specific
phenomena where 
colour non-leading contributions dominate.  Thus, if the $q$ and
$\overline{q}$ 
in three-jet events are close to each other, the colour-suppressed 
term (negative antenna $(\widehat{+ -})$ in
Eq.\ (\ref{4.1})) 
is kinematically enhanced, and is no longer negligible (see
subsection 6.4).  
A discriminative test between the string picture and perturbative
scenario can be
provided by the dependence of the multiplicity flow on the
particle mass $m_h$ 
or $p_h^{out}$ (momentum out of the event plane)
\cite{adkt2,kll}.  First of all, 
let us notice that the canonical string scenario predicts a {\it
larger} drag 
for heavier hadrons and for particles carrying a larger momentum
component 
perpendicular to the event plane, $p_h^{out}$.  The reasons are
the following.  
The Lorentz transformation of momentum $\vec{p}_i$ along the
boost direction is 
given by
$$
\vec{p}_i \vec{n}_B \; \rightarrow \; \gamma_B (\vec{p}_i
\vec{n}_B \: + \: 
\beta_B E_i),
$$
\noindent where $\vec{n}_B$ is a unit vector in the boost
direction, $\gamma_B$ 
and $\beta_B$ denote the $\gamma$-factor and the boost speed, and
$E_i$ stands 
for the particle energy.  
Whereas in the $cms$ the momentum components $\vec{p}_i \vec{n}_B $ have
both signs with equal probability, the positive sign dominates after the
boost and more particles are dragged into one hemisphere.
Since $E_i = \sqrt{p_i^2 + m_i^2}$ and
$p_i^2 = 
(p_i^{in})^2 + (p_i^{out})^2$, the effect becomes larger as $m_h$
or $p_h^{out}$ 
gets larger.  In addition, massive resonance decays are expected
to influence 
the particle flow in the out-of-plane direction.  For low
$p_h^{out}$, and at 
lower {\it cms} energies, the decay products do not always follow the
direction of the 
parent particle, and the drag effect may be diluted.

Within the perturbative approach there are no reasons to expect
any substantial 
enhancement for subsamples with large $p_h^{out}$ or $m_h$ since,
provided the 
energy is high enough for QCD cascades to dominate, the yields of
hadrons with 
different masses should be similar.  In what concerns the
$p_h^{out}$ dependence, 
even an opposite statement could be made:  with increase of
$p_h^{out}$ one 
leaves the event plane and the ratio of QCD motivated particle
flows in $q-g$ 
and $q-\overline{q}$ valleys can be easily shown to approach
unity instead of 
becoming larger \cite{adkt2}.

The first lower energy data \cite{jade,tpcc} demonstrated that
the string effect 
was enhanced by increasing $m_h$ or $p_h^{out}$.  If interpreted
in terms of a 
Lorentz boosted string, this enhancement would be of the same
strength at all 
c.m.\ energies, given fixed angles between jets.  As discussed
above, within the 
perturbative picture no enhancement is expected at high energies
for the 
subsamples with large $m_h$ or $p_h^{out}$.  While the
subasymptotic effects 
may play an important role at low energies, ultimately the
perturbative picture 
prevails at higher energies.  For illustration, we present in
Fig.\ 6.4 the energy evolution of the particle flow ratios
calculated within the 
ARIADNE model (\cite{ARIADNE}), which contains both soft-gluon
emission and 
string fragmentation.  One can easily see that the perturbative
regime here 
becomes more important than the string fragmentation
already at LEP-1 energies.  Such a test reflects
mainly the 
cascading property of the perturbative picture.  

A somewhat
different way to 
distinguish the (semihard) perturbative and non-perturbative
predictions was 
proposed in Refs.\ \cite{ts4,ts5}.  The idea is as follows. 
Consider a \lq\lq 
symmetric" 3-jet event and denote by $E_{qg} (E_{q\overline{q}})$
the energy flow 
midway between the $q$ and $g$ (between the $q$ and
$\overline{q}$) directions.  
In the string picture, $E_{qg}$ and $E_{q\overline{q}}$ are
essentially 
independent of the {\it cms} energy.  In the perturbative framework, on
the other hand, 
because of gluon emission one expects $E_{qg}, E_{q\overline{q}}
\sim \alpha_s 
W$.  The energy difference
\begin{equation}
\Delta E \; = \; E_{qg} (W) - E_{q\overline{q}} (W) \; = \; C_1
\: + \: C_2 \:
\alpha_s \: W
\label{4.10}
\end{equation}
\noindent would illustrate the contrast between the two
approaches.  The 
presence of both these terms in the data would demonstrate the
coexistence of 
the complementary contributions.  Needless to say, the
experimental problems in 
the case of both tests discussed above are considerable. 
Nevertheless, such 
measurements are of salient importance since they could provide
one with some 
clue to how to discriminate between the perturbative and
non-perturbative stages.

As shown in Ref.\ \cite{dkt11}, a new manifestation of the QCD wave 
nature of hadronic flows arises from studying the double-inclusive 
correlations of the interjet flows in $e^+ e^- \rightarrow q\bar qg$.
The point here is that one faces such subtle effects as the mutual influence 
of different $q\overline{q}$ antennae.  As a result, one finds:
\begin{equation}
\frac{d^2 N}{d \Omega_{1+} d \Omega_{1-}} \: / \: \frac{d^2 N}{d \Omega_{+-} 
d \Omega_{1-}} \: < \: \frac{dN}{d \Omega_{1+}} \: / \: \frac{dN}{d \Omega_{+-}}
\label{4.11}
\end{equation}
\noindent (see \cite{dkt11} for further details).  Here $(ij)$ denotes the 
direction midway between the partons $i$ and $j$.  Such effects cannot be 
mimicked by the orthodox Lund string model.  Only more sophisticated dipole 
formalism \cite{gg1} for the Lund Monte Carlo may reproduce them.

Finally, let us mention that colour coherence also leads to a rich diversity 
of collective drag effects in high-$p_\perp$ hadronic collisions (see Refs.\ 
\cite{dkt1,dkmt1,dkmt2,dkt8,emw}) and in Deep-Inelastic scattering \cite{akl}.  
One of the brightest examples is the colour drag in high-$p_\perp$ processes, 
such $\gamma, \mu^+ \mu^-$ pair, $W, Z$ production, where a colourless 
object is used as a trigger \cite{dkmt1}.

\subsection{Particle Flow Structure in Multijet Ensembles}
\subsubsection{QCD Portrait of Multijet Events}

As we discussed above, the collimation of the intrajet QCD
cascade around the 
parent parton becomes stronger as the jet energy increases.  If
one keeps the 
angle between the two jets fixed, then with the increase of the total
energy these 
jets become more and more distinguishable experimentally. 
Moreover, the 
collimation of an energy flux grows much more rapidly as compared
to a 
multiplicity flow.  Therefore, at asymptotically high energies
each event 
should possess the clear geometry, that reflects the topology of
the final 
hadronic system in terms of partons which participated directly
in the hard 
interaction.  The inclusive space-energy portrait of events
represents a natural 
partonometer for registration of the kinematics of the energetic
parent partons.  
While the hard component of a hadron system (a few hadrons with
energy fraction 
$z \sim 1$) determines the partonic skeleton of an event, the
soft component 
(hadrons with $z \ll 1$) forms the bulk of multiplicity
\cite{dkmt2,dkt10,dkt11}.

Closely following the radiation pattern, associated with the
partonic skeleton, 
the soft component is concentrated inside the bremsstrahlung
cones of QCD jets.  
Theoretically, the opening angle of each cone $\Theta_0$ is
bounded by the nearest 
other jet, since at larger angles $\Theta > \Theta_0$ particles
are emitted 
coherently by the overall colour charge of both jets.  As the
result, the total 
perturbative multiplicity is given by the additive sum of the
contributions of 
the bounded individual jets.

The interjet hadrons which at present energies form a sizeable
part of the total 
event multiplicity are distributed according to the colour
properties of the 
event as a whole.  Therefore, as a matter of principle, they
cannot be associated 
with any particular jet.  As a result, any attempt to force
particles to be 
assigned to a certain jet in an event may cause some problems.

As higher energies are attained the purely inclusive and
calorimetric 
characteristics for quantitatively dealing with hard collisions
become preferable 
to organizing the event according to a certain number of jets
(see e.g.\ 
\cite{dkmt2}).  There is in general a direct correspondence
between the jet 
directions and energy flux directions, so that one may naturally
study the jet 
shapes and any other characteristics of the hadronic system by
introducing 
inclusive correlations among energy fluxes and multiplicity
flows.  In this case 
when studying the interjet effects one really does not need to
apply the event 
selection procedures or jet finding algorithms.  The normalized
inclusive
quantities are 
free from soft and collinear singularities, and are therefore
well controlled 
perturbatively. In subsection 5.2 we have presented in detail the
results for the azimuthal angle of two particles in one jet, the
$Energy-(Multiplicity)^2\; Correlation$ $(EM^2C)$.

As the simplest example of the
multiplicity 
flow in two-jet events of $e^+ e^-$ annihilation
consider the single-particle angular distribution.  Its study is
accessible 
through an $(Energy)^2 - Multiplicity\; Correlation$ $(E^2 MC)$
\begin{eqnarray}
\label{4.12a}
\frac{dN}{d\Omega_{\vec{n}}} & = & \sum_{a,b} \: \int \: dE_a
dE_b dE_{\vec{n}}
\; \frac{E_a E_b \: d \sigma_3}{\sigma_2 \: dE_a dE_b
dE_{\vec{n}} \: d\Omega_a 
d\Omega_b d\Omega_{\vec{n}}}, \\
\label{4.12b}
\sigma_2 & = & \sum_{a,b} \: \int \: dE_a dE_b \; \frac{E_a E_b
\: d\sigma_2}
{dE_a dE_b \: d\Omega_a d\Omega_b}, \;\; \vec{n}_b \: \approx \:
- \vec{n}_a,
\end{eqnarray}
\noindent where the sum goes over all particle types.  The energy
weighted 
integrals over $E_a$ and $E_b$, at fixed angular directions
$\vec{n}_a$ and 
$\vec{n}_b \approx - \vec{n}_a$, specify the \lq\lq jet
directions" about which 
one has an associated multiplicity distribution at variable
angular direction 
$\vec{n} (\Omega)$.  The cross section (\ref{4.12b}), describing
the correlation 
between two back-to-back energy fluxes ($EEC$), contains the known
double 
logarithmic form factor (see \cite{dkmt2} and Refs.\
\cite{ddt1,ddt2}), that 
reflects the natural disbalance of the jet direction, caused by
gluon 
bremsstrahlung\footnote{The $EEC$ and its asymmetry ($EECA$) have
proved to be very 
useful in testing fundamental quantities of perturbative QCD such
as the running 
coupling constant (see e.g.\ \cite{h,bp,ms}).  But because of
their resulting 
emphasis on the earliest branchings, they are not convenient for
the investigation 
of the colour coherence effects.}.

The same angular distribution may be studied in terms of a more
simple 
double-inclusive correlation between the energy flux and the
multiplicity flow 
($EMC$)
\begin{eqnarray}
\frac{dN_2}{d\Omega_{\vec{n}}} & = & \sum_a \; \int \; dE_a \: 
dE_{\vec{n}} \; \frac{E_a \: 
d \sigma_2}{\sigma_1 \: dE_a dE_{\vec{n}} \: d\Omega_a
d\Omega_{\vec{n}}}, \nonumber\\
\sigma_1 & = & \sum_a \; \int \; dE_a \; \frac{E_a \: d
\sigma_1}{dE_a \: 
d \Omega_a},
\label{4.13}
\end{eqnarray}
\noindent The point here is that the main contribution also comes
from the 
two-jet sample whose kinematics is practically fixed by the
choice of the 
direction $\vec{n}_a$.  The difference between the distributions
(\ref{4.12a}) and 
(\ref{4.13}) occurs only when the angular direction $\vec{n}$ is
taken 
parametrically close to the backward \lq\lq jet axis", $\vec{n}
\approx 
- \vec{n}_a$.  In this case the shape of the distribution
(\ref{4.13}) near 
$\vec{n} = \vec{n}_a$ becomes somewhat wider due to the natural
\lq\lq shaking" 
of the non-registered jet in QCD.  A typical \lq\lq shaking
angle" can be estimated 
\cite{wr} as $\Theta_{sh} \sim (\Lambda/W)^\xi$, where $\xi =
b/(b + 4C_F) 
\approx 0.64$ for $n_f = 3$ active flavours. A Monte Carlo study of the
$EMC$ has been presented in Ref. \cite{ow2}.

The drag effect physics becomes accessible through a more
complicated example, 
$E^3 MC$, which in terms of the inclusive approach proves to be a
proper 
characteristic of the spatial distribution of the multiplicity
flow in 
$q\overline{q}g$ events \cite{dkmt2,dkt11}, see also below.  Some
other examples 
of practical interest for the experimental studies of colour
coherence phenomena 
based on the inclusive particle angular correlations were discussed
in subsections (5.2-5.5).

 In the general case multiple correlations of the \lq\lq
$E^\alpha M^\beta C$-kind" can be introduced.  This means that one should
fix \#$\alpha$ jet directions with help of energy fluxes, and then
consider correlations between \#$\beta$ multiplicity flows.  The
procedure of normalizing to a given energy configuration will
ensure that results are finite and well-behaved.

\subsubsection{Particle Distribution within Restricted Cone}

A specific application of the inclusive particle correlation
approach concerns 
studying the particle spectra within restricted jet opening
angles 
\cite{dkt10,dkmt1}.  Recall that according to Eq.\ (\ref{2.51})
the energy 
$E_{max}$ at the peak of the inclusive particle spectrum grows
rather slowly 
with jet energy $E$ (for the limiting spectrum even at $E \approx
1.5$ TeV the 
value of $E_{max}$ reaches only about 3 GeV).  To explore the
intrajet coherence 
origin of the hump-backed particle spectrum and in an attempt to
observe the 
depletion in its soft part for jets produced in hadronic
collisions, it proves 
to be important to study the particles restricted to lie within a
particular 
opening angle with respect to the jet.  For example, one might
consider the 
energy distribution of particles accompanying the production of
an energetic 
hadron and lying within a cone of half-angle $\Theta$ about the
direction of the 
trigger particle momentum.
There are some important advantages of this type of studies.
First of all, one examines here the energy (hardness) evolution of the
inclusive spectrum in the same experiment, avoiding the potential
additional problems caused by the possible difference in normalization
of different experiments. Secondly, the flavour composition of the
primary quarks does not change here. These, for instance, allow
one to perform the searches of the possible energy dependence of the
normalization factors $K^h$ in particle spectra, see (\ref{3.1}).

Parton cascades in this situation will populate mainly the
region
\begin{equation}
\frac{m_h}{\sin \Theta} \; \lapproxeq \; E_h \; < \; E
\label{4.14}
\end{equation}
\noindent with $E_h$ and $m_h$ being the energy and the effective mass of
the observed 
particle $(m_h \simeq Q_o \gapproxeq \Lambda)$.  The maximum of the
distribution, in $E_h$, 
is now forced to larger energies
\begin{equation}
\ln \: \frac{E_{max}}{m_h} \; \approx \; \frac{1}{2} \: \ln \:
\frac{E}
{m_h \sin \Theta} \: - \: B \: \left ( \sqrt{\frac{b}{16 N_C} \:
Y_\Theta} \; - \; 
\sqrt{\frac{b}{16 N_C} \; \ln \: \frac{m_h}{\Lambda} \:} \right
).
\label{4.15}
\end{equation}
This equation corresponds to Eq. (\ref{xibnlo}) for the truncated cascade; 
a similar equation can be
easily obtained for the limiting spectrum if  $\tau=Y$ in
(\ref{2.51}) is replaced by $Y_\Theta$.

 By choosing moderately small values of angle
$\Theta$ and varying 
$E$, one arrives at the \lq\lq moving peak" in accordance with
the relation
\begin{equation}
\frac{E}{E_{max}} \: \frac{dE_{max}}{dE} \; \approx \;
\frac{1}{2} \: - \: 
B \: \sqrt{\frac{b}{16 N_C Y_\Theta} \:}; \;\;\;\;\; Y_\Theta \; = \; 
\ln \: \frac{E \sin \Theta}{\Lambda}.
\label{4.16}
\end{equation}
\noindent The angular cut $\Theta$ is especially useful for jets
produced in 
hadronic collisions, since one is able to eliminate much of the
soft background.

The \lq\lq restricted cone" physics could, in principle, be
studied without 
exploiting any specific jet-finding algorithm in terms of $EMC$. 
As discussed 
in \cite{dkmt2}, the distribution of particles which are emitted
into a cone 
of aperture $\Theta$ around the \lq\lq jet axis" can be defined
as
\begin{equation}
\overline{D}^\Theta (x, E) \; = \; \frac{1}{E} \: \int_0^\Theta
\: d\Theta_{12} 
\: \int \: E_1 dE_1 \: \int \: dE_2 \; \frac{d\sigma_2}{\sigma
dE_1 dE_2 
d\Theta_{12}} \; \delta \left ( x \: - \: \frac{E_2}{E} \right ).
\label{4.17}
\end{equation}

As was discussed in detail in sections 2 and 3, in the case of restricted
cone physics the actual hardness parameter at $\Theta\ll 1$ is 
$Q=E_{jet}\Theta$. This fact lies in the very heart of the angular ordered
parton branching. Recently the CDF collaboration at TEVATRON \cite{TEVAsp}
has presented the preliminary results of the first studies of the inclusive
momentum spectra of charged particles in jets for a variety of dijet masses
(80-600 GeV) and opening angles ($\Theta$=0.168-0.466). These new data agree
very well with the MLLA expectations.
In particular, it is clearly demonstrated experimentally that
the appropriate
evolution variable in the QCD cascades is indeed $Q=E_{jet}\Theta$.

\subsubsection{Multiplicity Flows in Three Jet Events in MLLA}

Let us now come back to the multiplicity flows corresponding to the 
$q\overline{q}g$ sample of $e^+ e^-$ annihilation.  The description of the QCD 
spatial portrait of this simplest multi-jet system including the next-to-leading 
MLLA-corrections of the order of $\alpha_s$ is presented in detail in Refs.\ 
\cite{dkmt2,dkt4,dkt11}.  Here we compare particle flows in
$q\overline{q}g$ and $q\overline{q}\gamma$ events and in the next subsection
the respective total event multiplicities.

Usually this type of analysis is carried out experimentally
with a sample of three-jet
events obtained from a particular jet algorithm for a given resolution
scale. Alternatively, one can follow the inclusive approach (see subsections
5.2 and 6.3.1) which avoids such a selection procedure.  
In terms of the inclusive approach a proper characteristic of the spatial 
distribution of the multiplicity flow is $E^3 MC$ :
\begin{eqnarray}
\frac{dN_4}{d\Omega_{\vec{n}}} & = & \sum_{a,b,c} \: \int \: dE_a dE_b dE_c 
dE_{\vec{n}} \; \frac{E_a E_b E_c \: d\sigma_4}{\sigma_2 \: dE_a dE_b dE_c 
dE_{\vec{n}} \: d\Omega_a d\Omega_b d\Omega_c d\Omega_{\vec{n}}}, 
\nonumber \\
\sigma_3 & = & \sum_{a,b,c} \: \int \: dE_a dE_b dE_c \; \frac{E_a E_b E_c 
d\sigma_3}{dE_a dE_b dE_c \: d\Omega_a d\Omega_b d\Omega_c},
\label{4.18}
\end{eqnarray}
\noindent where the sum runs over all particles.  This represents an angular 
correlation between the three registered energetic particles $(a,b,c)$, moving 
in the directions $\vec{n}_a, \vec{n}_b$ and $\vec{n}_c$ and the multiplicity 
flow around the direction $\vec{n}$ as shown in Fig.\ 6.5. When all three
vectors $\vec{n}_a, \vec{n}_b$ and $\vec{n}_c$ are in the same plane, the main 
contribution to $dN_4$ comes from the $q\overline{q}g$ configuration of the 
primary parton system.  

In the practical application of the inclusive method one would define
the three jet directions within some angular size $\delta$ (say, about
10-20 degrees around each direction) and select all triples of particles in
an event which hit these angular cones. All these configurations contribute
 with the
weight $E_aE_bE_c$ whereas a forth particle is counted in direction $\vec n$
without a weight. 

In what follows we do not refer to a particular analysis method but consider
the soft radiation 
in the $q\overline{q}g$ system for three given angular directions.

In the leading order in $\alpha_s$ the massless parton 
kinematics is unambiguously fixed as follows:
\begin{eqnarray}
\vec{n}_+ \; \approx \; \vec{n}_a, & \vec{n}_- \; \approx \; \vec{n}_b, & 
\vec{n}_1 \; \approx \; \vec{n}_c ; \nonumber \\
& & \nonumber \\
x_+ \; = \; 2 \: \frac{\sin \Theta_{1-}}{\sum \sin \Theta_{ij}}, & x_- \; = \; 
2 \: {\displaystyle \frac{\sin \Theta_{1+}}{\sum \sin \Theta_{ij}}}, & x_1 \; = \; 
2 \: \frac{\sin \Theta_{+ -}}{\sum \sin \Theta_{ij}},\label{4.19} \\
& & \nonumber \\
& x \: + \: x_- \: + \: x_1 \; = \; 2, & 
\nonumber
\end{eqnarray}
\noindent with $x_i = 2 E_i/W$ being the normalized parton energies and $\Theta_{ij}$ 
the angles between partons $i$ and $j (+, - \equiv q, \overline{q}; \: 1 \equiv 
g_1)$.  We emphasize here, that, owing to intrajet coherence, the radiation of a 
secondary gluon $g_2 (k_2 \ll E_i)$ at angles larger than the aperture of each 
parton jet is insensitive to the jet internal structure.  Note that, in principle, 
corrections due to the mass effects in the case of $b$-initiated jets should be 
introduced.  However, in practice, such corrections are found to be rather small, 
see e.g.\ \cite{delphi4}.

Let us start from the radiation pattern for $q\overline{q}\gamma$ events. 
In subsection 6.2.1 we considered the radiation at large angles to the
jet directions, here we include the small angle radiation in the MLLA.
The angular distribution of particle flow can be written as 
\begin{equation}
\frac{8 \pi d N_{q\overline{q}\gamma}}{d\Omega_{\vec{n}}} \; = \; \frac{2}{a_+} 
\; N_q^\prime (Y_{q+}, Y_q) \: + \: \frac{2}{a_-} \; N_q^\prime (Y_{\overline{q}-},
Y_{\overline{q}}) \: + \: 2 I_{+ -} \: N_q^\prime (Y),
\label{4.20}
\end{equation}
\noindent where
\begin{equation}
Y_{q (\overline{q})} \; = \; \ln \frac{E_{q (\overline{q})}}{\Lambda}, \;\; 
Y_{q+} \: = \; \ln \: \left ( \frac{E_q \sqrt{a_+ / 2}}{\Lambda} \right ), \;\; 
Y_{\overline{q}-} \; = \; \ln \: \left ( \frac{E_q \sqrt{a_- / 2}}{\Lambda} 
\right ), \;\;  Y \; \equiv \; \ln \: \frac{E}{\Lambda}
\label{4.21}
\end{equation}
\noindent and
\begin{equation}
I_{+ -} \; = \; (\widehat{+ -}) \: - \: \frac{1}{a_-} \: - \: \frac{1}{a_+} \; 
= \; \frac{a_{+ -} \: - \: a_+ \: - \: a_-}{a_+ a_-}.
\label{4.22}
\end{equation}
\indent The factor $N_A^\prime (Y_{i}, Y) \; \equiv \; (d/dY_i) \: N_A (Y_{i}, Y)$ 
takes into account that the final registered hadron is a part of cascade ({\it 
cascading factor}), cf.\ subsection 5.2.2.
$N_A (Y_i, Y)$ stands for the
multiplicity in a jet $A (A = q,g)$ with the hardness scale $Y$ of particles 
concentrated in the cone with an angular aperture $\Theta_i$ around the jet 
direction $\vec{n}_i$.  In the above $a_i \equiv 1 - \vec{n}\vec{n}_i, \;\; 
\vec{n}_+ \approx \vec{n}_a, \;\; \vec{n}_- \approx \vec{n}_b$.  To understand 
the meaning of the quantity $N_A (Y_i, Y)$ it is helpful to represent it as
\begin{eqnarray}
N_A (Y_i, Y) & = & \sum_{B = q,g} \: \int_0^1 \: dz \: z \: \overline{D}_A^B 
(z, \Delta \xi) \: N_B (\overline{Y}_i),\label{4.23} \\
\Delta \xi \; = \; \frac{1}{b} \: \ln (Y/\overline{Y}_i) & , & \overline{Y}_i 
\; = \; Y_i + \ln z \; = \; \ln \left (\frac{zE}{\Lambda} \sqrt{\frac{a_i}{2}} 
\right ). \nonumber
\end{eqnarray}
\noindent Here $N_B (\overline{Y}_i)$ is the multiplicity in a jet with the 
hardness scale $\overline{Y}_i$, initiated by a parton $B$ within the cone 
$\Theta_i$, and $\overline{D}_A^B$ denotes the structure function for parton 
fragmentation $A \rightarrow B$.

Eq.\ (\ref{4.23}) accounts for the fact that due to intrajet coherence the 
radiation at small angles $\Theta_i \ll 1$ is governed not by the overall colour 
of a jet $A$, but by that of a subject $B$, developing inside a much narrower 
cone $\Theta_i$.

For the emission at large angles $(a_+ \sim a_- \sim 1)$ when all the factors 
$N^\prime$ are approximately the same, Eq.\ (\ref{4.20}) reads, cf.\ Eq.\ 
(\ref{4.3})
\begin{equation}
8 \pi \; \frac{d N_{q\overline{q}}}{d\Omega_{\vec{n}}} \; = \; 2 (\widehat{+ -}) 
\: N_q^\prime (\ln E/\Lambda).
\label{4.24}
\end{equation}
\noindent The cascading factor here can be presented as, 
cf. Eqs.\ (\ref{mfl}), (\ref{nglu})
\begin{equation}
\frac{N_C}{C_F} \: . \: N_q^\prime \left ( \ln \: \frac{E}{\Lambda} \right ) \;
 \approx \; N_g^\prime \left ( \ln \: \frac{E}{\Lambda} \right ) \; = \; \int^E 
\: \frac{dE_g}{E_g} \: 4N_C \: \frac{\alpha_s (E_g)}{2 \pi} \: N_g \left ( \ln 
\: \frac{E_g}{\Lambda} \right ).
\label{4.25}
\end{equation}
\noindent One can easily see that for the radiative two-jet events the emission 
pattern is given by the $q\overline{q}$ sample Lorentz boosted from the quark 
{\it cms} to the {\it lab} system (i.e.\ the {\it cms} of $q\overline{q}\gamma$), 
and the corresponding particle multiplicity should be equal to that in $e^+ e^- 
\rightarrow q\overline{q}$ at $W_{q\overline{q}}^2 = (p_q + p_{\overline{q}})^2$.

Now we turn to the three-jet event sample when a hard photon is replaced by a 
gluon $g_1$.  For a given $q\overline{q}g_1$ configuration the particle flow can 
be presented
\begin{eqnarray}
\frac{8 \pi dN_{q\overline{q}g}}{d\Omega_{\vec{n}}} & = & \frac{2}{a_+} \; 
N_q^\prime (Y_{q+}, Y_q) \: + \: \frac{2}{a_-} \; N_q^\prime (Y_{\overline{q}-}, 
Y_{\overline{q}}) \: + \: \frac{2}{a_1} \; N_g^\prime (Y_{g1}, Y_g) \nonumber \\
& + & 2 \; \left [ I_{1+} \: + \: I_{1-} \: - \: \left ( 1 \: - \: \frac{2C_F}
{N_C} \right ) \; I_{+-} \right ] \; N_g^\prime (Y),
\label{4.26}
\end{eqnarray}
\noindent where, in addition to the definitions in Eq.\ (\ref{4.21}) one has
\begin{equation}
Y_g \; = \; \ln \: \frac{E_g}{\Lambda}, \;\;\;\; Y_{g1} \; = \; \ln \: \left ( 
\frac{E_g \sqrt{a_1/2}}{\Lambda} \right ).
\label{4.27}
\end{equation}
\noindent This formula accounts for both types of coherence:  the angular ordering 
inside each of the jets and the collective nature of the interjet flows.  The 
first three terms in Eq.\ (\ref{4.26}) are collinear singular as $\Theta_i 
\rightarrow 0$ and contain the factors $N^\prime$, describing the evolution of 
each jet initiated by the hard emitters $q, \overline{q}$ and $g_1$.  The last 
term accounts for the interference between these jets.  It has no collinear 
singularities and contains the common factor $N_g^\prime (Y, Y)$ independent of 
the direction $\vec{n}$.  Eq.\ (\ref{4.26}) predicts the energy evolution of 
particle flows in $q\overline{q}g$ events.  It seems to be of importance to test 
experimentally this energy dependence, for instance, in the case of multiplicity 
flow projected onto the event plane.

As follows from (\ref{4.20})-(\ref{4.26}) when replacing a hard photon by a 
gluon $g_1$, with otherwise identical kinematics, an additional particle flow 
arises
\begin{equation}
\left ( \frac{8 \pi dN}{d\Omega_{\vec{n}}} \right )_g \; = \; \frac{8 \pi dN_{q
\overline{q}g}}{d\Omega_{\vec{n}}} \: - \: \frac{8 \pi dN_{q\overline{q}\gamma}}
{d\Omega_{\vec{n}}} \; = \; \frac{2}{a_1} \: N_g^\prime (Y_{g1}, Y_g) \: + \: 
[I_{1+} \: + \: I_{1-} \: - \: I_{+-}] \: N_g^\prime (Y).
\label{4.28}
\end{equation}
\noindent Note that for the case of large radiation angles both cascading factors 
$N^\prime$ become approximately equal and one has, cf.\ Eq.\ (\ref{4.1})
\begin{equation}
\left ( \frac{8 \pi dN}{d\Omega_{\vec{n}}} \right )_g \; = \; \biggl [ (\widehat
{1+}) \: + \: (\widehat{1-}) \: - \: (\widehat{+-}) \biggr ] \:  N_g^\prime 
(Y).
\label{4.29}
\end{equation}
\noindent An instructive point is that this expression is not positive definite.  
One clearly observes the net destructive interference in the region between the 
$q$ and $\overline{q}$ jets.  The soft radiation in this direction proves to be 
less than that in the absence of the gluon jet $g_1$ (see subsection 6.2).

\subsubsection{Topology Dependence of Mean Event Multiplicity}

Let us now come to the connection between the mean particle multiplicities in the 
two-jet and three-jet samples of $e^+ e^-$ annihilation.  Recall that the particle 
multiplicity in an {\it individual} quark jet was formally defined from the process 
$e^+ e^- \rightarrow q\overline{q} \rightarrow hadrons$ by 
(see sections 2, 3)
\begin{equation}
N_{e^+ e^-}^{ch} (W) \; = \; 2 \:  N_q^{ch} (E) \:  \left [ 1 + O \: 
\left ( \frac{\alpha_s (W)}{\pi} \right ) \right ] \: , \; W = 2E.
\label{4.30}
\end{equation}
\indent As we have already discussed when three or more partons are involved in 
a hard interaction the multiplicity cannot be represented simply as a sum of the 
independent parton distributions, but it rather becomes dependent on the geometry 
of the whole ensemble.

So, the problem arises of describing the multiplicity in three-jet events, 
$N_{q\overline{q}g}$, in terms of the characteristics of the individual $q$ and 
$g$ jets discussed above.  The quantity $N_{q\overline{q}g}$ should depend on the 
$q\overline{q}g$ geometry in a Lorentz-invariant way and should have a correct 
limit when the event is transformed to the two-jet configuration by decreasing 
either the energy of the gluon $g_1$ or its emission angle.

The angular integral of Eq.\ (\ref{4.20}) can be easily checked to reproduce the 
total multiplicity.  One can write $N_{q\overline{q}\gamma}$ as
\begin{equation}
N_{q\overline{q}\gamma} \; = \; \int \: \frac{dN_{q\overline{q}\gamma}}{d\Omega_
{\vec{n}}} \; = \; N_q (Y_q) \: + \: N_q (Y_{\overline{q}}) \: + \: 2 \ln 
\sqrt{\frac{a_{+-}}{2}} \:  N_q^\prime (Y).
\label{4.31}
\end{equation}
\noindent Now let us transform this formula to the Lorentz-invariant expression.  
To do this we rewrite
\begin{equation}
Y_{q(\overline{q})} \; = \; Y \: + \: \ln x_{+(-)}, \;\;\;\;\; x_{+(-)} \; \equiv 
\; E_{q(\overline{q})}/E
\label{4.32}
\end{equation}
\noindent and use the expansion
\begin{equation}
N_q (Y_q) \; = \; N_q (Y) \: + \: \ln x_+ \:  N_q^\prime (Y) \: + \: O 
\left ( \frac{\alpha_s}{\pi} \: N_q \right ).
\label{4.33}
\end{equation}
\noindent Then
\begin{equation}
N_{q\overline{q}\gamma} \; = \; 2 N_q (Y) \: + \: \ln \frac{x_+ x_- a_{+-}}{2} 
 \: N_q^\prime (Y) \: + \: O (\alpha_s N) \; = \; 2 N_q (Y_{+-}^*) \: 
[1 + O (\alpha_s)]
\label{4.34}
\end{equation}
\noindent with
\begin{equation}
Y_{+-}^* \; = \; Y \: + \: \ln \sqrt{\frac{x_+ x_- a_{+-}}{2}} \; = \; \ln \sqrt
{\frac{(p_+ p_-)}{2 \Lambda^2}} \; = \; \ln \frac{E^*}{\Lambda}.
\label{4.35}
\end{equation}
\noindent Here $E^*$ is the quark energy in the $cms$ of $q\overline{q}$, i.e.\ 
the Lorentz-invariant generalization of a true parameter of hardness of the 
process.

The multiplicity $N_{q\overline{q}g}$ can be written by analogy as
\begin{eqnarray}
N_{q\overline{q}g} & = & \int \: \frac{dN_{q\overline{q}g}}{d\Omega_{\vec{n}}} \; 
= \; N_q (Y_q) \: + \: N_q (Y_{\overline{q}}) \: + \: N_g (Y_g) \: + \nonumber \\
& & \left [ \ln \sqrt{\frac{a_{1+} a_{1-}}{2a_{+-}}} \: + \: \frac{2 C_F}{N_C} \: 
\ln \sqrt{\frac{a_{+-}}{2}} \right ] \:  N_g^\prime (Y)
\label{4.36}
\end{eqnarray}
\noindent where $Y_g = Y + \ln x_1$.  Proceeding as before one comes finally to 
the Lorentz-invariant result
\begin{equation}
N_{q\overline{q}g} \; = \; \biggl [ 2 N_q (Y_{+-}^*) \: + \: N_g (Y_g^*) \biggr 
] \: \left ( 1 + O \left ( \frac{\alpha_s}{\pi} \right ) \right )
\label{4.37}
\end{equation}
\noindent with
\begin{equation}
Y_g^* \; = \; \ln \sqrt{\frac{(p_+ p_-) (p_- p_1)}{2 (p_+ p_-) \Lambda^2}} \; = 
\; \ln \: \frac{p_{1 \perp}}{2 \Lambda},
\label{4.38}
\end{equation}
\noindent with $p_{1 \perp}$ the transverse momentum 
of $g_1$ in the $q\overline{q}~cms$\footnote{
Within the APA the observation that a gluon adds multiplicity related to its
$p_\perp$ was done in \cite{dkt11}. Similar result was derived in the string
length approach to the dipole emission picture \cite{agu}.}.

Comparing (\ref{4.31}) with (\ref{4.37}), we see that replacement of a photon 
$\gamma$ by a gluon $g_1$ in this frame 
leads to the additional multiplicity
\begin{equation}
N_g (Y_g^*) \; = \; N_{q\overline{q}g} (W) \: - \: N_{q\overline{q}\gamma} (W)
\label{4.39}
\end{equation}
\noindent which depends not on the gluon energy but on its {\it transverse 
momentum}, i.e.\ on the hardness of the primary process.  Eq.\ (\ref{4.37}) 
reflects the coherent nature of soft emission and has a correct limit when the 
event is transformed to the two-jet configuration.  For comparison with the 
data it looks convenient to rewrite (\ref{4.37}) in terms of the observed 
multiplicity in $e^+ e^-$ collisions (see Eq.\ (\ref{4.30})) as
\begin{equation}
N_{q\overline{q}g} \; = \; \left [N_{e^+ e^-} (2E^*) + \frac{1}{2} C_q^g 
(p_{1 \perp}) N_{e^+ e^-} (p_{1 \perp}) \right ] \; (1 + O (\alpha_s)).
\label{4.37a}
\end{equation}
In this formula $C_q^g(W)$ denotes the ratio of multiplicities in gluon and
quark jets at $cms$ energies $W$ which is presented within MLLA 
in subsection 3.2 and
approaches asymptotically the value $N_C/C_F = 9/4$. Alternatively,
one could use this formula to study the behaviour of $C_q^g(W)$ 
experimentally.

The result (\ref{4.37a})
can be written in another form, which is found \cite{dkt11,vak1} to 
be convenient for studies of the properties of a gluon jet, namely
\begin{equation}
N_{q\overline{q}g} \; = \; [N_g (Y_{1+}) \: + \: N_g (Y_{1-}) \: + \: 2N_q (Y_{+-}) 
\: - \: N_g (Y_{+-})] \:  [1 + O (\alpha_s)]
\label{4.40}
\end{equation}
\noindent where $Y_{ij} = \ln (\sqrt{(p_i p_j)/2\Lambda^2}) = \ln (E_{ij}^*/
\Lambda)$.  Such a representation deals with multiplicities of two-jet events at 
appropriate invariant pair energies 
$W_{ij}=2E_{ij}^*$ in the $(ij)$ $cms$ frame.
Expression (\ref{4.40}) has a 
proper limit, $2N_g(W/2)$, when the $q\overline{q}g$ configuration is forced to a 
quasi-two-jet one, $g (8) + q\overline{q} (8)$, with a sufficiently small angle 
between the quarks.  The experimental tests of Eq.\ (\ref{4.40}) should be 
performed with the tagged heavy quarks, see also Ref.\ \cite{jwg}.  Analogously to
(\ref{4.37a}) Eq.\ (\ref{4.40}) could be presented as
\begin{eqnarray}
N_{q\overline{q}g} & \simeq & \biggl [ N_{e^+ e^-} \: (2E^*) \: \biggl ( 1 \: - 
\: \frac{C_q^g(E^*)}{2} \biggr ) \nonumber \\
& + & \frac{C_q^g \: (E_{1+}^*)}{2} \: N_{e^+ e^-} \: (2E_{1+}^*) \nonumber \\
& + & \frac{C_q^g \: (E_{1-}^*)}{2} \: N_{e^+ e^-} \: (2E_{1-}^*) \biggr ].
\label{4.40a}
\end{eqnarray}

\subsection{Azimuthal Asymmetry of QCD Jets}

As has been already discussed in subsection 6.2, the treatment of the structure
of final states given by the Lund string picture qualitatively reproduces the 
QCD radiation pattern only up to $O (1/N_C^2)$ corrections (the large $N_C$-
limit).  Such corrections are, normally, not taken into full account
in all the parton shower Monte
Carlo algorithms.  The reason is that these terms often appear with a negative 
sign, or correspond to events with an undefined colour flow, and so it is not 
clear how to handle them in Monte Carlo shower simulations, e.g.\ \cite{bba} 
(see, however, footnote 21).  Normally, the omission is not drastic, but under
special conditions $1/N_C^2$ terms may become sizeable or even dominant 
\cite{dkt2}.

The simplest topical example is the azimuthal asymmetry of a quark jet in 
$q\overline{q}g$ events.  The radiation pattern is given here by Eq.\ (\ref{4.1}).  
When all the angles are large the third term in (\ref{4.1}) (negative colour-
suppressed antenna $(\widehat{+-})$) leads to a small correction to the 
canonical string interpretation of the drag effect.  However, if the $q$ and 
$\overline{q}$ are close to each other, this term is kinematically enhanced, 
and is no longer negligible.

The azimuthal distribution of particles produced inside a cone of opening half-
angle $\Theta_0$ may be characterized by an asymmetry parameter:
\begin{equation}
A (\Theta_0) \; = \; \frac{N_{\rightarrow g} (\Theta < \Theta_0) - 
N_{\rightarrow \overline{q}} (\Theta < \Theta_0)}{N_{tot} (\Theta < \Theta_0)} 
\; = \; \frac{(\Delta N)_{as}}{N_{tot}}.
\label{4.41}
\end{equation}
\noindent For parametrically small $\Theta_0$, one obtains (see Ref.\ \cite{dkt2} 
for further details)
\begin{eqnarray}
A (\Theta_0) & \simeq & \frac{2 \Theta_0}{\pi} \; G \sqrt{4 N_C \: 
\frac{\alpha_s (E \Theta_0)}{2 \pi}} \; ,\label{4.42} \\
& & \nonumber \\
G & = & \frac{N_C}{2 C_F} \; \cot \: \frac{\Theta_{+1}}{2} \: + \: \frac{1}
{2N_C C_F} \; \cot \: \frac{\Theta_{+-}}{2} \; .
\label{4.43}
\end{eqnarray}
\indent The first colour-favoured term in Eq.\ (\ref{4.43}) describes the
asymmetry due to the \lq boosted string' connecting the $q$ and $g$ directions.  
The corresponding asymmetry vanishes with an increase of $\Theta_{+1}$ as the 
string straightens.  Here, however, the second term enters into the game, 
forcing the asymmetry to increase anew, as shown in Fig.\ 6.6. This behaviour
might be interpreted as an additional repulsion between particles from $q$ and 
$\overline{q}$ jets.

For a symmetric configuration $\Theta_{+1} = \Theta_{-1} = \pi - \Theta_{+-}/2$, 
the colour-suppressed term in Eq.\ (\ref{4.1}) prevails when $\Theta_{+-} \leq 2 
\arctan (1/N_C) \approx 37^0$.  To realize this effect, one has to select 
$q\overline{q}g$ events with kinematics, in which case the hard gluon moves in 
the opposite direction to the quasi-collinear $q\overline{q}$ pair.  The above 
mentioned phenomenon can be studied experimentally by looking at the azimuthal 
distribution of particles around the identified quark direction, which can be 
done by tagging heavy flavour quark jets.

An interesting vista (see Ref.\ \cite{vak1}) on the manifestation of the quantum 
mechanical effects concerns $e^+ e^-$ events with the initial state energetic 
radiation (the so-called ISR events).  These events correspond to the $Z^0$ 
\lq\lq radiative tail".  Here one would expect that perturbative QCD differs 
qualitatively from the string picture in predicting the asymmetry of the 
azimuthal quark jet profile when the additional soft gluon (particle flow) 
is emitted in the direction orthogonal to the $q\overline{q}\gamma$ plane, 
see Fig.\ 6.7. This leads to a characteristic manifestation of the QCD wave
nature of hadronic flows.

In the ISR events the jet topology is well defined since a hard photon is emitted 
in the direction along the colliding beams and the $q\overline{q}$ system arises 
from the $Z$ decay.  In the case of additional soft radiation the QCD antenna 
pattern is given by Eq.\ (\ref{4.1}) with the appropriate kinematics of the 
colour emitters as illustrated by Fig.\ 6.7b.

For small $\Theta_0$ values the contribution of the $(\widehat{+ -})$ antenna to 
$(\Delta N)_{as}$ is proportional to
\begin{equation}
\left [ - \; \frac{1}{N_C^2} \right ] \:  \frac{\Theta_0}{\pi} \; \cot 
\frac{\Theta_q}{2} \; N_g^\prime (E \Theta_0)
\label{4.44}
\end{equation}
\noindent $(\Theta_q \equiv \Theta_{+ -})$, while the nonsingular (colour-
favoured) antenna $(\widehat{1-})$ produces only a parametrically small 
correction $(\Delta N)_{as} \sim \Theta_0^3$ (antenna $(\widehat{1+})$ does not 
contribute the asymmetry at all).

Neglecting this contribution one obtains for the symmetric configuration (c.f.\ 
Eqs.\ (\ref{4.42}), (\ref{4.43})
\begin{equation}
A_{q\overline{q}g} (\Theta_0) \; \approx \; - \: \frac{1}{N_C^2} \: \frac
{2 \Theta_0}{\pi} \; \sqrt{4 N_C \: \frac{\alpha_s (E \Theta_0)}{2 \pi}} \: 
\cot \: \frac{\Theta_q}{2} \; \approx \; - \: \frac{1}{N_C^2} \; A_{q
\overline{q}} (\Theta_0) .
\label{4.45}
\end{equation}
\indent Here the parameter $A_{q\overline{q}} (\Theta_0)$ describes the 
asymmetry of a quark jet in the standard ISR events, see Fig.\ 6.7a.

Let us emphasise that just the colour-suppressed term (which is normally
not included in 
Monte Carlo simulation programs) governs the overall asymmetry $A_{q
\overline{q}g}$.  It is of importance that the $O (1/N_C^2)$ term produces a 
negative asymmetry opposite to the small 
positive one due to the colour-favoured
contribution.  Therefore, the perturbative scenario differs here qualitatively
from the canonical string picture.

Accounting for both negative and positive contributions to $A_{q
\overline{q}g} (\Theta_0)$ leads to the conclusion that for small angles $\Theta_0$ 
the asymmetry is maximal (of the order of 2\%) when $\Theta_0/\Theta_q \approx 
1/N_C$.  In the general case when a gluon is not orthogonal to the $q\overline{q}g$ 
plane one can consider the $q$-jet asymmetry with respect to the $qg$ plane.

Experimental study of the asymmetry $A_{q\overline{q}g}$ could be performed, 
e.g.\ by looking at the energy-energy-multiplicity-multiplicity 
(energy-multiplicity-multiplicity) correlations \cite{dkmt2,dkmw1} in 
the ISR events.
It is worthwhile to mention that the underlying physics here is the same as
in the case of ``boosted" $Z^0$ events. However, in the latter case one will
need to deal with the nontrivial phase space for the registered particles.

The azimuthal asymmetry of produced jets is certainly not specific to 3-jet 
events of $e^+ e^-$ annihilation.  Similar phenomena should be observed, e.g.\ 
in high-$p_\perp$ hadronic processes (see Ref.\ \cite{dkt2,dkt4} for further 
details).  The qualitative difference between the predictions of QCD and its 
large-$N_C$-limit proves to be most spectacular in the case of $p\overline{p}$ 
scattering with identification of the scattered $q$-jet.  The asymmetry, predicted 
for the case of tagged $q$-jet is shown in Fig.\ 6.8. As is easily seen, in the
region of {\it cms} scattering angles, $70^o < \Theta_S < 110^o$ QCD predicts 
$A = +(4 : 7)\%$ at $\Theta_0 = 30^o$ unlike the opposite sign effect $A = 
-(2.5 : 1.5)\%$, originated from the large $N_C$-treatment of QCD formulae 
(\lq string-motivated' approach).

\section{Experimental Tests of Colour-Related Phenomena}
\setcounter{equation}{0}

In this section we focus mainly on the new experimental evidence of the interjet 
colour coherence effects from the recent LEP-1 experiments 
\cite{opal8,l34,opal12,aleph6,aleph5,delphi4}, for reviews  
see also \cite{wjm1,my1,ada1}.  The 
main lesson from these impressive studies is that now we have (quite successfully) 
entered the stage of quantitative tests of the details of colour drag 
phenomena.  Owing to the massive LEP-1 statistics one may perform the
further detailed analysis allowing discriminations between the non-perturbative 
(Lund string \cite{ags,agis}) and perturbative \cite{adkt2} interpretation of 
the celebrated string/drag effect.

\subsection{Comparison of the Particle Flow in $q\overline{q}g$ 
and $q\overline{q}\gamma$ events}

It has long been known \cite{adkt2} that comparing the particle flows in three-
jet multihadronic events with those in events with two jets and a hard isolated 
photon provides one with a very powerful method for studying the interjet 
collective phenomena.  As was discussed in subsection 6.2, a reduced particle
density is anticipated in the region between the two quark jets in $q\overline{q}g$ 
events, relative to the density between the two jets of $q\overline{q}\gamma$ 
events.  The obvious advantage of this method is that both the interquark regions 
are compared directly, the only difference being that the other hemisphere of 
the event contains either a gluon or a photon.  Unlike other methods, this one 
is not sensitive to differences between narrow $q$ and broad $g$ jets.  The event 
selection is assumed to be done in such a way that the two samples are kinematically 
similar.

The first experimental studies of the interjet colour coherence using this model 
independent approach were successfully performed at PEP/PETRA energies about a 
decade ago \cite{tpcm,wh,fos}.  Here we shall concentrate only on the new 
development based on the recent analysis 
of the LEP-1 data \cite{l34,opal12,delphi4}.

\subsubsection{Further Evidence for Particle Drag}
All experiments have performed rather similar analyses.  They compare the particle 
flow of $q\overline{q}g$ and $q\overline{q}\gamma$ events in the region between 
the $q$ and $\overline{q}$.  Care is taken that the $q\overline{q}$ systems be 
directly comparable.  The present data clearly demonstrate the predicted 
significant decrease in particle density in the angular region between quark 
and antiquark jets for $q\overline{q}g$ events as compared with $q\overline{q}
\gamma$ events.  Meanwhile, the incoherent shower Monte Carlo models show a large 
disagreement with the data.

Fig.\ 7.1, taken from the OPAL Collaboration
\cite{opal12}, illustrates the results of direct
comparison of the particle flow in the interquark region for $q\overline{q}g$ 
and $q\overline{q}\gamma$ events.  In this experiment the jets of the 
$q\overline{q}g$ and $q\overline{q}\gamma$ samples were ordered in decreasing 
energy. The jet of lowest energy in the 3-jet event is then taken as the 
gluon jet which is correct for 80\% of the events according to Monte Carlo
studies. 
The gluon jets in Fig.\ 7.1 are seen as a hump above $180^o$ (the position of the
hump depends on the value for the jet parameter $y_{cut}$ used).
The relative density of particles in the $q\overline{q}$ valley (in 
particular around $50^o - 110^o$) is markedly smaller in the case of 
$q\overline{q}g$ events (open points) than in the case of the $q\overline{q}\gamma$ 
events (solid points). The ratio of densities $R_{12}$ between jets 1 and 2
is found to be 0.71 $\pm$ 0.03.

Lepton-tagged $q\overline{q}g$ events were  used to enrich the 
purity of the gluon jet sample.  This technique provides a clear discrimination 
between the gluon and quark jets (at the cost of a large loss in statistics).
In this case the suppression effect becomes larger with
$R_{12}=0.60\pm 0.03$. Correcting for misidentified gluon jets with the help
of O($\alpha_s^2$) matrix element calculations \cite{ert} as implemented in
the JETSET \cite{JETSET}
 Monte Carlo yields $R_{12}=0.51\pm 0.03$.

The L3 Collaboration \cite{l34} 
has performed the comparison of the particle flows in the 
$q\overline{q}$ centre of mass frame and quantified the effect of particle drag by the 
ratio $R_N$ of the integrals of the flow distributions between $54^o$ and $135^o$.  
It was found that, in agreement with expectations, the ratio $R_N$ is significantly 
less than unity and decreases with increasing gluon identification purity.  
The numerical values of $R_N$ are well reproduced by the WIG'ged Monte Carlo 
models.

WIG'ged event generators describe the experimental findings perfectly well when 
the colour flow is respected both in the perturbative and the non-perturbative 
phases.  Monte Carlo shower models fail if either or both of these requirements 
are relaxed, for example, by assuming incoherent parton evolution
or applying independent jet fragmentation models
\cite{l34,opal12,delphi4,ts1}.

 Finally, let us note that both L3 \cite{l34} and OPAL \cite{opal12} have 
studied the dependence of the colour drag on out-of-plane momentum $p^{out}$.  
In
agreement with the predictions of Refs.\ \cite{adkt2,kll} (see also subsection 
6.2) the dependence on $p^{out}$ was found to be significantly weaker than at
lower energies. 
 Recall that the dependence of the magnitude of the
string/drag effect on $p^{out}$ (and registered particle mass) has to vanish 
asymptotically in the perturbative approach.

\subsubsection{Comparison with Analytical Calculations}
OPAL \cite{opal12} has performed the first fit of the observed charged particle 
flow distributions to the analytical perturbative formulae, 
see Fig.\ 7.2. The experimental data were corrected for detector effects and
for gluon purity using a Monte Carlo model \cite{JETSET,ert}.
In this comparison with the data
the large-angle emission 
formulae of Ref.\ \cite{adkt2} (see also subsection 6.2) were used and not
the complete expressions of Ref.\ \cite{dkt11,dkmt2,dkt4} 
for particle flow distributions at arbitrary angle. 
  To account for the intrajet cascading effects
Gaussian distributions of particles around the quark jet positions 
have been added to the perturbative formulae 
(\ref{4.4a}), (\ref{4.5}). In this way 
 a  qualitative description of 
the experimental results has been obtained. However, 
the separation of  the $q\overline{q}
\gamma$ and $q\overline{q}g$ data points appears to be larger in the data than in 
the fits. The comparison between theory and experiment will certainly be
most transparent if the jets are identified by experimental methods
and if not too different 
jet topologies are averaged over.

A step in this direction has been done recently 
by the DELPHI Collaboration \cite{delphi4} 
which has performed the first quantitative verification 
of the perturbative QCD prediction \cite{adkt2} for the ratio $R_\gamma$
\begin{equation}
R_\gamma \; = \; \frac{N_{q\overline{q}} (q\overline{q}g)}{N_{q\overline{q}} 
(q\overline{q} \gamma)}
\label{5.1}
\end{equation}
\noindent of the particle population densities in the interquark valley.  For a 
clearer quantitative analysis a comparison was performed for the $Y$-shaped 
symmetric events using the double vertex method for the $q$-jet tagging.  The 
observed normalized differential particle flows are 
depicted in Fig.\ 7.3 as a
function of the angle $\Psi$ of the particle to  the $q$-jet direction.  The 
ratio $R_\gamma$ of the charged particle flows in the $q\overline{q}$ angular 
interval $[35^o, 115^o]$ was found to be
\begin{equation}
R_\gamma^{\exp} \; = \; 0.58 \: \pm \: 0.06.
\label{5.2}
\end{equation}
\noindent This value is in a fairly good agreement with the expectation following 
from Eqs.\ (\ref{4.4a}) and (\ref{4.5}) at $N_C = 3$, for the same angular
interval
\begin{equation}
R_\gamma^{th} \; \approx \; \frac{0.65 N_C^2-1}{N_C^2-1} \; \approx \; 0.61.
\label{5.3}
\end{equation}
(Note that this ratio is slightly larger than the one predicted for Mercedes
type events in (\ref{4.6})).
The string effect is quantitatively explained by the perturbative prediction
and the above ratio does not appear to be affected by
hadronization effects in an essential way.

\subsection{Studies of the Properties of $Z^0 \rightarrow$ 3-Jet Events}

As was shown in subsection 6.2 interjet colour coherence causes destructive
interference in the region between the $q$ jet and the $\overline{q}$ jet and 
constructive interference in the region between the $q$ jet and the $g$ jet.  
Experimentally this phenomenon is studied by exploiting the same techniques 
that are used in the measurements of quark-gluon differences, see section 3.  
An observation of the larger particle flows between the quarks and the gluon 
than between the two quarks was the original method of demonstrating the string/
drag effect (see \cite{jade} and further references in \cite{h,ms,wjm1}).

The phenomenon has been studied vigorously by all four LEP-1 collaborations.  In 
particular, OPAL \cite{opal8} was the first to report on a model independent
evidence for the particle colour drag, using quark tagging in the symmetrical 
3-jet events.  The method of anti-tagging gluon jets in the symmetric 
$q\overline{q}g$ event topologies has been successfully applied since then 
(see e.g.\ \cite{delphi4} and reviews \cite{wjm1,wjm2,my1}).

\subsubsection{Phenomenological Results and Monte Carlo Models}
Recently ALEPH \cite{aleph6} has performed a detailed study using all 
3-jet events selected from a large sample of hadronic $Z$-decays by the $k_\perp$/
Durham \cite{rep,cdotw} algorithm.
A surplus of the particle 
population in the $q - g$ valley has been demonstrated quite convincingly.
The ratio $R_g=N_{qg}/N_{q\overline q}$ of the particle yield 
in the $qg$ and $q\overline{q}$ interjet regions has been measured for
different regions in the Dalitz plot of the 3-jet events. Kinematic regions
with the highest gluon jet purity gave the largest effect
$R_g \sim 2$. 

The results \cite{aleph6} are found to be in good 
quantitative agreement with the WIG'ged Monte Carlo models.  Moreover, the data 
allow us to conclude that colour drag is present already at the 
perturbative stage.  In some models the parton level results on their own
without hadronization already come close to the data, in others the
perturbative stage can be responsible only for part of  the effect. Here one should
remember that the parton level calculations refer to a cut-off 
parameter $Q_0$ which is 
 larger than that required in LPHD applications
($Q_0 \sim 1$ GeV in JETSET \cite{JETSET,aleph6}, for example).
 One may expect that
the string effect would increase for decreasing $Q_0$ as it does for
decreasing mass of the final state hadron (see next item). 
The incoherent shower Monte Carlo models give much too low 
$R_g$ values.  

Identification of charged hadrons $(\pi^\pm, K^\pm$ and $\overline{p})$ has 
allowed ALEPH \cite{aleph6}
to study mass dependence of the interjet $R_g$ values.  In full 
agreement with the perturbative expectations \cite{adkt2,kll} (see also subsection 
6.2) there is no strong mass dependence at LEP-1 energies. However, a 
systematic increase of the string effect (increase of $R_g$) with decreasing
hadron mass is observed (see Fig.7.4). This trend is not expected from the
 string hadronization models which would rather predict
 a decrease or no effect for 
the bulk of the particles (with momenta $p<500$ MeV \cite{aleph6}),
see also Fig. 6.4.

The string effect and colour coherence not only affect the particle
production in between the jets but also inside. 
Thus it was  shown \cite{aleph6} that the \lq\lq jets are crooked" \cite{ts1}:  
if a quark jet direction is reconstructed in $q\overline{q}g$ events, then there 
is a tendency for high-momentum particles to be found on the side away from the 
$g$ jet and low-momentum ones on the side towards the $g$.  The latter is what 
could be expected from the soft emission by the colour antenna/dipole spanned 
between the $q$ and the $g$ (see subsection 6.4 and Ref.\ \cite{dkt2}); the
former then follows as a consequence of the jet direction being found as an 
average.

If one allows for 
arbitrary 3-jet kinematic configurations
 new information can be obtained about the evolution of the event 
portrait with the variation of event topology, see subsection 
6.3.  In particular, ALEPH \cite{aleph5} has demonstrated that, in agreement
with the QCD radiophysics \cite{dkmt2}, the mean event multiplicity
in  three jet events depends 
both on the jet energies and on the angles between the jets.  These 
ALEPH results clearly show
the topological dependence of jet 
properties which are predicted  \cite{dkt11}
in  formulae (\ref{4.37}) and (\ref{4.40}).
  

\subsubsection{Comparison with Analytical Predictions}

An instructive analysis has been performed by DELPHI \cite{delphi4} who studied 
the threefold symmetric (Mercedes-like) events using the double vertex tagging 
method.  It is shown that the string/drag effect is clearly present in these 
fully symmetric events and it cannot be 
an artefact due to kinematic selections. The azimuthal angle dependence 
of particle density in the event plane is shown in Fig. 7.5.  
Quantitatively, comparing the minima located at $\pm [50^o, 70^o]$, 
the particle population 
ratio $R_g=N_{qg}/N_{q\overline q}$ 
in the $q - g$ and $q - \overline{q}$ valleys is measured to be
\begin{equation}
 R_g^{exp} = 2.23 \pm 0.37 
\labl{rgmer}
\end{equation}
for the $k_\perp$/Durham \cite{rep,cdotw} jet definition.

This is to be compared with the asymptotic prediction $R_g=2.46$ 
for projected rates at central angles in (\ref{rphiqg}) 
whereas for the above angular interval one finds
\begin{equation}
 R_g^{th} \approx 2.4.
\labl{rgmerth}
\end{equation}
in good agreement with the experimental value.
This number (\ref{rgmerth}) is obtained from the 
prediction for the full angular distribution \cite{dkt11}, see section
6.3.3,
which is also shown in Fig. 7.5 with the normalization adjusted. 
The calculation
takes into account both the interjet and intrajet coherence.
As can be seen, the relative depth of the two valleys between the jets is
well reproduced. Also reasonable is the distribution around the gluon jet
direction. 
On the other hand, more particles than expected are found within the quark jets. 
In part this
is a consequence of using in the calculation the asymptotical value
 $4/9$ for $1/C_q^g$, which, as we have already discussed, is approached 
rather slowly from above.
Despite these shortcomings, the main effect, the ratio $R_g$ of particle
densities in the valleys 
between the two types of jets, has been correctly predicted by the
APA.

Another prediction concerns the beam energy dependence. At a higher energy the
angular density in Fig.~7.5 gets essentially multiplied by an overall
factor \cite{dkt11}: the particle density in between the jets rises at a
rate comparable to the density within the jets.

\medskip
\subsection{Colour Coherence in Multi-jet Events from $p\overline{p}$ 
Collisions}

Hadronic high-$p_\perp$ processes are very rich in the collective drag phenomena 
(see e.g.\ \\
\cite{dkmt1,dkmt2,dkt8,emw,eks2}).  
Varying the experimental conditions 
(triggers) one may separate the dominant partonic subprocesses and switch from 
one subprocess to another.  Recall also that the length and height of the hadronic 
\lq\lq plateau" depend here on different parameters:  the length is determined 
by the total {\it energy} of the collision, and the height and the plateau 
structure depend on the {\it hardness} of the process governed, as a rule, by 
the {\it transverse} energy-momentum transfer.  
Thus, information becomes available 
that is inaccessible in $e^+ e^-$ annihilation where both energy and hardness 
were given by the value of $\sqrt{s}$.  
Therefore, the high-$p_\perp$ hadronic collisions 
could provide us with a very prospective laboratory for the detailed studies of 
the colour-related phenomena.  The first experimental results from the 
$p\overline{p}$ Tevatron collider \cite{pb1,cdf1,d0,nv} look rather promising.  
The studies in the $\overline{p}p \rightarrow 3 \: {\rm jet} \: + X$ events of 
the spatial correlations between soft and leading jets in multi-jet events have 
clearly demonstrated the presence of initial-to-final state interference effects 
in $p\overline{p}$ interactions.  Moreover, D0 \cite{d0} results indicate that 
the observed coherence phenomena are in agreement with the tree-level
parton level calculations \cite{dkt8,emw,ggk}. 
Recently the first data on $W$ + jet production from D0 \cite{mel} have
been presented. The hadronic antenna patterns for such process are entirely
analogous to that in the string/drag effect in $e^+e^- \to q\bar q g$
events. The colour coherence effects are clearly seen in the data and in our
view these studies have a very promising future.
Recall that the
detailed studies of the colour-related ``portrait'' of the high-$p_\perp$
events could provide a useful tool to discriminate the new physics signals
($H,Z,W'$, compositness $\ldots$), see, e.g. \cite{eks2}.

\section{Particle Spectra in Deep Inelastic Scattering}
\setcounter{equation}{0}

As is well known, Deep Inelastic Scattering (DIS) has always been a wonderful 
testing ground for the quantitative tests of the basic predictions of QCD.  The 
new era in the detailed investigation of QCD dynamics has been started with the 
advent of HERA.  New impressive QCD physics results continue to pour out from 
both the H1 and ZEUS collaborations at HERA, see e.g.\ \cite{al1,xfe}.

One of the important areas in which quantitative progress in probing QCD could 
be made is the comparative study of properties of jets produced in different 
reactions.  However, despite all their fundamental importance for QCD studies, 
the measurements of the final state particle distributions so far have not 
become a topic of primary interest for the experimentalists\footnote{To the best 
on our knowledge, the first QCD-based analysis of hadronic distributions in 
high energy fixed target experiment has been performed by the EMC \cite{emc}.}.  
Only recently the first HERA results on the inclusive particle spectra in DIS 
and their comparison with the $e^+ e^-$ data have been reported, see Refs.\ 
\cite{pl1,zeus,h1}.

\subsection{Particle Distributions in the Breit Frame}

A hard lepton-hadron interaction with high $-q^2 = Q^2 \gg \Lambda^2$ and fixed 
Bjorken variable $x = \frac{Q^2}{2 (p.q)}$ knocks a quark with longitudinal 
momentum $k = xp$ at virtuality level $k_\perp^2 \lapproxeq Q^2$ out of the 
initial partonic fluctuation.  The probability of finding the appropriately 
prepared quark-parton inside the target proton determines the DIS cross section 
(structure function).

The reader is reminded that in deeply inelastic scattering the picture of the 
development of the final state parton system is different from that in $e^+ e^-$ 
annihilation\footnote{The first theoretical treatment of this subject within 
perturbative scenario has been performed in Refs.\ \cite{dkt1,dkmt1,mc1,gdkt,cfm}.  
A recent comprehensive discussion has been given in Refs.\ \cite{gm1,yd}.}.  In 
particular, in DIS one needs to consider the structure of the parton wavefunction 
of the target hadron, which was prepared long before scattering and corresponds 
to a space-like bremsstrahlung cascade.  As was proved in Refs.\ \cite{dkt1,mc1,gdkt},
the effects of colour coherence here, as in the case of a timelike cascade, 
leads to a picture of soft bremsstrahlung with angular ordering.

The brief discussion of the theoretical issues below follows that of Ref.\ 
\cite{gdkt}.  Our main concern is with the distribution of particles associated 
with small $x$ deeply inelastic events.

\subsubsection{Arguments in Favour of the Breit Frame}
When discussing the final-state hadron distributions in DIS with large 
momentum transfer $Q^2$ and fixed value of Bjorken variable $x$ it is instructive 
to compare them with the jets produced in $e^+ e^-$ annihilation.  In the latter 
process two point-like colour quarks $(3 + \overline{3})$ moving apart cause 
QCD bremsstrahlung which populates the final state ($q$ and $\overline{q}$ jets).  
In DIS one deals with the struck quark (3) and the colour \lq\lq hole" 
($\overline{3}$).  The 
latter has its own internal structure (e.g.\ $\overline{3} + 8$ at $x \ll 1$) 
which complicates the properties of the target fragmentation (TF) region.  
As in $e^+ e^-$ collisions, the first natural step in studying the final-
state properties of hadrons in DIS is to examine the inclusive characteristics 
such as single-particle energy spectrum, rapidity distribution, etc.  As we 
know, these inclusive quantities are typically not Lorentz covariant.  The 
crucial question concerns the proper identification of the reference frame 
which is the most appropriate for QCD treatment of the final state structure 
in DIS.  For instance, since in DIS the hard collision occurs between the off-
mass-shell intermediate boson and the target proton (not between the electron 
and the proton), the laboratory frame does not look natural for studying the 
final state properties.  As was mentioned in \cite{yd}, analysis of the 
distribution of particles in this frame reminds us of an attempt to establish the 
laws of classical mechanics from a carousel laboratory.

In the golden age of the parton model \cite{rf,vng2} the exclusive role of the 
Breit (\lq\lq brick wall") frame was commonly accepted, see also 
\cite{swz,abas,ilk}.
The Breit frame is the one in which the momentum transfer $q^\mu$ has only a 
$z$-component:  $q^\mu (0, 0, 0, -Q)$.  In the quark-parton model, in this frame 
the struck quark enters from the left with $z$-momentum $p_z = \frac{1}{2} Q$ 
and sees the exchanged virtual boson as a \lq\lq brick wall" from which it simply 
rebounds with $p_z = - \frac{1}{2} Q$.  Thus, the left-hand hemisphere of the 
final state should look just like one hemisphere of an $e^+ e^-$ annihilation 
event at $E_{cm} = Q$.  The right-hand (\lq\lq beam jet") hemisphere, on the 
other hand, is different from $e^+ e^-$ because it contains the proton remnant, 
moving with momentum $p_z = (1 - x) Q/2x$ in this frame.  The standard argument 
in favour of this frame was that here the two sources of final particle production 
were expected to be best spatially separated, see Fig.\ 8.1a.  Namely, the current
fragmentation (CF) region due to the time-like evolution of the struck quark and 
the TF region populated with the evolving remnants of the disturbed space-like 
partonic system related to the relativistic proton should correspond to two 
opposite hemispheres in the Breit frame.  In this frame $x_p = \frac{2 p_h}{Q}$ 
is the proper equivalent of the $e^+ e^-$ definition of a fragmentation variable.  
As advocated in Refs.\ \cite{swz,dkt1,gdkt}, within the perturbative scenario all the 
advantages of the Breit frame as the centre-of-mass frame for the \lq\lq quark + 
hole" pair are still there.  In \cite{yd} Dokshitzer presented a new 
(perturbative) argument in favour of the Breit frame based on the analysis of the 
distribution of the accompanying large angle gluon radiation over pseudorapidity $\eta$ 
(relative to the virtual boson-proton axis).

According to \cite{yd} the proper choice of the reference frame is 
asymptotically dictated by 
fixing the point at which the intensity of the perturbative radiation (i.e.\ 
the hardness of the underlying process) is the highest.  Thus, in $e^+ e^-$ 
annihilation the pseudorapidity distribution of the final-state particles is 
symmetric and peaks at $\eta = 0$ in the $q\overline{q}$ centre-of-mass.  
Contrary to all other \lq\lq collinear frames", where $q$ and $\overline{q}$ 
have different energies, here the maximum of the $\eta$-distribution is 
positioned exactly at $\eta = 0$.  In the case of DIS the $\eta$-spectrum is 
asymmetric, but in a close similarity with $e^+ e^-$ collisions, the position 
of its maximum \lq\lq measures" the location of $\eta = 0$ in the Breit frame 
of the hard collision.

\subsubsection{Feynman-Gribov Puzzle}
Thus, in the Breit frame of reference the DIS process looks like the abrupt 
spreading of two colour states:  3 (the struck quark $q$) and $\overline{3}$ 
(the disturbed 
proton), moving in the opposite directions.  Here one conceptual remark is in 
order, see for more details \cite{ilk,gdkt,dkmt2}.  An important r\^{o}le in 
justification of the celebrated Feynman hypothesis \cite{rf} about a universal 
hadronic plateau (see Fig.\ 8.1a) has been assigned to the argument that the
continuously distributed hadrons connecting the two fragmentation regions are 
necessary to compensate for the fractional charges of quark-partons.  At the same
time, some serious doubts have been expressed about the possibility to organize 
dynamically such a state if one proceeds from the idea to consider the successive 
decays of outgoing partons as the only source of multiple production of final 
particles.  The problem has been formulated by Gribov in \cite{vng2}.  It was 
argued that in DIS the offsprings of the fragmentation of an initially prepared 
fluctuation had to leave the rapidity interval $0 < \ln \omega < \ln Q$ 
{\it depopulated} as shown in Fig.\ 8.1b ($\omega$ being the energy of registered 
hadron $h$).  The lack of hadrons with $\omega \ll Q$ in TF is, according to 
Gribov, related to the fact that the coherence of the partonic wave function 
remains undisturbed by the hard knocking out of a parton with momentum $xp 
\approx Q/2$.  As a result the upper (low momentum) part of the partonic 
fluctuation in Fig.\ 8.1b \lq\lq collapses" as if there were no hard scattering
at all.

The way in which this puzzle\footnote{In the old times it was sometimes called 
\lq\lq Feynman-Gribov puzzle" or the problem of \lq\lq Gribov's gap".} is 
resolved within the perturbative QCD context \cite{gdkt} appears to be in a 
certain sense mixed.  On the one hand, here Gribov's phenomenon does occur:  
the energy spectrum of the offsprings of the parton dissociation --- elements of 
the \lq\lq ladder" fluctuations determining the DIS cross-section --- proves to 
be concentrated mainly in the region of large energies $Q < \omega < p$.  On 
the other hand, there is a specific mechanism responsible for bridging the 
regions of target and current fragmentation in the spirit of the Feynman picture.  
This r\^{o}le is taken on by coherent bremsstrahlung of soft gluons which is 
insensitive to details of the structure of partonic wave function of the target 
hadron, being governed exclusively by the value of the total colour 
charge transferred in the scattering process.  As far as we are aware the 
importance of the $t$-channel colour transfer for filling in Gribov's gap 
was first emphasized in \cite{ilk}.  In fact, Gribov's phenomenon provides 
one with an additional support for the choice of the Breit frame.

\subsubsection{Current Fragmentation Region}
As it has been demonstrated in \cite{dkt1,dkmt1,gdkt}, the bremsstrahlung 
processes accompanying the emission of a struck quark lead to the formation in 
the CF region of a jet identical to the light quark jet in $e^+ e^-$ annihilation 
at $W^2 = Q^2$.  In particular, the appearance of a hump-backed plateau was 
predicted for the scaled momentum spectrum.  Therefore, it always looked quite 
instructive to compare the DIS data in the current region directly 
with $e^+ e^-$ results.  One can benefit here from the possibility to vary $Q^2$ 
quite smoothly and to perform some new quantitative tests of cascading picture 
of multiple hadroproduction in jets.  DIS measurements have the advantage over 
$e^+ e^-$ studies since they allow us to cover a large range in $Q$ in a single 
experiment.  Measurements with particle identification are especially interesting.  
However, we should warn the reader that in the $e^+ e^-$ case the data correspond 
to the mixture of the primary quarks with different flavours.  Corrections caused 
by the effects induced by heavy quarks should be taken into account at a certain 
state of the analysis.

\subsubsection{Target Fragmentation Region}
DIS also opens a new domain for studying charged particle spectra, namely the 
TF region.  Here the situation proves to be much more complicated, especially at 
$x \ll 1$ \cite{gdkt}.  The DIS occurs in this case on a \lq sea' quark from 
bremsstrahlung of soft $(q\overline{q})_g$ pairs in colour octet state ($g$ 
exchange in $t$-channel).  The dominant structure of the appropriate fluctuation 
can be characterized in terms of the multirung ladders determining the small $x$ 
behaviour of the structure function.  It is of importance to mention that sets of 
graphs which cancel in the structure function no longer cancel in the inclusive 
spectrum.  In addition to the offsprings of decaying subjets --- remnants of the 
ordered \lq ladders' (structural contribution), the collective coherent accompaniment 
arises, which is determined by the overall colour topology of the partonic system 
($t$-channel contribution).

In the TF there also arises the hump-backed particle distribution evolving with 
$\ln Q^2$ and $\ln 1/x$.  The resulting spectrum of hadron $h$ in the TF is depicted 
in Fig.\ 8.2 where the CF region is added to show the full hadronic distribution 
due to gluon emission.  For both CF and TF regions $\ln \frac{\omega}{\Lambda}$ 
is plotted separately in this figure.  Certainly, this plot could be used only 
for illustrative purposes since in \cite{gdkt} some practically important 
phenomena were not taken into account (for applications at more realistic 
conditions, see e.g.\ \cite{ava}).

One may anticipate that at HERA some specific perturbative predictions for the 
TF region \cite{gdkt} should be visible.  Thus, as in $e^+ e^-$ annihilation, 
coherence in DIS should lead to hardening of the hadronic spectra:  the 
population of particles with finite energies $\left ( \ln \frac{\omega}{\Lambda} 
\sim 1 \right )$ is expected to remain practically constant with increase of the 
hardness of the process.  It looks quite challenging to test experimentally the 
predicted evolution of the spectrum in TF with $Q^2$ and $x$.  Note also that at 
small values of $x$ the accompanying bremsstrahlung in the target fragmentation 
region is expected to be roughly twice as intense as that in the current fragmentation.

One final theoretical comment is here in order\footnote{We are grateful to Yu.\ 
L.\ Dokshitzer for elucidating discussion of this topic, see also Ref.
\cite{gdkt}.}.  In principle, one 
may wonder whether the transition at very low $x$ from the standard DGLAP 
evolution \cite{gl,ap,yld1} to the so-called BFKL \cite{bfkl} dynamics could 
strongly affect the predictions of Ref.\ \cite{gdkt}.  To our understanding the 
structure of the accompanying soft gluon radiation does not undergo any essential 
change.  In short, the reason is that even in the BFKL case the angular ordering 
of the \lq\lq structural gluons" (the ladder rungs which determine the cross 
section) is still there.

\subsection{Comparative Study with the Results from $e^+e^-$
Annihilation}

The first studies at HERA of the charged particle energy distributions in the 
Breit frame \cite{zeus,h1} look rather encouraging.  Recall that compared to the 
fixed target measurements the available values of $Q^2$ are much larger and $x$-
values are much smaller.  This results in far more intensive QCD radiation activity 
and opens a wide domain for examining the evolution of the particle spectra with 
$Q^2$ and $x$.

So far the main experimental efforts have concerned the comparative study of the 
current jet hemisphere in DIS in the Breit frame of reference with one hemisphere 
of an $e^+ e^-$ annihilation event at $\sqrt{s} = Q$.  It is of importance that 
the detailed comparison of the inclusive particle distributions over the natural 
scaled variable $x_p = \frac{2p_h}{Q}$ can be carried out without the 
necessity of
using a jet-finding algorithm, see e.g.\ \cite{dkt1,dkmt1,gdkt}.

The experimental analysis \cite{zeus,h1} conducted in the Breit frame shows 
(see Figs.\ 8.3--8.6) that the charged hadron spectrum not only has the same shape
as that seen in a single hemisphere of an $e^+ e^-$ event but also that this 
shape evolves in $Q^2$ in the same way as the latter does in terms of the $e^+ e^-$ 
centre-of-mass energy $\sqrt{s}$.  Even more, the absolute normalization, the 
mean charged multiplicity (see Fig.\ 8.3), follows the same curve except at low
$Q$, or $\sqrt{s}$, where there is a systematic shortfall \cite{zeus,h1}.  
It seems that this 
multiplicity discrepancy can be attributed to the contributions from the mechanisms 
at DIS that do not have an equivalent in $e^+ e^-$ collisions.  This concerns 
the proton-remnant hadronization, initial state QCD radiation etc., see e.g.\ 
\cite{gt}.  In our view, the string/drag effect can be partly responsible for a 
certain depopulation of particles in the current fragmentation.  Note that the 
difference in the flavour composition between the DIS and the $e^+ e^-$ case is 
unlikely to be responsible for this discrepancy.  This effect gives rise to 
differences in the results that are within the quoted experimental errors \cite{zeus}.

The HERA data \cite{zeus,h1} on the scaled charged particle distribution in 
$\xi_p = \ln \frac{1}{x_p}$, where $x_p = \frac{2p_h}{Q}$ in the Breit frame, 
exhibit the predicted hump-backed plateau, see Figs.\ 8.4, 8.5. Its
evolution is in agreement with the perturbative expectations.  The measured area, 
peak position and the width $\sigma$ of the spectrum confirm that the evolution 
variable, equivalent to the $e^+ e^-$ cms energy $\sqrt{s}$, is in the Breit 
frame $Q$.  The variation of the peak position $\xi_p^*$ with $Q$ follows the 
$e^+ e^-$ curve, see Fig.\ 8.6. A straight line fit to $\xi_p^*$ as a function
of $\ln Q$ gives a gradient of $0.75 \pm 0.05$ which is in agreement with Eqs.\ 
(\ref{2.51}) and (\ref{3.3}).  As expected \cite{gdkt}, 
see also Fig. 8.2, the data indicate no 
observable Bjorken $x$ dependence of the peak position $\xi_p^*$ 
as can be seen in Fig.\ 8.7 \cite{h1}.

In conclusion, HERA results on universality of the basic features of the charged 
particle distributions in the struck quark hemisphere provide us with new 
experimental support for the APA picture of multiparticle production in hard 
processes.  In particular, it is confirmed experimentally that the comparison 
with $e^+ e^-$ data can be made properly using as the evolution variable in the 
Breit frame the four-momentum transfer $\sqrt{Q^2}$.

\bigskip
\section{Inclusive Properties of Heavy Quark Initiated Events}
\setcounter{equation}{0}

In section 3 we considered mainly the properties of jets in $e^+ e^-$ collisions
obtained by analyzing the full hadron data sample with contributions from all 
flavours.  In the last years the intensive experimental studies have been 
performed with the heavy quarks $Q (c, b)$.  Experiments at $Z^0$ have produced 
a wealth of new interesting results on the profiles of jets initiated by heavy 
quarks, see e.g.\ \cite{my1,ada2,wjm2,pm1,osoaa}.  The accuracy of measurements 
started to be comparable with that in the $q\overline{q}$ events.  This is mainly 
related to the steady improvement in the heavy quark tagging efficiencies.  
Further progress is expected from the measurements at LEP-2 and, especially, at 
a future linear $e^+ e^-$ collider.  The principal physics issues of these 
studies are related not only to testing the fundamental aspects of QCD, but also 
to their large potential importance for measurements of heavy particle 
properties:  lifetimes, spatial oscillations of flavour, searching for 
CP-violating effects in their decays etc.  Properties of $b$-initiated jets are 
of primary importance for analysis of the final state structure in 
$t\overline{t}$ production processes.  A detailed knowledge of the 
$b$-jet profile is also essential for the Higgs search strategy.

The physics of heavy quarks has always been considered as one of the best 
testing grounds for QCD. 
In this section we will restrict ourselves to the discussion of the
inclusive energy spectrum of hadrons containing a heavy quark and of the
associated mean multiplicity which are related to our previous
considerations of soft particle production and LPHD.
These properties of heavy quark jets have recently been studied 
within the perturbative approach in a self-consistent way. We
 follow here mainly the results 
obtained by the Leningrad/St.\ Petersburg QCD group 
 \cite{dkt1,dfk1,dkt3,yld2,yld3,dkt7,dkt12,vak4}.  
In the calculations of the heavy quark energy spectrum -- unlike the previous
sections of the review -- the two-loop effects are included which allows one to
determine the coupling $\alpha_{\overline{MS}}$ which can be compared with
the results from other processes. In order to include the soft effects 
an effective coupling $\alpha_s^{\rm eff}(k)$ is introduced which remains finite
down to $k=0$. This procedure may be viewed as an alternative to the
previously considered fixed cut-off $k>Q_0$.

\medskip
\subsection{Energy Spectrum of Heavy Quarks}
\subsubsection{Leading Particle Effect in Heavy Quark Fragmentation}

 We first recall the main ideas of the conventional evolution 
approach to the description of the energy spectra of heavy flavoured hadrons 
$H_Q$. One starts from a phenomenological fragmentation function for the 
transition
\begin{equation}
Q (x) \; \rightarrow \; H_Q (x_H)
\label{7.1}
\end{equation}
\noindent and then traces its evolution with the annihilation energy 
$W$ by means of perturbative QCD.  Realistic fragmentation functions \cite{cpmb} 
exhibit a parton-model-motivated maximum at (hereafter $M$ is the heavy quark 
mass)
\begin{equation}
1 \; - \; \frac{x_H}{x} \: \sim \: \frac{{\rm const}}{M}.
\label{7.2}
\end{equation}
\noindent Taking gluon radiation at the perturbative stage of evolution into 
account would then induce scaling violations that soften the hadron spectrum by 
broadening (and dampening) the original maximum and shifting its position to 
larger values of $1 - x_H$ with increasing $W$.  Such an approach has been 
successfully applied in Ref.\ \cite{mn} where the effects of multiple soft gluon 
radiation have been examined.  The realistic fragmentation functions have been 
used there to describe the present day situation with heavy particle spectra and 
to make reliable predictions for the future.  Being formally well justified, 
this approach, however, basically disregards the effects that the finite quark 
mass produces on the accompanying QCD bremsstrahlung pattern, since the $W$-
evolution by itself is insensitive to $M$.

The so-called Peterson fragmentation function \cite{cpmb} is often used in the 
literature to describe the $Q \rightarrow H_Q$ transition,
\begin{equation}
C_Q (y) \; = \; N \: \frac{\sqrt{\epsilon_Q}}{y} \; \left [ \frac{1 - y}{y} \: 
+ \: \frac{\epsilon_Q}{1 - y} \right ]^{-2}, \;\; y = x_H/x,
\label{7.3}
\end{equation}
\noindent where $N$ is the normalization constant 
\begin{equation}
\int_0^1 \: dy C_Q (y) \; = \; 1; \;\;\; N \; = \; 
\frac{4}{\pi} \: [1 + {\cal O} 
(\epsilon_Q^{1/2}) ].
\label{7.4}
\end{equation}
\noindent The pick-up hadronization picture predicts that the small parameter 
$\epsilon_Q$ in (\ref{7.3}) should scale with heavy quark mass as
\begin{equation}
\epsilon_Q \; \approx \; \left ( \frac{m_q}{M} \right )^2 \; \propto \; M^{-2},
\label{7.5}
\end{equation}
\noindent (where $m_q$ is the quantity of the order of constituent light quark 
mass), in accordance with (\ref{7.2}).  A sharp peak of the energy distribution 
at large $y$,
\begin{equation}
1 - y_{\rm max} \; \propto \; \sqrt{\epsilon_Q} \; \propto \; M^{-1} \; \ll \; 1,
\label{7.6}
\end{equation}
\noindent manifests the leading particle effect in the heavy quark fragmentation 
\cite{afk}.

At the same time, assuming $M$ a sufficiently large scale, it 
is tempting to carry out a program of deriving the predictions that would keep 
under perturbative control, as much as possible, the dependence of $H_Q$ 
distributions on the quark mass.  As shown in detail  in Ref.~\cite{dkt3} 
the effects of perturbative gluon radiation 
are capable of reproducing the shape of the Peterson fragmentation function 
provided one feels courage enough to continue the perturbative description down 
to the region of small gluon transverse momenta.  The Sudakov form factor 
suppression of the quasi-elastic region $x \rightarrow 1$ results in the 
distribution which is qualitatively similar to the parton model motivated 
expression (\ref{7.3}).

As we discussed in detail above the dominant r\^{o}le of the perturbative 
dynamics has been very successfully tested in studies of light hadron 
distributions in QCD jets.  As far as the heavy quark spectrum is concerned, 
based on LPHD one may expect that a purely perturbative treatment is dual to 
the sum over all possible hadronic excitations.  Therefore, without involving 
any fragmentation function at the hadronization stage, one could attempt to 
describe the energy fraction distribution averaged over heavy-flavoured hadron 
states, the mixture that naturally appears, e.g., in the study of inclusive 
hard leptons.

Although such a purely perturbative approach might look rather naive, it is at 
least free from the problem of \lq\lq double counting" which one faces when 
trying to combine the effects of perturbative and hadronization stages.

\subsubsection{Perturbative Results for the Inclusive Spectrum}

In Refs.\ \cite{dkt12,dkt3} an expression was derived for the perturbative 
energy spectrum $D (x; W, M)$ of a heavy quark $Q$ produced in $e^+ e^- 
\rightarrow Q (x) + \overline{Q} +$ light partons as a function of the energy 
fraction $x = \frac{2E_Q}{W}$, $W$ and the heavy quark mass $M \equiv mW$.
%
The inclusive energy spectrum is normalized as
\begin{equation}
\int_{2m}^1 \: dx \: D (x; W, M)  =  1
\label{7.8b}
\end{equation}
and can be derived conveniently from its moment representation
\begin{equation}
D (x; W, M) \; \equiv \; \sigma_{\rm tot}^{-1} \: \frac{d \sigma}{dx} = 
\int_\Gamma \: \frac{dj}{2 \pi i} \: x^{-j} \: D_j (W, M).
\label{7.8a}
\end{equation}
\noindent Here the contour $\Gamma$ 
runs parallel to the imaginary axis in the complex 
moment $j$ plane to the right of all singularities. 
The integrand in (\ref{7.8a}) can be written as
\begin{equation}
\ln D_j \; = \; \int_{2m}^1 \: dx \: \left [x^{j - 1} - 1 \right ] \; 
\frac{dw (x; W, M)}{dx}.
\label{7.11}
\end{equation}
This type of relation is obtained for the distribution of the valence quark
in a quark jet in LLA \cite{gl,ap,yld1} (see also \cite{dkmt2}) where
$\frac{dw}{dx} \sim \Phi_F^F(x)$ and is calculated from 
the transition $q\to qg$ in the 
lowest order. In the present application 
 the \lq\lq radiator" $\frac{R}{2 \pi}
= C_F^{-1} v \frac{dw}{dx}$ is computed to the accuracy 
${\cal O}(\alpha_s^2(W))$
and takes into account the finite mass and two-loop effects.
The corresponding expression reads
\begin{eqnarray}
C_F^{-1} \: v \frac{dw}{dx} & = & \int_{\kappa^2}^{Q^2} \: \frac{dt}{t} \left \{
a (t) \left [ \frac{2 (x - 2m^2)}{1 - x} \: + \: \zeta^{-1} (1 - x) \right ] \: 
- \: a^\prime (t) (1 - x) \: + \: a^2 (t) \: \Delta^{(2)} (x) \right \} \nonumber 
\\
& & \label{7.12}\\
& + & \beta (x) \; \left \{ - \frac{2x}{1 - x} \: [a (Q^2) + a (\kappa^2) ] \: 
+ \: \zeta^{-1} \: \frac{x (x - 2m^2)}{2 (1 - x)} \: \left ( \frac{1 - x}
{1 - x + m^2} \right )^2 \: a (Q^2) \right \}. \nonumber
\end{eqnarray}
Here the integral over the gluon virtuality $t$ is limited by the 
two characteristic momentum scales 
\begin{eqnarray}
Q^2 \; \equiv \; Q^2 (x) & = & W^2 \: \frac{(1 - x)^2}{1 - x + m^2} \; z_0,
\nonumber \\
& & \label{7.13}\\
\kappa^2 \; \equiv \; \kappa^2 (x) & = & M^2 \: \frac{(1 - x)^2}{z_0}, 
\nonumber
\end{eqnarray}
\noindent with
$$
z_0 \; \equiv \; \frac{1}{2} \: (x - 2m^2 + \sqrt{x^2 - 4m^2} ) \; = \; x + 
{\cal O} (m^2).
$$
\noindent The following notation is also used:
\begin{eqnarray}
m \; \equiv \; \frac{M}{W} \leq \frac{1}{2}; & v \; \equiv \; \sqrt{1 - 4m^2}; & 
\beta (x) \; \equiv \; \sqrt{1 - 4m^2/x^2}, \;\; 0 \leq \beta \leq v; \;\;\; \zeta 
\; = \; 1 + 2m^2 \nonumber \\
& &  \label{7.14}\\
& a^\prime (k^2) \; = \; \frac{d}{d \ln k^2} \: a (k^2); & a (k^2) \; \equiv \; 
\frac{\alpha_s (k)}{2 \pi}. \nonumber
\end{eqnarray}
\noindent The radiator (\ref{7.12}) vanishes at
\begin{equation}
x \; = \; x_{\rm min} \; = \; 2m; \;\; \left ( \beta (x_{\rm min}) \; = \; 0, 
Q^2 \; = \; WM \: \frac{(1 - 2m)^2}{1 - m} \; = \; \kappa^2 \right ),
\label{7.15}
\end{equation}
\noindent thus justifying the lower kinematical limit in Eq.\ (\ref{7.11}).
In (\ref{7.12}) the 
two-loop effects are included in the $a'$ and $\Delta^{(2)}$
terms. It turns out in practical applications that 
$\Delta^{(2)}=0$ holds to a good approximation provided the physical effective
coupling $a_{\rm eff}=a_{\overline{MS}}\,(1+a{\cal K})$ is used with
\begin{equation}
{\cal K}=\left[ N_C\left( \frac{67}{18}-\frac{\pi^2}{6}\right)\;-
   \; \frac{5}{9} \; n_f \right].
\label{Kconst}
\end{equation}
Strictly speaking, Eq. (\ref{7.12}) applies for the vector channel only
 ($\zeta\equiv \zeta_V$), but the difference between the spectra in the
vector and axial vector channels vanishes like ($1-v$) in the relativistic
case and so the equation applies also for a $V+A$ mixture, in particular, at
the $Z^0$ peak. 

In the relativistic approximation, $m \ll \frac{1}{2}$, one arrives at the 
simplified expression
\begin{eqnarray}
C_F^{-1} \: \frac{dw}{dx} & = & \int_{\kappa^2}^{Q^2} \: \frac{dt}{t} \; \left 
\{ a (t) \: P (x) - a^\prime (t) (1 - x) + a^2 (t) \: \Delta^{(2)} (x) \right 
\} \nonumber \\
& + & a (Q^2) \; \left \{ \frac{- 2x}{1 - x} \: + \: \frac{x^2}{2 (1 - x)} 
\right \} \; + \; a (\kappa^2) \: \left \{ \frac{- 2x}{1 - x} \right \}; 
\label{7.16}\\
Q^2 & = & W^2 \: x (1 - x), \;\; \kappa^2 \; = \; M^2 (1 - x)^2/x, \nonumber
\end{eqnarray}
\noindent where
\begin{equation}
P (x) \; = \; \frac{1 + x^2}{1 - x}.
\label{7.17}
\end{equation}
\noindent The integration variable $t$ determines the physical hardness scale 
of the running coupling, and is related to the transverse momentum of the 
radiation.  In the dominant integration region,
\begin{equation}
k_\perp^2 \; \ll \; W^2, \;\;\; t \; = \; x  k_\perp^2.
\label{7.18}
\end{equation}
\noindent The lower limit $t_{\rm min}= \kappa^2$ sets the boundary for the essential 
gluon emission angles,
\begin{equation}
t \sim E_Q \: \frac{2}{W} \: (\omega_g \Theta)^2 \; \geq \; \kappa^2 \sim 
M^2 \: \frac{\omega_g^2}{E_Q} \: \frac{2}{W} \; \Longrightarrow \; \Theta \geq 
M/E_Q \equiv \Theta_0.
\label{7.19}
\end{equation}
\noindent This restriction manifests the \lq\lq Dead Cone" phenomenon 
characteristic for bremsstrahlung off a massive particle. Namely the forward
gluon radiation is suppressed 
\cite{dkt1,dkt7,dkt12,imd2}.
The latter is largely responsible for the differences between radiative particle 
production in jets produced by a light and a heavy quark, by a $c$-quark and a 
$b$-quark (see \cite{dkt7,yld2,bas}).

The analytical perturbative result for the inclusive spectrum $D (x; W, M)$ has 
the following properties \cite{dkt3}:
\begin{itemize}
\item it embodies the exact first order result \cite{bk} $D^{(1)} = \alpha_s  
\cdot f (x; m)$ and, as a consequence,

\item it has the correct threshold behaviour at $1 - 2m \ll 1$; 

\item in the relativistic limit $m \ll 1$, it accounts for all significant 
logarithmically enhanced contributions at high orders, including
\begin{itemize}
\item[1.] running coupling effects,
\item[2.] the two-loop anomalous dimension and
\item[3.] the proper coefficient function with exponentiated Sudakov-type logs, 
which are essential in the quasi-elastic kinematics, $(1 - x) \ll 1$;
\end{itemize}
\item it takes into full account the controllable dependence on the heavy quark 
mass that makes possible a comparison between the spectra of $b$ and
(primarily produced) $c$ quarks.
\end{itemize}
\noindent By \lq\lq threshold behaviour" we mean here the kinematical region of 
non-relativistic quarks, $|W - 2M| \sim M$, in which gluon bremsstrahlung 
acquires additional dipole suppression.  At the same time we will not account 
for the Coulomb effects that would essentially modify the production cross 
section near the actual threshold, $W - 2M \ll M$ \cite{fk}.
Note that in the derivation of the perturbative formulae 
(\ref{7.8a}-\ref{7.12}) the 
production of additional $Q\overline{Q}$ pairs has been disregarded.


 When $x$ is taken as close to 1 as
\begin{equation}
(1 - x) \; \lapproxeq \; \frac{\Lambda}{M}
\label{7.21}
\end{equation}
then one observes that  $\kappa^2\lapproxeq \Lambda^2 $ and the integration 
in Eq.\ (\ref{7.16}) leaves the perturbatively controlled 
domain, $\kappa^2 \gg \Lambda^2$.  This shows that the 
quark spectrum appears to be \lq\lq infrared sensitive" 
in the kinematical region 
(\ref{7.21}).  A large quark mass $M \gg \Lambda$ 
does not guarantee an infrared 
stability of the quark distribution at large $x$.  
Note that
a finite quark mass suppresses the 
emission of gluons at small {\it angles}, 
$\Theta \leq \Theta_0 \equiv 2M/W$ ("Dead Cone"), but not 
necessarily that of gluons with small {\it transverse momenta}. 
So, in the \lq\lq quasi-elastic" region (\ref{7.21}) 
the non-perturbative 
confinement physics inevitably enters the game.

One may single out effects of the non-perturbative momentum region by splitting 
the radiator into two pieces corresponding to large and small transverse 
momentum regions ($t \approx k_\perp^2$ for $x$ close to 1),
\begin{equation}
\frac{dw}{dx} \; = \; \frac{dw}{dx} \; [k_\perp^2 > \mu^2] \: + \: \left \{ 
\frac{dw}{dx} \right \}^{(C)} \; [k_\perp^2 \leq \mu^2].
\label{7.22}
\end{equation}
\noindent Correspondingly, the spectrum in the moment representation factorizes 
as 
\begin{equation}
D_j \; = \; F_j [k_\perp^2 > \mu^2] \: \times \: D_j^{(C)}.
\label{7.23}
\end{equation}
\noindent This formal separation becomes 
informative if one is allowed to choose 
the boundary value $\mu$ well {\it below} the quark mass scale 
(e.g.\ $\mu = {\cal O}(1$ 
GeV) providing $\mu/M \ll 1$ for the $b$ quark case).  
Within such a choice only 
$x$ close to 1 would contribute to $w^{(C)}$.  Keeping track of the leading 
$\frac{\mu}{M}$ effects, it is straightforward \cite{dkt3} to obtain the 
following formal expression for the \lq\lq confinement" part,
\begin{equation}
\ln D_j^{(C)} \; \approx \; - 2C_F \: \int_0^\mu \: 
\frac{dk_\perp}{k_\perp} \: 
\frac{\alpha_s (k_\perp)}{\pi} \; 
\ln \: \left [ 1 + \frac{k_\perp}{M} \: (j - 1) \right ].
\label{7.24}
\end{equation}
For moments of moderate order $j$ (that is, for $x$ not specifically 
close to 1) one can expand the logarithm in (\ref{7.24}) and arrives at
\begin{equation}
\ln D_j^{(C)} \; \approx \; - 2C_F \: (j - 1) \: \frac{\mu_\alpha}{M},
\label{7.25a}
\end{equation}
\noindent with
\begin{equation}
\mu_\alpha \; \equiv \; \int_0^\mu \: dk \: \frac{\alpha_s (k)}{\pi},
\label{7.25b}
\end{equation}
\indent Let us emphasize that the perturbatively based expressions (\ref{7.24}), 
(\ref{7.25a}) for the non-perturbative 
fragmentation function are meaningful \lq\lq if 
and only if\," the effective running coupling is free from a formal infrared 
singularity.  This is how the notion of an \lq\lq infrared regular coupling" 
appears in the context of the heavy quark inclusive spectra (for further details 
see Refs.\ \cite{dkt3,yld2,yld3}.

Note that neither the Peterson-like functions nor the perturbatively-motivated 
\lq\lq confinement" distribution are unambiguously defined quantities.  The 
former as an \lq\lq input" for the evolution are by themselves 
affected by gluon 
radiation effects at the hard scale $t \sim M^2$ that are present even at 
moderate $W \gapproxeq 2M$.  On the other hand, $D [k_\perp^2 \geq \mu^2]$ 
crucially depends on the arbitrarily introduced separation scale $\mu$ which 
disappears only in the product of the factors responsible for the perturbative 
and non-perturbative stages (see Eq.\ (\ref{7.23})).  Nevertheless, one may 
attempt to perform a direct correspondence 
between these two quantities.  Namely the phenomenological function
$C_Q (x)$ from (\ref{7.3})
in the $j$-representation can be confronted with $D_j^{(C)}$ as given 
by Eq.\ (\ref{7.24}), provided that  at small scales a phenomenologically
motivated, infrared finite $\alpha_s$ 
is incorporated into the perturbative analysis.

The formal asymptote $D (x \rightarrow 1)$ depends on details of the behaviour 
of $\alpha_s$ near the origin, eventually, on the convergence of the characteristic 
integral
\begin{equation}
\Xi_0 \; \equiv \; \int_0^\mu \: \frac{dk}{k} \: \frac{\alpha_s (k)}{\pi}.
\label{7.26}
\end{equation}
\noindent If $\alpha_s$ vanished at the origin, then $\Xi_0 < \infty$ and,
by evaluating the inverse Mellin transform, 
it is a simple exercise to derive 
the power-like behaviour
\begin{equation}
D (x) \: \propto \: (1 - x)^{-1 + 2C_F \Xi_0}.
\label{7.27a}
\end{equation}
\noindent For $\alpha_s (0) > 0$ one obtains instead a stronger suppression,
\begin{equation}
D (x) \: \propto \: (1 - x)^{-1} \:  \exp \left \{ - C_F \: \frac{\alpha_s (0)}
{2 \pi} \: \ln^2 (1 - x) \right \}.
\label{7.27b}
\end{equation}
\noindent In practice, in both cases the gross features of the popular Peterson 
fragmentation function are easily reproduced.

An important message comes from comparing the {\it mean energy losses} that 
occur at the hadronization stage.  Recall that the mean $x$ corresponds to the 
second moment, $D_{j = 2}$.  The distribution (\ref{7.25a}) results in
\begin{equation}
\ln D_2^{(C)} \; \equiv \; \ln \langle x \rangle^{(C)} \; = \; -2C_F \: 
\frac{\mu_\alpha}{M} \: \left [ 1 + {\cal O} \left ( \frac{\mu}{M} \right ) 
\right ], \;\;\; \langle x \rangle^{(C)} \; = \; \exp \left \{ -2C_F 
\frac{\mu_\alpha}{M} \right \}.
\label{7.28}
\end{equation}
\noindent This immediately leads to
\begin{equation}
1 - \langle x \rangle^{(C)} \; \approx \; 2C_F \: \frac{\mu_\alpha}{M},
\label{7.29}
\end{equation}
\noindent which justifies the expected \cite{afk} scaling law\footnote{A 
similar 
behaviour was advocated in Ref.\ \cite{rlj} which treated the difference between 
the hadron and the heavy quark masses as a small expansion parameter.}
\begin{equation}
1 - \langle x \rangle_{\rm fragm} \; \sim \; \sqrt{\epsilon_Q} \propto M^{-1}.
\label{7.30}
\end{equation}
\noindent Thus, instead of convoluting a phenomenological fragmentation function 
$C_Q (x)$ with the $W$-dependent \lq\lq safe" evolutionary quark distribution 
one may try to use consistently the \linebreak perturbatively-motivated description that 
would place no artificial separator between the two stages of the 
hadroproduction.  At first sight, one does not gain much by substituting one non-perturbative 
quantity --- the phenomenological fragmentation function $C_Q (x)$ --- by another 
unknown function, namely the effective long-distance interaction strength 
$\alpha_s^{\rm eff} (k)$ (at, say, $k \lapproxeq 2$ GeV).  An essential 
difference between the two approaches is that $\alpha_s^{\rm eff}$
is supposed to be a universal, 
process independent quantity and, therefore, fits naturally into 
the perturbative LPHD approach.  
Therefore, for instance, quite substantial 
differences between the inclusive spectra of $c$- and $b$-flavoured hadrons are 
expected to be under full perturbative control.  Moreover, the same notion 
of the infrared-finite effective coupling can be tried for a good many interesting 
problems in the light quark sector (for a list of such phenomena, see Ref.\ 
\cite{dkt3}).

\subsubsection{Comparison with Experimental Data}
The first comparison \cite{dkt3,yld3} of the perturbative predictions with the 
experimentally measured mean energy losses in charm and beauty sectors have 
demonstrated the consistency of the perturbative approach.

To examine the infrared sensitivity of the perturbative results various recipes 
of how to extrapolate the characteristic function
\begin{equation}
\xi (Q^2) \; = \; \int^{Q^2} \: \frac{dk^2}{k^2} \: 
\frac{\alpha_s (k)}{\pi} \: 
+ \: {\rm const}
\label{7.31}
\end{equation}
\noindent down to the confinement region of low $Q^2$ were tried in Ref.\ 
\cite{dkt3}.

The simplest prescription (the so-called $F$-model) reduces to freezing the 
running coupling near the origin.  One follows the basic perturbative 
dependence of $\alpha_s (k)$ down to a certain point $k_c^2$ where the coupling 
reaches a given value
\begin{equation}
\frac{\alpha_s (k_c)}{\pi} \; = \; A, 
\label{7.32}
\end{equation}
\noindent and then keeps this value down to $k^2 = 0$.  $\xi$ then takes the 
form
\begin{equation}
\xi_F (k^2)  = \left\{
    \begin{array} {l l}
       \xi (k^2)- \xi (k_c^2) &   {\rm for} \quad k^2 > k_c^2, \\
        A \ln (k^2/k_c^2) &  {\rm for} \quad   k^2 < k_c^2 
     \end{array}
       \right.                                                                                                                              
      \label{7.33}
\end{equation}
\noindent with $k_c$ related to $A$ by (\ref{7.32}).

A set of the so-called $G_p$-models gives another example of the trial 
effective coupling.  It appears when one regularizes the one-loop evolution 
function $\xi^{(1)}$
\begin{equation}
\xi^{(1)} (k^2) \; = \; \frac{4}{b} \: \ln \ln \: \frac{k^2}{\Lambda^2} \: + \: 
{\rm const}
\label{7.34}
\end{equation}
\noindent as follows:
\begin{equation}
\xi^{(1)}_G (k^2) \; = \; \frac{4}{b} \: \ln \ln \: \left ( \frac{k^{2 p}}
{\Lambda^{2 p}} + C_p \right ) \: + \: {\rm const}, \qquad   C_p \geq 1
\label{7.35}
\end{equation}
with $p=1,2,\ldots$
This corresponds to the effective coupling
\begin{equation}
\frac{\alpha_{s,G}^{(1)} (k)}{\pi} \; \equiv \; 
\frac{d \xi^{(1)}_G (k^2)}{d \ln k^2} 
\; = \; \left [ \frac{k^{2 p}}{k^{2 p} + C_p \Lambda^{2 p}} \right ] \: 
\frac{4}{b} \; \frac{p}{\ln (k^{2 p}/\Lambda^{2 p} + C_p)}.
\label{7.36}
\end{equation}
\noindent Such an expression preserves the asymptotic form for $\frac{\alpha_s^{(1)} 
(k)}{\pi}$ up to power corrections $\Lambda^{2 p}/Q^{2 p}$.  Notice that the 
effective coupling (\ref{7.36}) with $C_p = 1$ has a finite limit $\alpha_s (0)/
\pi = 4p/b$, while for $C_p > 1$ it vanishes at the origin.  The two-loop 
coupling results in
\begin{equation}
\frac{\alpha_{s,G}^{(2)} (k)}{\pi} \; \equiv \; 
\frac{d \xi^{(2)}_G (k^2)}{d \ln k^2} 
\; = \; \left [ \frac{k^{2 p}}{k^{2 p} + C_p \Lambda^{2 p}} \right ] \: 
\frac{4p}{b L_p} \; \left (1 - \frac{b_1}{b^2} \; \frac{\ln L_p}{L_p} \right ),
\label{7.37}
\end{equation}
\noindent with
\begin{equation}
L_p \; = \; \frac{1}{p} \: \ln \: \left (\frac{Q^{2 p}}{\Lambda^{2 p}} + C_p 
\right ), \;\;\; b_1 \; = \; \frac{34}{3} \: N_C^2 \: - \: \left (\frac{10}{3} \: 
N_C + 2C_F \right ) \; n_f.
\label{7.38}
\end{equation}
\indent After comparing the experimental data on the mean values of $\langle x_Q 
\rangle$\footnote{The world average values of $\langle x_Q \rangle$ presented 
in Ref.\ \cite{pm1} were taken as the input.  Within the errors in the numerical 
procedure of calculations in \cite{dkt3} updating of these values is unlikely to 
cause any essential difference.} with the generalized perturbative predictions 
which embody the infrared regular effective coupling one arrives at the following 
conclusions (for further details see Ref.\ \cite{dkt3}):
\begin{itemize}
\item[1.] The scaling violations in $\langle x_{a,b} \rangle (W)$, that is the 
ratios
$$
\frac{\langle x_a \rangle (W)}{\langle x_c \rangle (W_0)}, \;\;\; \frac{\langle 
x_b \rangle (W)}{\langle x_b \rangle (W_0)}
$$
\noindent viewed as a function of $W$, prove to be insensitive to our ignorance 
about the small momentum region in the effective coupling.
\item[2.] The same kind of stability has been observed with respect to the scaled 
positions of the {\it maxima} in the energy spectra,
$$
\frac{x_{\rm max} (W)}{x_{\rm max} (W_0)}.
$$
\item[3.] The {\it absolute} values of $c$ and $b$ energy losses, in particular, 
the ratios
$$
\frac{\langle x_c \rangle (W)}{\langle x_b \rangle (W)}
$$
\noindent appear to be strongly \lq\lq confinement-dependent".
\item[4.] This \lq\lq confinement-dependence" reduces to an integral quantity 
$\mu_\alpha$, defined above in (\ref{7.25b}), as the only relevant non-perturbative 
parameter.  A consistent {\it simultaneous} description of $\langle x_{a,b} 
\rangle (W)$ is achieved by constraining the value of the integral
\begin{equation}
(2 \: {\rm GeV})^{-1} \: \int_0^{2 \: {\rm GeV}} \: dk \: \frac{\alpha_s^{\rm eff} 
(k)}{\pi} \; = \; 0.18 \pm 0.01 (\exp) \pm 0.02 ({\rm theor}).
\label{7.39}
\end{equation}
\end{itemize}
\indent As was first suggested by M\"{a}ttig (see first reference in 
\cite{pm1}), the $W$ evolution in the quark energy losses allows one to 
determine the scale parameter $\Lambda$.  The value of the latter proves to be 
practically insensitive to the adopted scheme of extrapolation of 
$\alpha_s^{\rm eff}$, see \cite{dkt3}.  The numerical analysis of the heavy quark 
energy losses carried out in \cite{dkt3} leads to 
\begin{equation}
\alpha_{\overline{MS}} (M_Z) \; = \; 0.125 \pm 0.003 (\exp) \pm 0.004 
({\rm theor}).
\label{7.40}
\end{equation}
This value  is in a good agreement with the
results of the (more) traditional methods of determination of
$\alpha_{\overline{MS}} (M_Z)$ (for a recent review, see \cite{pnb}).

 One remark is here in order.  The notion of the infrared finite 
$\alpha_s^{\rm eff}$ exploited here differs from that used in the 
renormalization-scheme-invariant approach to the $e^+ e^-$ annihilation 
cross-section (see e.g.\ \cite{ckl,ms2}) and the $\tau$-lepton hadronic width 
\cite{ckl}.  Nevertheless, it is worth mentioning that the value of the couplant 
$\frac{\alpha_s}{\pi} (0)$ and the integral measure derived in \cite{ms2} are 
both consistent with (\ref{7.39}).  Notice also, that there is a natural theoretical 
scale to which the \lq\lq measurement" of $\alpha_s^{\rm eff}$ below 1-2 GeV is to 
be compared.  As shown by Gribov \cite{vng1}, in the presence of light quark colour 
confinement occurs when the effective coupling (parameter $A$ of the $F$-model) 
exceeds rather {\it small} critical value
\begin{equation}
A > \left \{ \frac{\alpha_s}{\pi} \right \}^{\rm crit} \; = \; C_F^{-1} \: \left [1 - 
\sqrt{\frac{2}{3}} \right ] \; \approx \; 0.14.
\label{7.41}
\end{equation}
\noindent Then within the Gribov's confinement scenario an interesting 
possibility arises.  Namely, if the phenomenological $\alpha_s^{\rm eff}$ extracted 
from the data does exceed $\alpha_s^{\rm crit}$ but remains numerically small, 
this would provide a better understanding of the perturbative approach to 
multiple hadroproduction in hard processes.

Thus, it is empirically confirmed \cite{dkt3} that the characteristic integral 
(\ref{7.25b}) turns out to be a fit-invariant quantity which one has to keep 
fixed in order to describe the absolute values of energy losses.  As pointed out by 
Gribov, it can be looked upon as the long-distance contribution to the 
QCD field energy of a heavy quark.  It is worthwhile to notice that such an 
integral appears in the relation between the running heavy quark mass at scale 
$\mu$ and the pole mass \cite{bsuv}
\begin{equation}
M^{\rm pole} - M (\mu) \; = \; \frac{8 \pi}{3} \: 
\int_{|k| < \mu} \: 
\frac{d^3 k}{(2 \pi)^3} \: \frac{\alpha_s (k)}{k^2} \; = \; C_F \: \int_0^\mu \: 
dk \: \frac{\alpha_s (k)}{\pi} \; \equiv \; C_F \: \mu_\alpha.
\label{7.42}
\end{equation}
\indent To illustrate the phenomenological advances of the perturbative approach 
we present in Fig.\ 9.1 the expected $W$ evolution of the inclusive $b$ and $c$-quark
spectra\footnote{The results presented in Fig. 9.1 and in Eq. (\ref{7.43})
below are taken from Ref. \cite{dkt3}, except for those at $W=200$ GeV which
have been prepared for LEP-2 in application of the same approach. 
We thank Yu. Dokshitzer
for providing us with these results.}.  
These plots were calculated within the $F$-model with the values of 
$A$ and $\Lambda$ which allow the best fit to the mean energy losses.  For 
comparison the best-fit $G_2$-model results are also shown for LEP-1 energy.  It 
is easily seen that the $F$- and $G$-model predictions are quite close to each 
other.  One may conclude that it suffices to fix the integral parameter 
(\ref{7.25b}) 
together with the value of $\Lambda$ to predict the 
differential energy distributions 
with the reasonable accuracy\footnote{As pointed out in \cite{dkt3}
it is desirable to design a stable computation procedure for numerical
evaluation of the inverse Mellin transform (\ref{7.8a}).}.  
Note that the same parameters of $F$-model as in 
Fig.\ 9.1 lead to the following values of $\langle x_Q \rangle$:
\begin{eqnarray}
\langle x_b \rangle \; \simeq \; 0.698, \;\; \langle x_c \rangle \; \simeq \; 
0.511 & {\rm at} & W \: = \: 91 \: {\rm GeV}, \nonumber \\
& & \labl{7.43}\\
\langle x_b \rangle \; \simeq \; 0.648, \;\; \langle x_c \rangle \; \simeq \; 
0.475 & {\rm at} & W \: = \: 200 \: {\rm GeV}. \nonumber
\end{eqnarray}
As the energy losses at 91 GeV have been used 
to fix the parameters only the
results at 200 GeV should be considered as a prediction.

The duality arguments could be best applied to the description of the energy
fraction distributions averaged over the heavy-flavour states. 
Such a situation appears naturally e.g. in measurements of the inclusive
hard leptons, distributions over the heavy particle decay paths etc.
The inclusive quantities presented in this section can also 
be applied to describe 
 the exclusive heavy hadron spectra  such as those of $D^*$, for
example, although a precise description of these distributions cannot 
be fully achieved 
by the present formalism. 
 Within the accuracy of the  data now
 and in the foreseeable future this is unlikely to cause
 any essential difference.
 We stress that the perturbative approach provides one with a
 fully predicted (and
 well controllable) evolution of the spectra with both the overall
 energy and the heavy quark mass.



\subsection{On the Multiplicity of Events Containing Heavy Quarks} 

As we discussed above, the perturbative scenario predicts a suppression of soft 
gluon radiation off an energetic massive quark $Q$ inside the forward cone of 
aperture $\Theta_0 = \frac{M}{E_Q}$ (Dead Cone, e.g. \cite{dkt1,dkt7}).  This 
phenomenon is responsible for the perturbative mechanism of the leading particle 
effect and, at the same time, induces essential differences in the structure of 
the accompanying radiation in light and heavy quark initiated jets.  According 
to the LPHD concept, this should lead to corresponding differences in 
\lq\lq companion" multiplicity and energy spectra of light hadrons, see for 
further details \cite{dkt1,bas} (see also \cite{dfk1,yld2,dkt7}).

In particular, it is a direct consequence of the perturbative approach
 that the difference of companion mean 
multiplicities of hadrons, $\Delta N_{Qq}$ from equal energy (hardness) heavy 
and light quark jets should be $W$-independent (up to power correction terms 
${\cal O} \left ( (M^2/W^2) \right )$.  This constant is different for $c$ and 
$b$ quarks and depends on the type of light hadron under study (e.g.\ all charged, 
$\pi^0$, etc.).  This is in marked contrast with the prediction of the so-called 
naive model based on the idea of reduction of the energy scale \cite{yia1,yia2,pcr}, 
$N_{Q\overline{Q}} (W) = N_{q\overline{q}} ((1 - \langle x_Q \rangle ) W)$, so 
that the difference of $q$- and $Q$-induced multiplicities grows with $W$ 
proportional to $N (W)$.  Below we shall briefly elucidate this bright prediction 
of the APA  \cite{dkt13}.

When examining the perturbatively-based yield of light hadrons accompanying the 
$Q\overline{Q}$ production one has to understand first how does the development 
of the parton cascades initiated by $Q$ depend on the quark mass $M$.  Particle 
multiplicity is an infrared sensitive quantity dominated by the emission of 
relatively soft gluons.  In this domain the only difference between the heavy and 
massless quark cases comes from the Dead Cone phenomenon:  soft bremsstrahlung in 
the forward region
\begin{equation}
\Theta_{1+} \; \lapproxeq \; \Theta_0 \; \equiv \; M/E_Q
\label{7.44}
\end{equation}
\noindent is suppressed (e.g.\ \cite{dkt1,dkt7}).  Here $\Theta_{1+}$ is the 
angle between a primary gluon $(g_1)$ and the quark $(+)$.  In the region of 
large gluon radiation angles
\begin{equation}
\Theta_{1+} \; \gg \; \Theta_0
\label{7.45}
\end{equation}
\noindent the finite mass effects are power suppressed and do not affect the 
picture of strictly angular ordered evolution of $g_1$ as a secondary jet.  It 
is straightforward to verify that within the MLLA accuracy the finite mass 
induces only a small integral correction to the $AO$ prescription for the next 
generation gluon $g_2$.  As a result, in the region (\ref{7.45}) the internal 
structure of secondary gluon jets is identical to that for the light $q$ case,
and the emission angle $\Theta_{1+}$ should be taken as an evolution parameter 
to restrict the subsequent cascading.  As usual, the smaller the radiation 
angle $\Theta_{1+}$, the less populated with the offspring partons the gluon 
subjet $g_1$ is.

The situation changes, however, when $\Theta_{1+}$ becomes smaller than $\Theta_0$.  
The Dead Cone region (\ref{7.44}) gives a sizeable (though nonleading) 
contribution, of order $\sqrt{\alpha_s}$, and should be taken into account 
within the MLLA accuracy.  Here the opening angle of the jet $g_1$ \lq\lq freezes" 
at the value $\Theta_0$ and no longer decreases with $\Theta_{1+} \rightarrow 0$.  
The reason for this is rather simple.  Normally, in the \lq\lq disordered" angular 
kinematics $\Theta_{21} \geq \Theta_{1+}$ the destructive interference between 
emission of a soft gluon $g_2$ by quark and $g_1$ cancels 
the independent radiation 
$g_1 \rightarrow g_2$.

Meantime, in the massive quark case the interference contribution enters the game 
only when $\Theta_{2+} > \Theta_0$, so that the cancellation of the independent
$1 \rightarrow 2$ term inside the $\Theta_0$-cone does not occur.  In physical 
terms what happens is the loss of coherence between $+$ and $1$ as emitters of 
the soft gluon $2$ due to accumulated longitudinal separation $\Delta z_{1+}$ 
between massless and massive charges $(v_1 = 1, v_+ \approx 1 - \Theta_0^2/2 
< 1)$ \cite{dkt13}.  It is the ratio $\rho = (\Theta_{1+}^2 + 
\Theta_0^2)/\Theta_{21}^2$ which determines whether interference is essential 
or not.  When this ratio is larger than unity, $Q$ and $g_1$ are separated far 
enough for $g_2$ to be able to resolve them as two separated classical charges.  
Then the gluon $g_1$ acts as an independent source of the next generation 
bremsstrahlung.  Otherwise, no additional particles triggered by $g_1$ emerge on 
top of the yield determined by the quark charge (which equals the total colour 
charge of the $Q g_1$ system).  In the massless limit $(\Theta_0 = 0)$ this 
reproduces the standard $AO$ picture, $\Theta_{21} < \Theta_{1+}$.  In the 
massive case one concludes that the upper limit of the relative gluon angle 
remains finite when $\Theta_{1+}$ falls inside the Dead Cone:  $\Theta_{21} < 
\Theta_0$ for arbitrarily small $\Theta_{1+} \ll \Theta_0$.

Thus, the proper evolution parameter for the subsequent parton cascading of the 
primary gluon $g_1$ (generalization of an \lq\lq opening angle" of jet 1) may 
be chosen as 
\begin{equation}
\overline{\Theta}_{1+}^2 \; \equiv \; \Theta_{1+}^2 \: + \: \Theta_0^2.
\label{7.46}
\end{equation}
\noindent Another comment is in order concerning generalization of the argument 
of the running coupling to the massive quark case.  Here again the substitution 
similar to (\ref{7.46}) is applicable.  Namely, for the effective coupling, that 
determines inclusive probability of the emission $Q \rightarrow Q + g_1$, one has 
to use as an argument (see Ref.\ \cite{yia2})
\begin{equation}
k_\perp^2 \; = \; \omega_1^2 \: 
\left [ \left (2 \sin \frac{\Theta_{1+}}{2} \right )^2 
\; + \; \Theta_0^2 \right ] \; \approx \; \omega_1^2 \: \overline{\Theta}_{1+}^2.
\label{7.47}
\end{equation}
\indent Within the MLLA the expression for the multiplicity of light particles 
accompanying the production of a heavy quark pair, $N_{Q\overline{Q}} (W)$, can 
be obtained by convoluting the probability of a single gluon bremsstrahlung off 
a heavy quark $Q$ with the parton multiplicity 
initiated by the gluon subjet with
the hardness parameter $k_\perp$ (see Ref.\ \cite{dkt13}),
\begin{equation}
N_{Q\overline{Q}} (W) \; = \; 2 \: \int_0^4 \: \frac{\left (2 \sin 
\frac{\Theta_{1+}}{2} \right )^2 \: 
d \left ( (2 \sin \frac{\Theta_{1+}}{2} \right )^2}
{\left [ \left (2 \sin \frac{\Theta_{1+}}{2} \right )^2 \: 
+ \: \Theta_0^2 \right ]^2} 
\: \int_{[k_\perp > Q_0]} \: dz \: 
\Phi_F^G (z) \: \frac{\alpha_s (k_\perp)}{4 \pi} \: 
N_G (k_\perp)
\label{7.48}
\end{equation}
\noindent with $z = \frac{\omega_1}{E} \: \left (E = \frac{W}{2} \right )$ and 
$k_\perp$ given by Eq.\ (\ref{7.47}); $Q_0$ denotes a transverse momentum
cut-off as used in the previous sections.  
Here $\Phi_F^G$ stands for the standard 
DGLAP kernel.

The main contribution to $N_{Q\overline{Q}}$ comes from the DL phase space region.  
Meanwhile, the expression (\ref{7.48}) keeps track of significant SL effects as 
well, provided that the multiplicity factor $N_G$ is calculated with the MLLA 
accuracy.

Introducing a convenient variable
\begin{equation}
\kappa^2 \; = \; \frac{E^2}{\omega^2} \: k_\perp^2 \; = \; E^2 \: \left [ \left ( 
2 \sin \frac{\Theta_{1+}}{2} \right )^2 \: + \: \Theta_0^2 \right ],
\label{7.49}
\end{equation}
\noindent one can rewrite Eq.\ (\ref{7.48}) as
\begin{equation}
N_{Q\overline{Q}} \: (W) \; = \; 2 \: \int_{M^2}^{W^2} \: \frac{d\kappa^2}
{\kappa^2} \: \left [ 1 - \frac{M^2}{\kappa^2} \right ] \: \int_{Q_0}^\kappa \: 
\frac{dk_\perp}{\kappa} \: \Phi_F^G (z) \: \frac{\alpha_s (k_\perp)}{4 \pi} \: N_G 
(k_\perp),
\label{7.50}
\end{equation}
\noindent with $k_\perp = z \kappa$.  In the massless limit $M \lapproxeq Q_0$ the 
contribution of the $M^2/\kappa^2$ term vanishes and (\ref{7.50}) reproduces 
the known result for the $q$-jet multiplicity.  The integration can be performed 
with the use of the relation corresponding to the evolution equation for light 
quark jet multiplicity:
\begin{equation}
N_q^\prime (Q) \; \equiv \; \frac{\partial}{\partial \ln Q^2} \: N_q (Q/2) \; = 
\; \int_{Q_0}^{Q} \: \frac{dk_\perp}{Q} \: \Phi_F^G (z) \: \frac{\alpha_s (k_\perp)}{4 \pi} 
\: N_G (k_\perp),
\label{7.51}
\end{equation}
\noindent with
\begin{equation}
N_{q\overline{q}} \: (Q) \; = \; 2 N_q \: \left ( \frac{Q}{2} \right ),
\label{7.52}
\end{equation}
\noindent see Eqs.\ (\ref{2.21}), (\ref{4.30}).

Then one can rewrite the expression for the companion multiplicity in the heavy 
quark jet in the form
\begin{equation}
N_Q \: (E) \; = \; N_q \: (E) \; - \; N_q \: \left ( \frac{M}{2} \right ) \: - 
\: N_q^\prime \: \left ( \frac{M}{2} \right ) \: + \: {\cal O} (\alpha_s N_q; 
\Theta_0^2 N_q^\prime),
\label{7.53}
\end{equation}
\noindent where $2E = W$.

Notice that the factor of $2$ in the argument of $N_q$ generates a $\sqrt{\alpha_s} 
N_q (M)$ correction and is under control in the present analysis, whereas it 
could be omitted in the $N_q^\prime$ term as producing $\alpha_s N_q (M)$ terms 
which we systematically neglect.  Within this accuracy the term $N^\prime \sim 
\sqrt{\alpha_s} N_q (M)$ can be embodied into the multiplicity factor by shifting 
its argument, namely
\begin{equation}
N_q \: \left ( \frac{M}{2} \right ) \: + \: N_q^\prime \: \left ( \frac{M}{2} 
\right ) \; \approx \; N_q \: \left ( M \: \frac{\sqrt{e}}{2} \right ).
\label{7.54}
\end{equation}
\noindent As a result, one arrives at a formula allowing the expression of 
the companion 
multiplicity in $e^+ e^- \rightarrow Q\overline{Q}$ in terms of that in the light 
quark production process $e^+ e^- \rightarrow q\overline{q}$ (assuming $M \gg 
\Lambda)$ which reads as 
\begin{equation}
N_{Q\overline{Q}} \: (W) \; = \; N_{q\overline{q}} \: (W) \; - \; N_{q\overline{q}} 
\: (\sqrt{e} M) \; \left [ 1 \: + \: {\cal O} (\alpha_s (M)) \right ].
\label{7.55}
\end{equation}
\noindent The total particle multiplicity in $Q\overline{Q}$ events then reads as
\begin{equation}
N^{e^+ e^- \rightarrow Q\overline{Q}} \: (W) \; = \; N_{Q\overline{Q}} \: (W) \; 
+ \; n_Q^{dk},
\label{7.56}
\end{equation}
\noindent where $n_Q^{dk}$ stands for the constant {\it decay} multiplicity of 
the heavy quarks $(n_Q^{dk} = 11.0 \pm 0.2$ for $b$-quarks, $n_Q^{dk} = 5.2 \pm 
0.3$ for $c$-quarks, see Ref.\ \cite{bas} for details).

The main consequence of (\ref{7.56}) is that the {\it difference} between particle 
yields from $q$- and $Q$-jets at fixed annihilation energy $W$ depends on the 
heavy quark mass and remains $W$-independent \cite{dfk1,dkt7,bas}
\begin{eqnarray}
\delta_{Qq} & = & N^{e^+ e^- \rightarrow Q\overline{Q}} \: (W) \; - \; 
N^{e^+ e^- \rightarrow q\overline{q}} \: (W) \; = \; {\rm const} \: (W), \nonumber 
\\
& & \label{7.57}\\
\delta_{bc} & = & N^{e^+ e^- \rightarrow b\overline{b}} \: (W) \; - \; 
N^{e^+ e^- \rightarrow c\overline{c}} \: (W) \; = \; {\rm const} \: (W). \nonumber 
\end{eqnarray}
\noindent Let us emphasize that it is the QCD coherence which plays a fundamental 
role in the derivation of this result.  Due to this the gluon bremsstrahlung 
off {\it massive} and {\it massless} quarks differ only at parametrically small 
angles $\Theta \lapproxeq \Theta_0 \equiv M/E$ where, due to the $AO$, cascading 
effects are majorated by the $N^\prime (M)$ factor.  The relative accuracy of the 
result (\ref{7.57}) is
 $\sqrt{\alpha_s (M)} \: \frac{M^2}{W^2}$ \cite{dkt13}.  

The results of the recent experiments on multiplicities in $b\overline{b}$ and 
$c\overline{c}$ events (see for details \cite{ada2,wjm2}) are 
summarized in Tables 3,4 and in Fig. 9.2.
\begin{table}[ht]
\begin{center}
\begin{tabular}{|c|c|c|} \hline
& Value & Exp \\ \hline
$\langle n^{ch} \rangle_{c\overline{c}}$ & 21.44 $\pm$ 0.62 & OS \\ \hline
$\delta_{bq}^{ch}$ & 2.90 $\pm$ 0.30 & DOS \\ \hline
$\delta_{cq}^{ch}$ & 0.89 $\pm$ 0.62 & OS \\ \hline
\end{tabular}
\caption{Results on the average multiplicity of charged particles
in $Z \rightarrow c\overline{c}$ 
and on the average differences $\delta_{bq}^{ch}, \delta_{cq}^{ch}$ 
(see Eq.\ (\protect\ref{7.57})) from DELPHI (D),
OPAL (O) and SLD (S)
\protect\cite{ada2,delphi1,osoaa}.}
\end{center}
\end{table}
\begin{table}[ht]
\begin{center}
\begin{tabular}{|c|c|c|c|} \hline 
Particle & Multiplicity & From $B$ Decay & Exp \\ \hline
Charged & 23.43 $\pm$ 0.48 & 5.72 $\pm$ 0.38 & DOS \\ \hline
$\pi^0$ & 10.1 $\pm$ 1.2 & 2.78 $\pm$ 0.53 & D \\ \hline
$K^+$ & 2.74 $\pm$ 0.50 & 0.88 $\pm$ 0.19 & D \\ \hline
$K^0$ & 2.16 $\pm$ 0.12 & 0.58 $\pm$ 0.06 & D \\ \hline
$\phi$ & 0.126 $\pm$ 0.023 & 0.032 $\pm$ 0.011 & D \\ \hline
$p$ & 1.13 $\pm$ 0.27 & 0.141 $\pm$ 0.059 & D \\ \hline
$\Lambda$ & 0.338 $\pm$ 0.047 & 0.059 $\pm$ 0.011 & D \\ \hline
\end{tabular}
\caption{Average multiplicities in $Z \rightarrow b\overline{b}$ and in the 
decay of single $B$ hadrons from DELPHI (D), 
OPAL (O) and SLD (S)
\protect\cite{ada2,delphi1,delphix4}.}
\end{center}
\end{table}
To the available accuracy, the results on $\delta_{Qq}$ are seen to be 
independent of $W$, in marked contrast to the steeply rising total multiplicity, 
and are thus consistent with the MLLA.  The \lq\lq naive" hypothesis appears to 
be disfavoured by the data\footnote{On recent modifications of the naive model, 
see Ref.\ \cite{vap}.}.  We turn now to the absolute values of the charged 
multiplicity differences $\delta_{Qq}^{ch}$.  Neglecting in Eq.\ (\ref{7.55}) 
the subleading ${\cal O} (\alpha_s)$ corrections, the following MLLA 
expectations were found in Ref.\ \cite{bas} by inserting the appropriate
experimental numbers into the $r.h.s$ of Eqs. (\ref{7.55})-(\ref{7.57}):
\begin{equation}
\delta_{bq}^{ch} \; = \; 5.5 \: \pm \: 0.8, \;\;\;\; \delta_{cq}^{ch} \; = \; 
1.7 \: \pm \: 0.5 .
\label{7.58}
\end{equation}
\noindent These exceed the experimental values given in Table 3 
showing the essential role of the next-to-MLLA (order $\alpha_s (M)\cdot  
N (M)$) terms. 
 Recently in Ref.\ \cite{vap} an attempt was made to improve Eq.\ 
(\ref{7.55}).  This was taken as a welcome news by the experimentalists, e.g.\ 
\cite{wjm2}.  However, we need to mention here that the very picture of 
accompanying multiplicity being induced by a single cascading gluon, 
implemented in \cite{vap}, is not applicable at the level of subleading effects 
(see e.g.\ \cite{dkmt2}).  Therefore a self-consistent reliable theoretical 
improvement of the MLLA predictions for the absolute values of $N_{Qq}$ 
remains to be achieved \cite{dkt13}.

Further detailed experimental results could provide stringent tests of the 
perturbative predictions.  Thus, the study of $b$-quark events at the energies 
of LEP-2 looks rather promising, e.g.\ \cite{vak2}.  For instance, in this 
region the modified naive model \cite{vap} predicts a negative value of $\delta_{bq}$ 
(about -1) in marked contrast with the perturbative expectation $\delta_{bq} \approx 3$.  
It will be very interesting to check whether the differences $\delta_{bq}^h$ of 
mean multiplicities for the identified particles remain energy independent.  
Results from the $Z^0$ (see Table 4) are available as reference values. In 
particular, the DELPHI results \cite{delphi1} suggest 
 $\delta_{bq}^{\pi^0} \simeq 0$ and therefore, using again Eqs. 
 (\ref{7.55})-(\ref{7.57}),
one may expect at LEP-2 
the relation
\begin{equation} 
N_{\pi^0}^{e^+ e^- \rightarrow b\overline{b}}
/N_{\pi^0}^{e^+ e^- \rightarrow q\overline{q}} \; \simeq \; 1.
\label{7.59}
\end{equation}
The energy behaviour of $\delta_{Qq}^h$ should be watched closely at future 
linear colliders.

Finally, let us mention that the spectra of light particles from the heavy quark 
jets are predicted to be depopulated in the hard momentum region compared with 
the light-$q$-jets \cite{dkt1,dkt7,kdt}.  Meanwhile, the particle yield in the 
soft momentum region should remain unaffected.  This is a direct consequence of 
the Dead Cone phenomenon.

At $W \gg M \gg \Lambda$ the perturbative expression for the $Q$-jet companion 
particle distribution can be approximately given by \cite{dkt7} 
\begin{equation}
D_Q^h \: (x, W) \; = \; D_q^h \: (x, W) \: - \: 
D_q^h \: \left( \frac{x}{\langle y \rangle}, M \sqrt{e} \right),
\label{7.60}
\end{equation}
\noindent where $D_q^h (x, W)$ 
is the standard spectrum of particles 
in a $q$-jet (see section 2) and $\langle y \rangle$ being the averaged scaled energy 
of the $Q$.  This problem needs further detailed studies.

 \section{Summary and Outlook}

Perturbative QCD proves  to be a very successful theory in
its application to hard processes.
Still, the problem of the soft limit of the theory and 
of colour confinement is not
 solved. Therefore, at present, multihadron production phenomena
cannot be derived  systematically 
solely from  perturbation theory without additional
model-dependent
assumptions. Most popular and successful in reproducing the details are  
hadronization schemes which combine perturbative parton cascades with 
explicit parametric models for the production of the final hadrons.
The complexity of these models requires Monte Carlo methods to derive
their predictions.

In this review we have concentrated on the question of the extent to which the
semisoft phenomena in particle production in hard processes reflect 
the properties
of the perturbative parton cascade. 
According to the preconfinement idea colour singlet clusters are formed
already during the perturbative phase of the jet evolution 
and they then smoothly transform
into hadronic clusters in such a way that there is a similarity in global
characteristics between parton and hadron final states.
According to the hypothesis of Local  Parton Hadron Duality
the parton cascade can be evolved down 
to a small scale $Q_0$ of the order of the
hadronic masses and  one
compares directly the 
inclusive hadronic observables with the corresponding results 
computed at the parton level. 
 The attractive feature of this scenario is the small number
of parameters, namely the QCD scale $\Lambda$, the 
transverse momentum cut-off $Q_0$ and the overall normalization.
This approach describes a large variety of  observables such as inclusive
particle spectra and correlations inside jets as well as 
the bright interjet
coherence phenomena
which reflect the QCD wave nature of hadroproduction. 
At the moment there are various remarkable  successes
of this approach. In some cases apparent disagreement with experiment
could well result from the approximations made in the analytical
calculations.
In the present exploratory phase  it is difficult to design
a crucial test, rather it
is left to experiment to decide
for which observables the LPHD approach will work well 
and what are the limits  
for its  application  
(certainly one cannot expect that such a scenario will reproduce 
the resonance bumps!).

The hope is that the further  detailed
predictions of such a duality picture 
and their experimental tests will eventually not only
provide a simple and predictive approach for the 
description of  multiparticle phenomena
but will also 
lead to some progress in the study of the confinement problem. 
The duality between partonic and hadronic final states suggests a rather
soft confinement mechanism.
In the following
we summarize the main theoretical and experimental results and present a
brief discussion of future possibilities.

\subsection{Theoretical approximations and accuracy of predictions}

The partons participating in
the primary hard process generate gluon
cascades by brems\-strahlung processes which yield the 
multiparton final state.
The hard processes are calculated from the corresponding matrix elements,
the partonic cascades from the evolution equation of the multiparton
generating functionals which include energy conservation and the full parton
splitting functions. If only the contributions from the leading soft and
collinear singularities are kept (Double Log Approximation) one obtains
 predictions for asymptotically high energies, in particular 
one finds finite scaling
limits in logarithmic variables after appropriate rescaling. In the
Modified Leading Log Approximation 
one obtains the next-to-leading  corrections of relative order
$\sqrt{\alpha_s}$ 
which turn out to be essential and only slowly decrease
with energy. This leads to the ``preasymptotic'' scaling laws, i.e. an apparent 
scaling behaviour in the present energy range, the simplest example being 
KNO scaling.

Further improvement in accuracy is obtained 
by summing up the perturbative series
which solves the evolution equation with the proper initial condition
and turns into the asymptotic solution. This
yields the results for the parton cascade in the full energy 
region starting from
threshold up to the highest energies. Such calculations have been performed,
for example, for the energy spectra and their moments
and for the subjet multiplicities.

These methods yield analytical results which provide 
us with an understanding 
of how the phenomena are related to the theoretical ideas and 
therefore can lead to characteristic predictions, sometimes in terms of
 basic QCD parameters.
Alternatively, one can also apply Monte Carlo methods to
derive the properties of the parton cascade. In principle, such computations
promise a higher accuracy, for example, 
they take into full account the
energy-momentum conservation. However, when using such calculations for
LPHD tests one has to take care of the proper parameters, especially
a small $Q_0 \gsim\Lambda$, 
also the cut-off procedures are sometimes different from those 
considered in the standard analytical calculations.
Alternatively, one can restrict the applications 
of Monte Carlo calculations to observables which
are not sensitive to the $Q_0$ parameter in a sufficiently large range.

\subsection{Tests of the parton hadron duality}
First, there is a class of quantities which are infrared safe, i.e.
are not sensitive to the cut-off $Q_0$. For the intrajet cascade
such characteristics are:
a) energy evolution of particle multiplicity,
b) moments of multiplicity distributions,
c) azimuthal angle correlations,
d) angular correlations for large relative polar angles.
The DLA predictions for scaling properties of distributions 
are well confirmed by the data but
in general their  shape is not reprodeced. On the other hand,
when the MLLA calculations are available they are  usually
in a good agreement with the data.  
If this is not the case,
the results from the parton level Monte Carlo give 
a satisfactory description, indicating that the accuracy of the analytical
calculation was not sufficient.

Another example is the  particle flow at large angles
to the jet directions, such as the string/drag effect 
which is well reproduced experimentally. 

Secondly, there are observables whose
distributions depend explicitly 
on the cut-off, especially particle energy spectra
and correlations. For the energy spectrum a full solution 
within the MLLA accuracy taking into account
the initial condition at threshold is available
and the shape of the predicted
hump-backed plateau and especially the 
position of the maximum and the first few moments 
are in a very good agreement with
the data choosing $Q_0 \gsim\Lambda\approx$ 250 MeV. The leading (DLA) and 
next-to-leading (MLLA) contributions alone are not
sufficient to describe the data in the full available energy range. 
Also the shape of the 2-particle energy correlation function is well
described by the MLLA, but not
the absolute normalization.
Here the results from calculations which include higher order 
contributions have 
to be awaited before
the final judgement.

Ideas for the description of observables for 
identified particle species are still very
crude. The data, especially on particle spectra,
 confirm the trend that the mass of the particle is related to
the cut-off $Q_0$, but a systematic description is not yet available.

Thirdly, there are the subjet observables which connect the infrared safe jet
properties at large mass scale (large $y_c$) 
with the particle properties at low scale.
Whereas the perturbative calculations 
generally describe well the jet properties 
at large scale, 
they fail in some cases to reproduce the transition to small scale
(an example is the ratio of subjet
multiplicities in quark and gluon jets). This failure
could well be due to the unsufficient accuracy of the 
analytical calculations as is suggested by the related Monte Carlo results.

\subsection{Sensitivity to specific properties of perturbative QCD}
An important question is whether the agreement of the data with
the perturbative predictions is significant or merely 
reflects some  general properties 
of final states already present in the phenomenological 
nonperturbative models. We emphasize
here some  results which are characteristic of the perturbative QCD dynamics.

First there are the various consequences from the coherence of the soft
gluon emission in the QCD bremsstrahlung cascade.
This yields the particular shape of the hump-backed plateau for the
distribution in the variable $\xi = \ln( E_{jet}/E)$ 
with the suppression
of the soft particle production. Models without coherence would yield a
flatter parton distribution. The coherence also implies an
approximate energy independence of the low energy
particle yield whereas  the overall central rapidity density rises with
the energy. Furthermore, coherence 
leads to nontrivial negative interference
phenomena in the azimuthal and
polar angle correlations. 

A spectacular phenomenon is the celebrated string/drag effect
which suppresses particle production opposite to the gluon jet in 3 jet events
despite the overall increase of multiplicity. Other counter-intuitive negative
interference effects are met in the subjet analysis of 3-jet events in $e^+e^-$
annihilation.

Another interesting effect is the strong variation of the coupling
$\alpha_s$ at
small scales. This has an important impact on the prediction 
for the energy spectra.
A perturbative model with fixed coupling in the same overall approximation
scheme would fail badly.

\subsection{Prospects of future measurements}

LEP-1 has performed an important task in studying QCD physics.
It has benefited from the record statistics and the substantial
lack of background. We have learned much interesting physics
but the need for further detailed QCD analyses of the
data recorded at LEP-1 has not decreased. In particular,
these studies may teach us about the most interesting transition
regime from the perturbative to the non-perturbative regime in
multiparticle production.
On October 31st 1995 LEP successfully collided electrons and positrons at
130 GeV. This was the first stage en route to the full operation of
LEP-2, and results from the higher energies became available immediately.
At LEP-2 QCD studies are more challenging, first of all, because of
unfavourable statistics. In the meantime, there are also some 
advantages over the lower-energy $e^+e^-$ colliders for examining 
QCD physics (for example, $\gamma\gamma$ and heavy quark jet physics).

The  program of QCD studies at future linear $e^+e^-$ colliders
looks quite interesting. Here the anticipated higher luminosity gives a
definite advantage over LEP-2 in performing the very precise tests of
QCD predictions. New areas for QCD studies at future linear colliders are
opened by the high energy $\gamma\gamma$ and $e\gamma$ colliding facilities.
Polarization of electrons and photons will provide a useful tool in helping
to separate signals from different processes and to reduce
background contributions.

The semisoft QCD physics will remain one of the important
topics for investigation in the TEVATRON and HERA experiments.
Of special interest here are further quantitative tests of the
perturbative universality of parton jets appearing in different
processes. The detailed studies of the rich diversity of
colour-related phenomena in hadronic high $p_\bot$ processes
looks very promising. In DIS further quantitative analyses
(including particle identification) of the current jet hemisphere
are very important. A new domain for testing the perturbative approach
could be opened by the studies of the target fragmentation region
in DIS.

We present below a brief list of the possible areas of the
analyses  concerning the semisoft QCD physics.
\begin{enumerate}
\item There are clear  predictions for the {\it spectra of particles}
from the perturbative approach. By measuring the spectra 
 within restricted opening
angles with respect to the jet one can explore the coherent
origin of the hump-backed plateau thus sharpening the
influence of angular ordering on the particle branching.
Of special interest is the energy evolution of the spectrum from the lowest
to the highest energy, also there is the question
to what extent the soft particle production
rates are energy independent as expected. 
\item Further detailed examinations of the {\it differences between quark
and gluon jets} are important to establish the 
perturbative QCD origin of the 
particle production phenomena
beyond the phase space effects
(examples include average multiplicities and exclusive production rates
of particles and particle clusters (subjets), the distribution in the 
$\xi_E= \log (E_{jet}/E)$ variable). 
A promising new source of gluon jets 
would become available at future $e^+e^-$ colliders through the 
 process $\gamma\gamma\to gg$.
\item The study of 
{\it jets initiated by heavy quarks} is of special importance. 
There are  predictions for the energy 
spectrum of light particles in heavy quark jets.
It will also be interesting to check whether the
differences of mean multiplicities between heavy and light quark jets
for various primary quarks remain energy independent.
From the study of the distribution of
particles inside fixed cones, in particular their energy
and angular spectra, one can learn about the expected 
depletion of the forward particle production for angles $\Theta<m_Q/E_Q$.

Also further studies of the inclusive energy
spectrum of the leading heavy flavoured hadrons and their detailed comparison
with the perturbative predictions look promising.
\item Detailed tests of {\it QCD coherence dynamics}
will require comprehensive
studies of the total three-dimensional pattern of particle flows in
three-jet events in $e^+e^-$ annihilation and in $W$ + jets events in 
hadronic collisions. 
Of special interest here are the angular structure
and the energy dependence of the multiplicity flow and its dependence
on $m_h$ and $p_{out}$. 
\item 
An important question concerns the {\it onset of the 
perturbative regime}, whether the relevant scale really is as low as a few
hundred MeV. A clue to this question can be obtained from the study of 
inclusive soft
particle production perpendicular to the primary $q\bar q$ or $gg$-like
parton antennae. Tests are proposed for various hard processes
($e^+e^-$, $ep$ and hadron-hadron collisions).
Another approach to the same question is the study of jet or subjet
observables at high resolution (small $y_{cut}$ parameter). The relevant
scale in both cases may be different.
\item
The study of {\it asymptotic scaling behaviour} of distributions in
logarithmic 
energy and angular variables 
with the jet energy and jet opening angle
provides a test of the underlying 
perturbative picture based on the parton cascade dominated by
bremsstrahlung processes. 
\item 
To investigate the role of {\it QCD motivated
$1/N^2_c$ interference terms} one can measure the
azimuthal asymmetry of a jet in $q\bar q g$ events in
which the gluon moves in the direction opposite to the quasicollinear
$q\bar q$ pair. The comparison with the analytical results (accounting
for both interjet and intrajet phenomena) should make it possible
to distinguish reliably the perturbative predictions from those of
existing fragmentation schemes. Another test can be obtained 
by comparing the particle production in  $q\bar q$ and
$q\bar q g$ events perpendicular to the 
jet axis or production plane respectively.

\item
An interesting manifestation of the {\it QCD wave nature of
hadronic flows} arises from the double-inclusive correlations 
$d^2N/d\Omega_\alpha d\Omega_\beta$ ($\alpha,\beta$ denote the
interjet valleys) of the
interjet flows in $q\bar q g$ events. Here one faces such tiny effects
as the mutual influence of different antennae.
\item
{\it Energy and momentum correlations} are potentially affected by resonance
effects and it is therefore important to further test the perturbative
approach in these observables. In future  studies of angular dependent
spectra or correlations it would be 
advantageous to consider {\it Energy$^n$-Multiplicity$^m$-Correlations
(E$^{\ n}$M$^{\ m}$C)}
which avoid any reference to  a jet axis.
This allows better treatment of the observables in
analytical calculations.
\item 
The distributions of {\it identified hadrons} are
not well understood within the MLLA-LPHD scenario.
 Here  more detailed analyses of spectra of different hadrons 
from different jets could be helpful. 
In particular, one needs to find some clue to
the observed correlations between the peak position
$\xi^*_p,\xi^*_{prim}$ and particle masses.
In the angular correlations the dependence on the particle masses is
predicted to disappear above a certain critical angle.
A hot topic here is also the comparatative study
of $\eta,\eta'$ production in quark and gluon jets.
The universality of the fragmentation into the various particle species
should be carefully examined;
this universality has been questioned recently for strange particles.

\end{enumerate}

The need for QCD studies
in the semisoft region has not decreased. On the contrary, there
are many open questions that require further investigation in
their own right, and for their importance to other subjects.
The colour-related effects are becoming a phenomenon of
large potential value as a new tool for studying dynamics of
particle production in multijet systems.
Knowledge about the  applicability 
of the perturbative calculations in
 multiparticle physics and the determination of their limits
could be an important help in the understanding of
the colour confinement problem which is still 
an important challenge after many years 
of studying  strong interaction physics.

\section*{Acknowledgements}
\addcontentsline{toc}{section}{Acknowledgements}

We are grateful to A. Akeroyd, N. Brook, I. Dremin, S. Lupia,
J. Wosiek and especially to Yu. Dokshitzer and T. Sj\"ostrand for 
numerous fruitful discussions, useful comments and suggestions.
We are very thankful to Sharon Fairless and Rosita Jurgeleit for invaluable
assistence.
VAK thanks the Theory Group of the
Max-Planck-Institute for the kind hospitality during the course
of this work. This work was supported in part by the UK Particle Physics
and
Astronomy Research Council.




\newpage

\section*{Figure Captions}
\addcontentsline{toc}{section}{Figure Captions}
%
\begin{itemize}
\item[Fig.\ 3.1] Modifications of the asymptotical leading-order
particle distribution caused by the next-to-leading order
effects.

\item[Fig.\ 3.2] (a) $Q_0$ dependence of the maximum of the MLLA
truncated
distribution at $\Lambda = 150$ MeV \cite{dkt5}.  The left
end-points of the curves correspond to the limiting spectrum
$(\lambda = 0)$;\\
 (b) Energy evolution of the peak position for
different values of $Q_0$ \cite{dkt5}.  The results of the
straightforward numerical calculation in the limiting case
(solid
line) are compared with the asymptotic formula (\ref{2.51}).

\item[Fig.\ 3.3] Energy evolution of the $\xi$ spectrum at $Q_0=2\Lambda$.
The
shape parameters are shown for $W = 2 * 250$ GeV \cite{dkt5}.

\item[Fig.\ 3.4] $Q_0$-dependence of the kurtosis at LEP-1 energy
\cite{dkt5}.

\item[Fig.\ 3.5] Shrinkage of the cones in which the fixed shares
of multiplicity ($\delta$) and energy ($\epsilon$) of a jet $A (A
= q, g)$ are deposited \cite{dkt10}.

\item[Fig.\ 3.6] (a) Plots of $\xi_p$ distribution of charged
particles \cite{opal1} together with the prediction of
(\ref{2.44}) and (\ref{2.54}) at $\Lambda_{ch} \simeq 250$
MeV;\\
(b) $\xi_p$ distribution of charged particles at LEP-1.5 measured
by the DELPHI 
\cite{delphiL15}, ALEPH \cite{alephL15}  and
OPAL \cite{opalL15} Collaborations
in comparison  with the prediction of the
limiting spectrum (\ref{2.44}) at $\Lambda_{ch} =270$~MeV (solid line).
The dashed line shows the prediction of the limiting spectrum after 
the correction
for kinematical effects at low momenta, see Ref.\  \cite{klo}.

\item[Fig.\ 3.7] Energy dependence of the mean charged
particle
multiplicity $N^{ch}$ together with the MLLA-LPHD prediction
(from Refs. \cite{ms,l3L15,delphiL15,alephL15,opalL15}).

\item[Fig.\ 3.8] Peak position $\xi^*$ of the inclusive
$\xi$-spectra of all charged particles as a function of
$Y= \log (\sqrt{s}/2\Lambda_{ch})$. Comparison between
experimental data collected in \cite{ms} and LEP-1.5 data
\cite{delphiL15,alephL15,opalL15} and theoretical predictions
numerically extracted \cite{klo} from the shape of the limiting
spectrum (solid line); the crosses correspond to $\sqrt{s}=200$ GeV
and 500 GeV.

\item[Fig.\ 3.9] 
The average multiplicity $\overline N_E$ and the first four 
cumulants of charged particles' energy spectra $E dn/dp$ 
as the functions of  $\xi_E$,
i.e., the average
value $\bar \xi_E$, the dispersion $\sigma^2$, the skewness $s$ and the
kurtosis $k$, 
are shown as the functions of $Y = \log (\sqrt{s}/2Q_0)$ for $Q_0$ = 270 MeV.
Data points from MARK I\protect\cite{mark1},
TASSO\protect\cite{tassox}
and LEP-1 (weighted averages of
ALEPH\protect\cite{alephx2}, 
DELPHI\protect\cite{delphix4}, L3\protect\cite{l31}
and
OPAL\protect\cite{opal1}).
Predictions for the limiting spectrum (i.e. $Q_0 = \Lambda$)
of MLLA with running $\alpha_s$
 and of MLLA with fixed $\alpha_s$ are also shown (for $n_f$ = 3).
Predictions for the average multiplicity refer to the two-parameter formula
$\overline{N}_E = c_1 2(4/9) \overline{N} + c_2$ (from Ref.~\cite{lo}).

\item[Fig.\ 3.10] (a) $\xi_p$ distribution for $\pi^0$ at the $Z^0$
resonance compared with the MLLA limiting spectrum at
$\Lambda_{eff} = 147$ MeV \cite{l32}.  These data are in a
good
agreement with the results from other LEP-1 experiments for
neutral
and charged pions, see Refs.\ \cite{ms,opal2,aleph1,delphi1};\\
(b)
 $\xi_p$ distribution of $K^0$ mesons
\cite{opal5} as compared to the numerical result of solution of
MLLA Evolution Equation for truncated cascade (solid curve) and
to the distorted Gaussian formula with the MLLA shape parameters
(dashed curve).  Taking into account of the finite mass effects
induces a certain decrease in the values of the free parameters, $Q_0
(m_K) = 250$ MeV, $K^{K^0} = 0.34$, see Ref.\ \cite{dkt9} for
details.

\item[Fig.\ 3.11] Measured $\xi_p$ distributions for (a)
$\pi^\pm$, (b) $K^\pm$, (c) $p/\overline{p}$ and (d)
$K^0/\overline{K}^0$ \cite{topaz}.  Both of the TOPAZ and PEP4/TPC
measurements are plotted with MLLA fits based on Eq.\
(\ref{2.34}) (Solid curve: TOPAZ, Dashed curve: PEP4).

\item[Fig.\ 3.12]  (a) Position of the peak $\xi_p^*$ as a function
of the centre-of-mass energy for pions, kaons and protons for the
inclusive spectra presented by ALEPH \cite{aleph1} (filled
points) and those of other experiments
\cite{topaz,opal2,tasso1,tasso2,tpc,gdc}.  The solid lines are
fits \cite{aleph1} from top to bottom, pion, kaon and proton
data according to Eqs.\ (\ref{2.51}), (\ref{2.76}).  The arrow
represents the estimated shift in $\xi^*$ for kaons at the $Z$
due to $b$ hadron decays, and is to be compared with the
extrapolation of Eqs.\ (\ref{2.51}), (\ref{2.76}) (dashed
line).  This shift brings the data into reasonable agreement with
an extrapolation of the fitted function.  The dot-dashed line is
the prediction for an incoherent parton shower;\\
(b) 
Energy evolution of the peak position
\cite{l31}
compared to the MLLA limiting result.  The data on $\pi^0$
production are presented as a function of $\sqrt{s} = W$ at
$\Lambda_{eff} = 150$ MeV.  The data on charged particles are
replotted as a function of $\sqrt{s} = W 
(\Lambda_{eff}/\Lambda_{ch})$ \cite{dkt9}.

\item[Fig.\ 3.13] Measured $\xi_p$ distributions for $\rho_0, 
\omega, K^{*0}$ and $\Phi$ in $Z^0$ decays.  The spectra are
fitted with a Gaussian over the ranges indicated by the dotted
lines \cite{aleph3}. 

\item[Fig.\ 3.14] Relation between the peak position of the
$\xi_p$
distribution at the $Z^0$ and the particle mass
\cite{delphi3}. 
The full lines are separated fits for the meson and baryon
distributions with $\xi_p^* \propto - \ln m_h$. 

\item[Fig.\ 3.15] 
Distribution in $\xi_p=\ln(1/x_p)$ from JETSET Monte Carlo \cite{JETSET}
at the parton level (``Gluons") with and without colour coherence
included, and for charged particles ($C^\pm$) of the final hadronic
state (from Ref. \cite{bcu}).

\item[Fig.\ 3.16]
Invariant density $E dn/d^3p$ of charged particles
in $e^+e^-$ annihilation 
as a function of the particle energy
$E=
(p^2+Q_0^2)^{1/2}$ at $Q_0$ = 270 MeV.
Data  at various $cms$ energies
are compared to MLLA predictions with the overall normalization adjusted
(from \protect\cite{klo2}).

\item[Fig.\ 3.17]
 Comparison  of experimental data on the invariant distribution 
$E dn/d^3p$ as function of particle energy $E$ from
MARK I\protect\cite{mark1} with predictions from the
MLLA with running $\alpha_s$ and
with fixed $\alpha_s$ at the same $cms$ energy, using  $Q_0$ = 270 MeV
(from \cite{lo}).

\item[Fig.\ 3.18] Ratio of mean charged particle multiplicities in
gluon and quark jets as a function of the jet energy $E$. Open symbols:
from symmetric, Y-shaped 3-jet 
events $(E\approx 24$ GeV) with natural quark flavor
composition at the $Z^0$ \cite{aleph3j,delphi4,opal10}; full symbols:
all data referring to $c$- and $b$-quark content as in $q\bar q g/q\bar q\gamma$
sample \cite{delphi4}.
\item[Fig.\ 4.1]
The KNO multiplicity distribution in $x=n/\bar n$ for
infinite energies (dashed line) and LEP-1 energies (full line),
calculated for a gluon jet with fixed $\alpha_s$ in the 
next-to-MLLA
\cite{yld}. The negative binomial distribution (open points)
with parameter $K=7$ is also shown for comparison.
\item[Fig.\ 4.2]
Two-particle multiplicity moment $R_2\equiv F_2=<n(n-1)>/<n>^2$
as a function of cms energy. The QCD results in the leading order
(DLA), next-to-leading order (MLLA)
 \cite{mw} and from a parton level Monte Carlo
(HERWIG \cite{HERWIG}) \cite{wo} in comparison with experimental data
(from \cite{aleph8,wo}).
\item[Fig.\ 4.3]
Ratio of cumulant to factorial moments $H_q$, as a function
of the moment rank $q$. Data from SLD collaboration
\cite{sld}.
\item[Fig.\ 4.4]
The mean jet multiplicity in $Z^0\to $ hadrons as a function of the resolution
scale $y_{cut}$. Prediction from the all order $\alpha_s$ and
next-to-leading $\ln y_{cut}$ computation, corrected for hadronization
according to the JETSET \cite{JETSET} and HERWIG \cite{HERWIG} models,
are fitted to data of the L3 Collaboration \cite{L3jmul}.
\item[Fig.\ 4.5]
Subjet multiplicity in 2-jet and 3-jet events at the $Z^0$ as a function of
subjet resolution $y_0$ for fixed jet resolution $y_1=0.01$. 
The data of the OPAL Collaboration \cite{opalsj} are compared with the
${\cal O} (\alpha_s)$ and the all order $\alpha_s$
calculation in the next-to-leading $(\ln y_{0,1})$ approximation.
\item[Fig.\ 4.6]
Ratio of subjet multiplicities in quark and gluon jets as measured by 
the ALEPH Collaboration \cite{alephsj} as a function of subjet resolution
$y_0$. The data at $y_0\sim 10^{-2}$ 
come close to the asymptotic prediction 9/4.

\item[Fig.\ 5.1]
Normalized two-particle momentum correlation $R_2(\xi_1, \xi_2)$ 
as measured by the OPAL
Collaboration
\cite{opalmc} for two bands in the $\xi_1, \xi_2$ plane,
compared with the analytical predictions 
\cite{fw} for 3 values of the parameter $\Lambda $
(band I: $|\xi_1 - \xi_2| < 0.1$,
band V: $6.9 < \xi_1 + \xi_2 < 7.1$). The dashed curves indicate the
asymptotic DLA predictions for $\Lambda$=255 MeV. 

\item[Fig.\ 5.2]
Normalized two-particle momentum correlation 
$R_2 (\xi_1, \xi_2)$ for $\xi_1 = \xi_2$ as a
function of $\xi/\xi^*$,
where $\xi \equiv \xi_1$ and $\xi^*$ is 
the peak position of the one-particle spectrum, at the 
hadron and parton level from the
HERWIG MC \cite{HERWIG}
together with the analytical results in linear
approximation
 \cite{fw} for $\Lambda $ = 255 MeV.
Also shown are the OPAL data
\cite{opalmc}. The dashed line indicates the asymptotic DLA prediction
(from Ref. \cite{wo}).

\item[Fig.\ 5.3]
(a)-(c) The contributions of elementary 
hard processes to the leading 
and next-to-leading two-particle
correlation $C_{\rm EM}$; (A)-(D) to 
the three-particle correlation $C_{\rm EMM}$.
Parton no.~1 corresponds to the energy 
counter, parton no.~2 to the recoiling jet, partons no.~3 and no.~4 to
the registered jets
(from Ref. \cite{dmo}).

\item[Fig.\ 5.4]
Azimuthal correlations $C(\phi)$ of partons in a jet, at leading order 
and next-to-leading order for different
values of $\Lambda $ and comparison to the Monte Carlo (HERWIG) 
\cite{HERWIG} result
 \cite{dmo}.

\item[Fig.\ 5.5]
Azimuthal correlations
 of hadrons in a jet as measured by the OPAL Collaboration
 in comparison with model calculations \cite{opalph}.

\item[Fig.\ 5.6]
Rescaled correlation in the relative angle $\hat r (\epsilon )$ in the forward cone of half opening
angle $\Theta $ from DELPHI (preliminary data) in comparison with the asymptotic DLA prediction,
the JETSET MC \cite{JETSET}
as well as with the HERWIG MC \cite{HERWIG} at
parton level for two primary energies $\sqrt{s}$ (from
\cite{mb}).

\item[Fig.\ 5.7]
(a) Correlation in the relative angle $r(\vartheta_{12})$ in the forward cone of half opening angle $\Theta $,
for different $\Theta $\\ 
(b) the same as (a) but in rescaled coordinates,
testing the angular scaling
 (preliminary data from DELPHI
\cite{mb}). Also shown are the DLA predictions for asymptotic and LEP-1
energies \cite{ow2}.

\item[Fig.\ 5.8]
Particle-Particle Correlation Asymmetry (PPCA) as function of
relative spherical angle $\chi \equiv \vartheta_{12}$
from L3 experiment
\cite{ppc3} in comparison with the JETSET MC\cite{JETSET}
with angular
ordering (``coherent") and without (``incoherent").

\item[Fig.\ 5.9]
(a) Factorial multiplicity moments $F^{(n)}$ 
for the symmetric ring at production angle $\Theta$ and half width 
$\delta$ (preliminary data from 
DELPHI Collaboration \cite{mb})
in comparison with the DLA predictions (\ref{cmomno});\\
(b) same moments as
in (a) after rescaling as in (\ref{rnall}) ($\Lambda$=0.15 GeV, E=45 GeV).


\item[Fig.\ 6.1] Kinematics of interjet radiation in three-jet
events.

\item[Fig.\ 6.2] Directivity diagram of soft gluon radiation,
projected onto the $q\overline{q} \gamma$ (dashed line) and
$q\overline{q} g$ (solid) event planes.  Particle flows of Eqs.\
(\ref{4.4a}) and (\ref{4.5}) are drawn in polar coordinates:
$\Theta = \phi_2, r = \ln 2W (\phi_2)$.  Dotted circles show the
constant levels of particle flow:  $W (\phi_2) = 1, 2, 4$
(from Ref.\ \cite{adkt2}).

\item[Fig.\ 6.3] String model (a) and perturbative QCD (b,c)
pictures for the $e^+ e^- \rightarrow q\overline{q}g$ events. 
Dashed lines show topology of colour strings.  Dash-dotted line
shows \lq\lq negative antenna".

\item[Fig.\ 6.4] The ratio of particle flow between $q$ and $g$
jets $(N_2)$ to that between $q$ and $\overline{q}$ jets $(N_1)$,
for Mercedes-like $q\overline{q}g$ configuration (using $60^o$
sectors) \cite{kll}.

\item[Fig.\ 6.5] Angular inclusive correlation between three
energetic $(a, b, c)$ and one soft particle $(\vec{n})$ in the
$e^+ e^-$ annihilation process.

\item[Fig.\ 6.6] Comparison of the QCD (solid) and large-$N_C$
limit (dashed) predictions for (a) the $G$ factor and (b) asymmetry 
parameter $A$ in the symmetric $q\overline{q}g$ events \cite{dkt2}.

\item[Fig.\ 6.7] The azimuthal asymmetry of the quark jet in the
ISR events without ($a$) and with ($b$) additional soft gluon
emitted in the direction orthogonal to the $\overline{q}q \gamma$
plane.  Dashed lines show topology of colour strings, dash-dotted
line shows \lq\lq negative antenna".

\item[Fig.\ 6.8] Asymmetry parameter $A (\Theta_0 = 30^o)$ of the scattered
tagged $q$-jet in $p\overline{p}$-collisions \cite{dkt2}.

\item[Fig.\ 7.1] Charged particle flow in the event plane for two-jet
radiative events, and three-jet multihadronic events \cite{opal12}.  
$\alpha$ is the angle in the event plane as measured from the most 
energetic $(q)$ jet towards the second $(\overline{q})$ jet, c.f.\ 
Fig.\ 6.1.  The jet definition is based on the Durham algorithm \cite{cdotw}
with $y_{cut} = 0.007$. The data are not corrected for detector effects.

\item[Fig.\ 7.2] Fits of the charged particle flows to the expressions
(\ref{4.4a}), (\ref{4.5}) \cite{opal12}.  Gaussians centred around jet 1 and
jet 2 were added to account for the intrajet cascading (dotted lines).

\item[Fig.\ 7.3] Charged particle flow in $q\overline{q}g$ and $q\overline{q}\gamma$
events \cite{delphi4}.  The dotted lines indicate the intervals of angular 
integration $\pm [35^o, 115^o]$.

\item[Fig.\ 7.4] The inter-jet particle flow ratio R($qg$/$q\overline q$)
as function of particle type  in  momentum intervals,
as measured by ALEPH Collaboration \cite{aleph6}.

\item[Fig.\ 7.5] Charged particle flow in $q\overline{q}g$ events
\cite{delphi4} in comparison with analytical prediction \cite{dkt11}.

\item[Fig.\ 8.l] Structure of the hadron plateau in DIS process according to
Feynman \cite{rf} (a) and Gribov \cite{vng2} (b) (shaded area shows target 
fragmentation region), $y = \ln \omega$ ($\omega$ being energy of registered 
hadron).

\item[Fig.\ 8.2] Evolution of hadron distribution for DIS in the Breit
frame.  (a) fixed $x$, three \linebreak different $Q^2 = 9,500, 11 000 \: {\rm GeV}^2$
and (b) at fixed $Q^2$ for three different $x = 0.14, 0.0067, \linebreak 0.00005$.  
$y$ is defined as $y = \ln (\omega/\Lambda_{eff}) \: {\rm with} \: \Lambda_{eff} 
= 150$ MeV for both current and target fragmentation regions.  
It is assumed that after identification 
of a particle as belonging to either one of these regions, it enters either 
the $lhs$ or the $rhs$ of the plot, respectively \cite{gdkt}.

\item[Fig.\ 8.3] Mean charged multiplicity in $ep$ current jet fragmentation as
a function of $Q^2$ ($K^0$ and $\Lambda$ decays excluded).  Twice the
 multiplicity measured by ZEUS 
\cite{zeus} is compared to the results from $\overline{\nu}p$ scattering 
at FNAL \cite{fnal} 
and from $e^+ e^-$ annihilation at $\sqrt{s} = Q$ (from Ref. \cite{brook}).

\item[Fig.\ 8.4] $\xi_p$-distributions of the charged particles in the current
region of the Breit frame as a function of $Q^2$ \cite{zeus}.  The ZEUS data are 
compared to the MLLA limiting fomula (\ref{2.44}).

\item[Fig.\ 8.5]H1 \cite{h1} data on $\xi_p$-distributions of the charged particles
in the current hemisphere of the Breit frame of reference for (a) $12 < Q^2 < 
80 \: {\rm GeV}^2$ and (b) $Q^2 > 100 \: {\rm GeV}^2$.

\item[Fig.\ 8.6] Evolution of the peak position compared with $e^+ e^-$ data
\cite{h1}.  The solid line is the MLLA-based fit to the H1 data.

\item[Fig.\ 8.7] Peak position of $\xi_p$-distribution as a function of
Bjorken $x$ showing no significant $x$-dependence; for intervals
$12<Q^2<25~GeV^2$ (solid circles), $25<Q^2<45~GeV^2$ (open triangles),
$45<Q^2<80~GeV^2$ (solid triangles), and $200<Q^2<500~GeV^2$ (open circles).
The dashed lines refer to the $\xi_{peak}$ expected from the fit at the
relevant mean $Q^2$ value.

\item[Fig.\ 9.1] The energy evolution of the inclusive $b$ and $c$ spectra
as calculated from (\ref{7.8a}) in
\cite{dkt3}.  The results of the best-fit $F$-model (solid) and $G_2$-model 
(dashed) are shown.  
Curve (1) refers to $c$-quarks ($M_c = 1.5$ GeV), curve (2) to $b$-quarks
($M_b = 4.75$ GeV), using parameters 
$\Lambda^{(3)} = 500$ 
MeV, $A = 0.19$ and $C_2 = 2.22$ (in Eqs. (\ref{7.32}) and (\ref{7.35})).

\item[Fig.\ 9.2] The differences between the mean charged multiplicities of
heavy- and light-quark events $\delta_{Ql}$ ($\equiv \delta_{Qq}$) 
for (a) charm and (b) bottom.  The area filled by 
diagonal lines is the prediction of a naive model (from Ref.\
\cite{wjm2}).
\end{itemize}

\end{document}